\documentclass[dvips]{article}
\usepackage{supertabular,lscape,epsfig}
\usepackage{amssymb}
\usepackage{amsmath}
\DeclareSymbolFont{ppa}{OT1}{ppl}{m}{it}
\DeclareMathSymbol{\vv}{\mathalpha}{ppa}{'166}

\begin{document}

\newcommand{\dd}{\,{\rm d}}
\newcommand{\ie}{{\it i.e.},\,}
\newcommand{\etal}{{\it et al.\ }}
\newcommand{\eg}{{\it e.g.},\,}
\newcommand{\cf}{{\it cf.\ }}
\newcommand{\vs}{{\it vs.\ }}
\newcommand{\zdot}{\makebox[0pt][l]{.}}
\newcommand{\up}[1]{\ifmmode^{\rm #1}\else$^{\rm #1}$\fi}
\newcommand{\dn}[1]{\ifmmode_{\rm #1}\else$_{\rm #1}$\fi}
\newcommand{\upd}{\up{d}}
\newcommand{\uph}{\up{h}}
\newcommand{\upm}{\up{m}}
\newcommand{\ups}{\up{s}}
\newcommand{\arcd}{\ifmmode^{\circ}\else$^{\circ}$\fi}
\newcommand{\arcm}{\ifmmode{'}\else$'$\fi}
\newcommand{\arcs}{\ifmmode{''}\else$''$\fi}
\newcommand{\MS}{{\rm M}\ifmmode_{\odot}\else$_{\odot}$\fi}
\newcommand{\RS}{{\rm R}\ifmmode_{\odot}\else$_{\odot}$\fi}
\newcommand{\LS}{{\rm L}\ifmmode_{\odot}\else$_{\odot}$\fi}

\newcommand{\Abstract}[2]{{\footnotesize\begin{center}ABSTRACT\end{center}
\vspace{1mm}\par#1\par
\noindent
{~}{\it #2}}}

\newcommand{\TabCap}[2]{\begin{center}\parbox[t]{#1}{\begin{center}
  \small {\spaceskip 2pt plus 1pt minus 1pt T a b l e}
  \refstepcounter{table}\thetable \\[2mm]
  \footnotesize #2 \end{center}}\end{center}}

\newcommand{\TableSep}[2]{\begin{table}[p]\vspace{#1}
\TabCap{#2}\end{table}}

\newcommand{\FigCap}[1]{\footnotesize\par\noindent Fig.\  %
  \refstepcounter{figure}\thefigure. #1\par}

\newcommand{\TableFont}{\footnotesize}
\newcommand{\TableFontIt}{\ttit}
\newcommand{\SetTableFont}[1]{\renewcommand{\TableFont}{#1}}

\newcommand{\MakeTable}[4]{\begin{table}[htb]\TabCap{#2}{#3}
  \begin{center} \TableFont \begin{tabular}{#1} #4 
  \end{tabular}\end{center}\end{table}}

\newcommand{\MakeTableSep}[4]{\begin{table}[p]\TabCap{#2}{#3}
  \begin{center} \TableFont \begin{tabular}{#1} #4 
  \end{tabular}\end{center}\end{table}}

\newenvironment{references}%
{
\footnotesize \frenchspacing
\renewcommand{\thesection}{}
\renewcommand{\in}{{\rm in }}
\renewcommand{\AA}{Astron.\ Astrophys.}
\newcommand{\AAS}{Astron.~Astrophys.~Suppl.~Ser.}
\newcommand{\ApJ}{Astrophys.\ J.}
\newcommand{\ApJS}{Astrophys.\ J.~Suppl.~Ser.}
\newcommand{\ApJL}{Astrophys.\ J.~Letters}
\newcommand{\AJ}{Astron.\ J.}
\newcommand{\IBVS}{IBVS}
\newcommand{\PASP}{P.A.S.P.}
\newcommand{\Acta}{Acta Astron.}
\newcommand{\MNRAS}{MNRAS}
\renewcommand{\and}{{\rm and }}
\section{{\rm REFERENCES}}
\sloppy \hyphenpenalty10000
\begin{list}{}{\leftmargin1cm\listparindent-1cm
\itemindent\listparindent\parsep0pt\itemsep0pt}}%
{\end{list}\vspace{2mm}}

\def\TYLDA{~}
\newlength{\DW}
\settowidth{\DW}{0}
\newcommand{\dw}{\hspace{\DW}}

\newcommand{\refitem}[5]{\item[]{#1} #2%
\def\REFARG{#3}\ifx\REFARG\TYLDA\else, {\it#3}\fi
\def\REFARG{#4}\ifx\REFARG\TYLDA\else, {\bf#4}\fi
\def\REFARG{#5}\ifx\REFARG\TYLDA\else, {#5}\fi.}

\newcommand{\Section}[1]{\section{#1}}
\newcommand{\Subsection}[1]{\subsection{#1}}
\newcommand{\Acknow}[1]{\par\vspace{5mm}{\bf Acknowledgements.} #1}
\pagestyle{myheadings}

\newfont{\bb}{ptmbi8t at 12pt}
\newcommand{\xrule}{\rule{0pt}{2.5ex}}
\newcommand{\xxrule}{\rule[-1.8ex]{0pt}{4.5ex}}
\def\thefootnote{\fnsymbol{footnote}}
\begin{center}
{\Large\bf Binary Lenses in OGLE-III EWS Database. Seasons 2002--2003}
\vskip1cm
{\bf M.~~ J~a~r~o~s~z~y~\'n~s~k~i$^1$,~~ A.~~ U~d~a~l~s~k~i$^1$,~~ 
M.~~ K~u~b~i~a~k$^1$,~~ M.~~ S~z~y~m~a~\'n~s~k~i$^1$, 
G.~~ P~i~e~t~r~z~y~\'n~s~k~i$^{1,2}$,~~ I.~~ 
S~o~s~z~y~\'n~s~k~i$^1$,~~ K.~~\.Z~e~b~r~u~\'n$^1$,
~~ O.~~ S~z~e~w~c~z~y~k$^1$~~ and~~ 
\L.~~ W~y~r~z~y~k~o~w~s~k~i$^1$}
\vskip3mm
{$^1$Warsaw University Observatory, Al.~Ujazdowskie~4,~00-478~Warszawa, Poland\\
e-mail:(mj,udalski,mk,msz,pietrzyn)@astrouw.edu.pl\\
(soszynsk,zebrun,szewczyk,wyrzykow)@astrouw.edu.pl\\
$^2$ Universidad de Concepci{\'o}n, Departamento de Fisica,
Casilla 160-C, Concepci{\'o}n, Chile\\
e-mail: pietrzyn@hubble.cfm.udec.cl}
\end{center}

\Abstract{We present 15 binary lens candidates from OGLE-III Early
Warning System database for seasons 2002--2003. We also found 15 events
interpreted as single mass lensing of double sources. The candidates were
selected by visual light curves inspection.  Examining the models of binary
lenses of this and our previous study (10 caustic crossing events of
OGLE-II seasons 1997--1999) we find one case of extreme mass ratio binary
(${q\approx0.005}$) and the rest in the range ${0.1<q<1.0}$, which may
indicate the division between planetary systems and binary stars. There is
no strong discrepancy between the expected and the observed distributions
of mass ratios and separations for binary stars.}{Gravitational lensing --
Galaxy: center -- binaries: general}

\Section{Introduction}
In this article we present the results of the search for binary lens 
events among microlensing phenomena discovered by the Early Warning
System (EWS -- Udalski \etal 1994b, Udalski 2003) of the third phase of
the Optical Gravitational Lens Experiment (OGLE-III) in seasons 
2002--2003. This is the continuation of the study of seasons 1997--1999
presented by Jaroszy\'nski (2002, hereafter Paper~I). 

As estimated by Mao and Paczy\'nski (1991) several percent of all
microlensing events in our Galaxy should be caused by binary systems of
stars acting as lenses. In the same paper the analysis of microlensing
events caused by planetary systems is proposed as a way of discovering
extra-solar planets. The sufficiently large database of binary and/or
planetary microlensing events may serve as an independent tool of
studying such systems. Some basic ideas for binary lens analysis can be
found in the review article by Paczy\'nski (1996).

Lensing by two point masses has been studied by Schneider and Weiss
(1986). Various aspects of binary lens modeling have been described
(among others) by Gould and Loeb (1992), Bennett and Rhie (1996), Gaudi
and Gould (1997), Dominik (1999),  Albrow \etal (1999c), and Graff and
Gould (2002).

The first microlensing phenomenon interpreted as being due to the binary
system was the event OGLE-7 (Udalski \etal 1994a). Several binary lens
events with good light curve coverage were used to study the atmospheres of
the source stars (\eg Albrow \etal 1999b) or to constrain the lensing
system parameters (\eg Alb\-row \etal 1999a). The first lens mass measurement
was obtained by An \etal (2002) based on a binary lens event with combined
effects of parallax motion and caustic crossing. The systematic study of 21
binary lens events found in MACHO data was presented by Alcock \etal
(2000).

Paper~I presents the analysis of 18 binary lens events found in OGLE-II
data reduced with difference photometry, DIA, (Alard and Lupton 1998) by
Wo{\'z}niak (2000) and Wo{\'z}niak \etal (2001). The aim of the present
study is similar. The combined sample of events showing caustic
crossings (10+15 events in OGLE-II and OGLE-III to date) is large
enough for a crude statistical analysis of the binary lens population. 

In the next Section we describe the selection of binary lens candidates.
In Section~3 we describe the procedure of fitting  models to the data. 
The results are described in Section~4, and the discussion follows in
Section~5. The extensive graphical material is shown in Appendices. 

\Section{Choice of Candidates}
The OGLE-III data are routinely reduced with difference photometry (DIA)
which gives high quality light curves of variable objects. The EWS
system of OGLE-III (Udalski 2003) automatically picks up candidate
objects with micro\-len\-sing-like variability.

There are 389/462 microlensing events candidates selected by EWS in
2002/ 2003 seasons. We visually inspected all candidate light curves looking
for features characteristic of binary lenses (multiple peaks, U-shapes,
asymmetry). We avoided light curves showing excessive noise. We selected
8/16 candidate events in 2002/2003 data for further study.

This work is a continuation of Paper~I, where we have introduced some
criteria to obtain a sample of high S/N ratio microlensing light curves among
transient events. Following this approach we find the average base flux of
each light curve and its standard deviation~$\delta$. It may include also a
small amplitude intrinsic source variability and is usually higher than the
averaged observational error obtained during photometric data reduction. We
rescaled all observational errors by the factor
$s \equiv \delta/\langle\sigma_i\rangle_{\rm base}$ calculated for the unlensed part
of the light curve. In Paper~I we considered only light curves with at
least 7 high flux (\ie ${\ge5\delta}$ above the base) measurements. We
relax this requirement here, when constructing a single mass lens sample,
which includes events with at least 5 points $5\delta$ above the base
flux. For the single lens sample we also require a reasonably good fit,
postulating ${\chi^2/{\rm DOF}\le2}$. There are 268 single lens events in
2002/2003 data passing these criteria.  Three of the binary lens candidates
do not have the required number of high flux measurements.

\Section{Fitting Binary Lens Models} 
The models of the two point mass lens were investigated by many authors
(Schneider and Weiss 1986, Mao and DiStefano 1995, DiStefano and Mao 1996,
Dominik 1998, to mention only a few). The effective methods applicable for
extended sources have recently been described by Mao and Loeb (2001). While
we use mostly the point source approximation, we extensively employ their
efficient numerical schemes for calculating the binary lens caustic
structure and source magnification.

We fit binary lens models using the $\chi^2$ minimization method for
the light curves. It is convenient to model the flux at the time $t_i$
as: 
$$F_i=F(t_i)=A(t_i)\times F_s+F_b\eqno(1)$$
where $F_s$ is the flux of the source being lensed, $F_b$ is the blended
flux (from the source close neighbors and possibly the lens), and the
combination ${F_b+F_s=F_0}$ is the total flux, measured long before or
long after the event. The lens magnification (amplification) of the
source $A(t_i)=A(t_i;p_j)$ depends on the set of model parameters $p_j$.
Using this notation one has for $\chi^2$:
$$\chi^2=\sum_{i=1}^N\frac{\left(A_i~F_s+F_b-F_i\right)^2}{\sigma_i^2}\eqno(2)$$
where $\sigma_i$ are the rescaled  errors of the flux measurement taken
from the DIA photometry. The dependence of $\chi^2$ on the binary lens
parameters $p_j$ is complicated, while the dependence on the
source/blend fluxes is quadratic. The subset of equations ${\partial
\chi^2/\partial F_s=0}$; ${\partial\chi^2/\partial F_b=0}$ can be solved
algebraically, giving ${F_s=F_s(p_j;\{F_i\})}$ and
${F_b=F_b(p_j;\{F_i\})}$ thus effectively reducing the dimension of the
parameter space. In some cases this approach may give unphysical
solutions with negative blended flux (${F_b<0}$). Using the fluxes
($F_s$, $F_b$) as independent parameters may help avoiding such
problems, and we do it in a part of calculations. In particular we
always use this method when modeling single lens events needed for
comparison.

Binary lens models are possibly and typically non unique (Dominik 1999).
The  presence of caustics and cusps in the lens theory (Schneider,
Ehlers and Falco 1992, Blandford and Narayan 1992) makes the $\chi^2$
dependence on the model  parameters complicated and discontinuous for
point sources. For extended sources the discontinuities in strict
mathematical sense are not present, but the $\chi^2$ surface remains
complex, possessing large number of local minima and deep narrow
valleys, which makes the search for minima complicated. For this reason
we combine the scan of  the parameter space and the random choice of
initial parameters  with  the standard minimization techniques in our
search. 

For most of the light curves we investigate, the caustic crossings are not
well sampled, and we are forced to use a point source approximation in
the majority of our models. In two cases (events OGLE 2003-BLG-170 and OGLE
2003-BLG-267) the caustic crossings are resolved, so the extended source
models can be fitted. In these cases the strategy resembling Albrow \etal
(1999c) for finding binary lens models can be used. It is based on the fact
that some of the parameters (source angular size, strength of the caustic)
can be fitted independently, so for an initial fit one can split the
parameter space into two lower dimensionality sub-manifolds.

A binary system consists of two masses $m_1$ and $m_2$, where by
convention ${m_1\le m_2}$. The Einstein radius of the binary lens is
defined as: 
$$r_{\rm E}=\sqrt{\frac{4G(m_1+m_2)}{c^2}\frac{d_{\rm OL}d_{\rm LS}}
{d_{\rm OS}}}\eqno(3)$$
where $G$ is the constant of gravity, $c$ is the speed of light, $d_{\rm
OL}$ is the observer--lens distance, $d_{\rm LS}$ is the lens--source 
distance, and ${d_{\rm OS}\equiv d_{\rm OL}+d_{\rm LS}}$ is the
distance between the observer and the source. The Einstein radius
serves as a length unit and the Einstein time: ${t_{\rm E}=r_{\rm
E}/\vv_\perp}$, where $\vv_\perp$ is the lens velocity relative to the line
joining the observer with the source, serves as a time unit. The passage
of the source in the lens background is defined by seven parameters:
${q\equiv m_1/m_2}$ (${0<q\le1}$) -- binary mass ratio, $d$ --
binary separation expressed in $r_{\rm E}$ units, $\beta$ -- angle
between the source  trajectory as projected onto the sky and the
projection of the binary axis, $b$ -- impact parameter, $t_0$ --
time of closest approach of the source to the binary center of
mass, $t_E$ -- Einstein time, and $r_s$ -- source radius. Thus we
are left with the seven or six  dimensional parameter space, depending
on the presence/absence of observations covering the caustic
crossings. 

We begin with a scan of the parameter space using a logarithmic grid of
points in $(q,d)$ plane (${10^{-3}\le q\le 1}$, ${0.1\le d\le 10}$) and
allowing for continuous variation of the other parameters. The choice of
starting points combines systematic and Monte Carlo searching of regions in
parameter space allowing for caustic crossing or cusp approaching
events. The $\chi^2$ minimization is based on downhill method and uses
standard numerical algorithms. When a local minimum is found we make a
small Monte Carlo jump in the parameter space and repeat the downhill
search. In some cases it allows to find a different local minimum.  If it
does not work several times, we stop and try next starting point.

Minimizations with fixed physical binary parameters serve mostly to obtain
$\chi^2(q,d)$ maps showing the preferred binary models. We perform also
minimizations in higher dimensions, including mass ratio, binary
separation, source flux and blended flux as independent parameters. This
improves the models, since now no variables are limited to grid values
only. Even more important is the fact that in a higher dimension space
there may exist a downhill path joining two points impossible in a lower
dimension subspace. The $\chi^2(q,d)$ maps are improved during high
dimension minimization: whenever the running value of $\chi^2$ is lower
than the value assigned to the closest grid point, we exchange them. The
maps shown in Appendix~1 are obtained as a result of such a procedure.

Only the events with characteristics of caustic crossing (apparent
discontinuities in observed light curves, U-shapes) can be treated as
safe binary lens cases. The double peak events may result from cusps
approaches, but may also be produced by double sources (\eg Gaudi and
Han 2004). In such cases we also check the double source fit of the
event postulating: 
$$F(t)=A(u_1(t))\times F_{\rm s1}+A(u_2(t))\times F_{\rm s2}+F_{\rm b}\eqno(4)$$
where $F_{\rm s1}$, $F_{\rm s2}$ are the fluxes of the source
components, $F_{\rm b}$ is the blended flux, and $A(u)$ is the single
lens amplification (Paczy\'nski 1986). The dimensionless source -- lens
separations are given as:
$$u_1(t)=\sqrt{{b_1}^2+\frac{(t-t_{01})^2}{{t_{\rm E}}^2}}\quad
u_2(t)=\sqrt{{b_2}^2+\frac{(t-t_{02})^2}{{t_{\rm E}}^2}}\eqno(5)$$
where $t_{01}$, $t_{02}$ are the closest approach times of the source
components, $b_1$, $b_2$ are the respective impact parameters, and   
$t_{\rm E}$ is the (common) Einstein time.

\Section{Results}
In Table~1 we show the results of binary lens fitting for a long list of
events. Some of them are "strong cases", which we denote by "b" (for
binary) in the third column. Other may be better interpreted as single
mass lensing of double sources, which we denote by "d". If binary lens
and double source models have similar formal quality, but the binary
light curve has caustic crossings and/or cusp approaches in places not
covered by observations, we choose the double source model.

Columns 1--2 give the event identification (year and EWS number), the
fourth column gives $\chi^2$ and DOF (degrees of freedom) number, and the
other columns give the parameters for the best models of each of the
events. In some cases we include also another solution if it belongs to a
distinct $\chi^2$ minimum and lies inside the confidence region. Our best
fits are also shown on the plots in Appendix~1. For each binary there are
three separate plots. The first shows the source trajectory as projected
onto the lens plane with caustic structure and binary components
included. The model light curves and observed magnitudes are shown in the
second plot. Third diagram shows $\chi^2$ confidence regions in the $\lg
q$--$\lg d$ plane.

\MakeTable{lc@{\hspace{6pt}}c@{\hspace{6pt}}rccccrrrl@{\hspace{7pt}}l}{12.5cm}{Parameters of binary lens modeling}
{\hline
\noalign{\vskip3pt}
Year & Event &  & $\chi^2/$DOF & $s$ & $q$ & $d$ & $\beta~~~$ & $b~~~$ & $t_0~~~$
&  $t_E~~~$ & ~~~$f$  & ~~~~$r_s$ \\
\noalign{\vskip3pt}
\hline
2002 & 051 & b & 110.0/ 58 & 1.45 & 0.943 & 1.390 &   96.71 & $-0.48$ &  2391.4 &   88.1 &  0.87 \\
2002 & 069 & b & 1014./ 95 & 1.18 & 0.721 & 0.497 &  110.41 & $-0.02$ &  2456.4 &   99.7 &  1.00 \\
2002 & 099 & d & 134.4/151 & 1.63 & 0.248 & 1.963 &   16.39 & $ 0.09$ &  2413.0 &   34.4 &  0.37 \\
2002 & 114 & b &  77.5/ 80 & 1.80 & 0.745 & 0.623 &   83.13 & $-0.04$ &  2412.5 &   75.8 &  0.12 \\
2002 & 135 & d & 170.3/159 & 1.34 & 0.144 & 0.398 &   42.28 & $ 0.06$ &  2441.5 &  148.2 &  0.06 \\
     &     & d & 171.8/159 & 1.34 & 0.128 & 1.089 &  297.78 & $-0.44$ &  2440.9 &   32.2 &  0.76 \\
     &     & d & 172.0/159 & 1.34 & 0.678 & 0.734 &   34.65 & $-0.33$ &  2441.0 &   39.3 &  0.51 \\
2002 & 158 & d & 115.7/103 & 1.08 & 0.051 & 0.872 &   45.80 & $ 0.18$ &  2467.9 &   56.5 &  0.69 \\
2002 & 256 & d & 171.3/110 & 1.22 & 0.081 & 1.066 &  315.79 & $ 0.15$ &  2485.9 &   40.1 &  0.25 \\
2002 & 321 & d &  94.1/ 97 & 1.16 & 0.251 & 0.708 &   30.72 & $ 0.08$ &  2523.9 &   50.9 &  0.96 \\
2003 & 021 & b & 131.2/121 & 1.42 & 0.799 & 0.941 &   57.91 & $-0.09$ &  2776.8 &   54.9 &  0.96 \\
2003 & 056 & b & 147.5/120 & 1.78 & 0.743 & 1.497 &  318.72 & $-0.02$ &  2764.8 &   40.3 &  0.79 \\
2003 & 084 & d & 110.1/103 & 1.65 & 0.794 & 0.794 &   85.34 & $-0.01$ &  2697.3 &  124.8 &  0.51 \\
2003 & 124 & b & 128.0/122 & 1.25 & 0.666 & 0.959 &   72.57 & $-0.11$ &  2768.7 &   73.0 &  0.27 \\
2003 & 135 & b &  59.5/110 & 3.55 & 0.129 & 0.847 &  123.00 & $-0.20$ &  2724.3 &  339.9 &  0.06 \\
2003 & 170 & b & 161.1/149 & 1.46 & 0.789 & 1.213 &  133.66 & $-0.35$ &  2794.1 &   15.6 &  0.75 & 0.0027 \\
2003 & 194 & d & 117.2/112 & 1.24 & 0.692 & 3.357 &  157.24 & $ 0.77$ &  2749.4 &   32.4 &  0.87 \\
     & 194 & d & 118.3/112 & 1.24 & 0.702 & 0.562 &  134.46 & $ 0.09$ &  2804.4 &   21.8 &  0.91 \\
2003 & 200 & b &  90.8/105 & 1.51 & 0.209 & 1.495 &  122.70 & $-0.06$ &  2836.4 &   46.0 &  0.54 \\
2003 & 235 & b & 151.1/175 & 1.19 & 0.005 & 1.128 &  318.91 & $-0.10$ &  2848.2 &   75.0 &  0.58 \\
2003 & 236 & b & 125.4/122 & 1.32 & 0.175 & 0.838 &  196.36 & $-0.13$ &  2801.5 &   73.8 &  0.12 \\
2003 & 260 & b & 127.2/120 & 1.38 & 0.112 & 2.269 &  288.48 & $-1.52$ &  2968.7 &  272.3 &  0.10 \\
     & 260 & b & 129.6/120 & 1.38 & 0.488 & 1.845 &  106.77 & $-0.42$ &  2840.7 &  139.4 &  0.05 \\
     & 260 & b & 129.8/120 & 1.38 & 0.450 & 0.624 &  256.90 & $-0.01$ &  2827.8 &  106.7 &  0.04 \\
2003 & 266 & d &  88.6/ 98 & 1.38 & 0.448 & 1.071 &  246.51 & $ 0.02$ &  2822.9 &   25.0 &  0.05 \\
2003 & 267 & b & 1042./250 & 1.23 & 0.628 & 0.352 &   87.88 & $-0.01$ &  2845.3 &   88.8 &  0.37 & 0.0008 \\
2003 & 291 & b & 325.3/159 & 1.45 & 0.837 & 3.457 &  198.53 & $ 0.48$ &  2956.6 &   39.8 &  0.49 \\
2003 & 340 & d & 179.4/148 & 1.54 & 0.001 & 1.105 &  265.20 & $ 0.03$ &  2900.9 &  208.4 &  0.15 \\
     & 340 & d & 179.9/148 & 1.54 & 0.049 & 1.447 &  153.99 & $-0.08$ &  2904.0 &  106.6 &  0.42 \\
2003 & 380 & b &  85.3/118 & 1.29 & 0.615 & 0.784 &  169.05 & $ 0.14$ &  2876.5 &   78.9 &  0.21 \\
\noalign{\vskip3pt}
\hline
\noalign{\vskip5pt}
\multicolumn{13}{p{12.4cm}}{Note: The table contains the year and the
EWS event number, the event classification according to this study ("b" --
for binary lens, "d" -- for double source), the rescaled $\chi^2$ value and
the DOF number, the error scaling factor $s$
($\chi^2=\chi^2_\mathrm{raw}/s^2$),
mass ratio $q$, binary separation $d$, source trajectory
direction $\beta$, impact parameter $b$, time of passing by the center of
mass $t_0$, Einstein time $t_E$, and blending parameter $f=F_s/F_0$.  The
source radius $r_s$ (in Einstein units), is given only in cases of events
with well resolved caustic crossings.}}

In Table~2 we show the results of double source modeling. We include the
ambiguous cases of binary lens / double source models as well as some
events showing well separated, smooth maxima in their light curves,
which may be safely treated as single mass lensing of double sources.
The comparison of two kinds of fits is given in Appendix~2, and the well
separated double source events are shown in Appendix~3. 

\MakeTable{ccrcccccrcc}{12.5cm}{Parameters of double source modeling}
{\hline
\noalign{\vskip3pt}
Year & Event & $\chi^2/$DOF & $s$ & $b_1$ & $b_2$ & $t_{01}$ & $t_{02}$ & 
~~~$t_{\rm E}$ &  $f_1$ & $f_2$ \\
\noalign{\vskip3pt}
\hline
2002 & 018 &  92.1/107 & 1.20 & 0.4495 & 0.4356 & 2351.09 & 2573.47 &  31.6 & 0.203 & 0.797 \\
2002 & 099 & 134.2/151 & 1.63 & 0.0821 & 0.0294 & 2402.93 & 2425.23 &  47.1 & 0.147 & 0.051 \\
2002 & 135 & 181.7/159 & 1.34 & 0.0394 & 0.0951 & 2432.39 & 2449.89 &  81.2 & 0.033 & 0.070 \\
2002 & 158 & 120.8/103 & 1.08 & 0.0690 & 0.1934 & 2456.42 & 2472.80 &  60.2 & 0.065 & 0.543 \\
2002 & 256 & 182.2/110 & 1.22 & 0.0025 & 0.0220 & 2471.92 & 2488.38 & 199.9 & 0.002 & 0.023 \\
2002 & 321 &  90.6/ 97 & 1.16 & 0.0099 & 0.0261 & 2517.96 & 2524.34 &  61.9 & 0.071 & 0.553 \\
2003 & 063 & 118.2/113 & 1.64 & 0.5878 & 1.6755 & 2748.84 & 2891.20 &  29.1 & 0.352 & 0.648 \\
2003 & 067 & 113.4/114 & 1.48 & 0.5433 & 0.5792 & 2772.01 & 3077.06 &  66.8 & 0.386 & 0.470 \\
2003 & 084 & 145.1/103 & 1.65 & 0.0894 & 0.0667 & 2717.20 & 2759.63 & 108.1 & 0.279 & 0.721 \\
2003 & 095 & 112.7/108 & 2.39 & 0.2313 & 0.3403 & 2775.44 & 2881.33 &  44.4 & 0.273 & 0.327 \\
2003 & 124 & 210.5/122 & 1.25 & 0.1068 & 0.0001 & 2751.10 & 2768.21 & 125.1 & 0.124 & 0.014 \\
2003 & 126 & 137.1/108 & 1.97 & 0.1480 & 1.0042 & 2774.57 & 2830.57 &  14.9 & 0.820 & 0.180 \\
2003 & 194 & 127.0/112 & 1.25 & 0.0000 & 0.2027 & 2802.66 & 2804.57 &  19.8 & 0.069 & 0.918 \\
2003 & 266 & 119.1/ 98 & 1.39 & 0.2285 & 1.1596 & 2810.27 & 2825.33 &  10.4 & 0.223 & 0.777 \\
2003 & 340 & 191.2/148 & 1.55 & 0.0649 & 0.0114 & 2900.11 & 2902.33 & 134.2 & 0.228 & 0.040 \\
\noalign{\vskip3pt}
\hline
\noalign{\vskip5pt}
\multicolumn{10}{p{12.0cm}}{Note: The table contains the year and event
number according to EWS, the rescaled $\chi^2$ value and the DOF number,
the error scaling factor $s$,
the impact parameters $b_1$ and $b_2$ for the two source components,
times of the closest approaches $t_{01}$ and $t_{02}$, Einstein
time $t_{\rm E}$, and blending parameters $f_1=F_{\rm s1}/(F_{\rm
s1}+F_{\rm s2}+F_{\rm b})$ and $f_2=F_{\rm s2}/(F_{\rm s1}+F_{\rm
s2}+F_{\rm b})$.}}

For the further study of binary mass lensing we choose only the "safe"
cases (denoted "b" in Table~1). We also use the binary lenses of
Paper~I, but we limit ourselves only to caustic crossing events. The
cusp approach events of Paper~I are typically not well constrained. Some
of them would probably be better interpreted as double source events.

\subsection{Distribution of Mass Ratios and Separations}

For statistical study of binary lens properties we use 10 caustic
crossing events of seasons 1997--1999 (Paper I)  and 15 "safe" events of
the present study. Three of the events considered have more than one 
model with similar quality of fits belonging to two or three different
classes of binary lenses (close/intermediate/ wide). In such cases we
assign statistical weights $w_{\rm 1st}$, $w_{\rm 2nd}$, etc.\ to the
models, assuming the relation:
$$\frac{w_{\rm 2nd}}{w_{\rm 1st}}
=\exp\left(-\frac{\chi^2_{\rm 2nd}-\chi^2_{\rm 1st}}{2}\right)\eqno(6)$$
and similarly for the 3rd model of the same event if applicable. The
histograms for the distributions of mass ratio $q$ and binary separation
$d$ are shown in Fig.~1. 

\begin{figure}[htb]
\includegraphics[height=60.0mm,width=60.0mm]{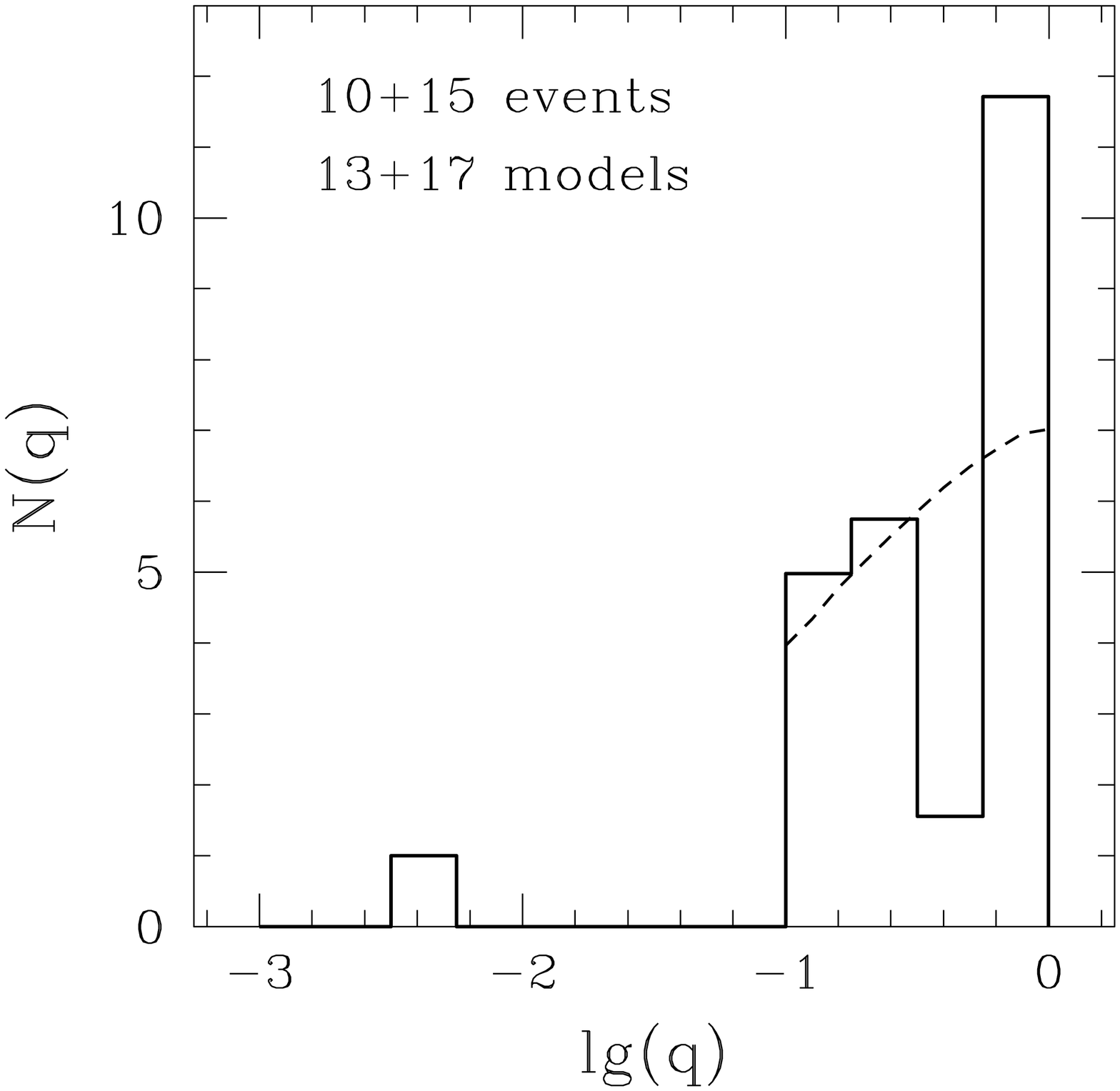}\hfill
\includegraphics[height=60.0mm,width=60.0mm]{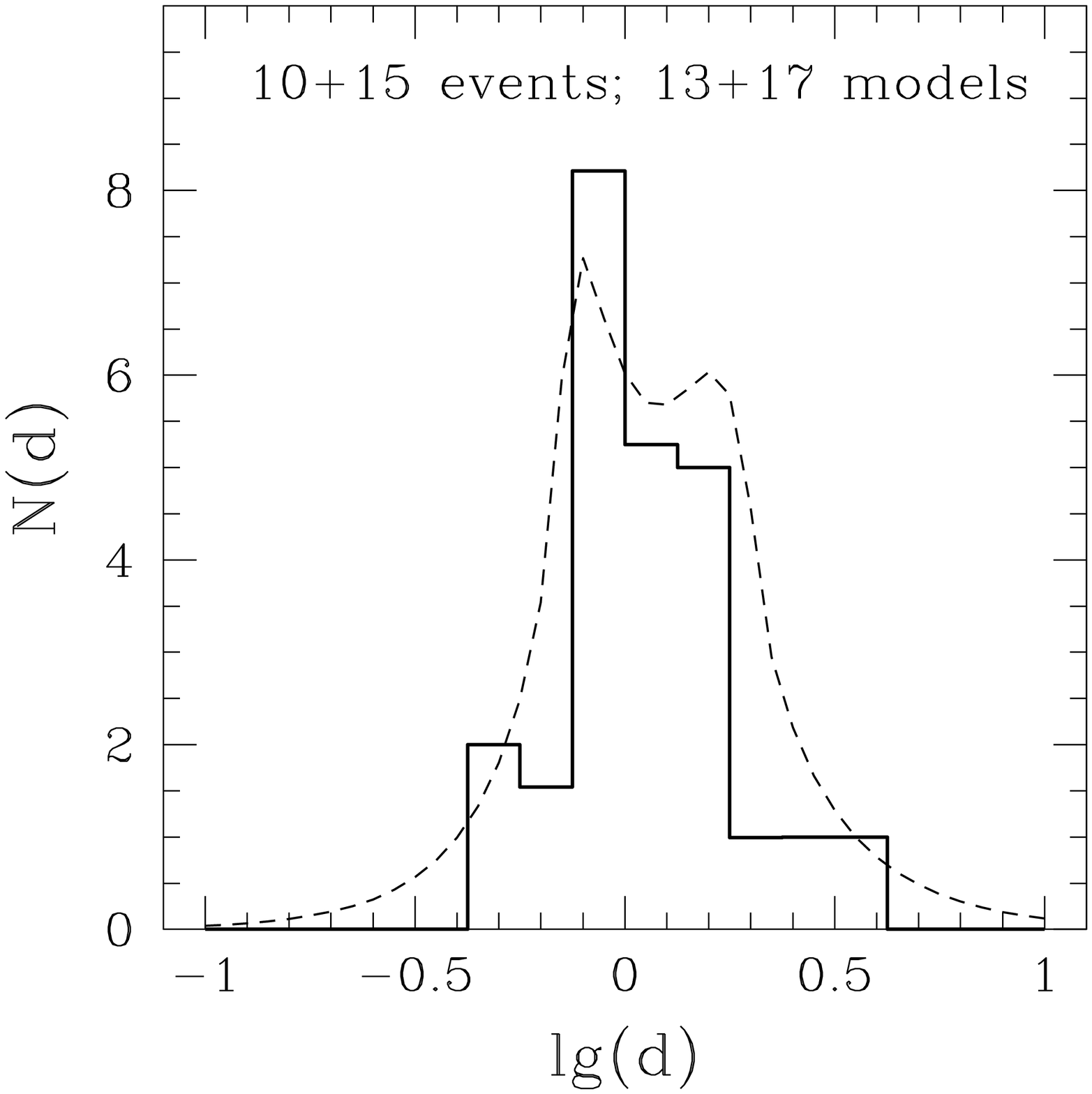}%
\FigCap{Histograms showing the distribution of the mass ratio $q$ (left)
and the binary separation $d$ (right) for the gravitational lens models
(solid). Crude theoretical predictions of the distributions are plotted
with dashed lines (see text for details). All binary lenses (seasons
1997--1999 and 2002--2003) are included.}
\end{figure}

As can be seen in the left panel the majority of binary lenses has mass
ratio in the range ${0.1<q<1.0}$, typical for binary stars. There is one
strong case of extreme mass ratio binary, which may be considered a
``planetary system'', the event OGLE 2003-BLG-235/MOA 2003-BLG-53
described in detail by Bond \etal (2004). In Paper~I we have reported 2
cases of extreme mass ratio binaries, but both belong to cusp approach
cases, not considered here. One of the models of OGLE 2003-BLG-340 has an
extreme mass ratio ${q=0.0014}$, but the alternative model has ${q=0.049}$,
and the double source model is not excluded.

We also show the predicted distributions of mass ratios and separations 
for the binary stars based on simplified assumptions. Following the
approach of Mao and Paczy\'nski (1991), based on Abt (1983) and Trimble
(1990) (compare Paper~I), we assume, that for the range ${0.1\le q\le
1}$ the logarithms of the mass ratios of the binary stars and the
logarithms of their semi-major axes are distributed uniformly:
$$P_{\rm bin}\left(\lg q,\lg d\right)\sim{\rm const}.\eqno(7)$$
The observed distributions are influenced by the probability of given
system to cause an observable binary lensing event. For a binary of
known parameters and a source moving in a given direction the chance of
crossing the caustic is proportional to its width perpendicular to the
source trajectory. The size of caustic along the source trajectory is
proportional to the time elapsing between caustic crossings, which also
influences the probability of classifying the event as binary, but we
neglect this effect as too difficult to model. Assuming that the
observability of binary lens event is proportional to the chance of
caustic crossing one has:
$$P_{\rm obs}\left(\lg q,\lg d\right)\sim 
P_{\rm bin}\left(\lg q,\lg d\right)\times w\left(\lg q,\lg d\right)
\sim w\left(\lg q,\lg d\right)\eqno(8)$$
where $w\left(\lg q,\lg d \right)$ is the caustic width averaged over all
possible source path directions. Averaging over one of the parameters we
obtain the one dimensional probabilities for finding a binary lens with
a given mass ratio or separation, shown in Fig.~1 with dashed lines. No
prediction is given for the planetary systems since there is no sufficient
observational evidence to model the mass ratio distribution as in the case
of binary stars.

The binary separation distribution has a strong peak around
${d\approx1}$. Theoretically we predict two peaks at ${d\approx0.9}$ and
${d\approx1.7}$.

We apply the Kolmogorov -- Smirnov test to the cumulative 1D
distributions of binary mass ratios and separations. The postulated
distribution of these parameters (Eqs.~7,~8) cannot be rejected. 

\subsection{Distribution of Events Duration}

The Einstein time ($t_{\rm E}$) can be found both for single and binary
lenses. For comparison we also fit single events from OGLE-II and III
databases. Not all fits are satisfactory. First we reject fits which
have too high $\chi^2/$DOF. Because of the parameter degeneracy in
single lens fitting (Wo\'zniak and Paczy\'nski 1997) there remains a
risk of including unphysical fits with formally good quality. Fits with
very small impact parameters (large amplification) small source flux
(small parameter $f$) and very long time scale belong to this category.
To reject them we ignore all models with ${f<0.01}$. We show histograms
for single and binary events durations in Fig.~2.

\begin{figure}[htb]
\includegraphics[height=60.0mm,width=60.0mm]{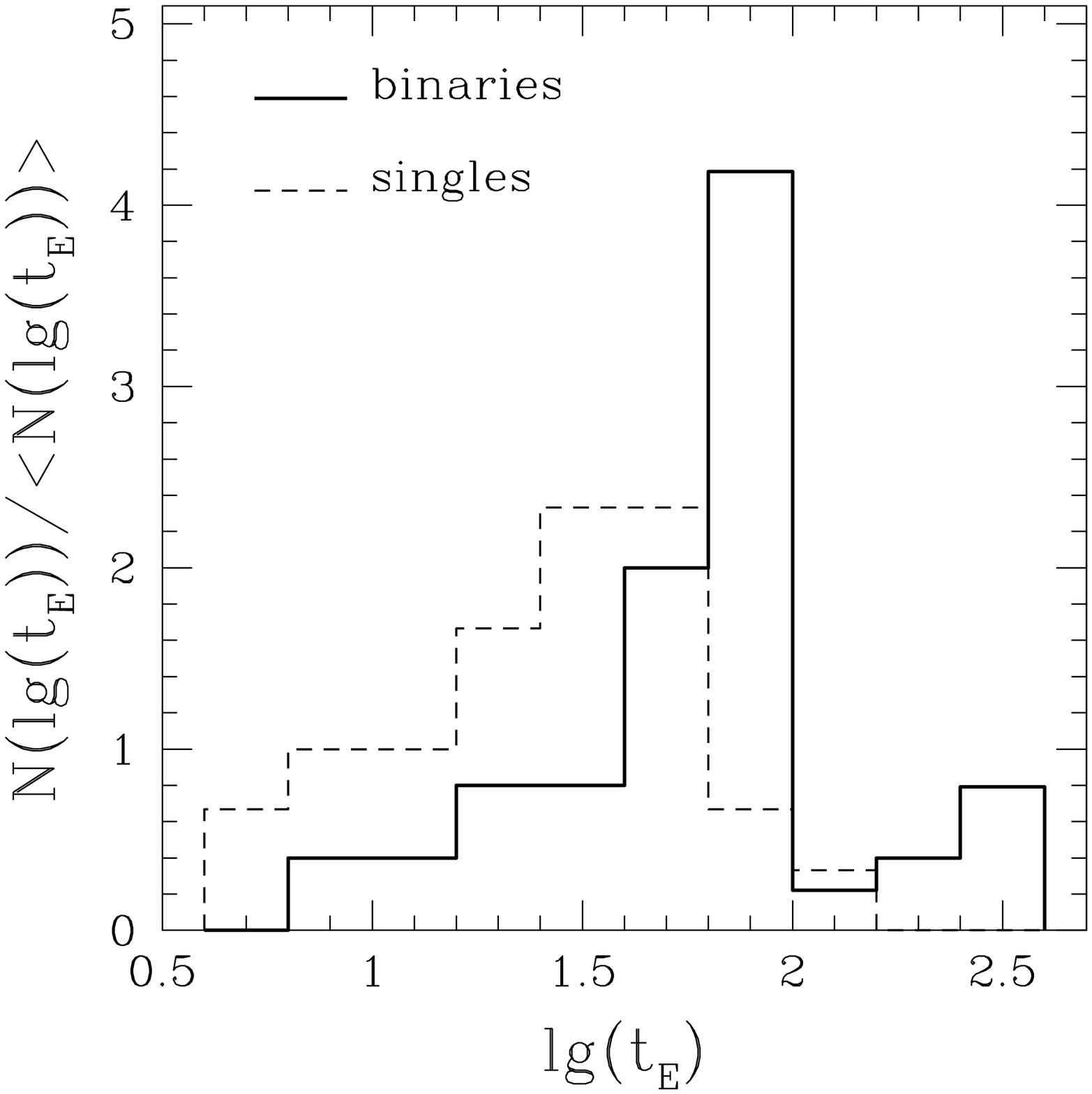}\hfill%
\includegraphics[height=60.0mm,width=60.0mm]{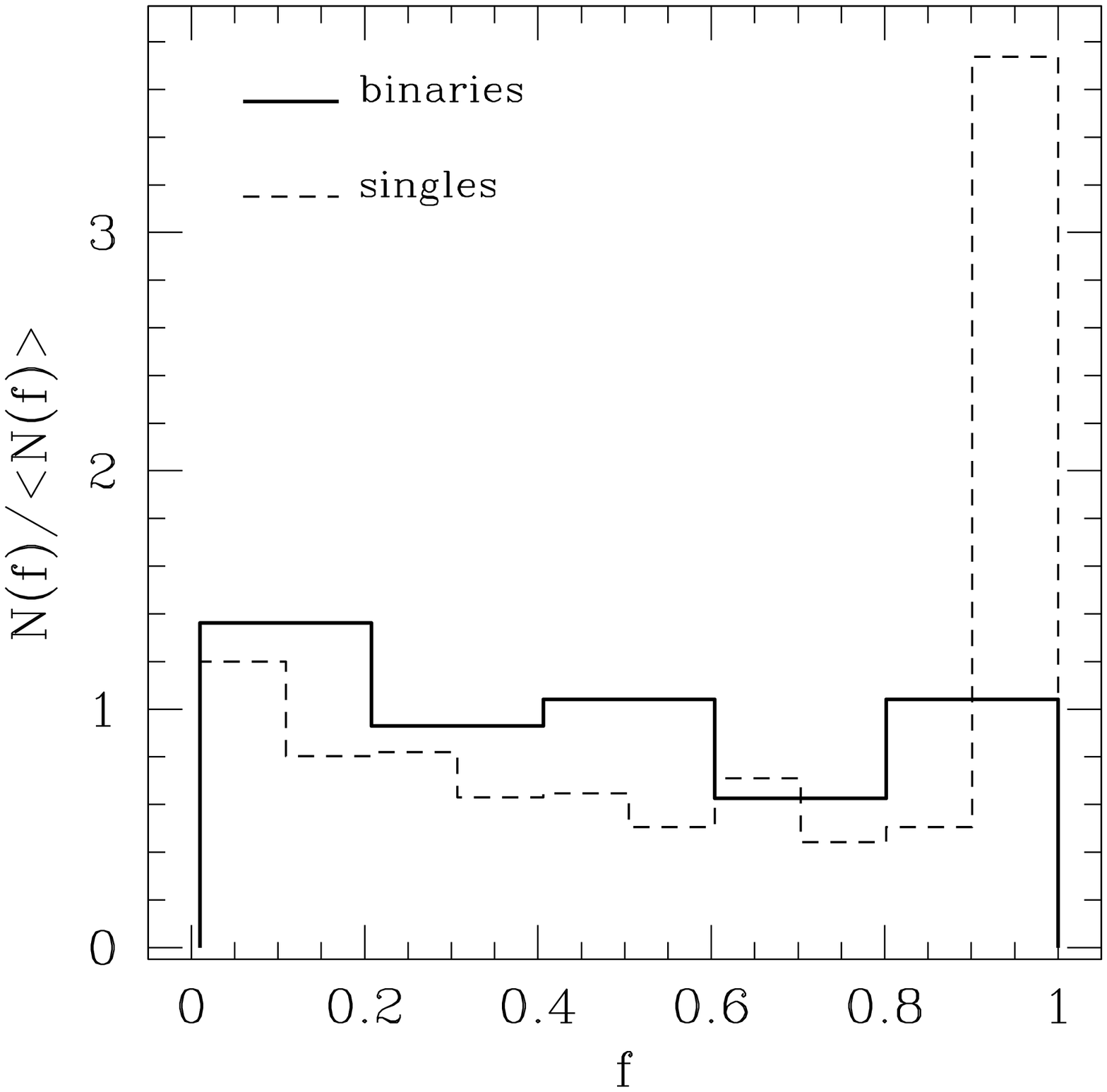}%
\FigCap{Histograms showing the distributions of the Einstein time (left)
and the blending parameter (right). Binary events (solid lines) and
single events (dashed) are shown.}
\end{figure}

The durations of binary events are systematically longer as compared
to single lens events. The binaries should be (on average) twice as
massive as single lenses, so the binary events should be $\sqrt2$ times
longer. The averaged logarithms of Einstein times for single/binary lens
events in our sample correspond to event durations of 32\upd/63\upd, in
disagreement with the above estimate. The longer events have better
chance of being classified as binary, but it is hard to estimate to
what extend such selection effect explains the difference. 

\subsection{The Blending Parameter $f$}
We show the distributions of blending parameter values for single and
binary events. Again we neglect single lens models with ${f<0.01}$, which
removes most of the artifacts related to parameter degeneracy. Another
problem -- crowding of single mass fits with ${f\approx1}$ and the
unphysical models with ${f>1}$ (compare Paper~I) -- is circumvented by
inclusion of fluxes $F_s$ and $F_b$ as independent parameters of the
fits. We put the histograms showing the distribution of blending
parameter $f$ in Fig.~2 for both single and binary lens models. 

We also show the positions of binary and single lens models on a
$\lg(t_{\rm E})$--$f$ diagram in Fig.~3. There is no apparent
correlation between the parameters in the diagram, which means that the
majority of parameter degeneracy artifacts have been removed. The
crowding of points near upper boundary ($f =1$) represents the bright
sources which are not affected by blending.

\begin{figure}[htb]
\vglue-3mm
\centerline{\includegraphics[height=123.0mm,width=123.0mm]{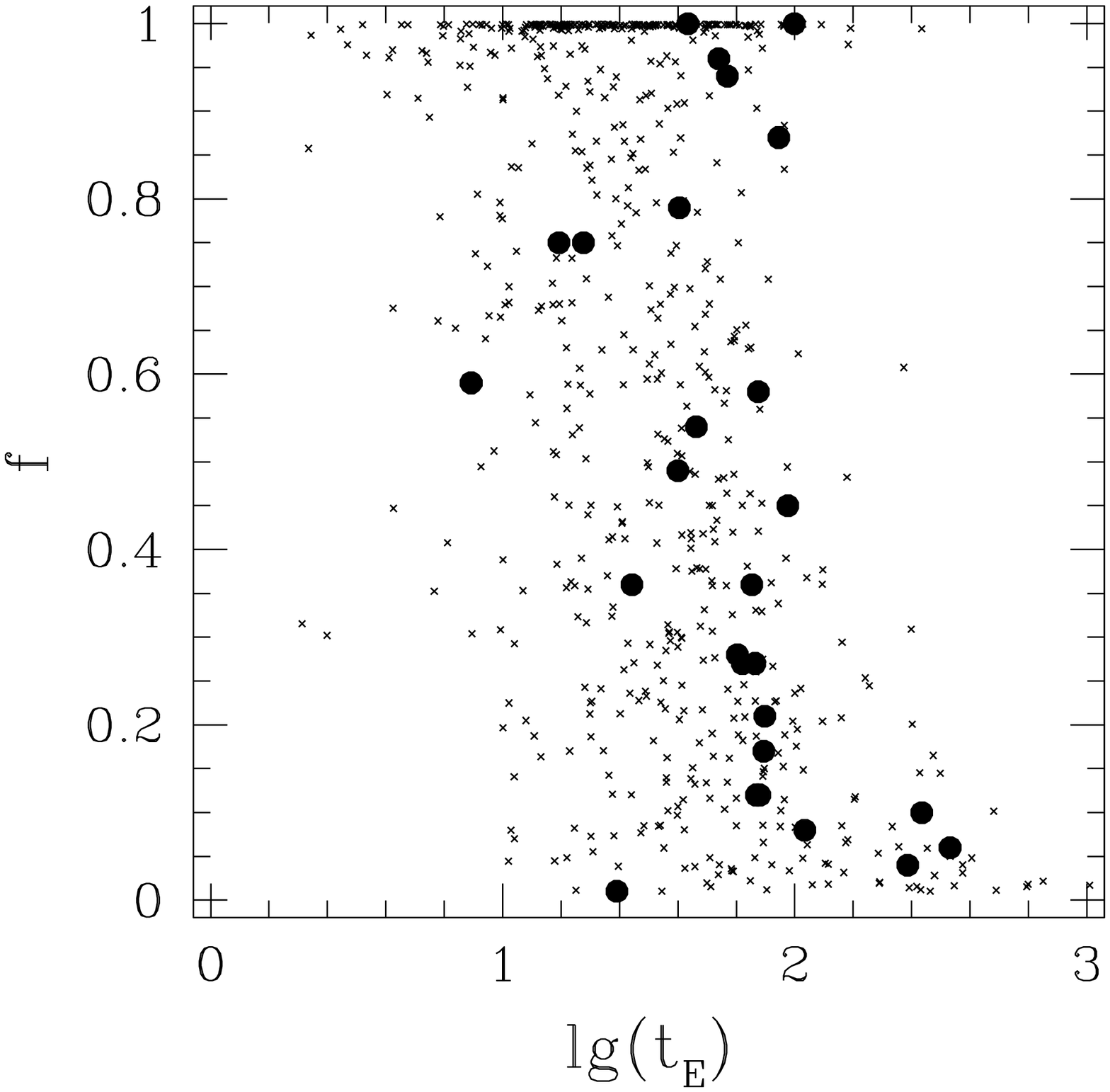}}
\vspace*{-5mm}
\FigCap{The location of single lens (crosses) and binary lens models
(large dots) on $\lg(t_{\rm E})$--$f$ plane. Events from seasons 
1997--1999 and 2002--2003 are included.}
\end{figure}

\Section{Discussion}
In this study (as in Paper~I) we use rescaled errors when estimating the
quality of fits {\it via} $\chi^2$ test. The rescaling allows more adequate
definition of confidence regions in parameter space and gives more reliable
comparison between different fits. The linear scaling of all errors may be
not adequate, since the relative accuracy of flux measurements is by far
better for bright sources, and estimates based on the epochs when the
source is faint may be not sufficient. Examining our models we also see
that fits for the brightest sources are formally the worst. 

For long lasting events the effects of binary and Earth orbital motions may
be important, but we neglect both of them in this study. In general fits
with more parameters require more observations during the event. In the
case of caustic crossing events it is rather time between the crossings,
not the Einstein time, which should be compared with the orbital period of
binary/Earth to check whether their motion can influence the quality of our
models. Finally, for strongly blended events (${f\ll1}$) the Einstein time
is not a good measure of their duration, since the measurable changes in
observed energy flux can be seen only when the amplification is really
high. For single lens events the rough estimate of relevant time scale is
given by $ft_{\rm E}$. For binary events the intra-caustic time plays the
role. All our models with extremely long duration (${t_{\rm E}>100\upd}$)
are also strongly blended, and the introduction of extra parameters in
these cases is not promising. The event which certainly deserves improved
modeling including parallax effect and/or binary rotation is OGLE
2003-BLG-267 (${t_{\rm E}=89\upd}$, ${f=0.37}$), a case with well sampled
caustic crossings and showing some systematic differences between
observations and model (see Appendix~1). Similarly it is impossible to
obtain a satisfactory model for OGLE 2003-BLG-291 without including some
extra parameters. We are going to describe this event with more details
elsewhere.

The main purpose of this study is the statistical characteristic of the
population of binary lenses. The total number of ``strong'' binary lens 
cases (including those of Paper~I) is still low (25). The distribution
of our models in $(\lg q,\lg d)$ plane is in agreement with the
hypothesis that binary stars are distributed uniformly in these
parameters for ${0.1\le q\le1}$. Since the probability of observing an
event caused by a binary with very small (${d\le0.1r_{\rm E}}$) or very
large (${d\ge10r_{\rm E}}$) separation is negligible, it is impossible
to check the wider range of separations using microlensing and the
present database. For a ``typical'' event including binary in the
Galactic disk the Einstein radius ${r_{\rm E}\approx1{\rm a.u.}}$, so our
approach probes the binary stars with separations of similar order of
magnitude. 

Of the planetary lenses only one (OGLE 2003-BLG-235/MOA 2003-BLG-53 --
compare Bond \etal 2004) is a ``strong case''. The other extreme mass
ratio models of Paper~I may be replaced by a less extreme mass ratio
models and/or by a double source models. The event OGLE 2002-BLG-055 is
not included in considerations because it is poorly constrained
(Jaroszy\'nski and Paczy\'nski 2002) and may also be modeled as a double
source event (Gaudi and Han 2004).

The distribution of binary events duration (Fig.~2) is systematically
shifted to longer $t_{\rm E}$ as compared to single lens events, which
can be partially explained as a result of the higher on average binary
masses. The well pronounced peak of the binary duration distribution
must be a fluctuation or some kind of selection effect, which makes the
binary events with $t_{\rm E}$ between 64 and 100 days more likely to be
discovered. 

Some of the events of this study were also reported elsewhere. The event
OGLE 2002-BLG-069 was observed by the PLANET collaboration and the caustic
exit was monitored spectroscopically by VLT (Cassan \etal 2004) to probe
the atmosphere of the source star -- a G5III giant in Galactic bulge. The
atmospheric study does not require the full binary lens model, which has
not been published yet, so the direct comparison is not possible.

The event OGLE 2003-BLG-135 was also observed by the MOA group (Bond \etal
2001) and is named MOA 2003-BLG-21. More interesting is the event OGLE
2003-BLG-235/MOA 2003-BLG-53. The combined observations of OGLE and MOA
were modeled by the two teams and reported as a planetary microlensing
event set including the coverage of the second caustic crossing (Bond \etal
2004). The parameters of the best fit based on large data set including the
coverage of the second caustic crossing (predicting ${q=0.0039}$,
${d=1.12}$, and ${t_{\rm E}=61.5}$) are in rough agreement with our
results.

The event OGLE 2003-BLG-095 (treated here as an example of double source
lensing) was modeled by Collinge (2004), who considers also the binary lens
model of the event and the influence of the parallax effect. The double
source models presented here and in his paper are similar.

The events OGLE 2003-BLG-170 and OGLE 2003-BLG-267 have well covered
caustic crossings. The latter is also showing some effects of parallax
and/or internal variability. The further study of these events may give
some limits on the possible distances and masses of the binary lenses.

\Acknow{We thank Bohdan Paczy\'nski for many helpful discussions and
Shude Mao for the permission of using his binary lens modeling 
software. This work was supported in part by the Polish KBN grants 
2-P03D-016-24 and 2-P03D-021-24, the  NSF grant AST-0204908, and NASA
grant NAG5-12212.}

\newpage

\centerline{{\bf Appendix~1: Binary Lens Models of Candidate Events}}

Below we present the plots for the 24 events for which the binary lens
modeling has been applied. Some of the models, especially cases interpreted
as cusp approach events, are not well constrained. The majority of events
modeled as cusp approaches have alternative double source models of similar
quality, and are shown in Appendix~2.

The events are ordered and named according to their position in the OGLE
EWS database for seasons 2002 and 2003. For two events we also give their
names in MOA database. Some events have more than one binary lens model of
comparable fit quality (compare Table~1) which we show as the 1st, 2nd etc
models.

Each case is illustrated with three panels. The most interesting part of
the source trajectory, the binary and its caustic structure are shown in
the left panel for the case considered. The labels give the $q$ and $d$
values. In the middle panel the part of the best fit light curve is
compared with observations. The labels give the rescaled $\chi^2$/DOF
values, the Einstein time $t_{\rm E}$ in days and the source flux / base
flux ratio. The diagram on the right shows the 68\% and 99\% confidence
regions in the $\lg q$--$\lg d$ plane. The location of the best fit is
marked with a large dot, and the position of the fit illustrated in a
given row -- with a cross. In some cases the confidence regions are
small and completely or partially hidden behind the dots.

\noindent\parbox{12.7cm}{
\noindent {\bf OGLE 2002-BLG-051} 

\vspace*{5pt}

 \includegraphics[height=46mm,width=42.5mm]{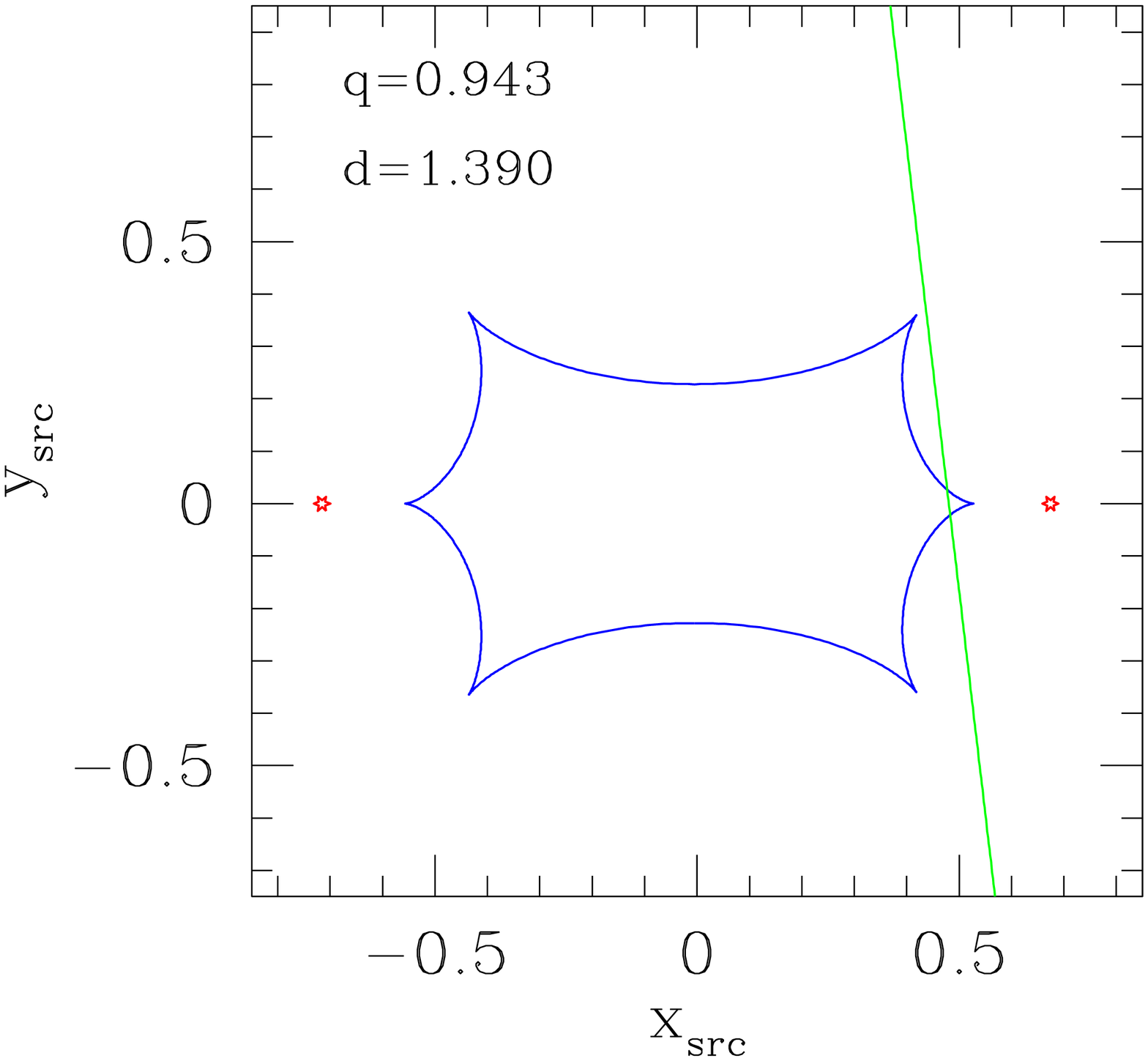}%
 \includegraphics[height=46mm,width=42.5mm]{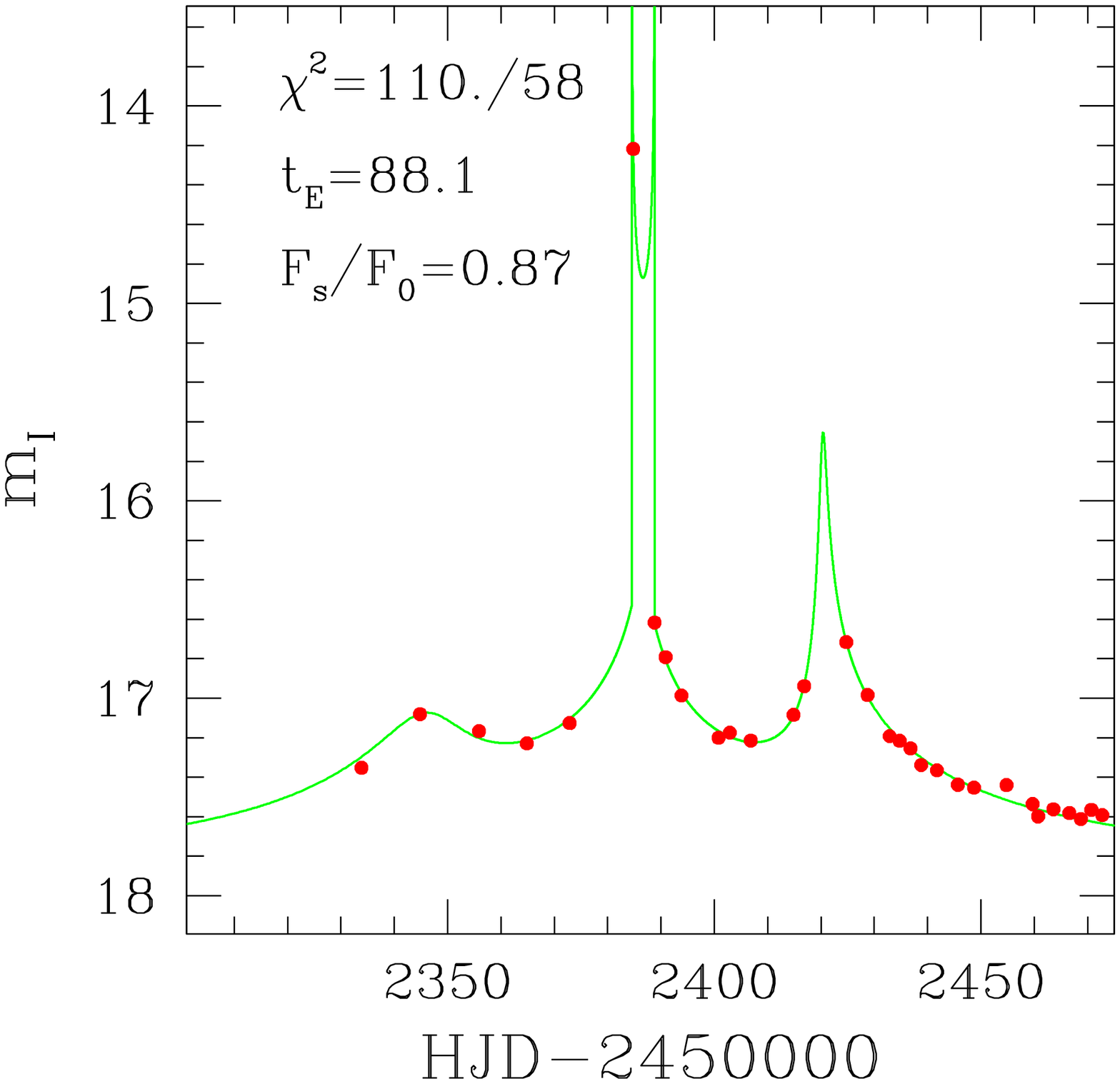}%
 \includegraphics[height=46mm,width=42.5mm]{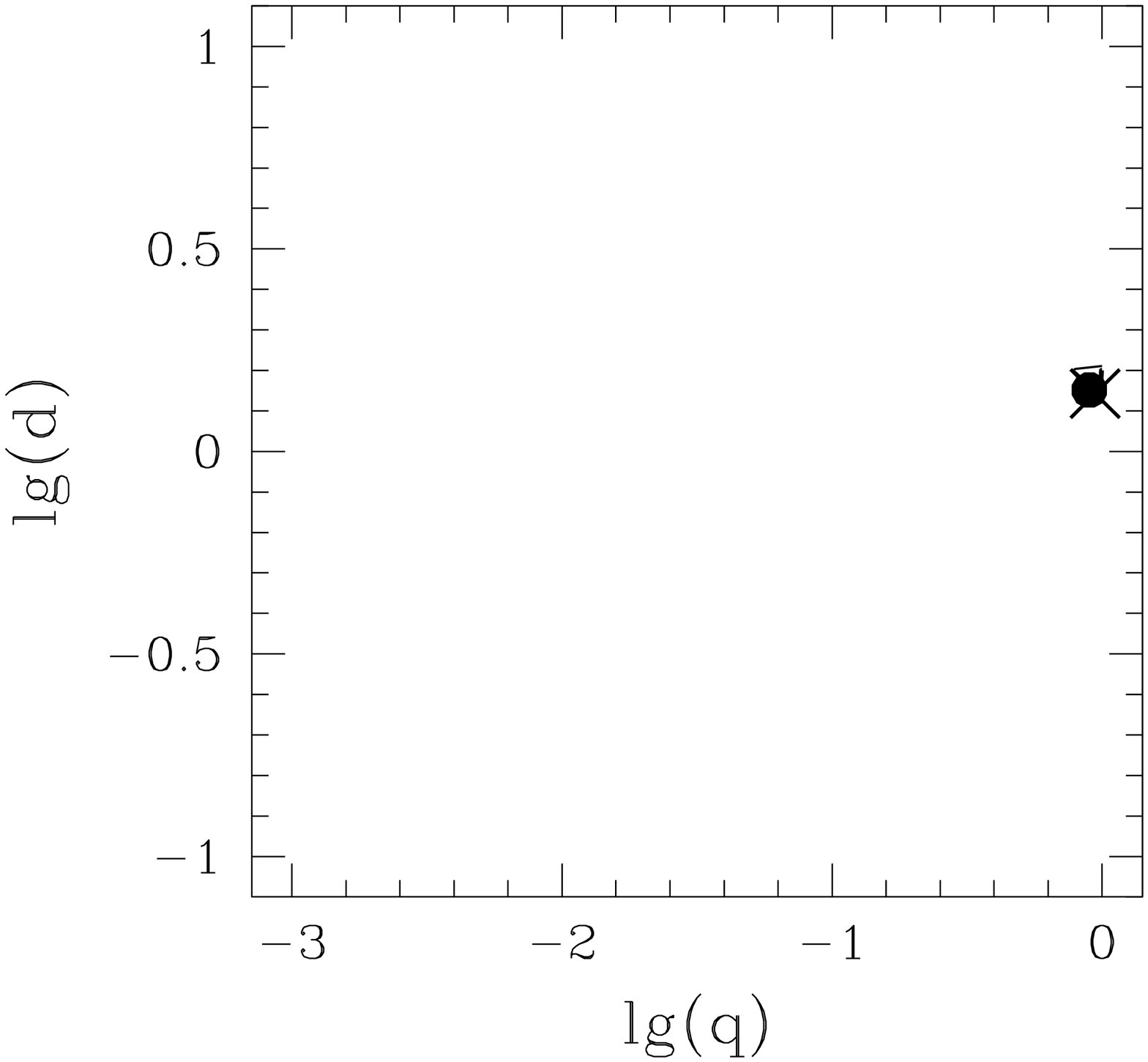}%
}

\noindent\parbox{12.7cm}{
\noindent {\bf OGLE 2002-BLG-069} 

\vspace*{5pt}

 \includegraphics[height=46mm,width=42.5mm]{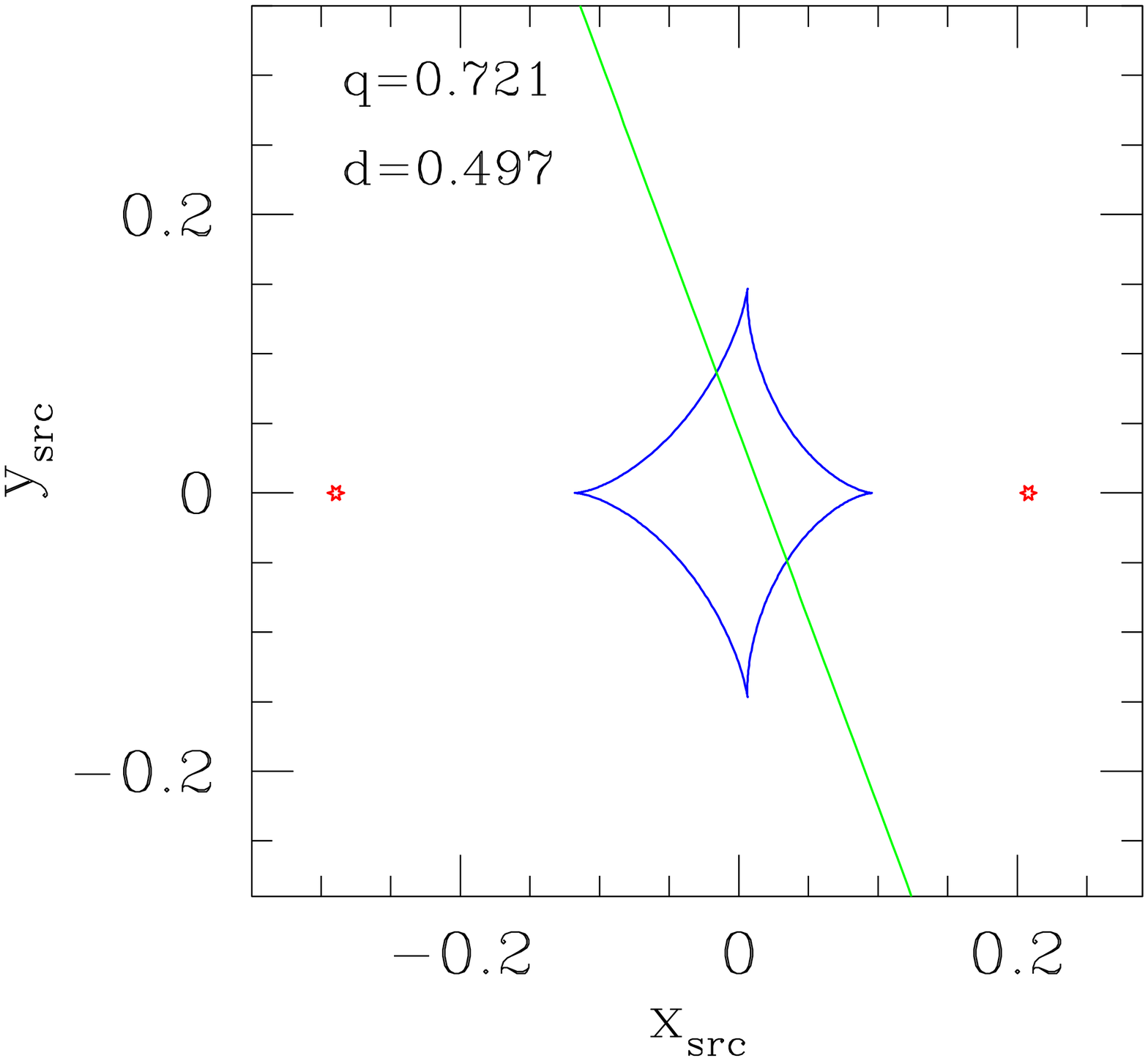}%
 \includegraphics[height=46mm,width=42.5mm]{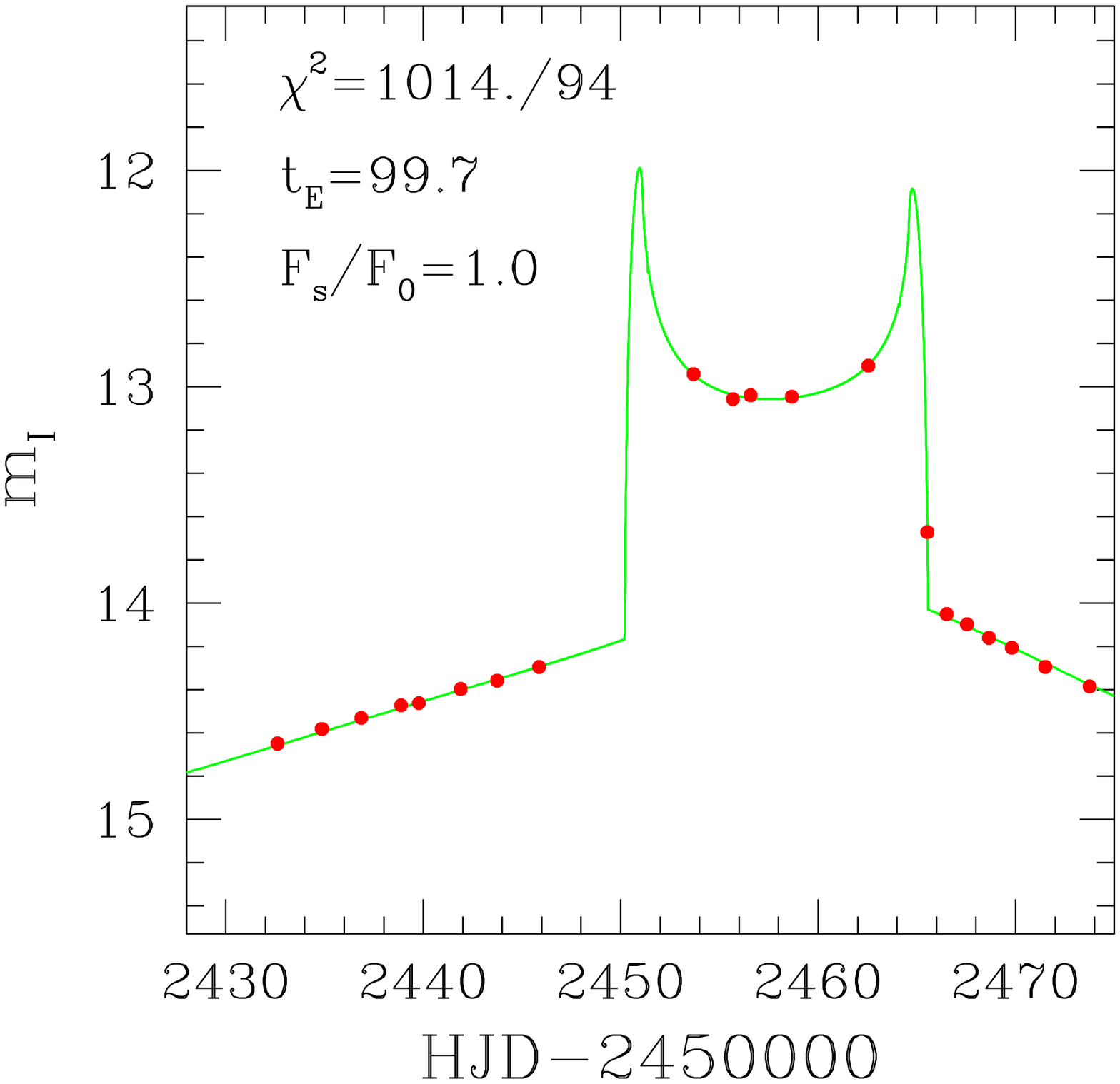}%
 \includegraphics[height=46mm,width=42.5mm]{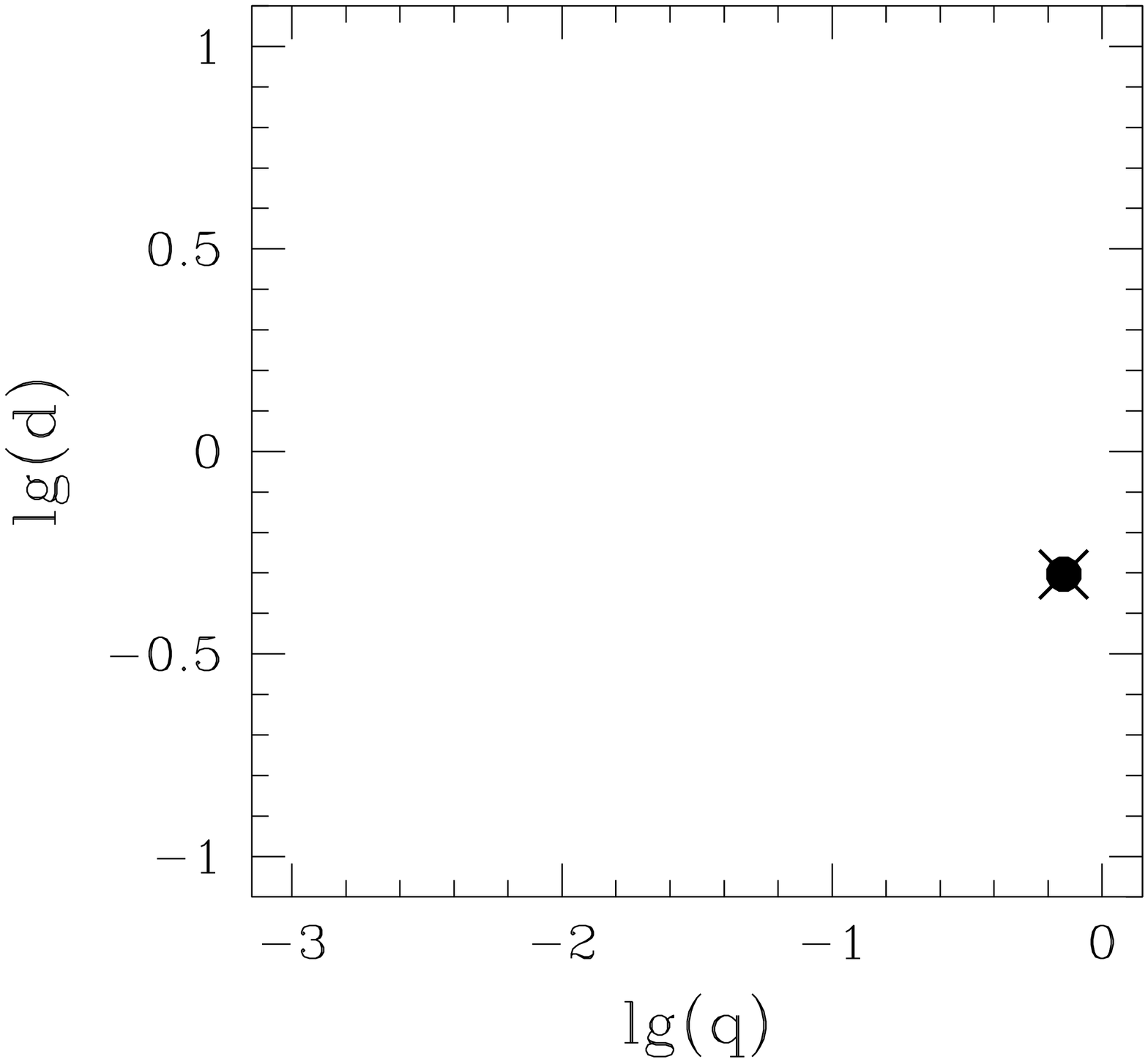}%
}

\noindent\parbox{12.7cm}{
\noindent {\bf OGLE 2002-BLG-099} 

\vspace*{5pt}

 \includegraphics[height=46mm,width=42.5mm]{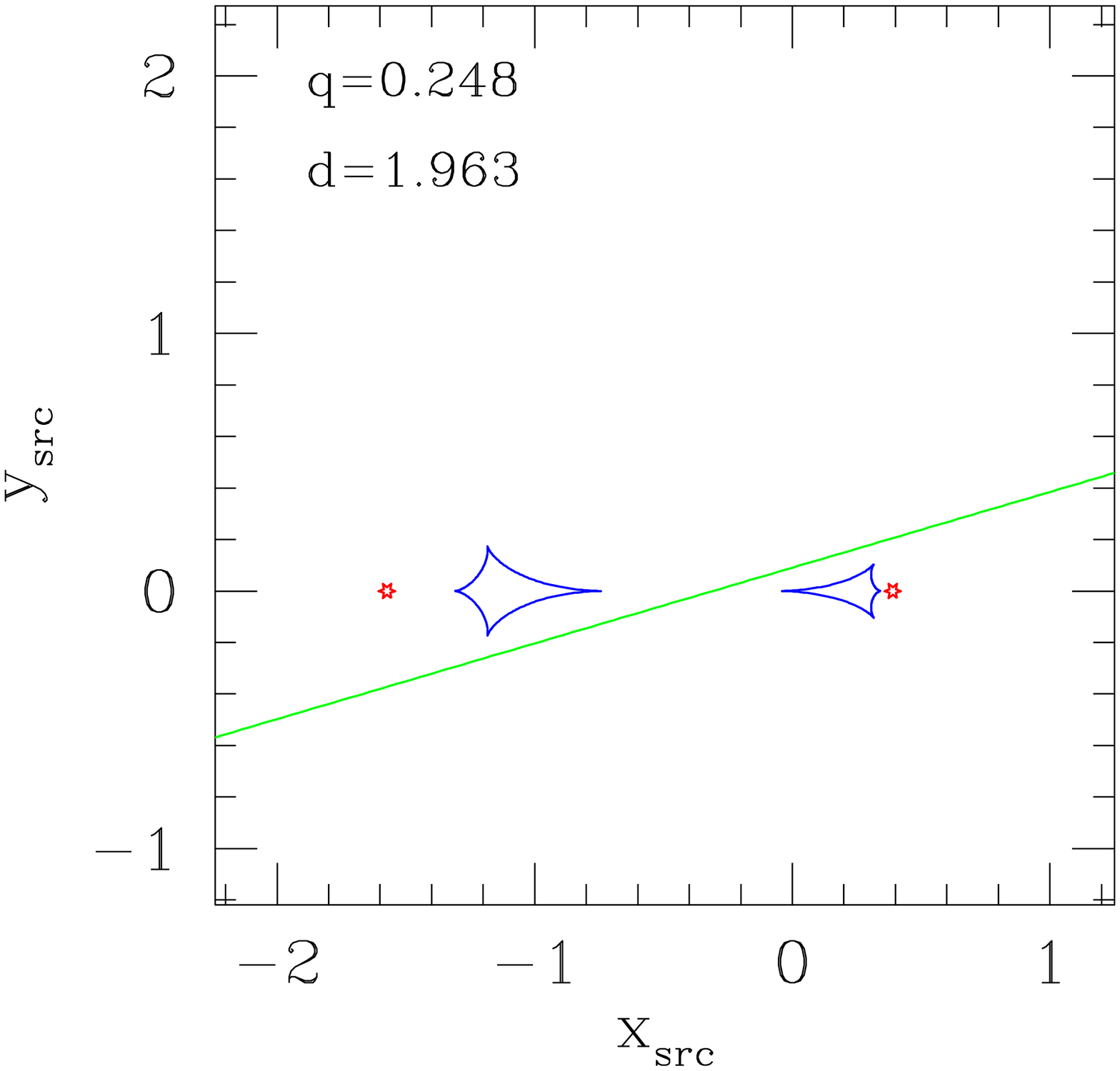}%
 \includegraphics[height=46mm,width=42.5mm]{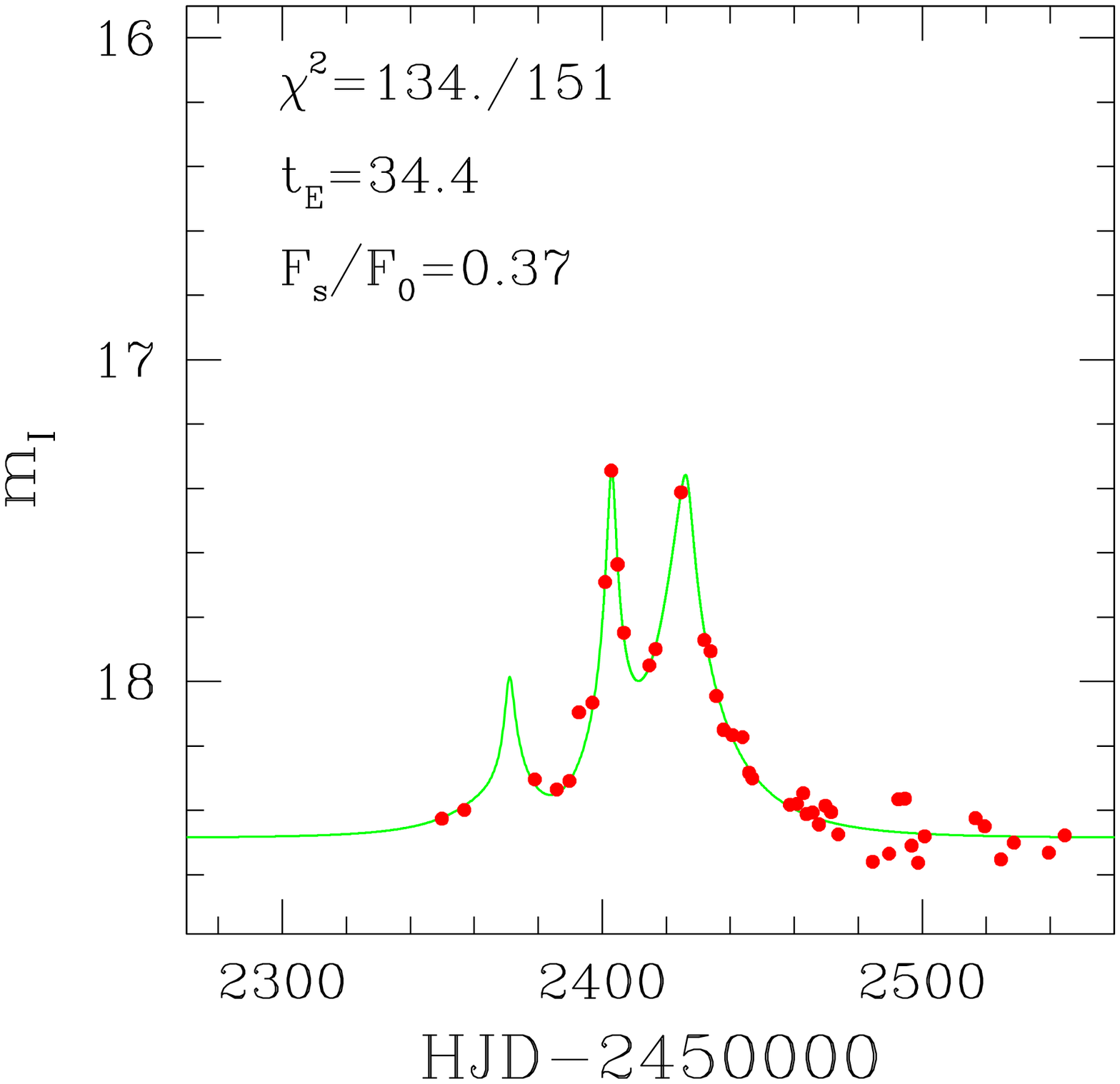}%
 \includegraphics[height=46mm,width=42.5mm]{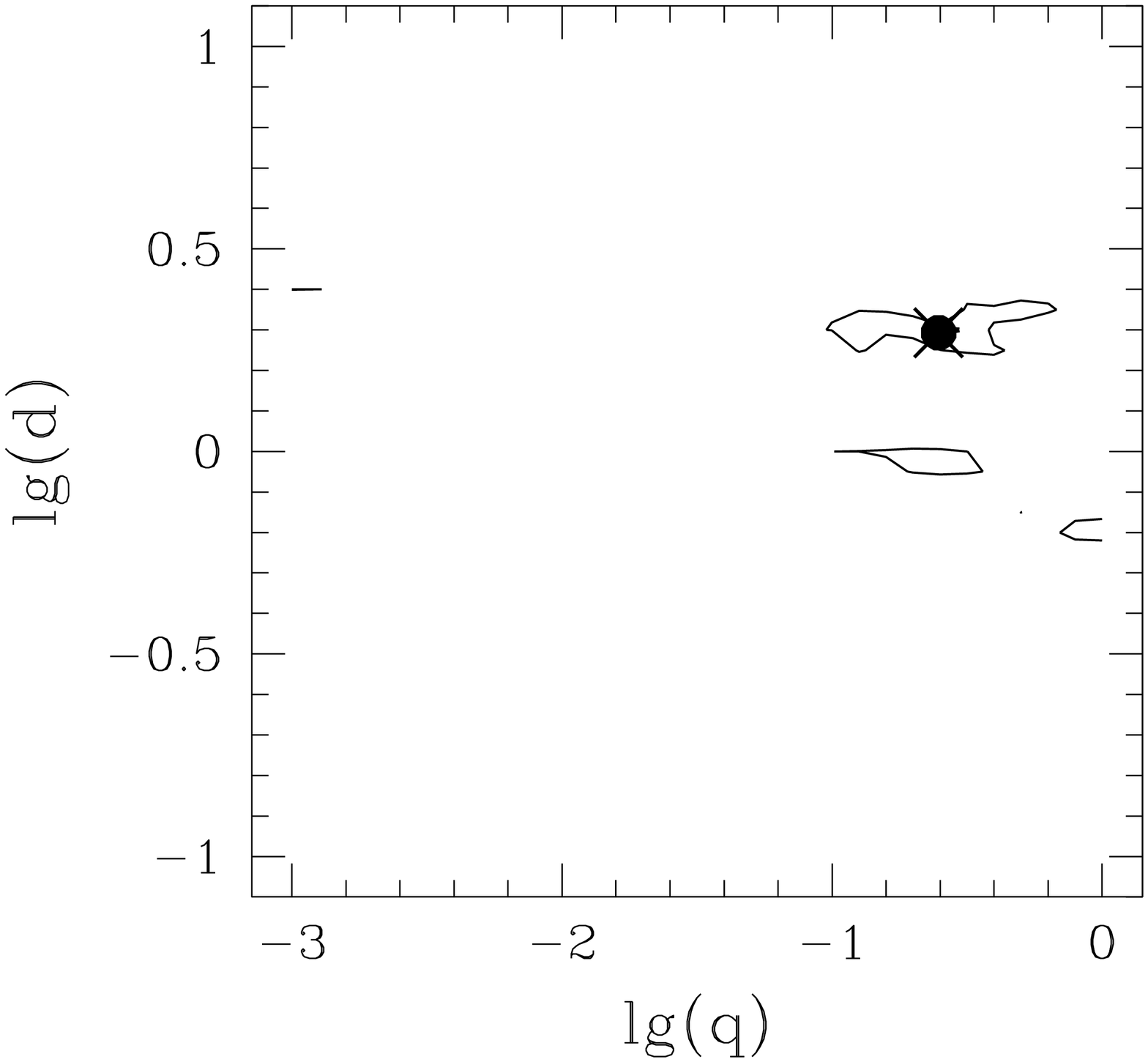}%
}

\noindent\parbox{12.7cm}{
\noindent {\bf OGLE 2002-BLG-114} 

\vspace*{5pt}

 \includegraphics[height=46mm,width=42.5mm]{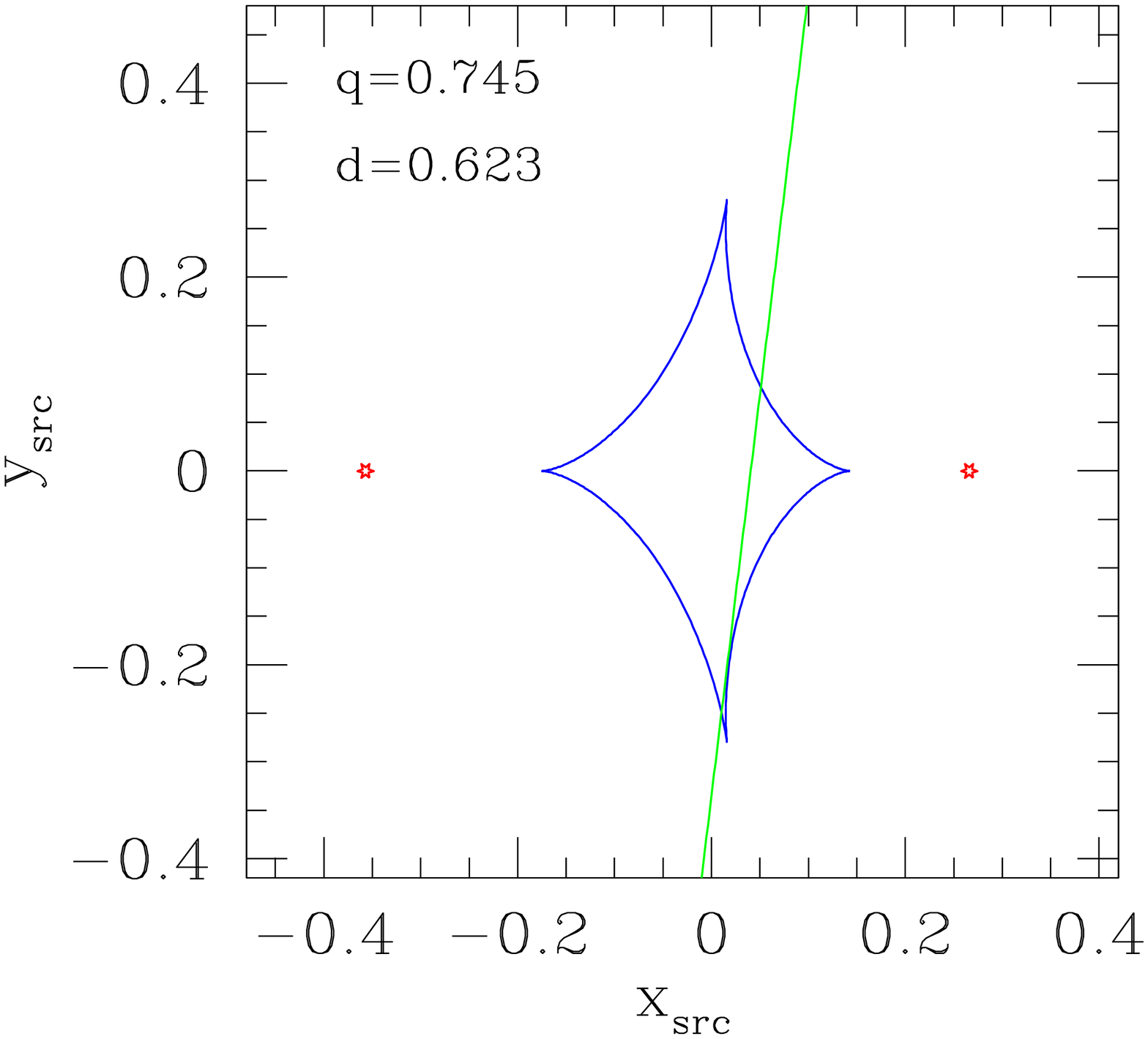}%
 \includegraphics[height=46mm,width=42.5mm]{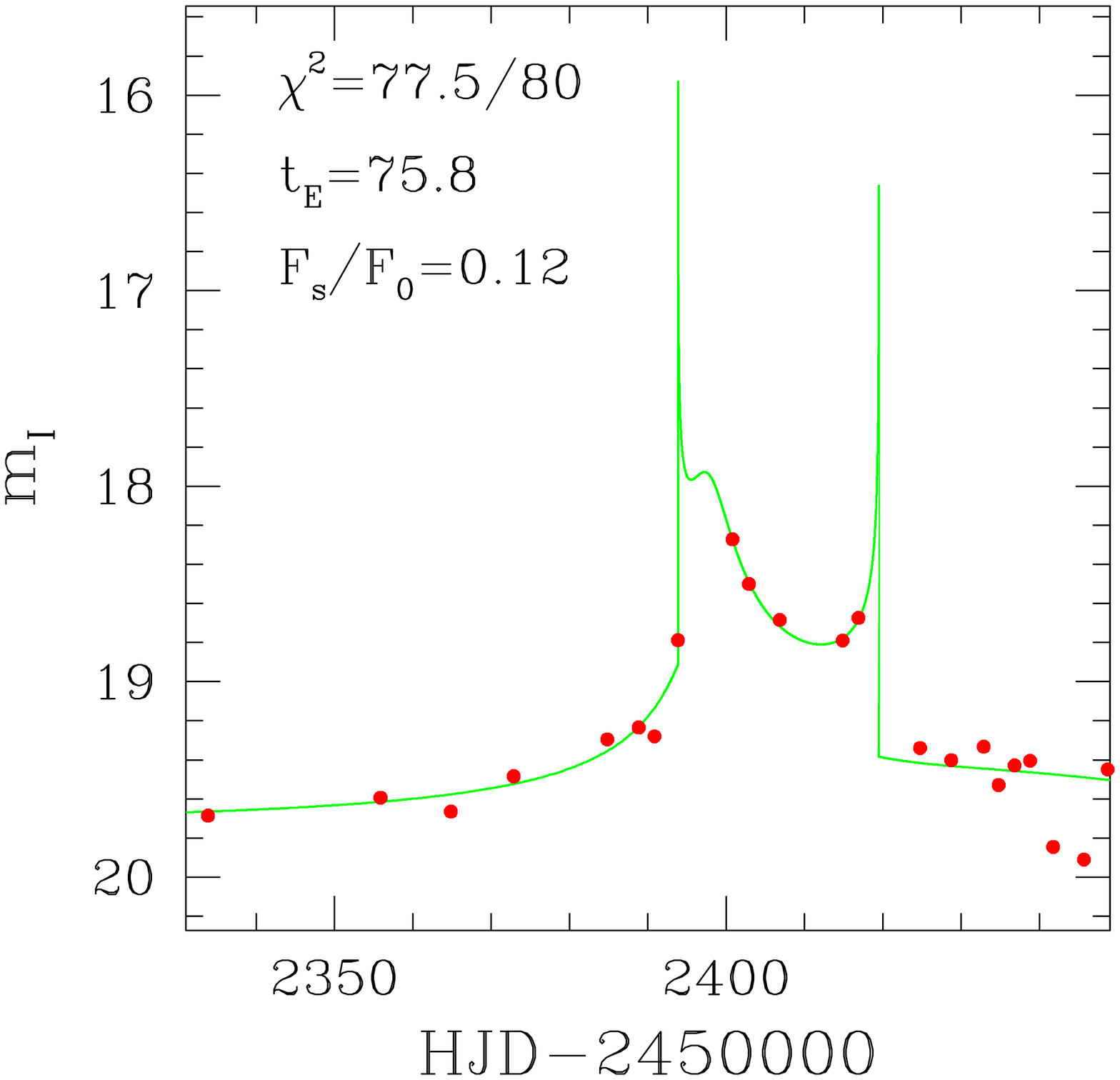}%
 \includegraphics[height=46mm,width=42.5mm]{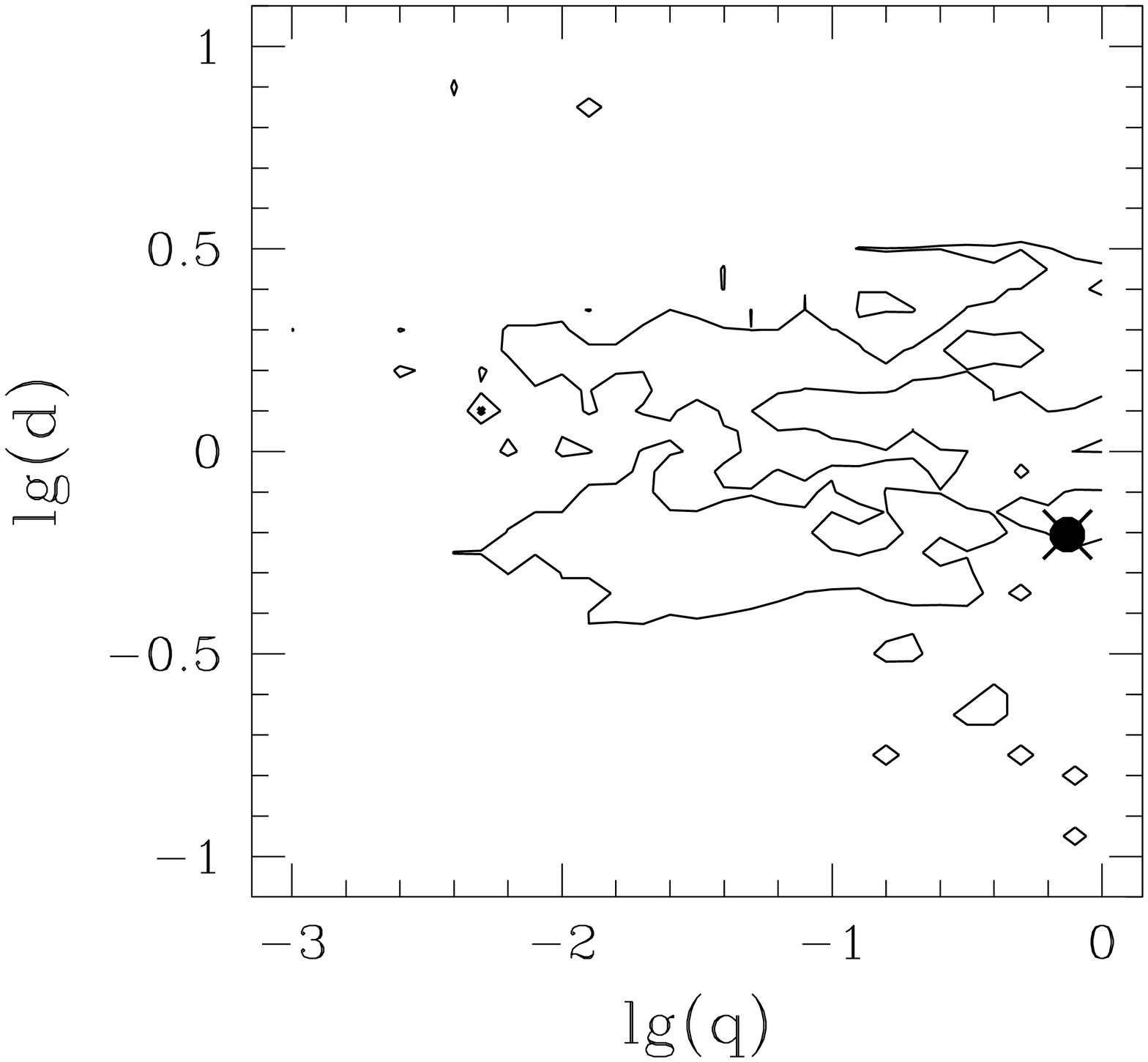}%
}

\noindent\parbox{12.7cm}{
\noindent {\bf OGLE 2002-BLG-135 (1st model)} 

\vspace*{5pt}

 \includegraphics[height=46mm,width=42.5mm]{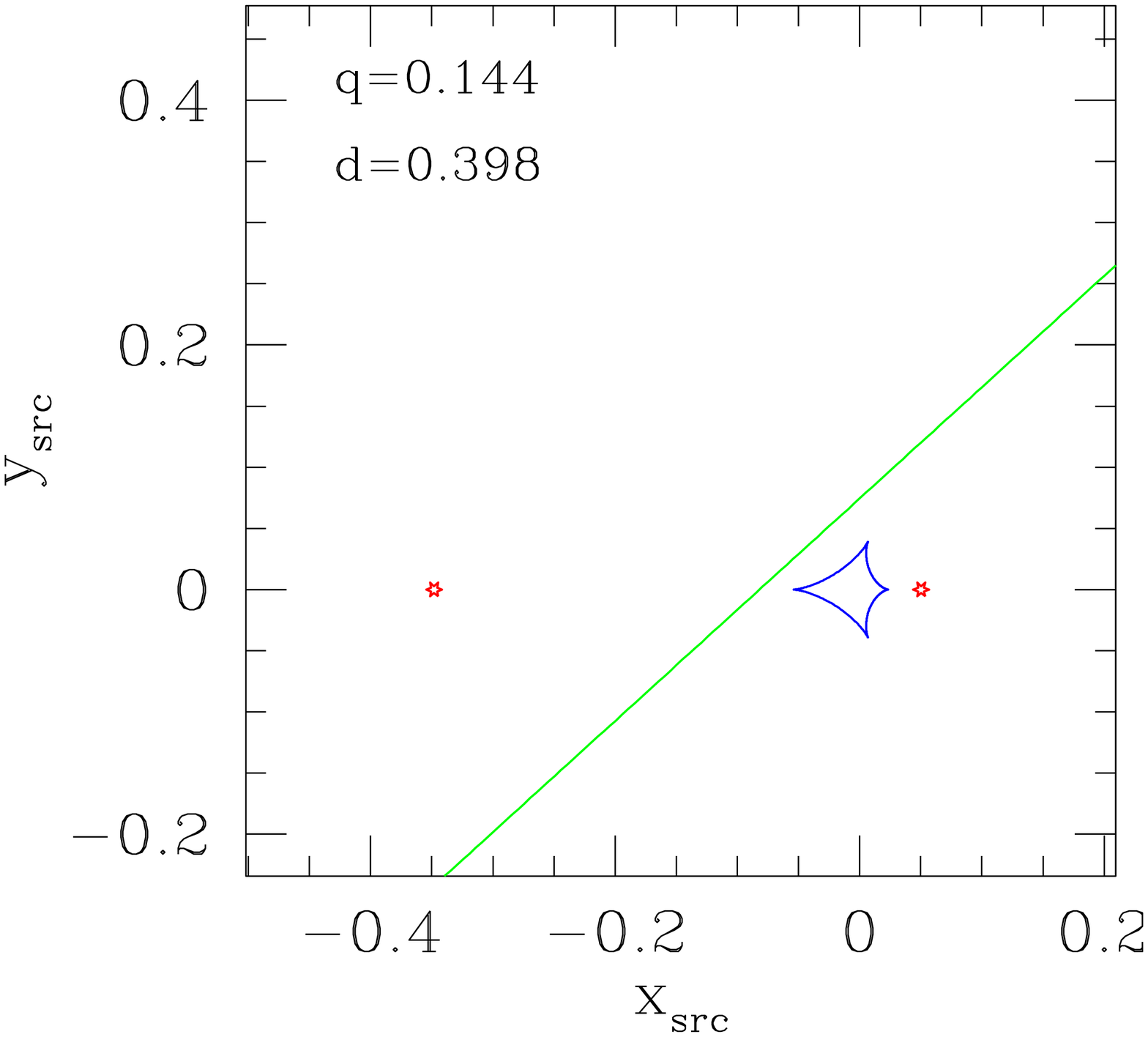}%
 \includegraphics[height=46mm,width=42.5mm]{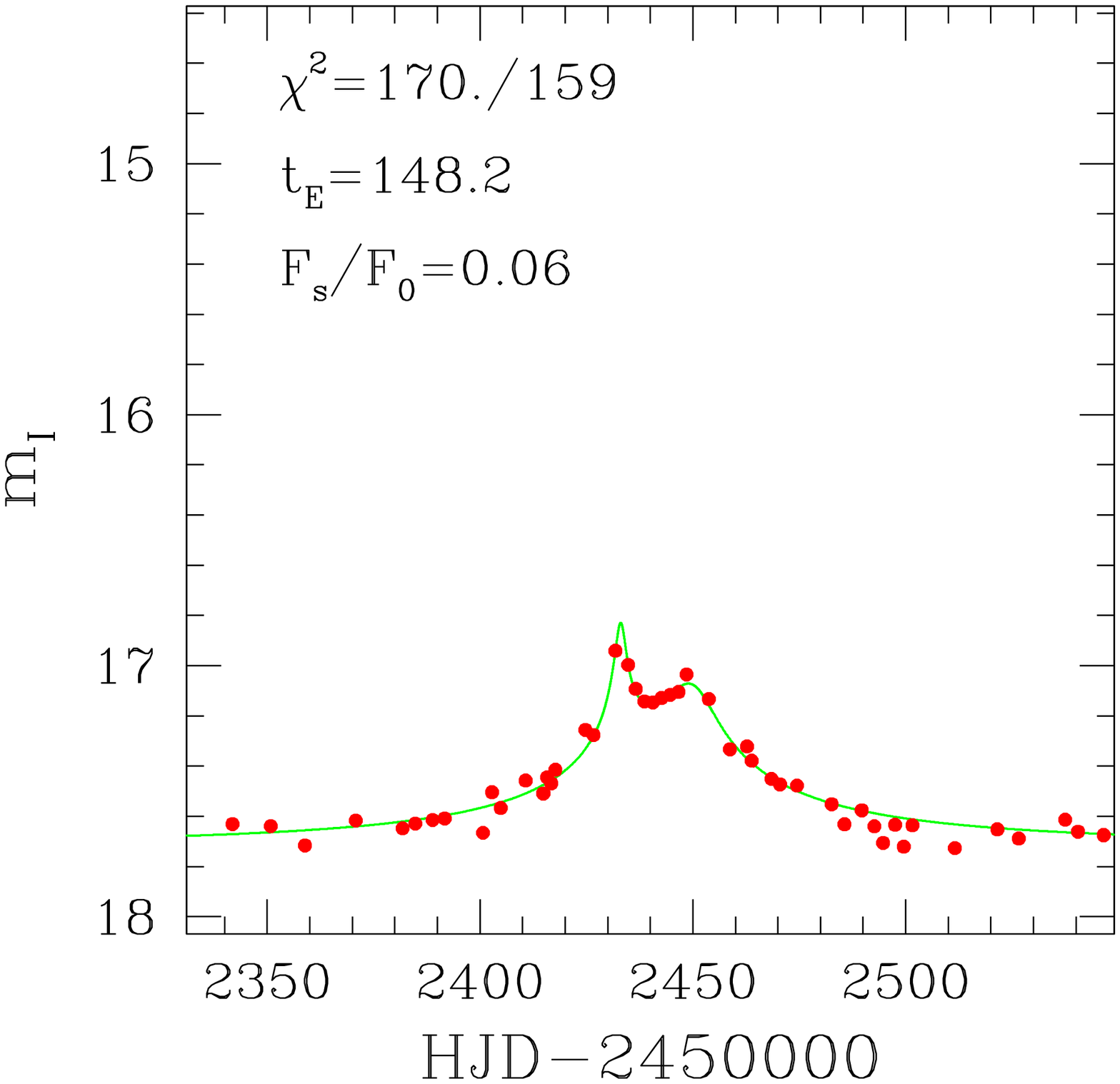}%
 \includegraphics[height=46mm,width=42.5mm]{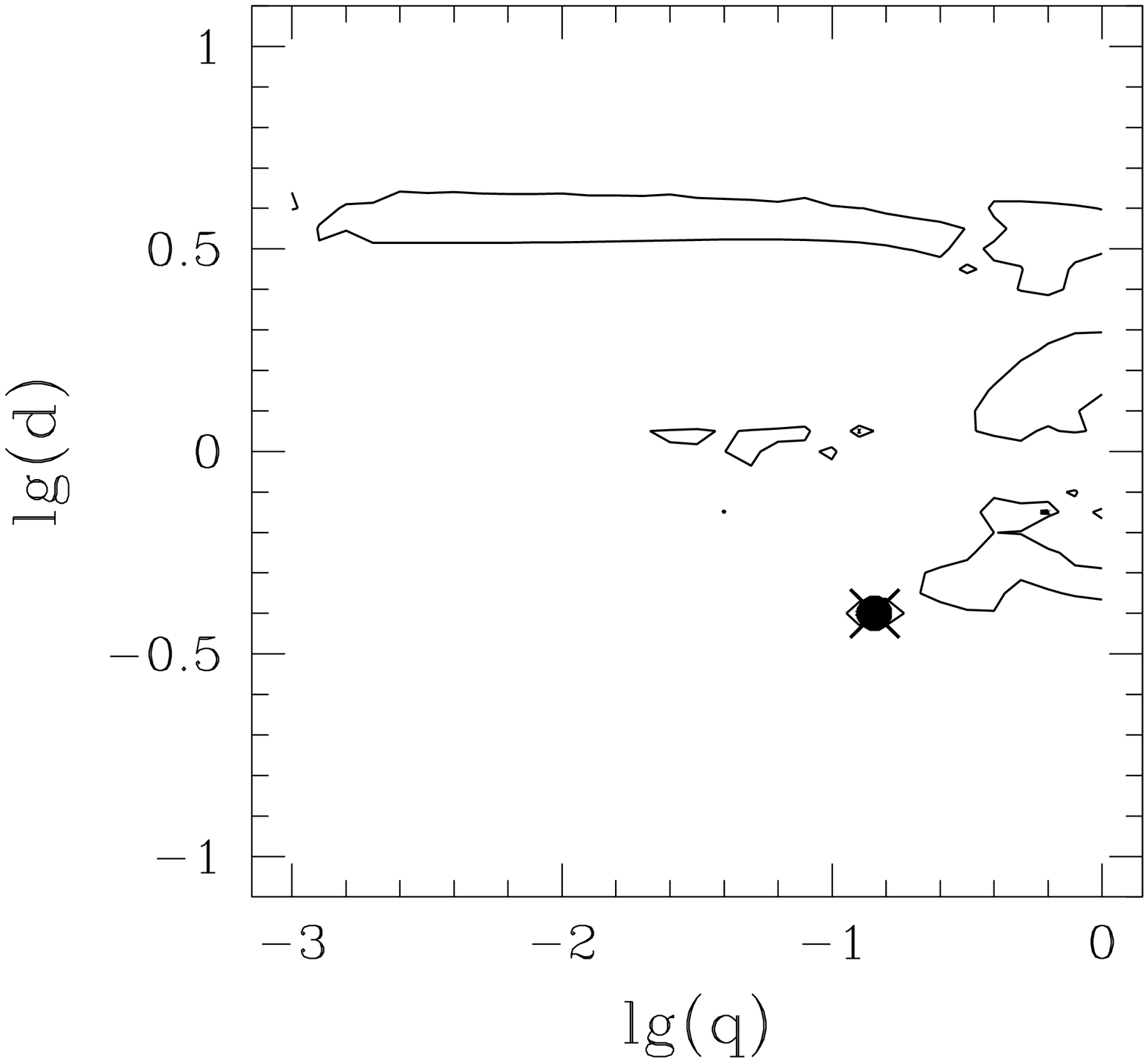}%
}

\noindent\parbox{12.7cm}{
\noindent {\bf OGLE 2002-BLG-135 (2nd model)} 

\vspace*{5pt}

 \includegraphics[height=46mm,width=42.5mm]{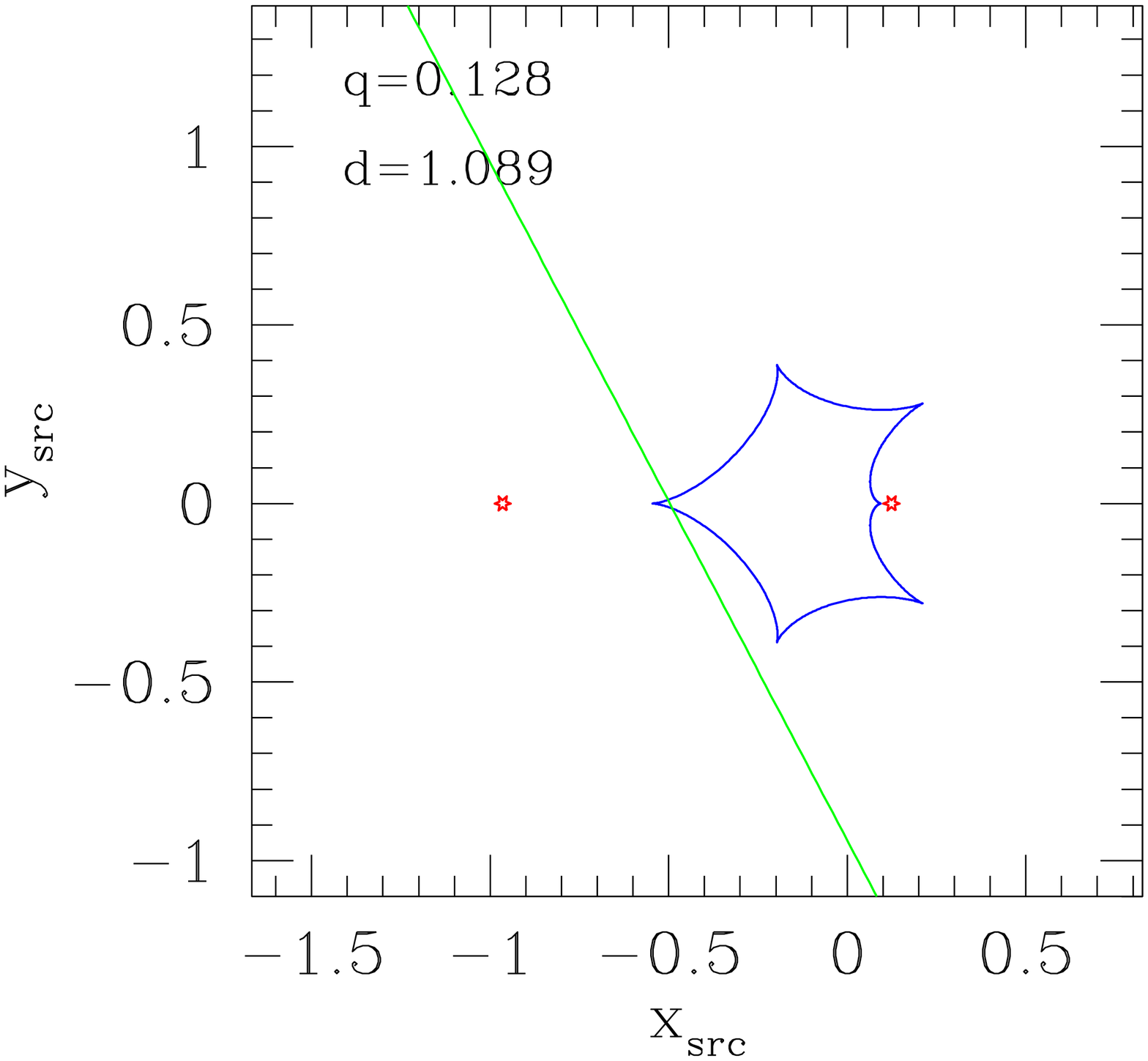}%
 \includegraphics[height=46mm,width=42.5mm]{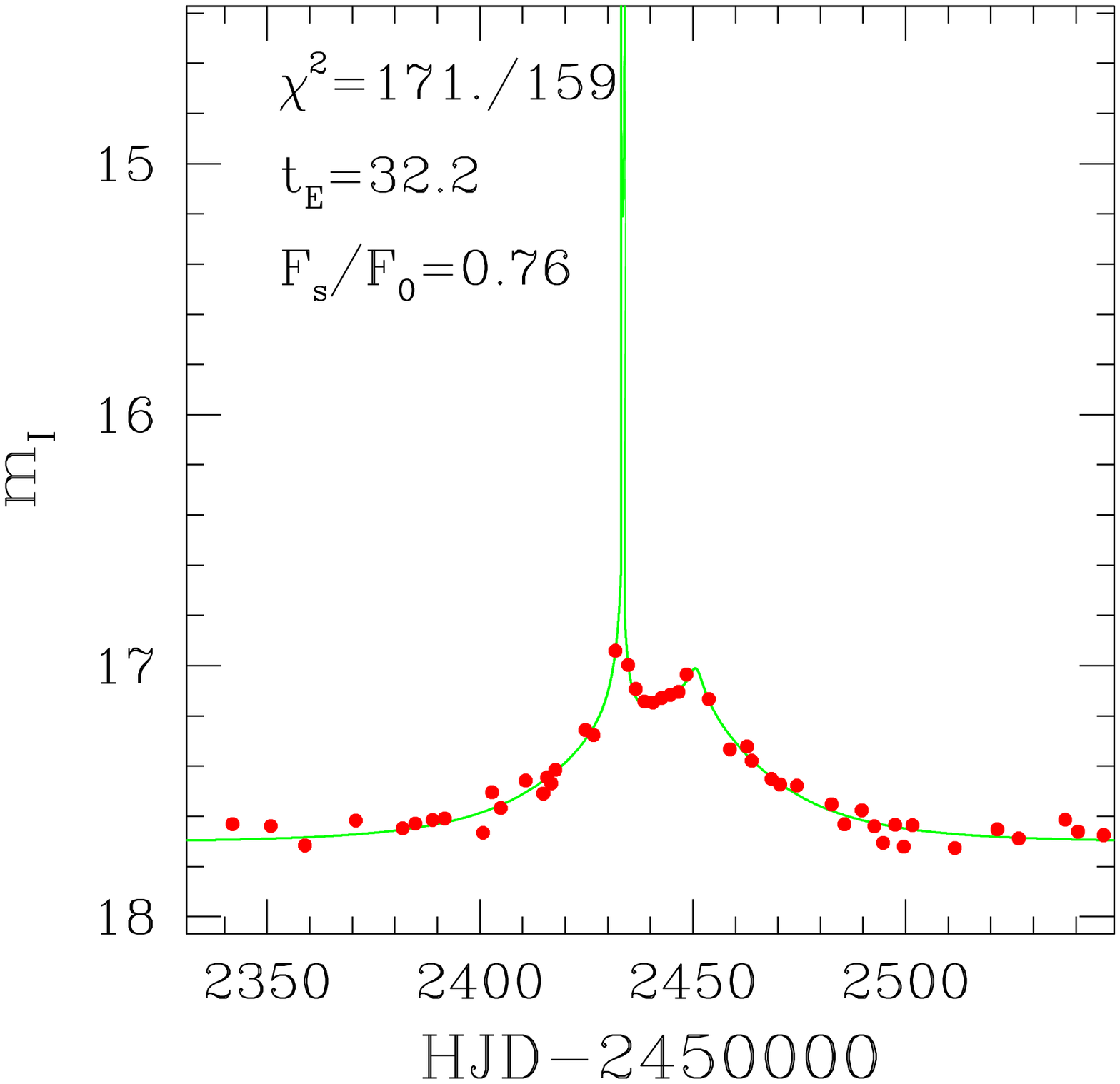}%
 \includegraphics[height=46mm,width=42.5mm]{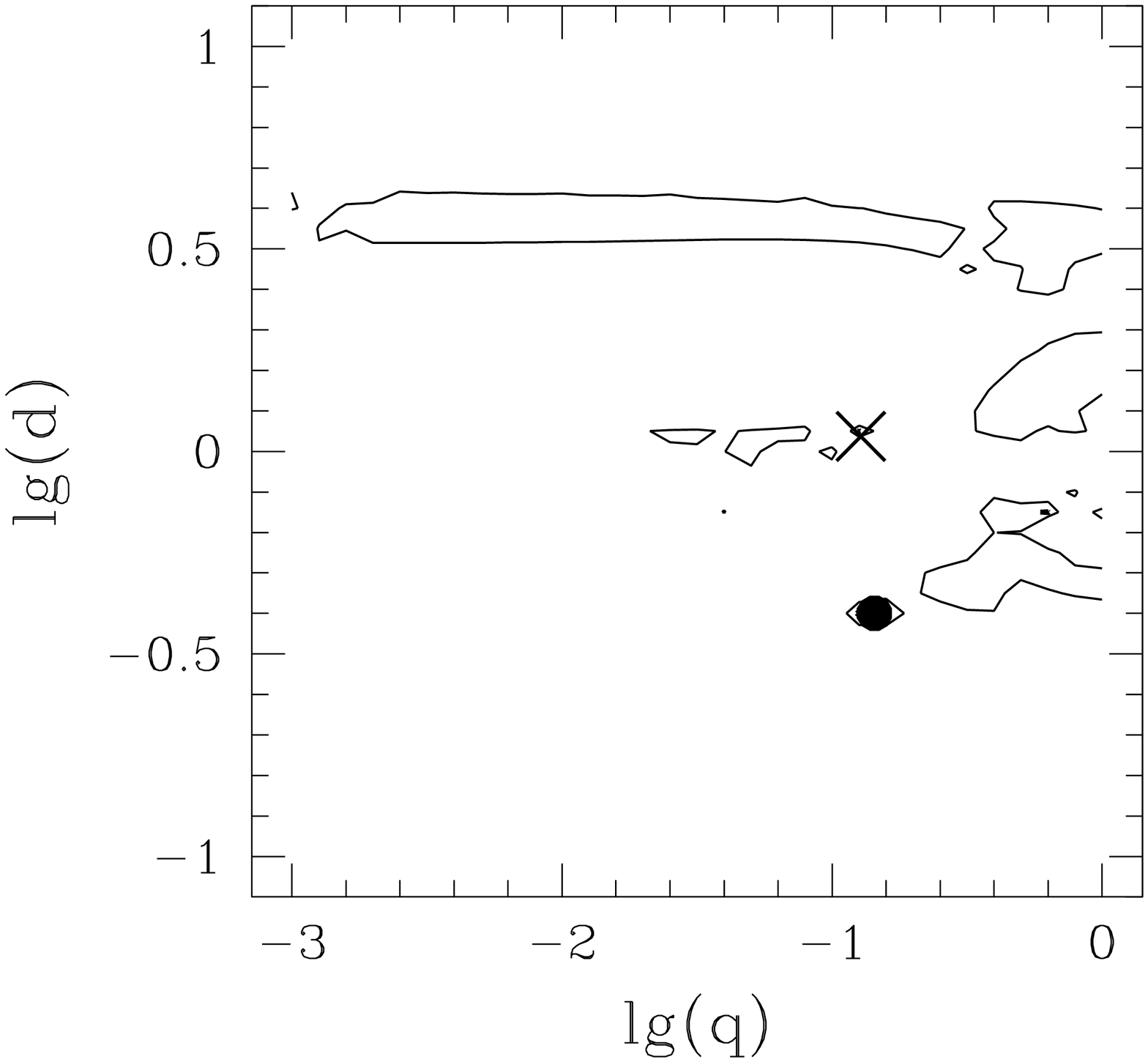}%
}

\noindent\parbox{12.7cm}{
\noindent {\bf OGLE 2002-BLG-135 (3rd model)} 

\vspace*{5pt}

 \includegraphics[height=46mm,width=42.5mm]{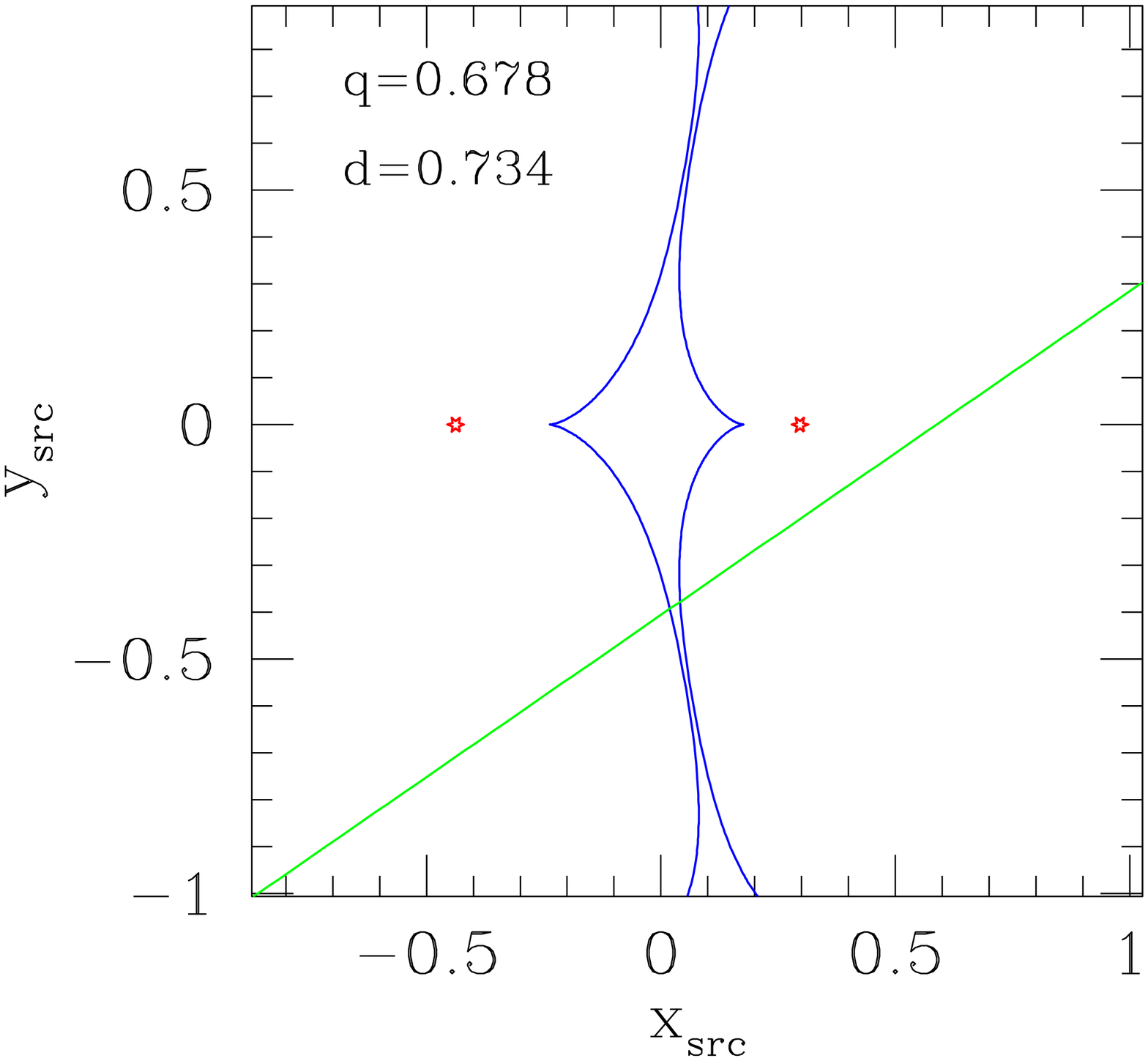}%
 \includegraphics[height=46mm,width=42.5mm]{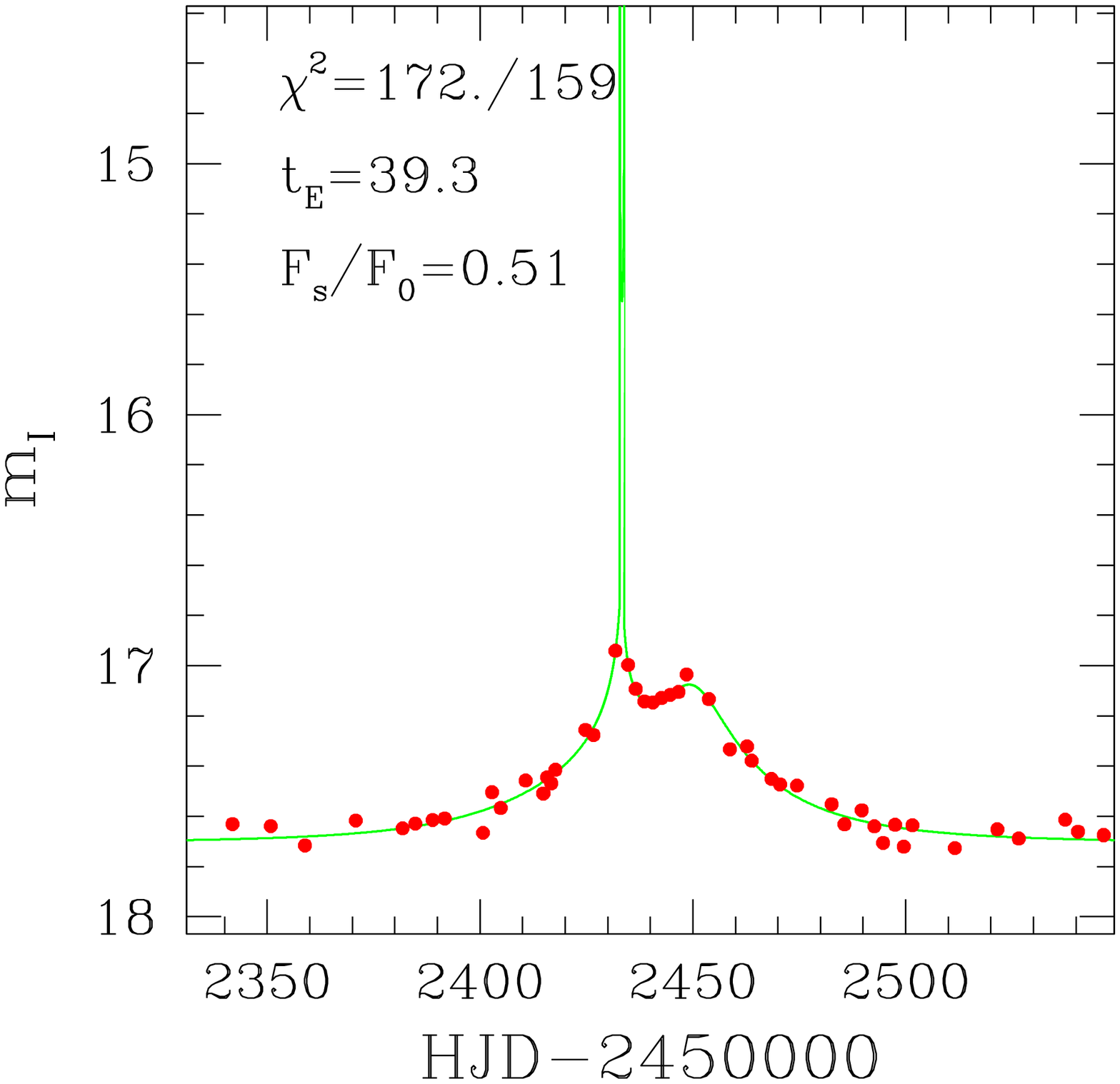}%
 \includegraphics[height=46mm,width=42.5mm]{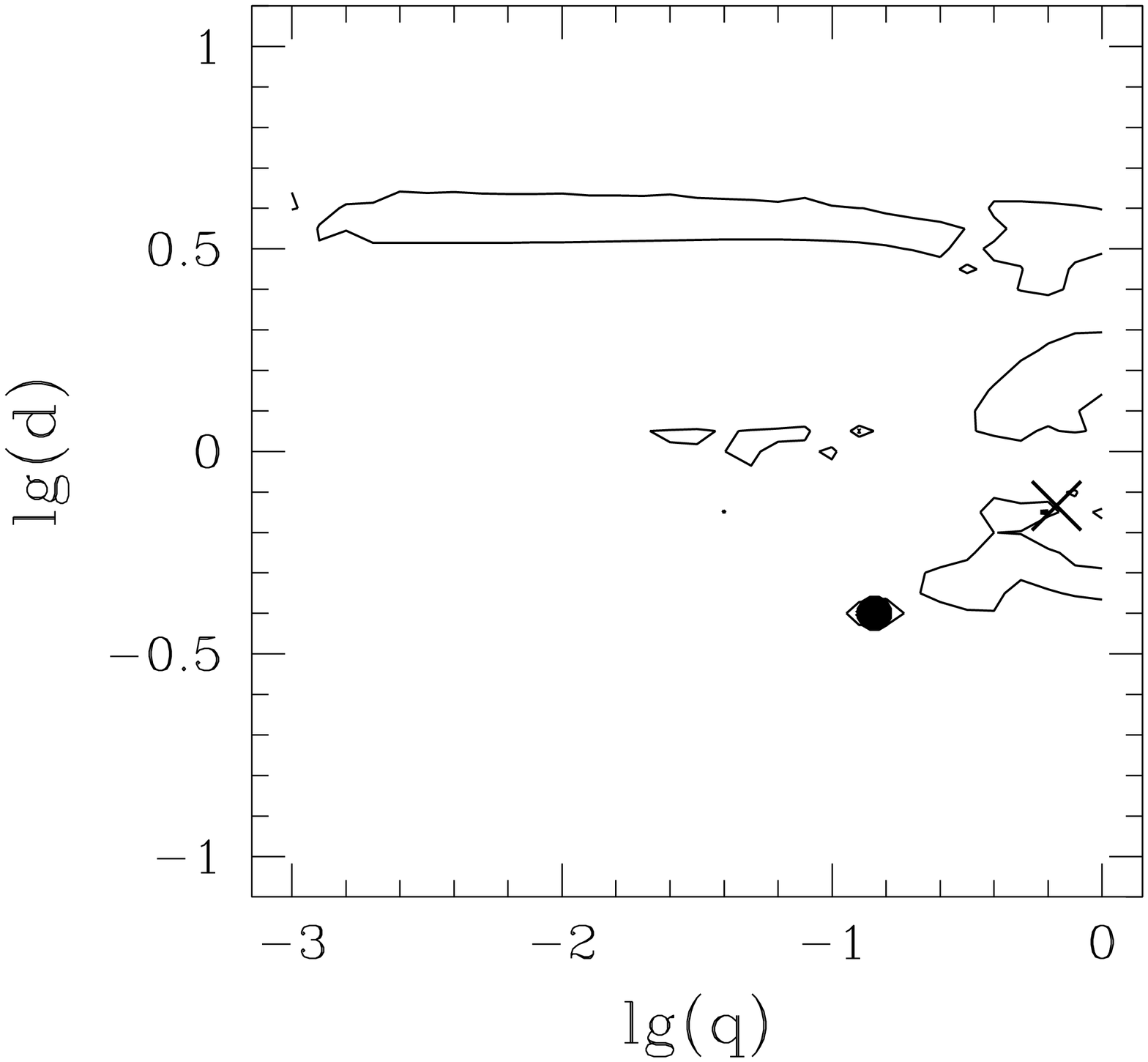}%
}

\noindent\parbox{12.7cm}{
\noindent {\bf OGLE 2002-BLG-158} 

\vspace*{5pt}

 \includegraphics[height=46mm,width=42.5mm]{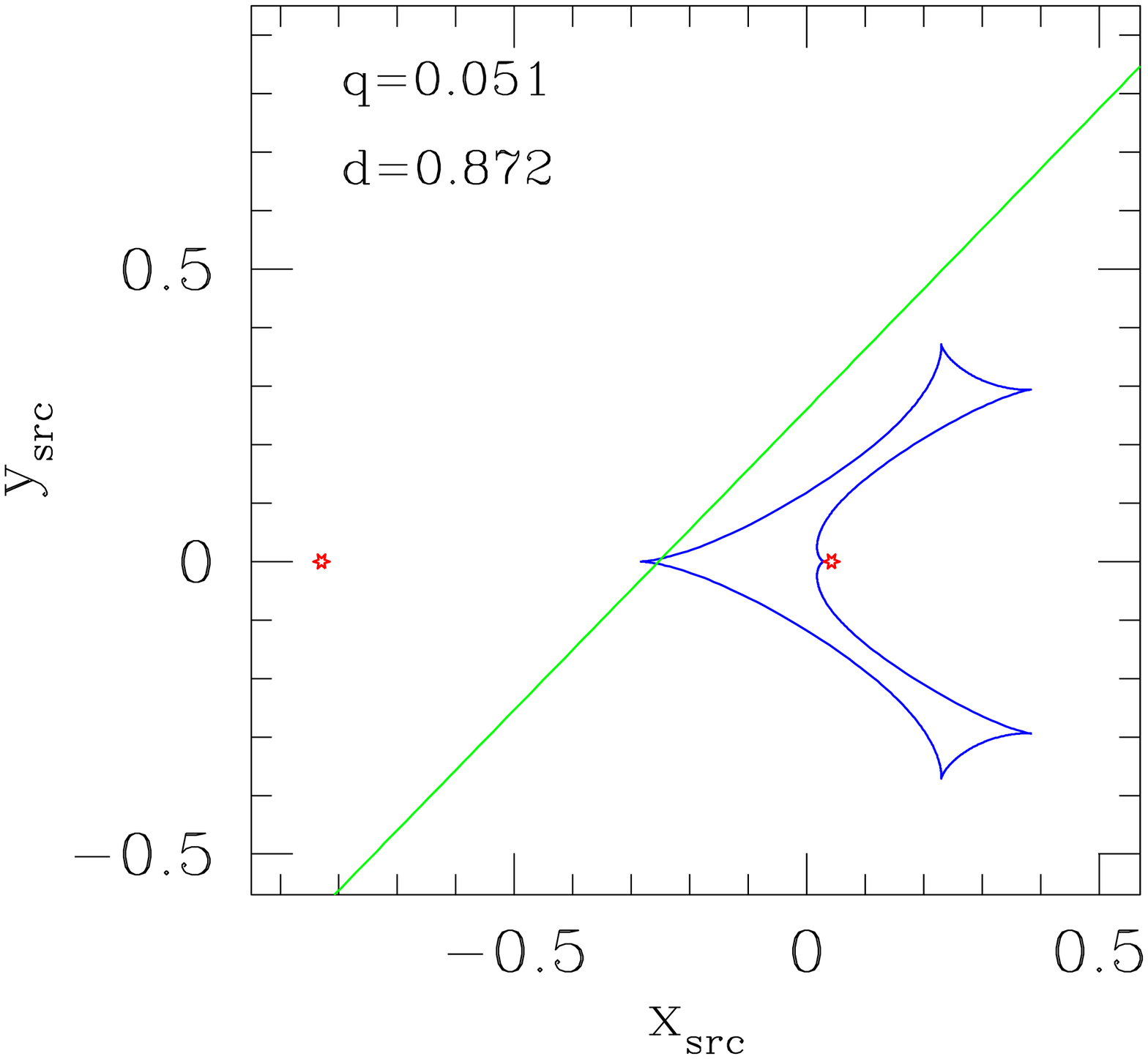}%
 \includegraphics[height=46mm,width=42.5mm]{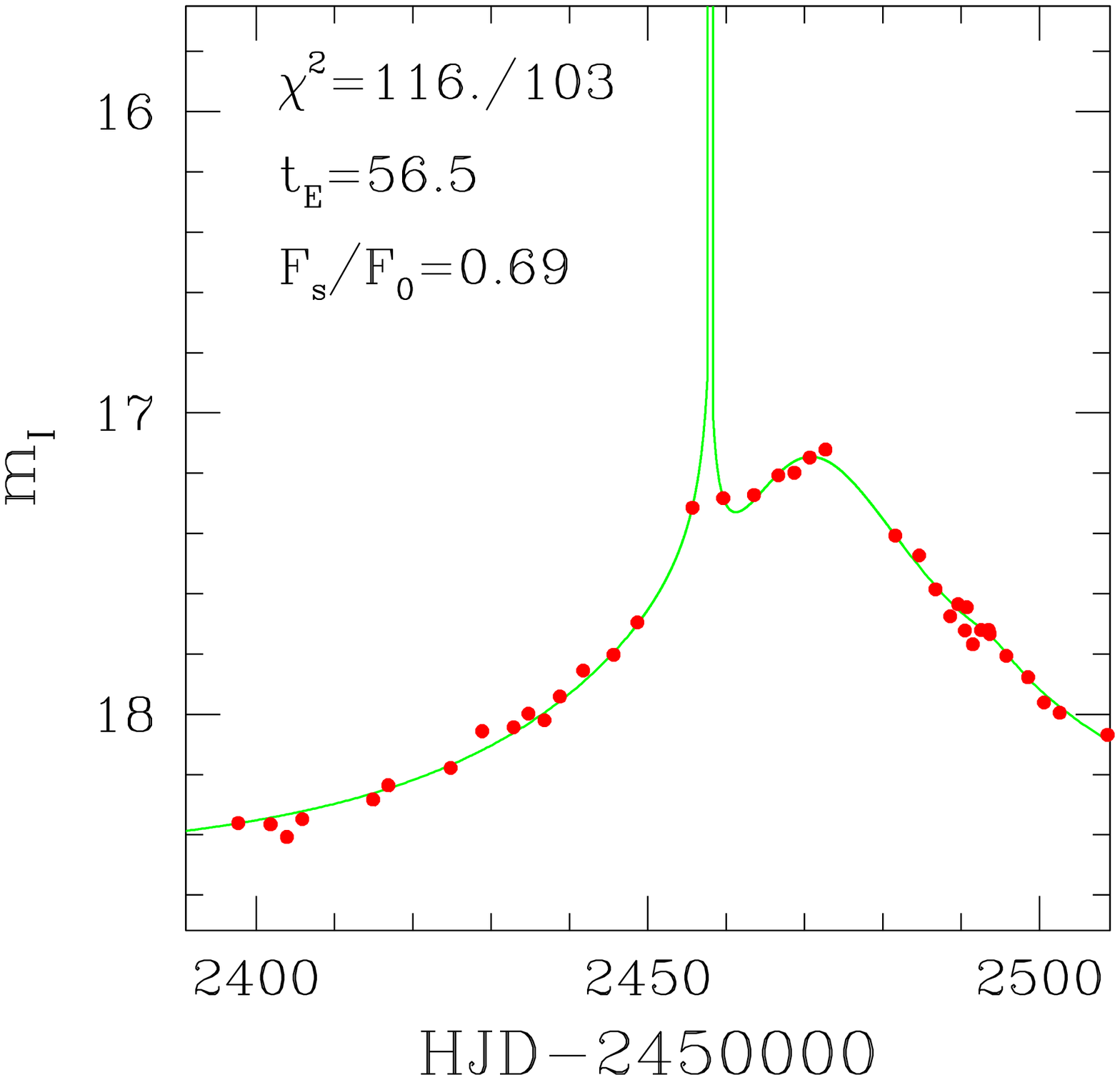}%
 \includegraphics[height=46mm,width=42.5mm]{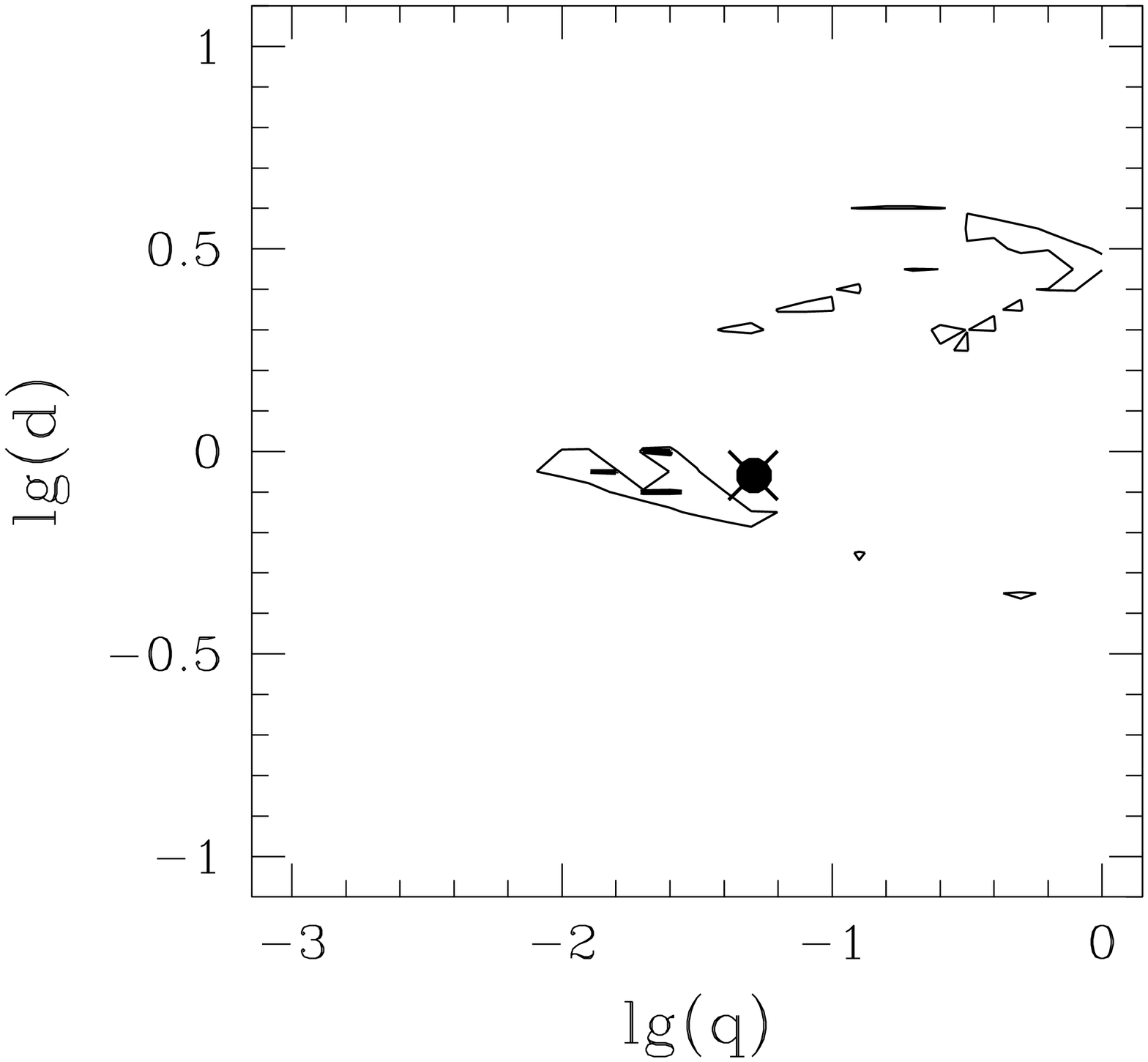}%
}

\noindent\parbox{12.7cm}{
\noindent {\bf OGLE 2002-BLG-256} 

\vspace*{5pt}

 \includegraphics[height=46mm,width=42.5mm]{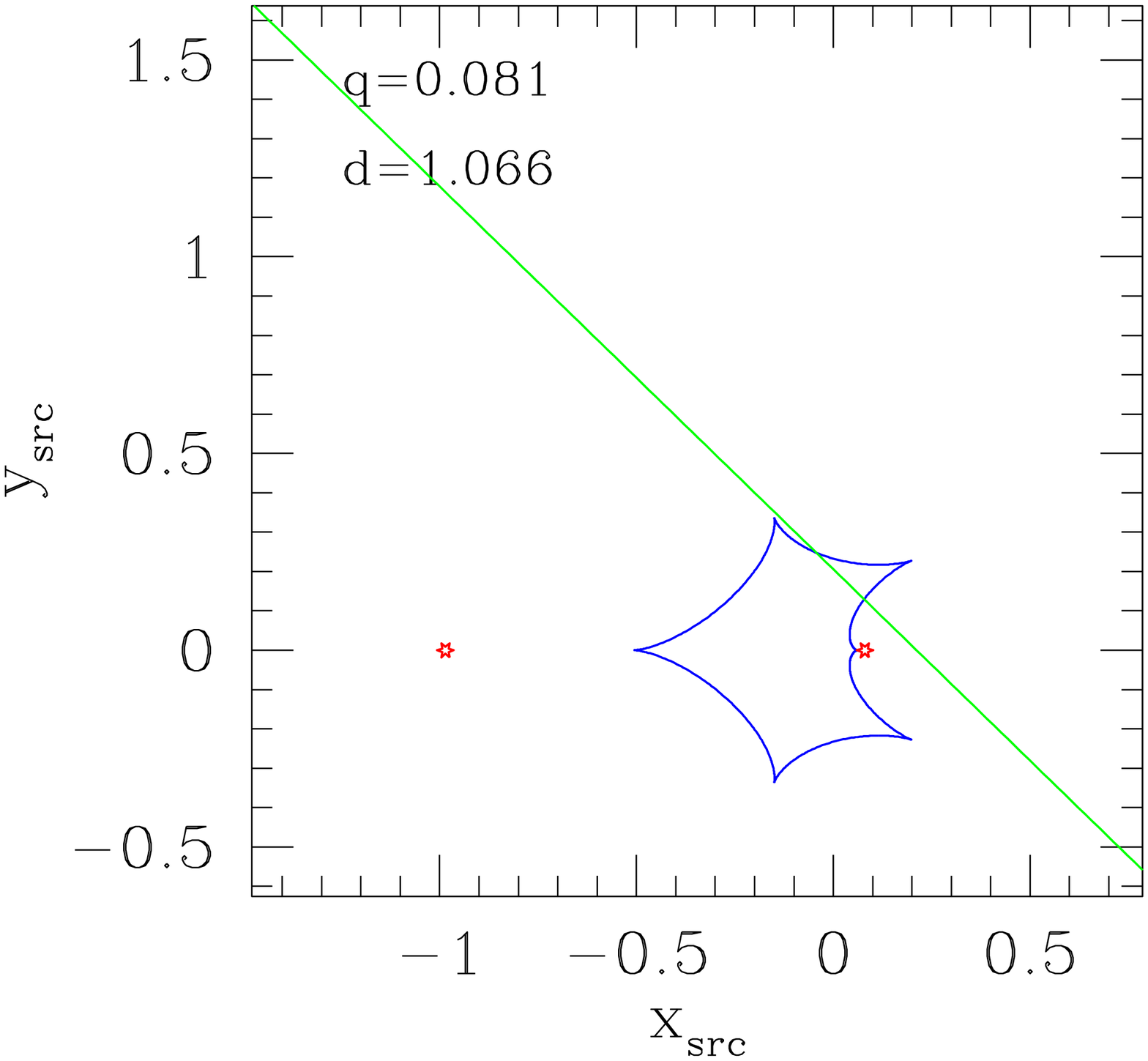}%
 \includegraphics[height=46mm,width=42.5mm]{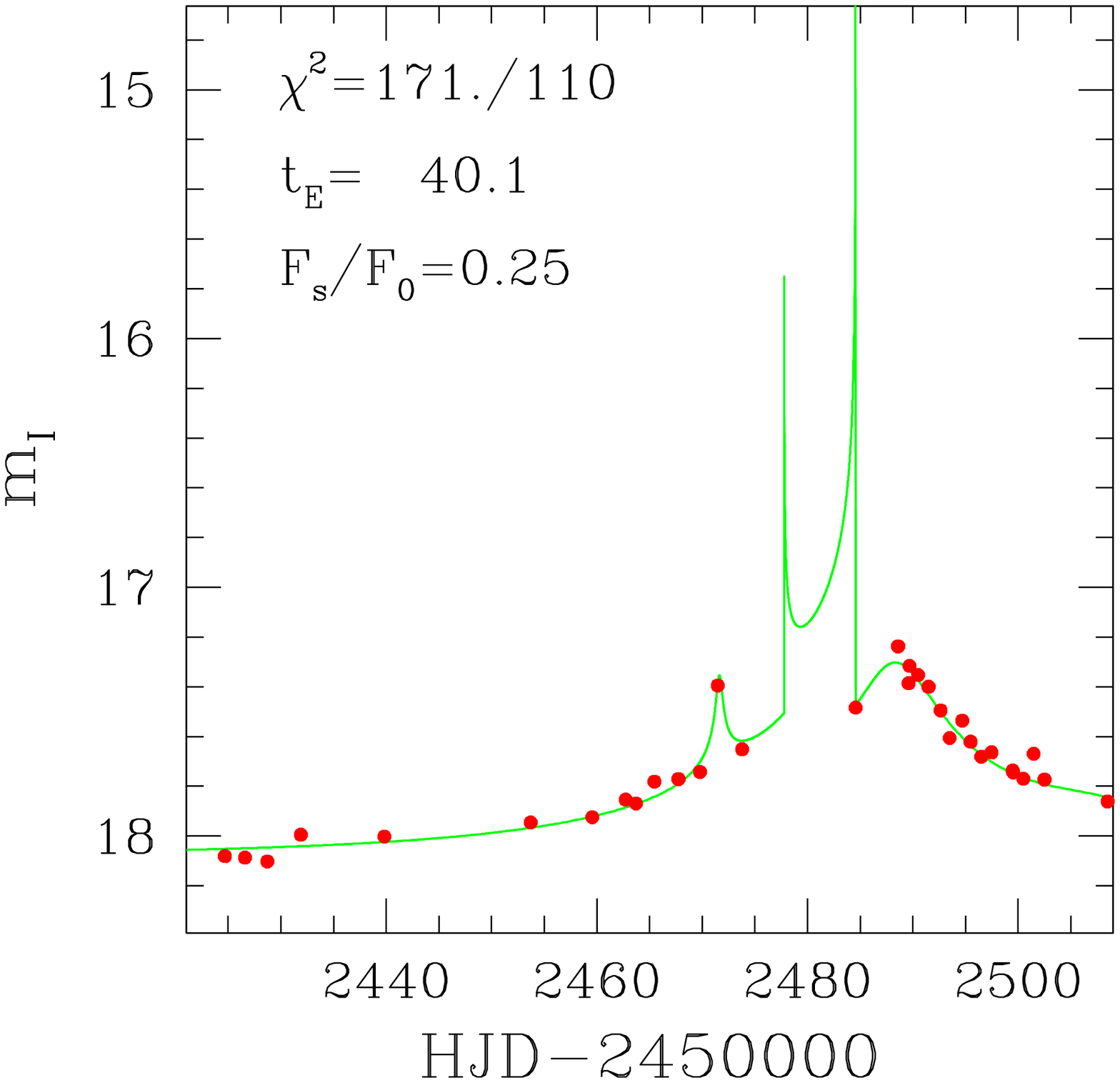}%
 \includegraphics[height=46mm,width=42.5mm]{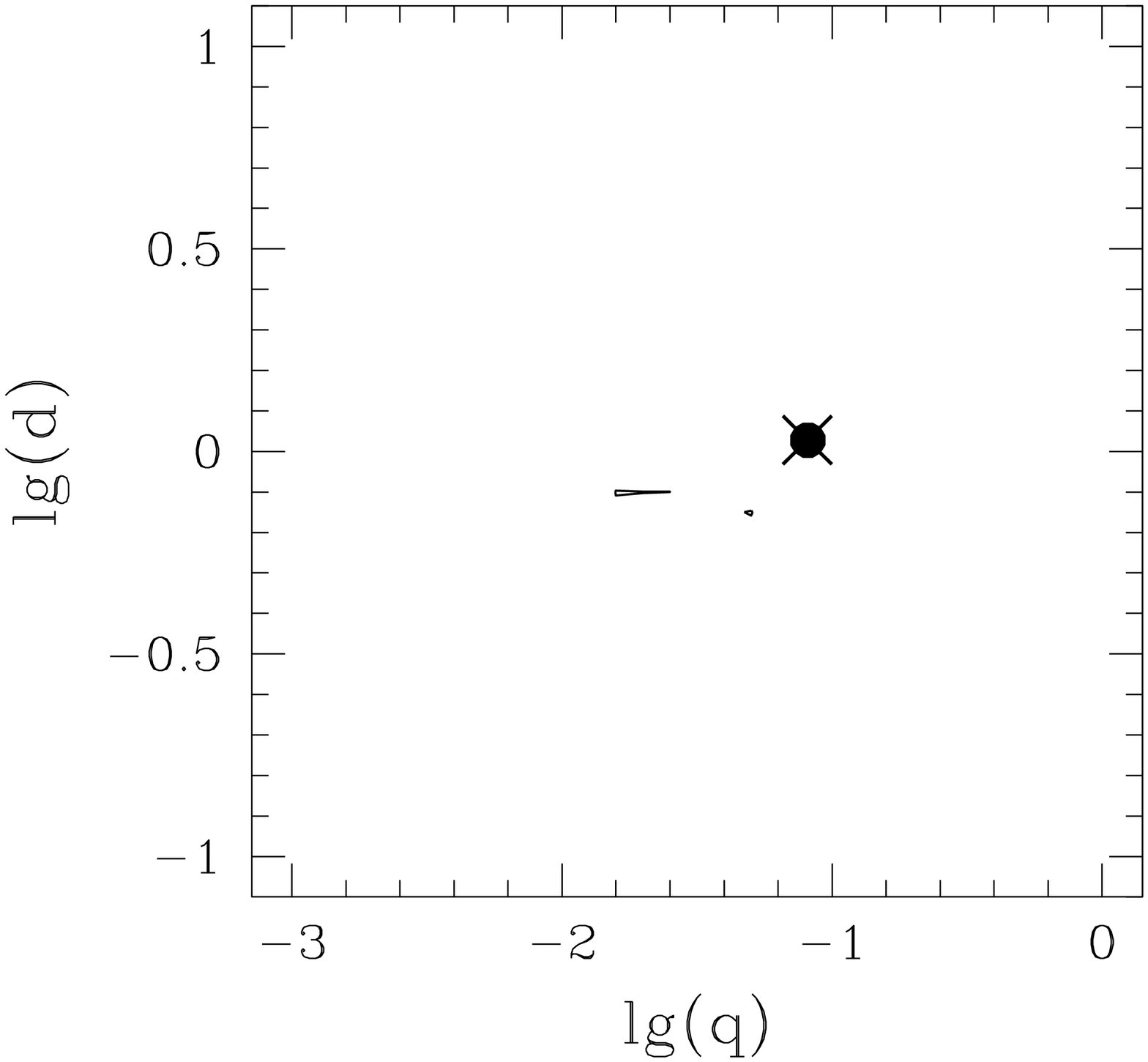}%
}

\noindent\parbox{12.7cm}{
\noindent {\bf OGLE 2002-BLG-321} 

\vspace*{5pt}

 \includegraphics[height=46mm,width=42.5mm]{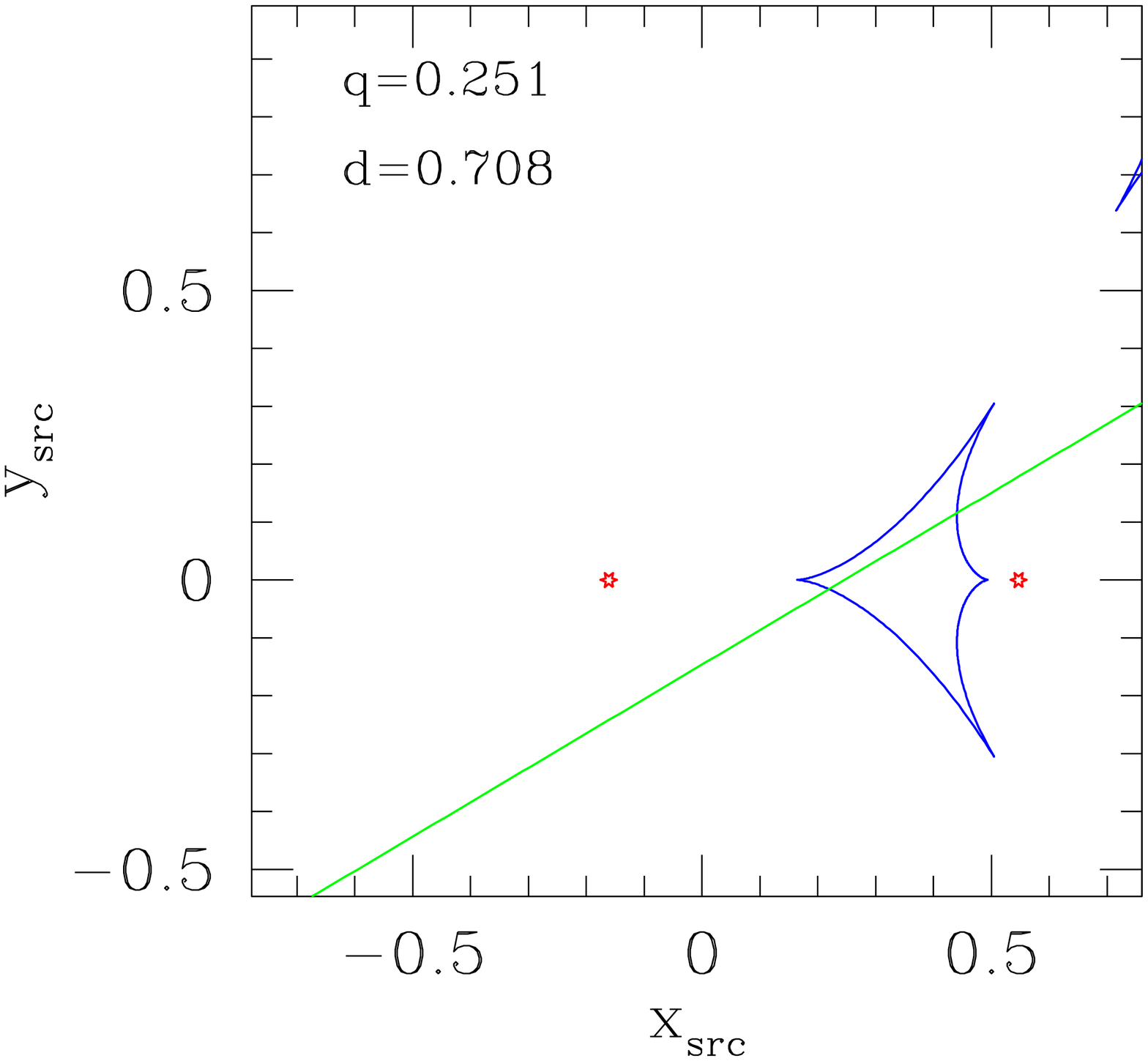}%
 \includegraphics[height=46mm,width=42.5mm]{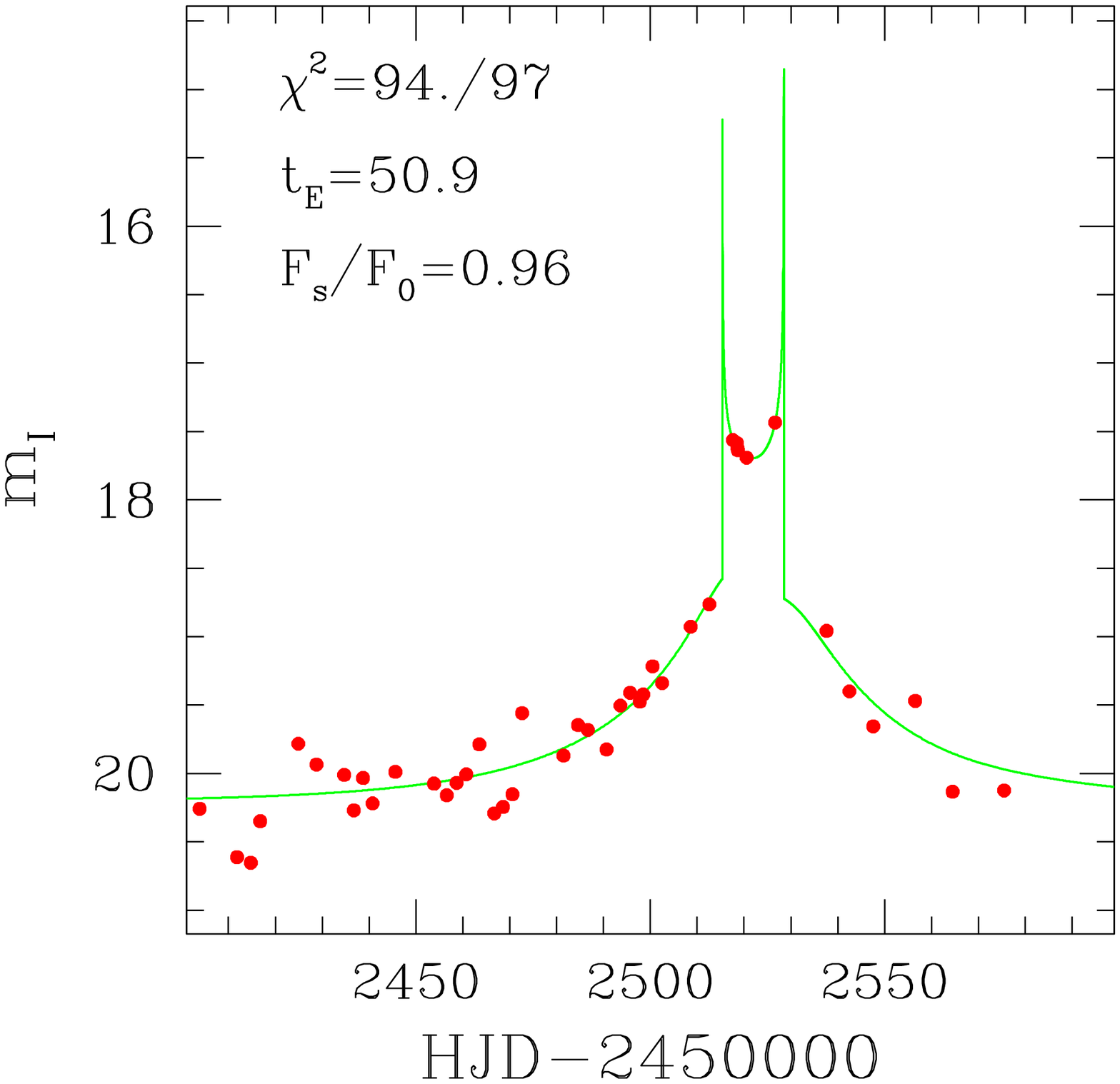}%
 \includegraphics[height=46mm,width=42.5mm]{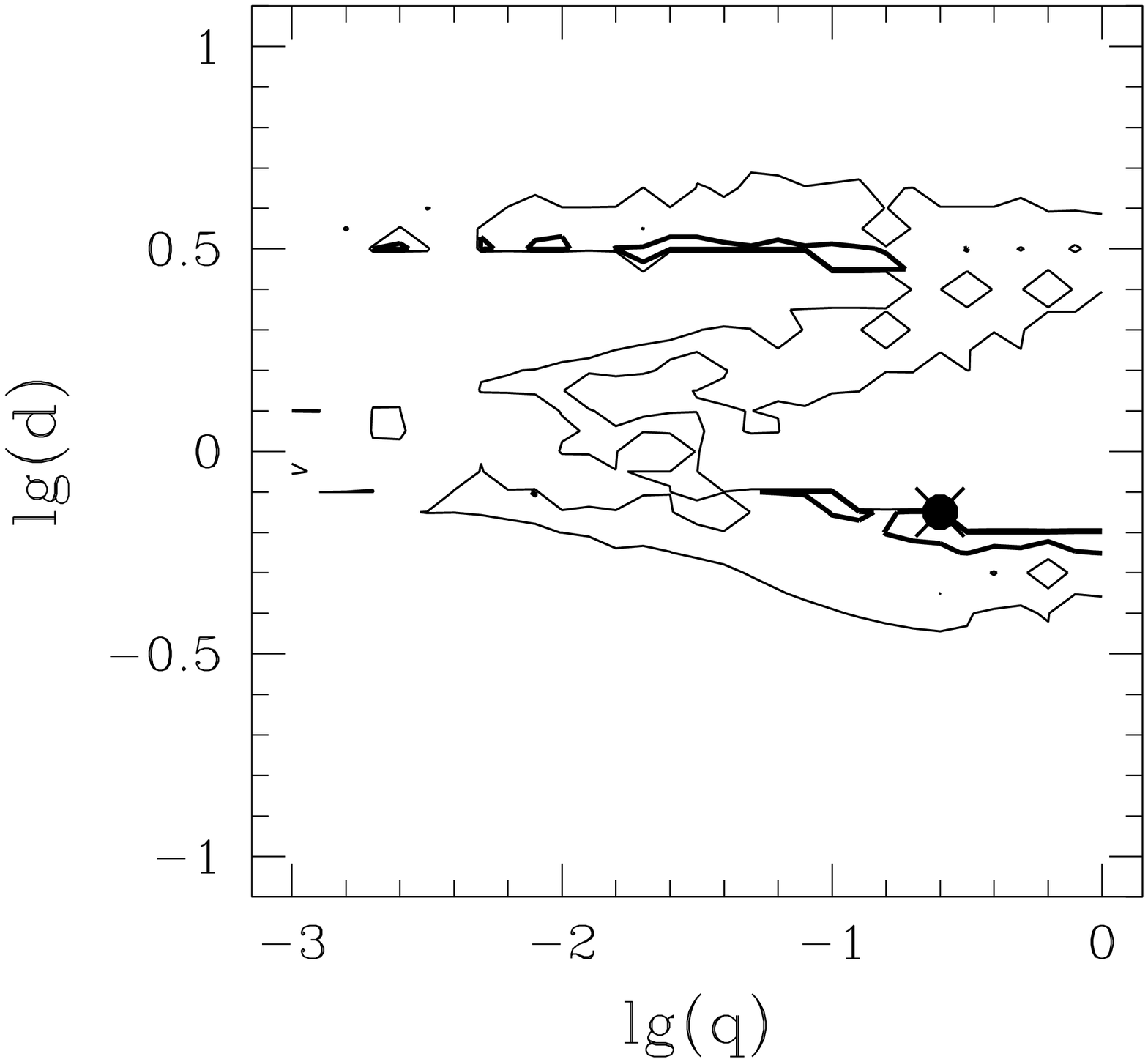}%
}

\noindent\parbox{12.7cm}{
\noindent {\bf OGLE 2003-BLG-021} 

\vspace*{5pt}

 \includegraphics[height=46mm,width=42.5mm]{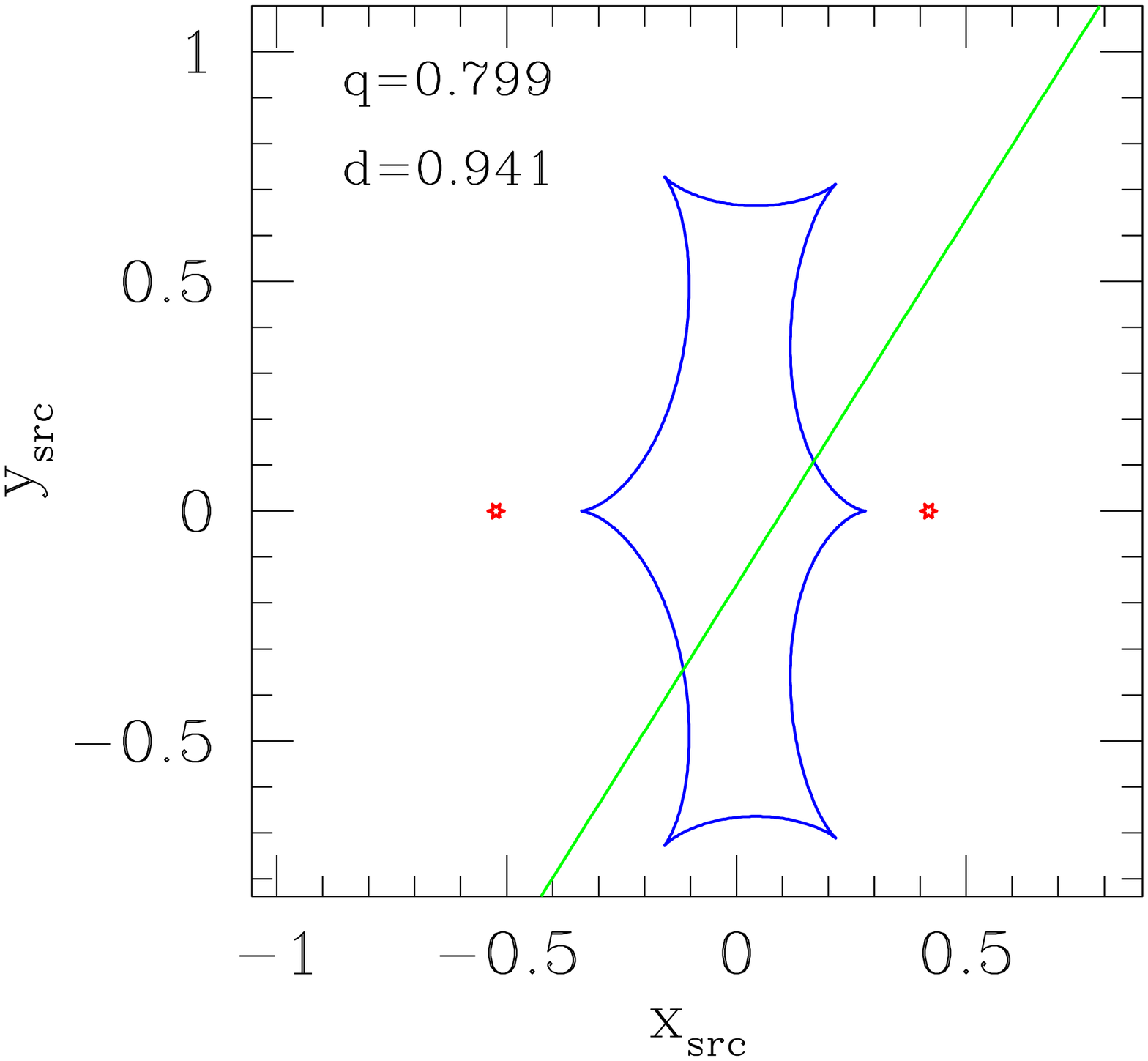}%
 \includegraphics[height=46mm,width=42.5mm]{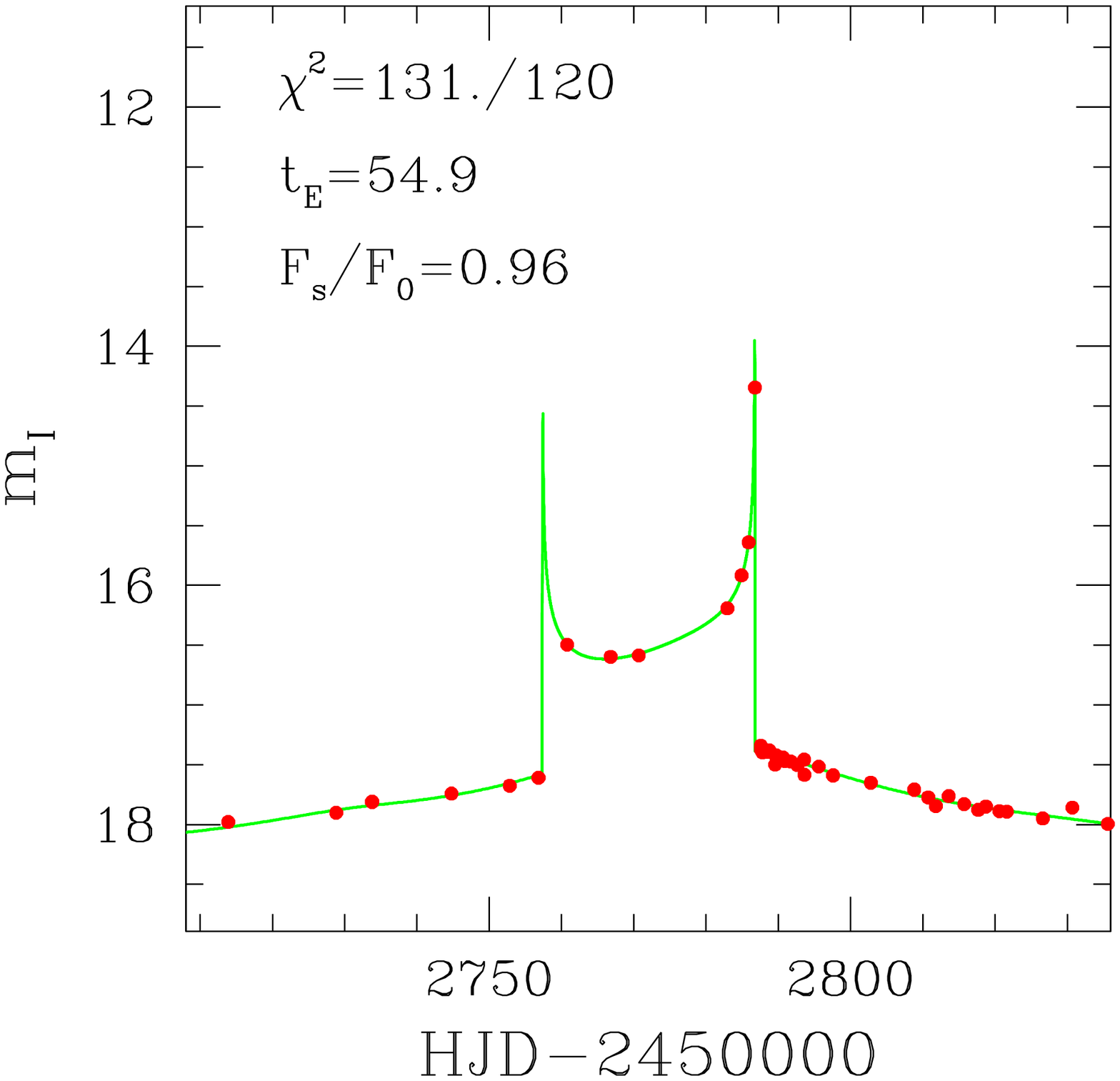}%
 \includegraphics[height=46mm,width=42.5mm]{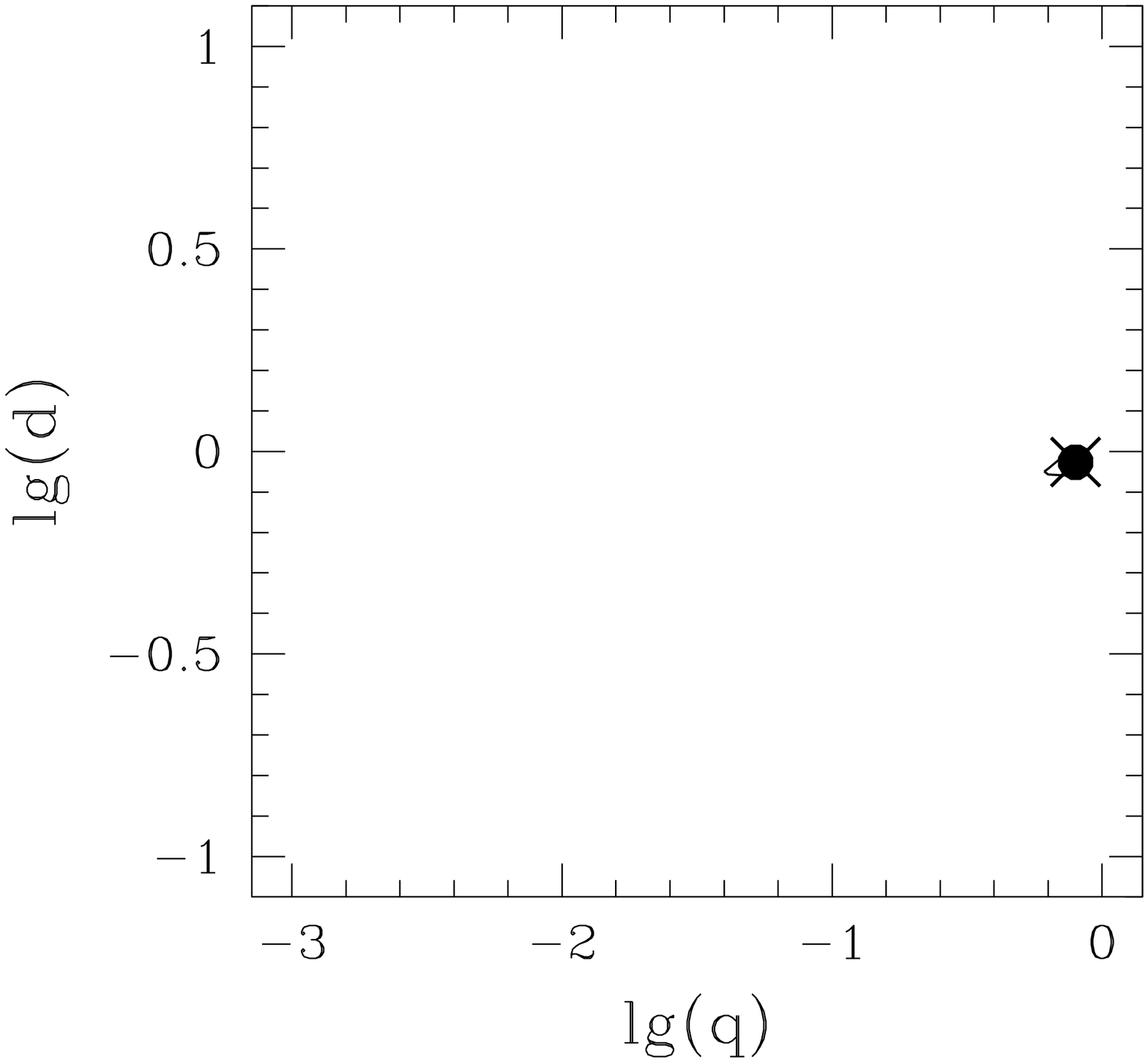}%
}

\noindent\parbox{12.75cm}{
\noindent{\bf OGLE 2003-BLG-056}                 

\vspace*{5pt}

 \includegraphics[height=46mm,width=42.5mm]{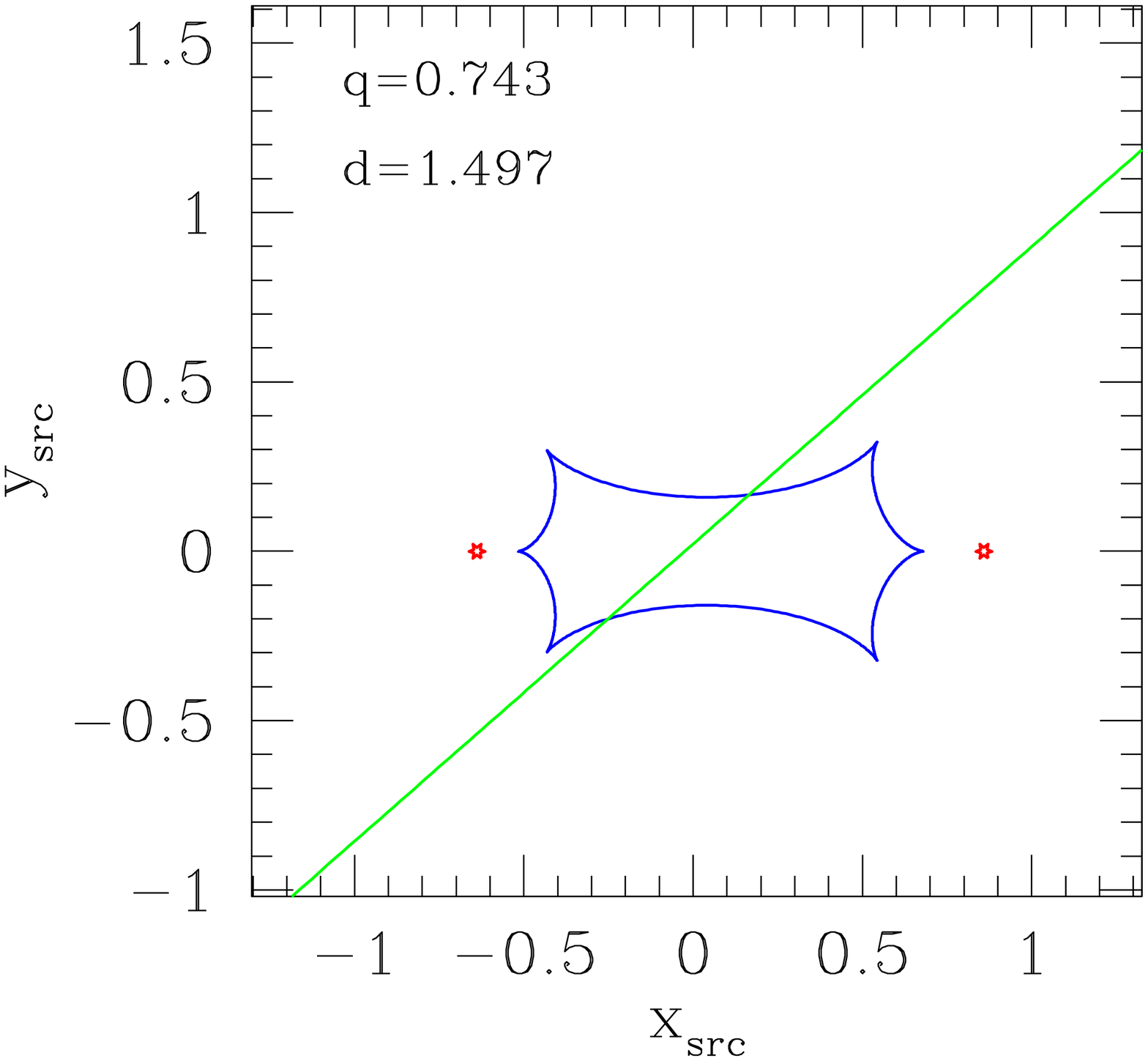}%
 \includegraphics[height=46mm,width=42.5mm]{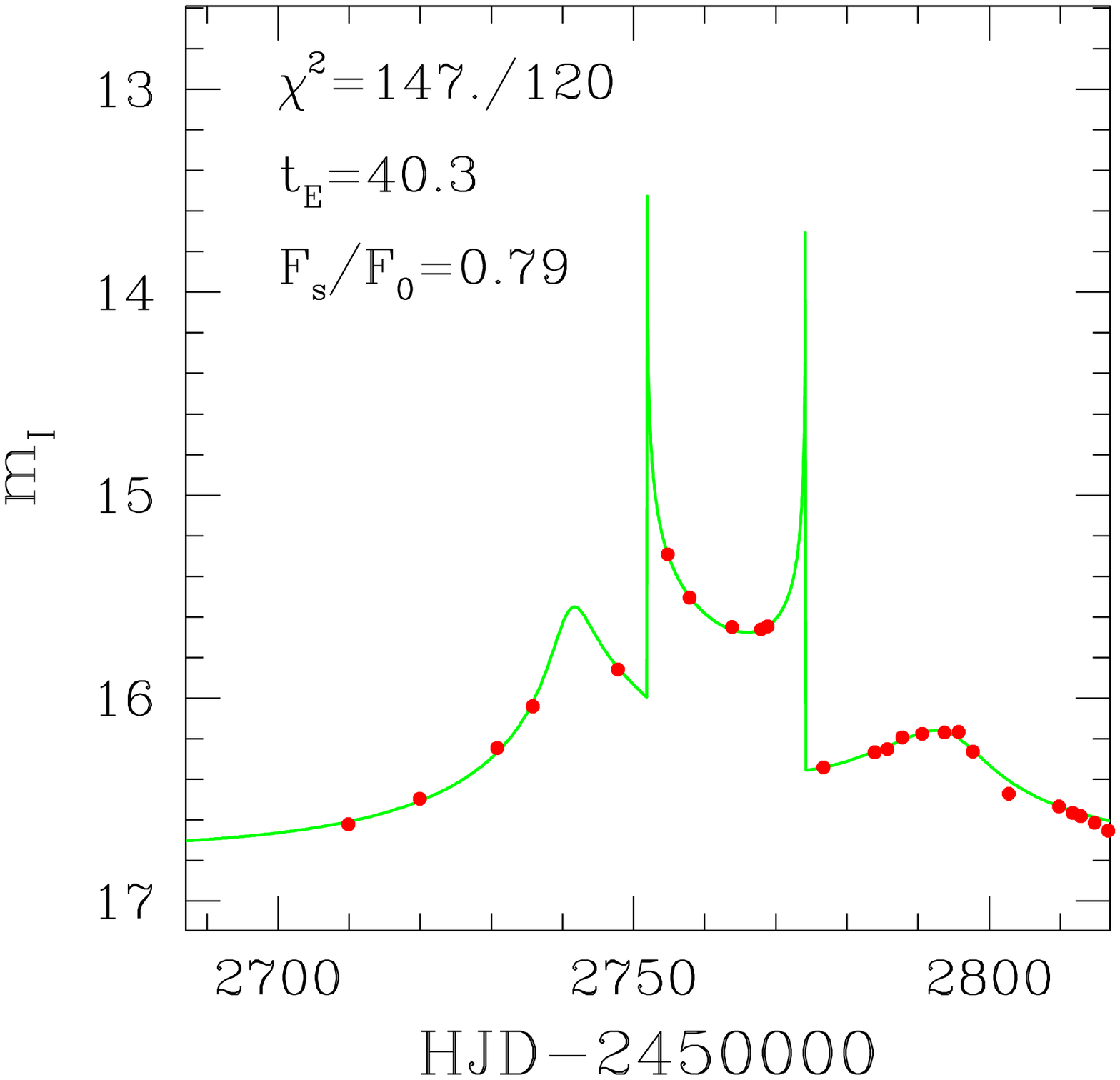}%
 \includegraphics[height=46mm,width=42.5mm]{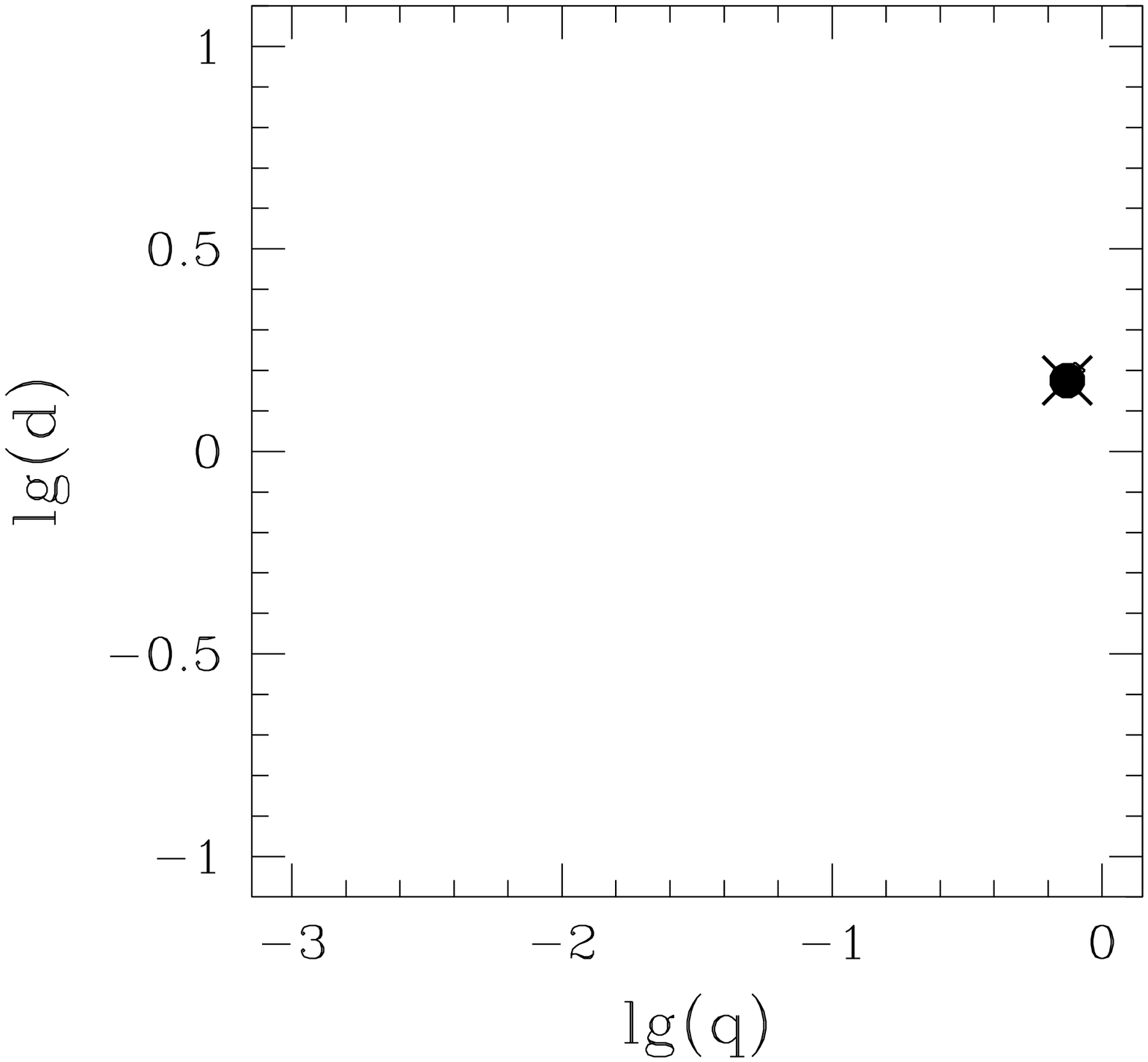}
}

\noindent\parbox{12.75cm}{
\noindent{\bf OGLE 2003-BLG-084}                 

\vspace*{5pt}

 \includegraphics[height=46mm,width=42.5mm]{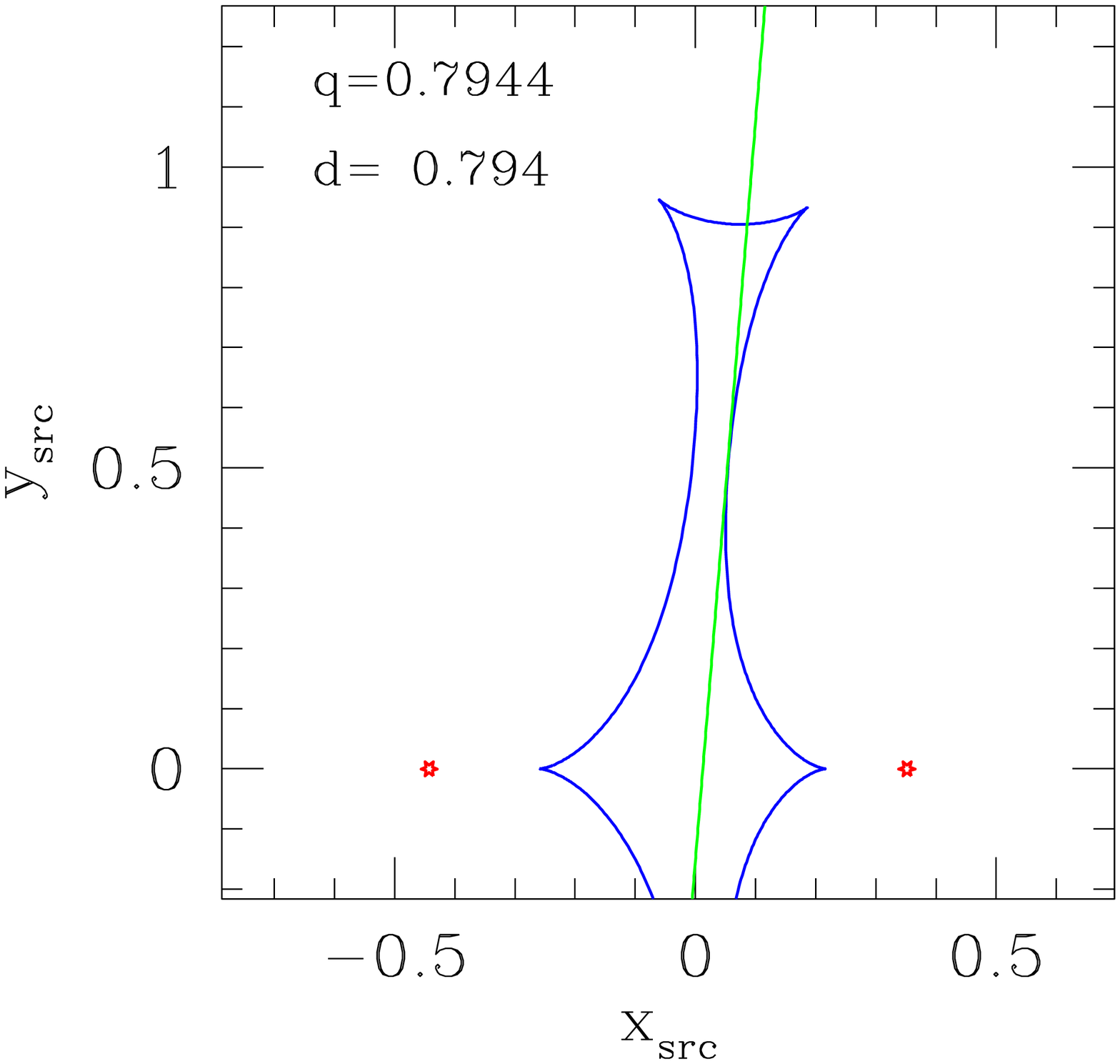}%
 \includegraphics[height=46mm,width=42.5mm]{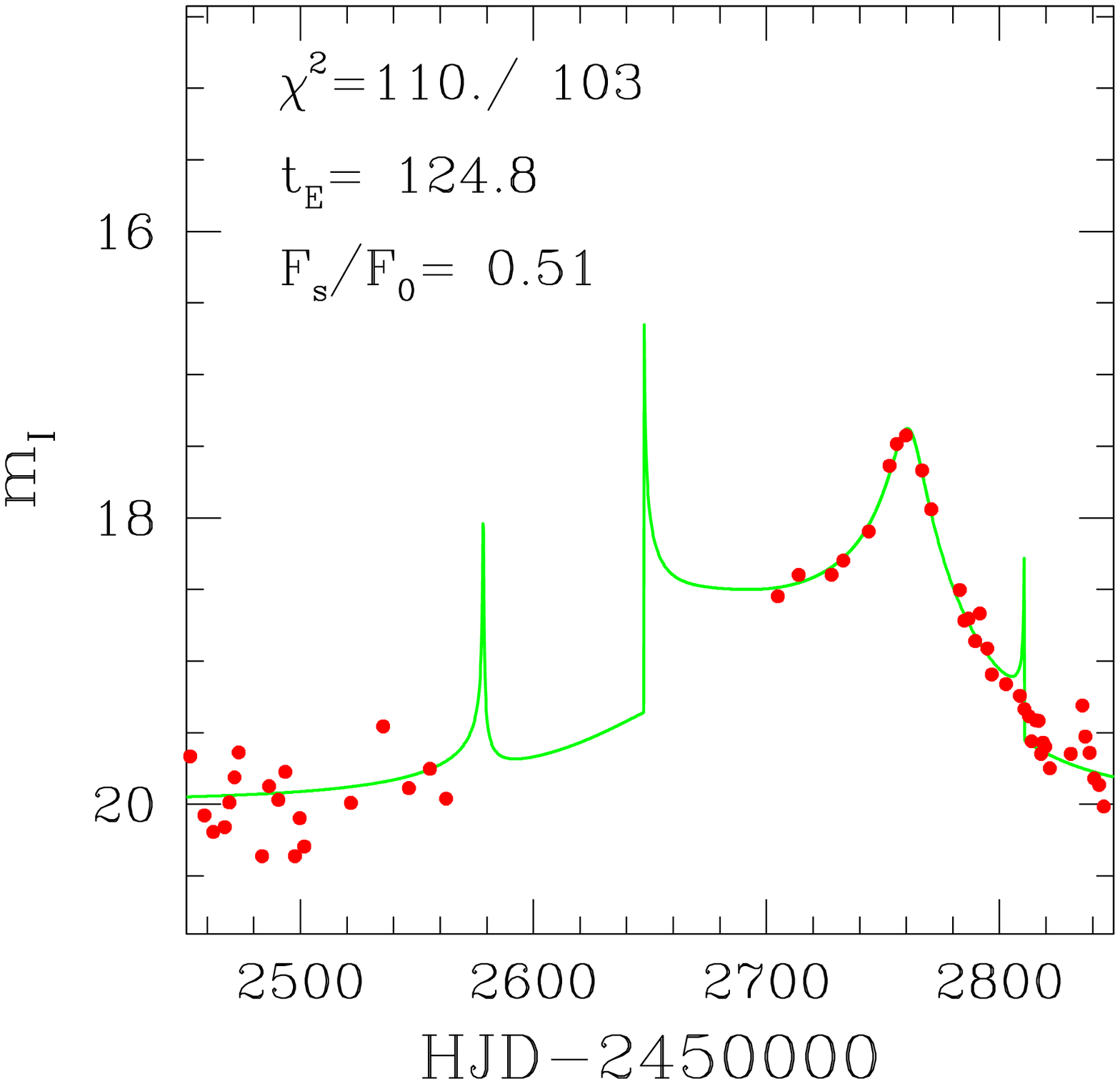}%
 \includegraphics[height=46mm,width=42.5mm]{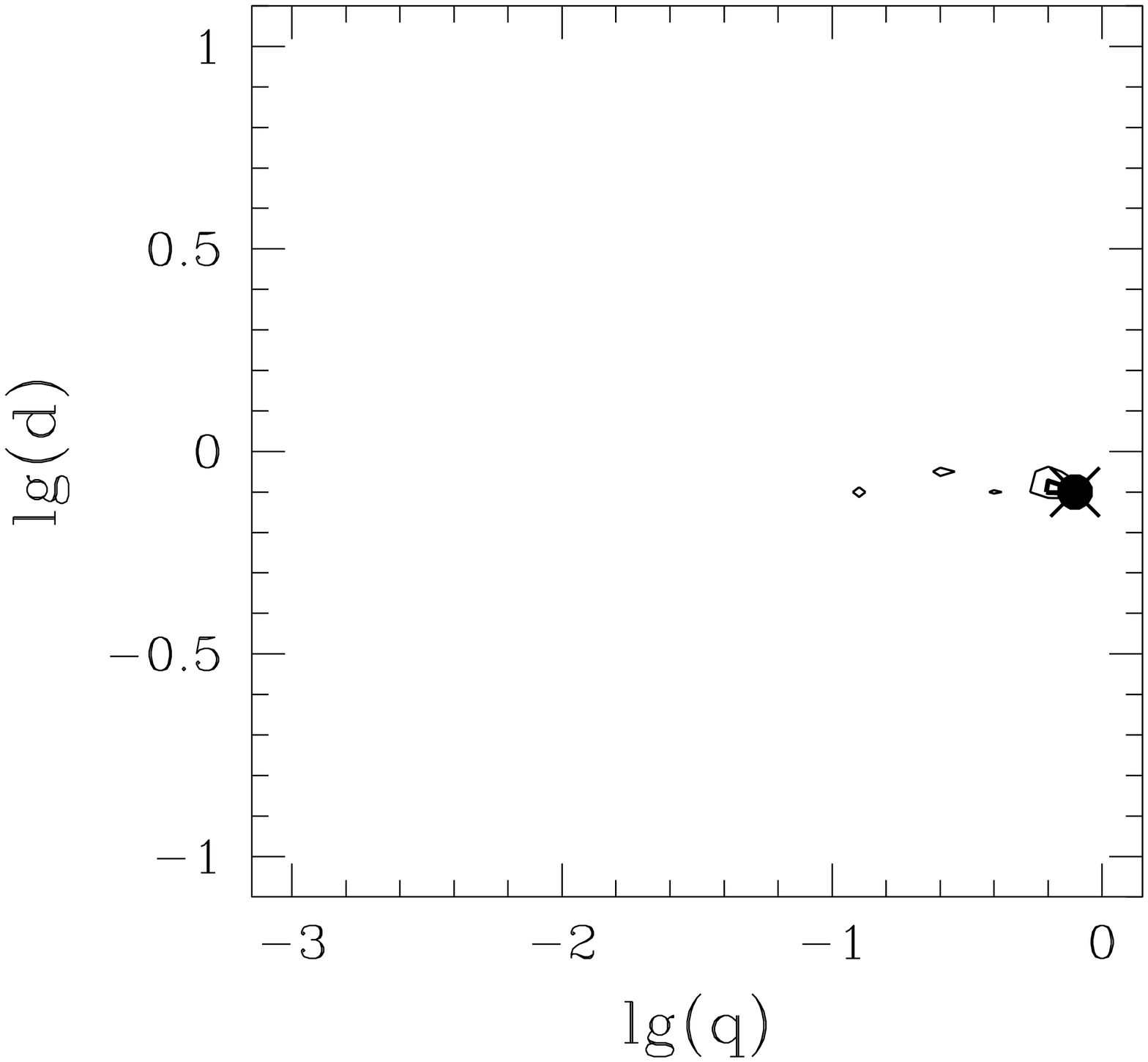}
}

\noindent\parbox{12.75cm}{
\noindent{\bf OGLE 2003-BLG-124}                                   

\vspace*{5pt}

 \includegraphics[height=46mm,width=42.5mm]{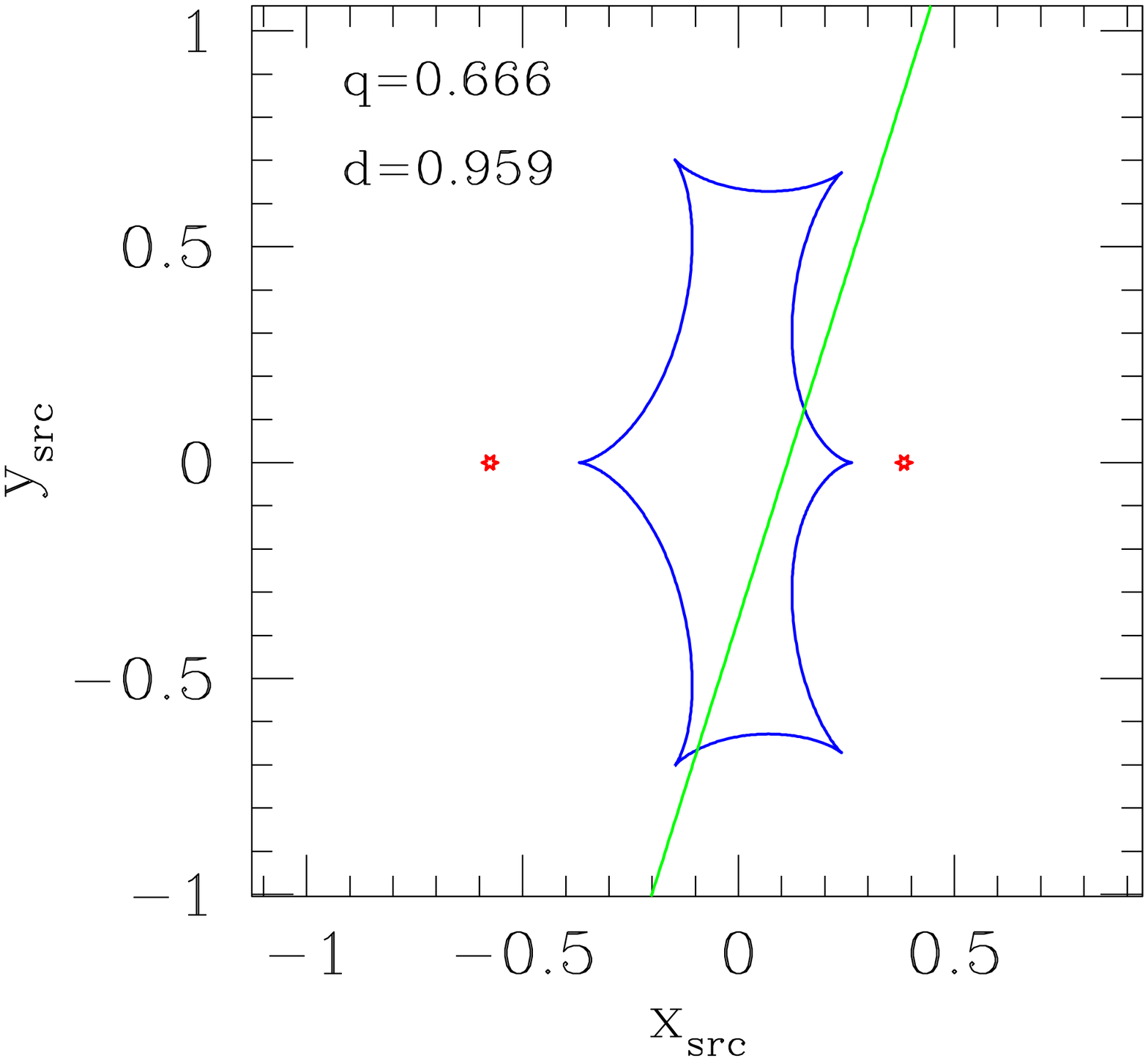}%
 \includegraphics[height=46mm,width=42.5mm]{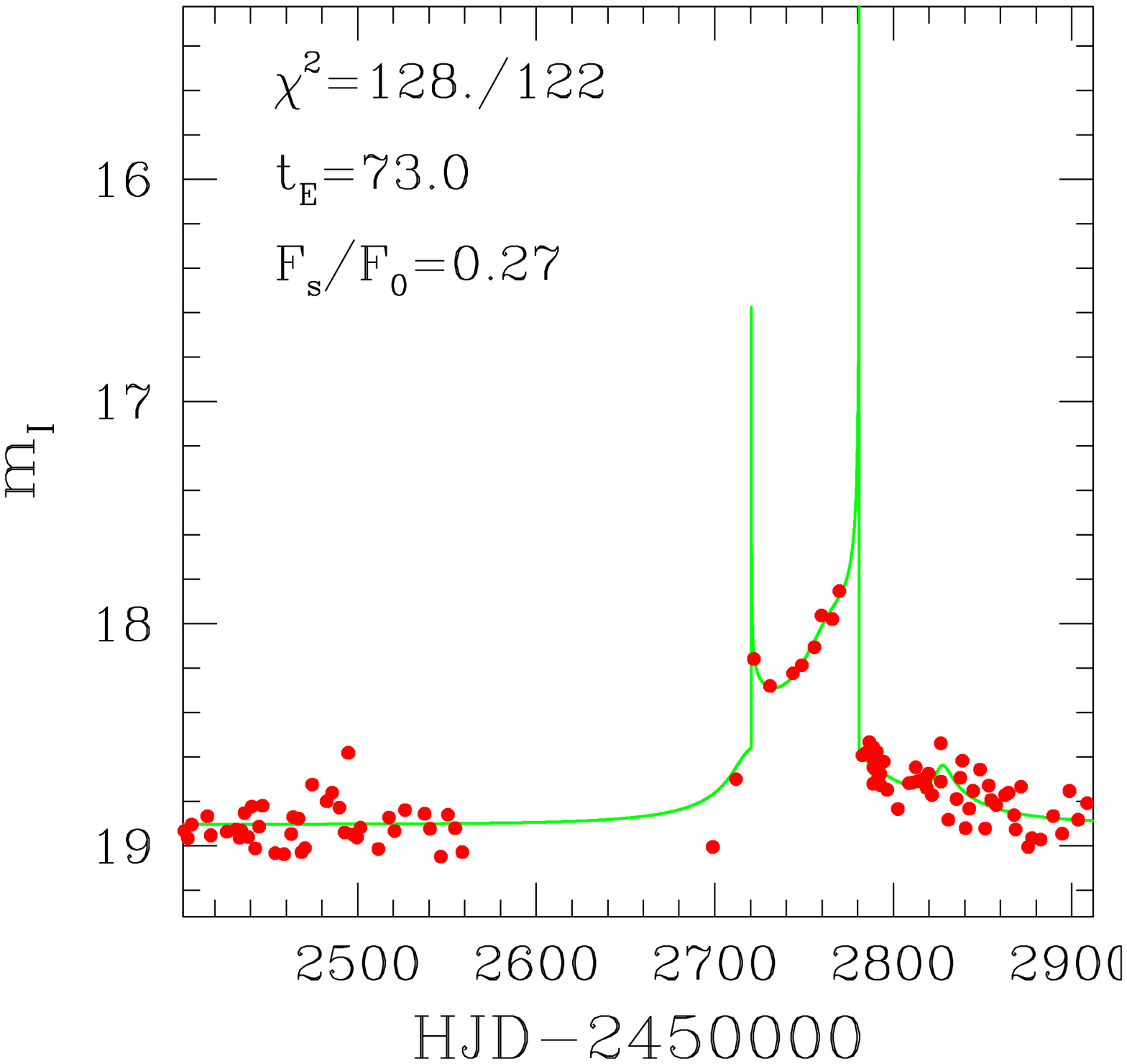}%
 \includegraphics[height=46mm,width=42.5mm]{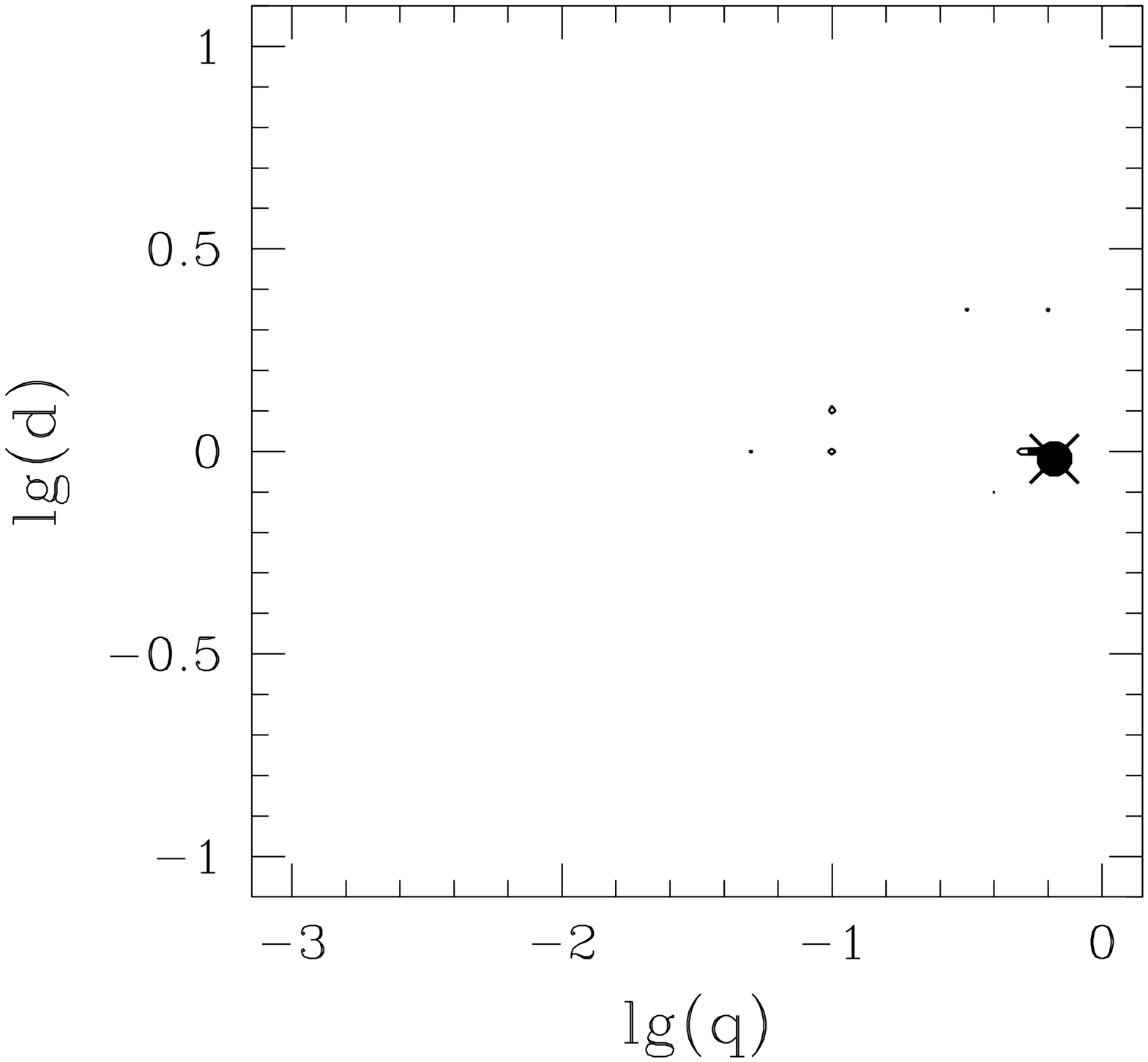}%
}

\noindent\parbox{12.75cm}{
\noindent{\bf OGLE 2003-BLG-135 / MOA 2003-BLG-21}

\vspace*{5pt}

 \includegraphics[height=46mm,width=42.5mm]{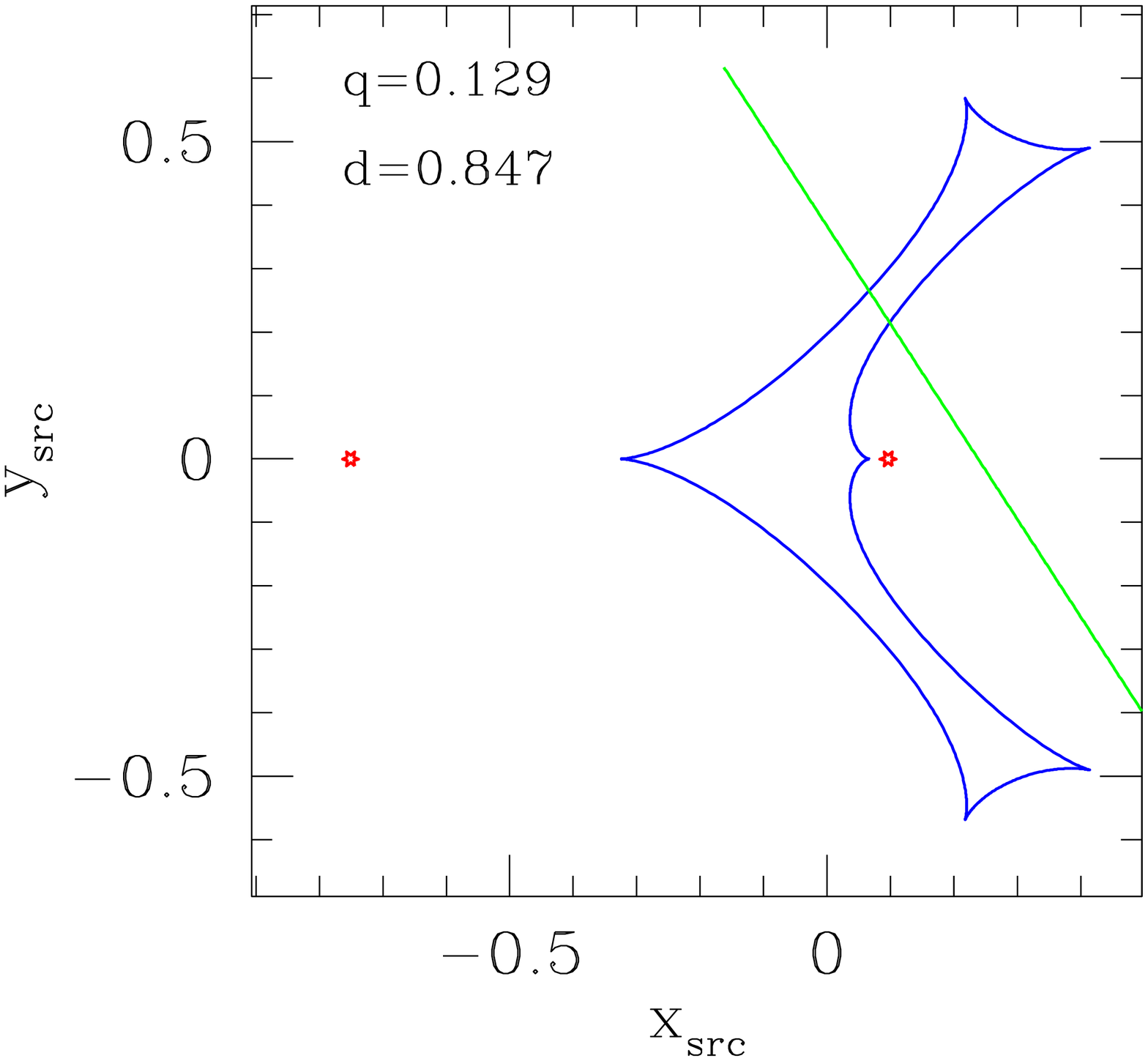}%
 \includegraphics[height=46mm,width=42.5mm]{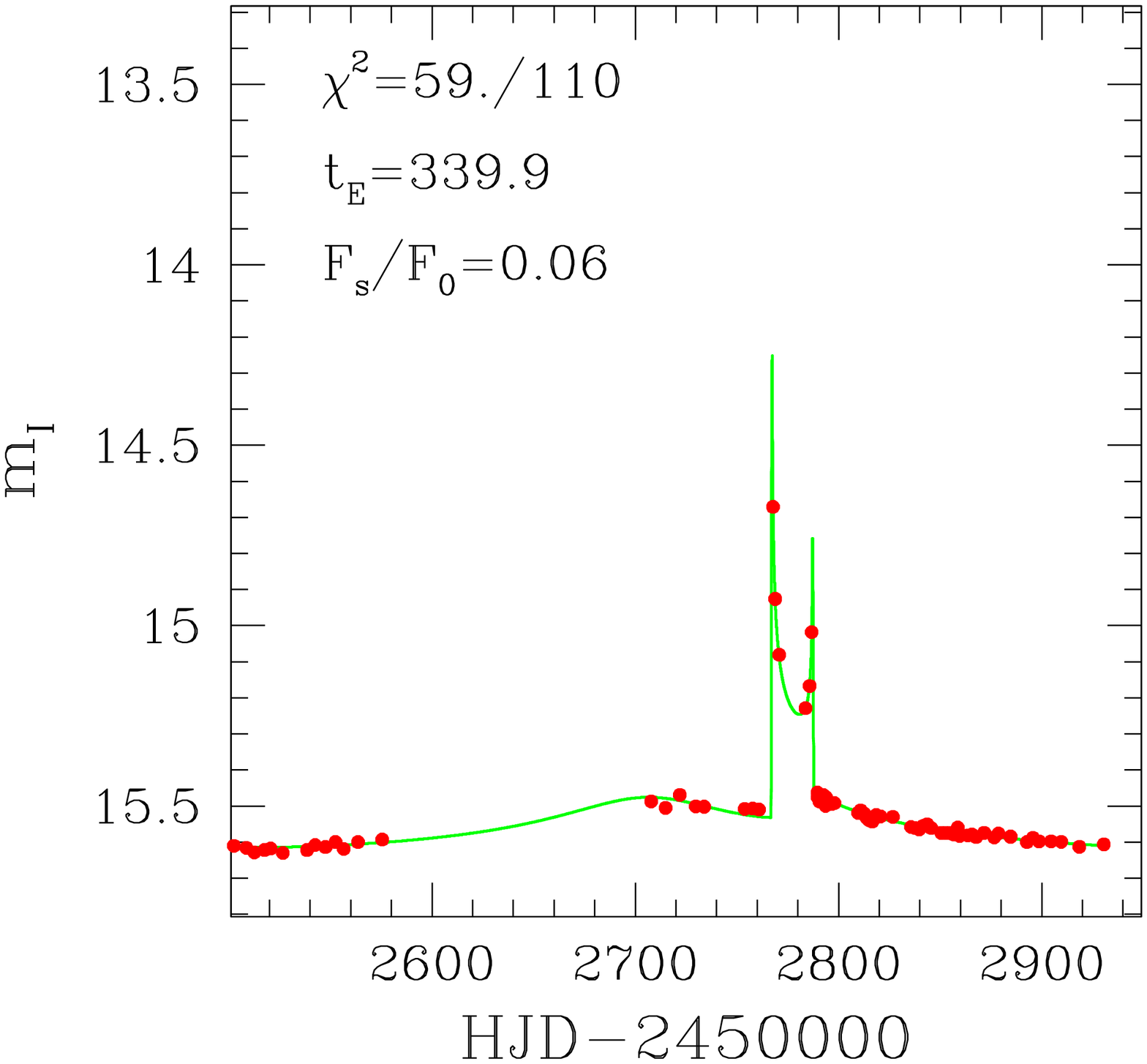}%
 \includegraphics[height=46mm,width=42.5mm]{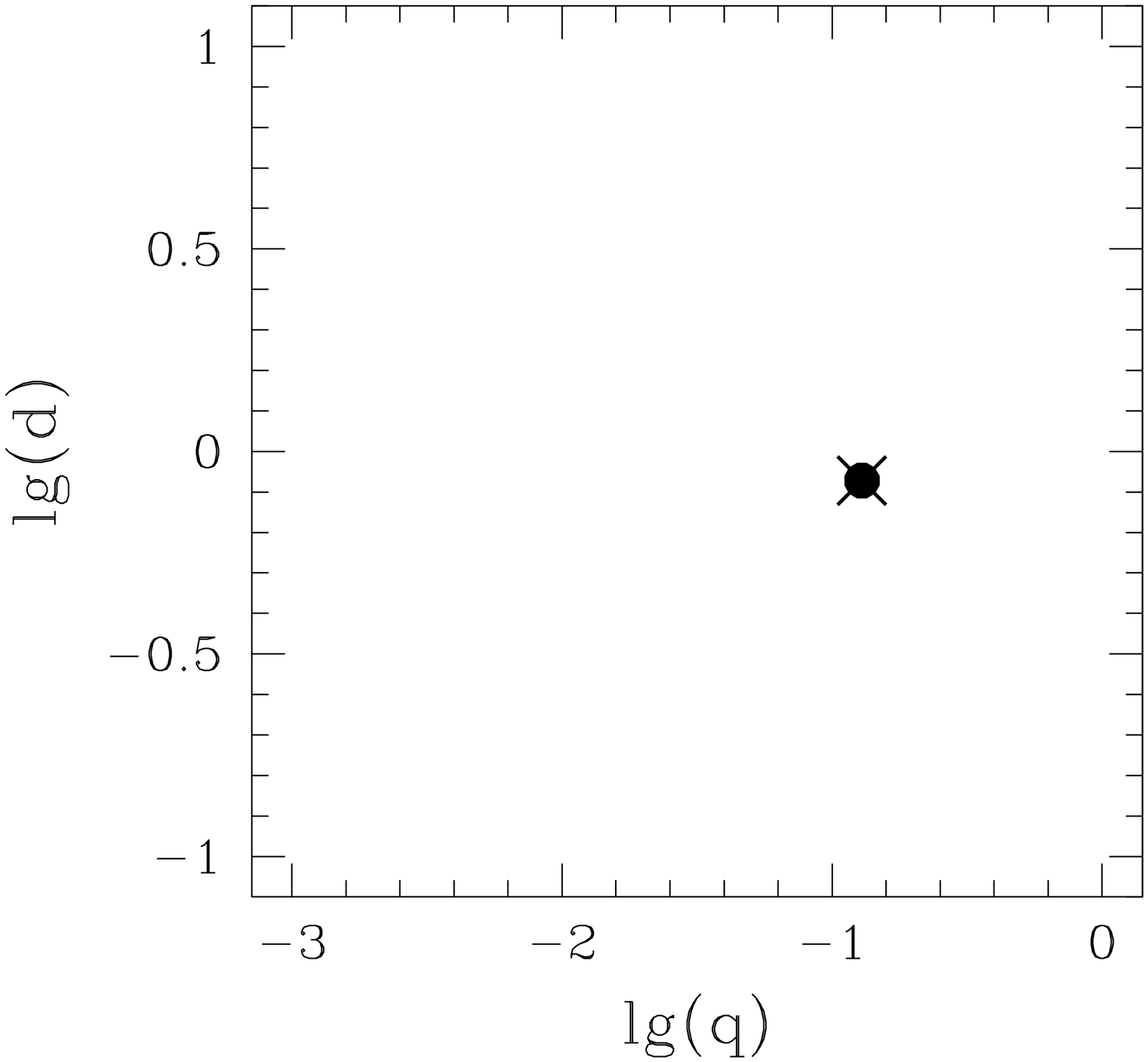}%
}

\noindent\parbox{12.75cm}{
\noindent{\bf OGLE 2003-BLG-170}               

\vspace*{5pt}

 \includegraphics[height=46mm,width=42.5mm]{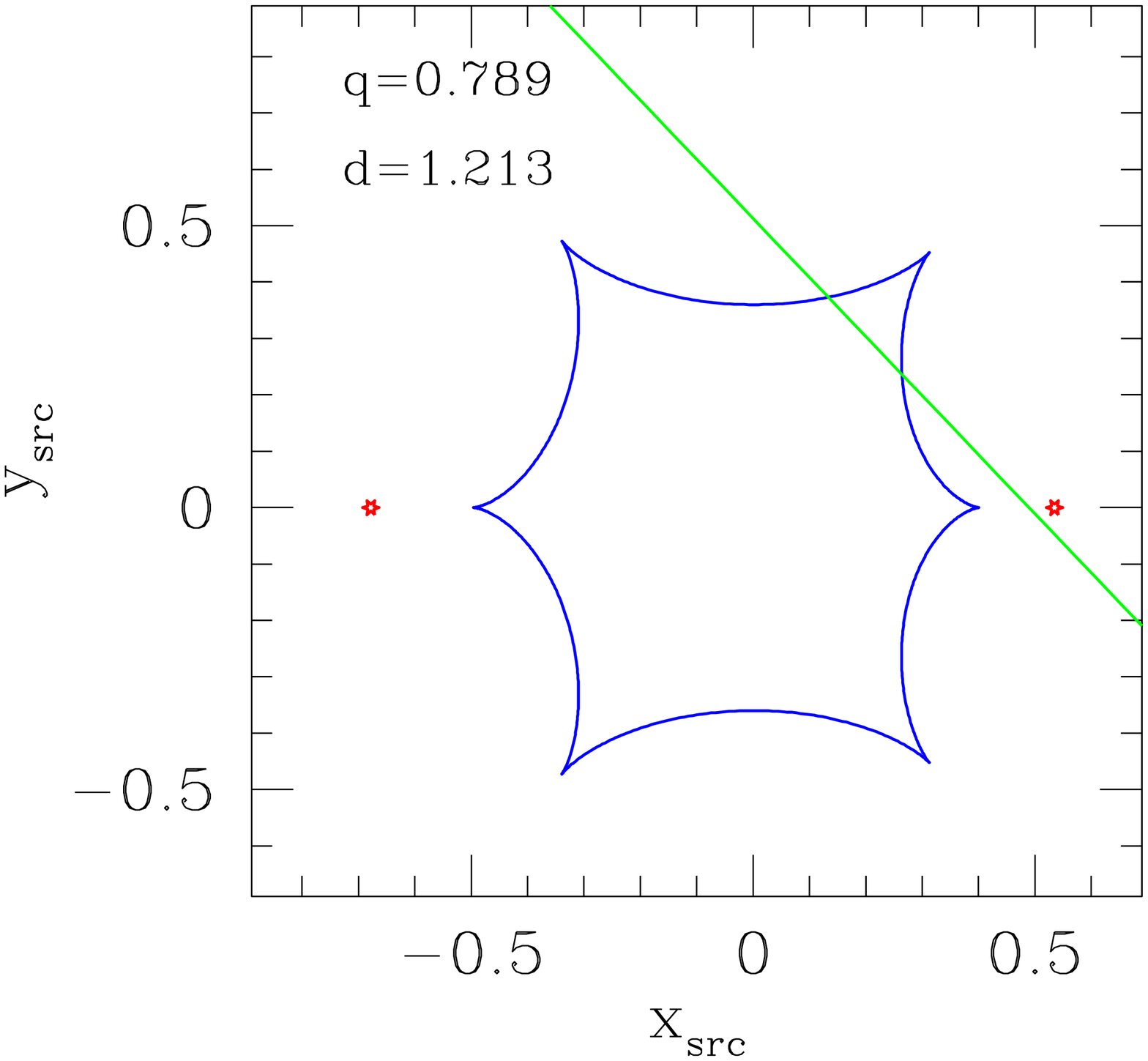}%
 \includegraphics[height=46mm,width=42.5mm]{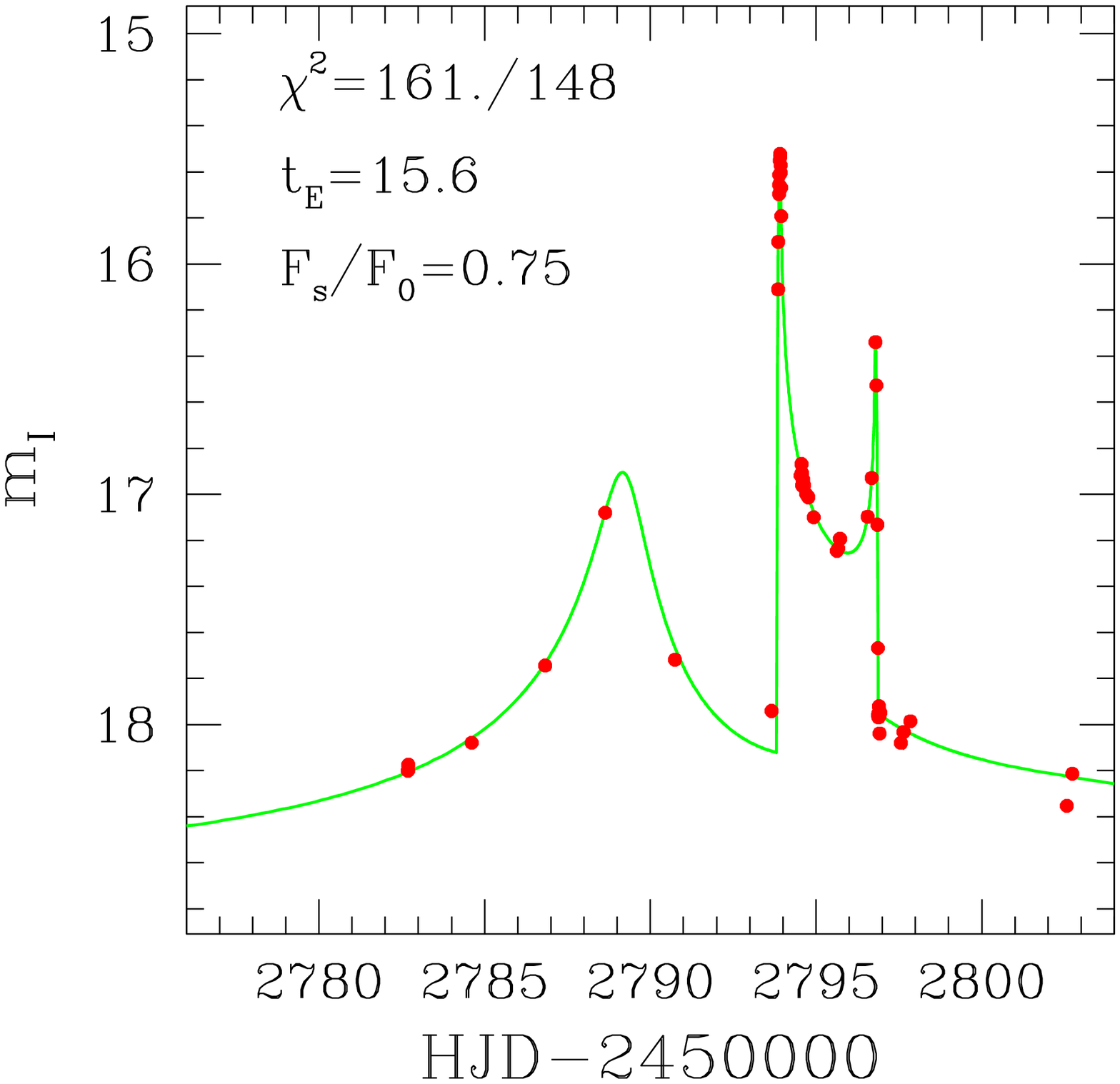}%
 \includegraphics[height=46mm,width=42.5mm]{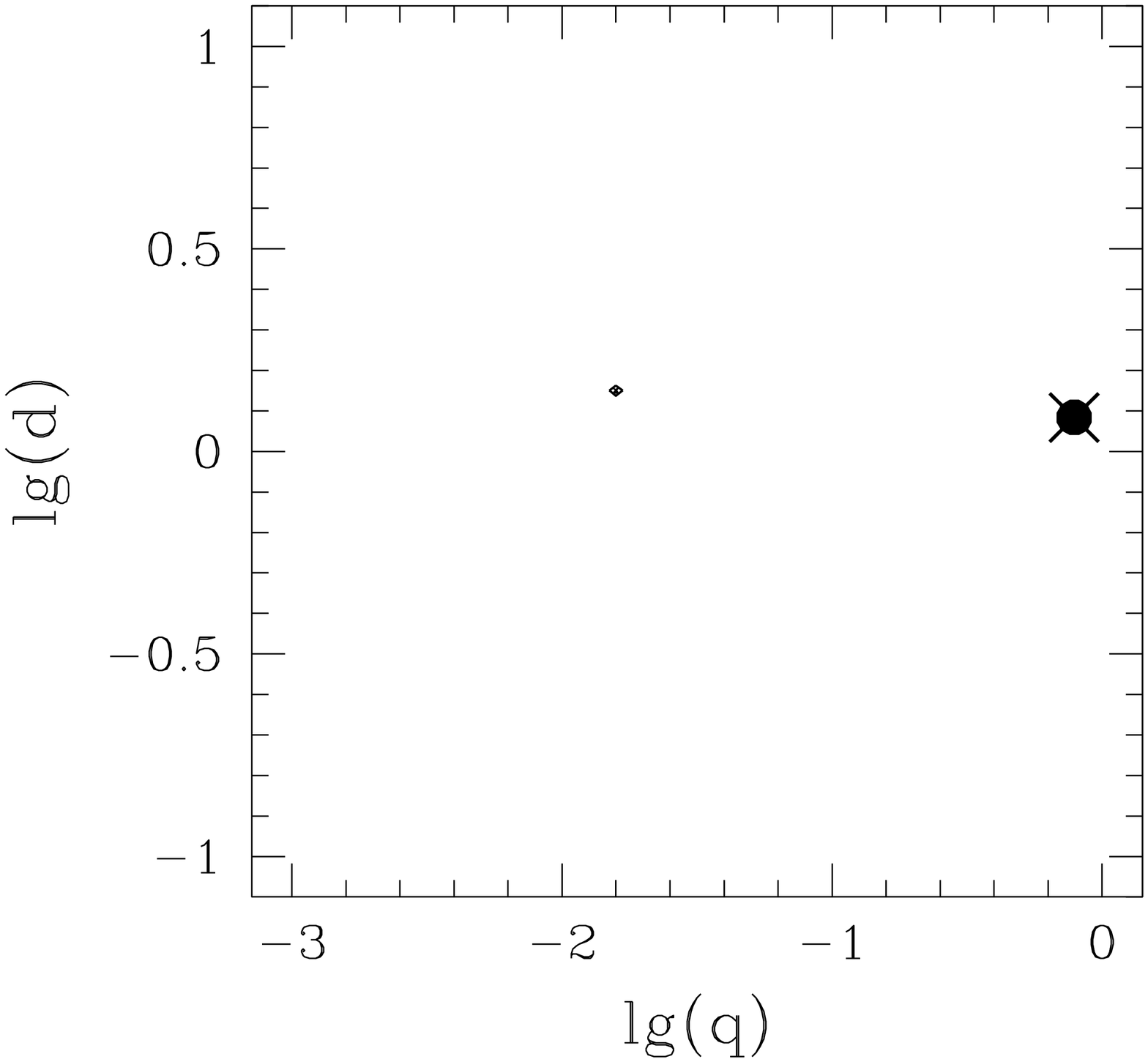}
}

\noindent\parbox{12.75cm}{
\noindent{\bf OGLE 2003-BLG-194 (1st model)} 

\vspace*{5pt}

 \includegraphics[height=46mm,width=42.5mm]{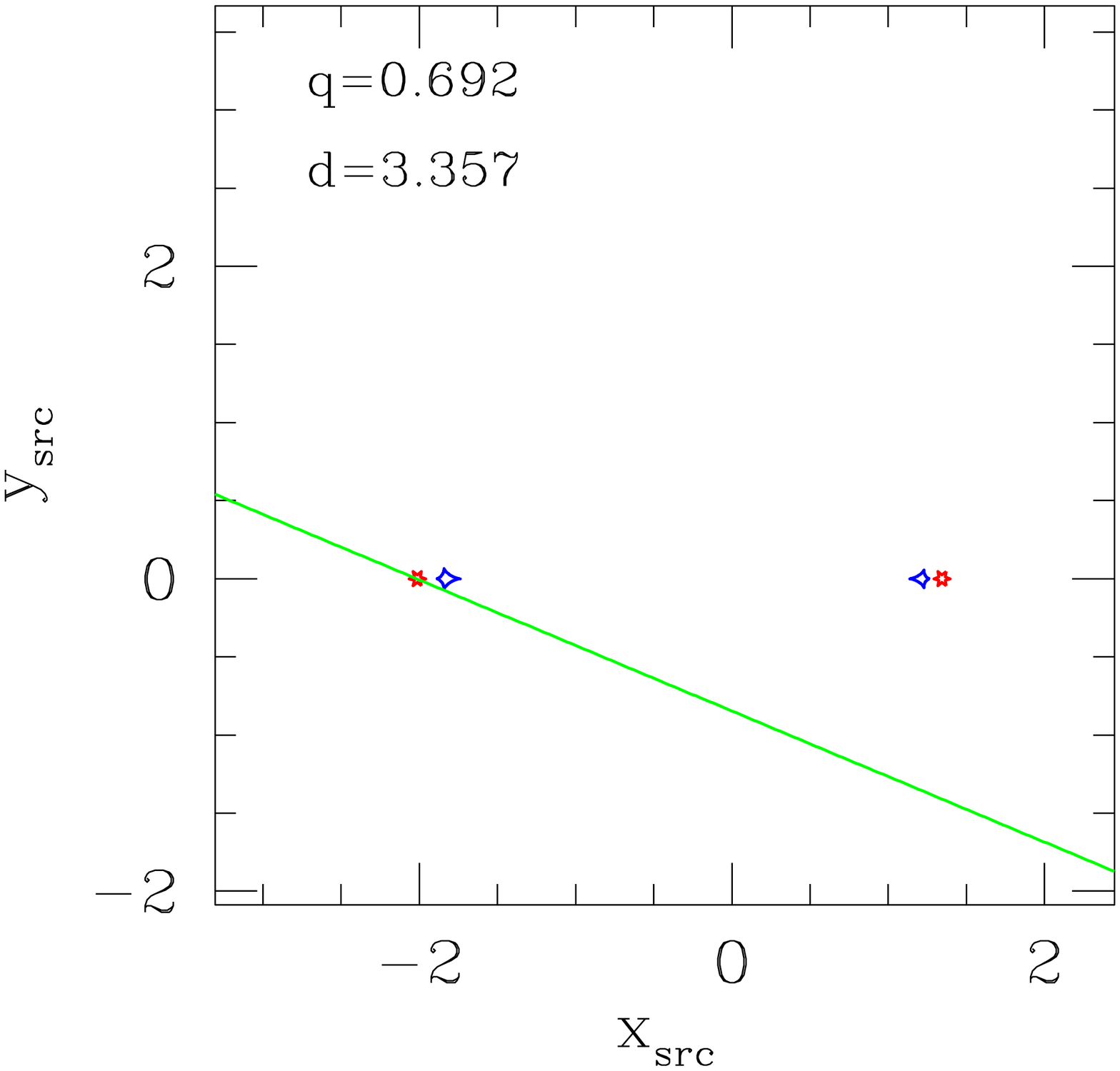}%
 \includegraphics[height=46mm,width=42.5mm]{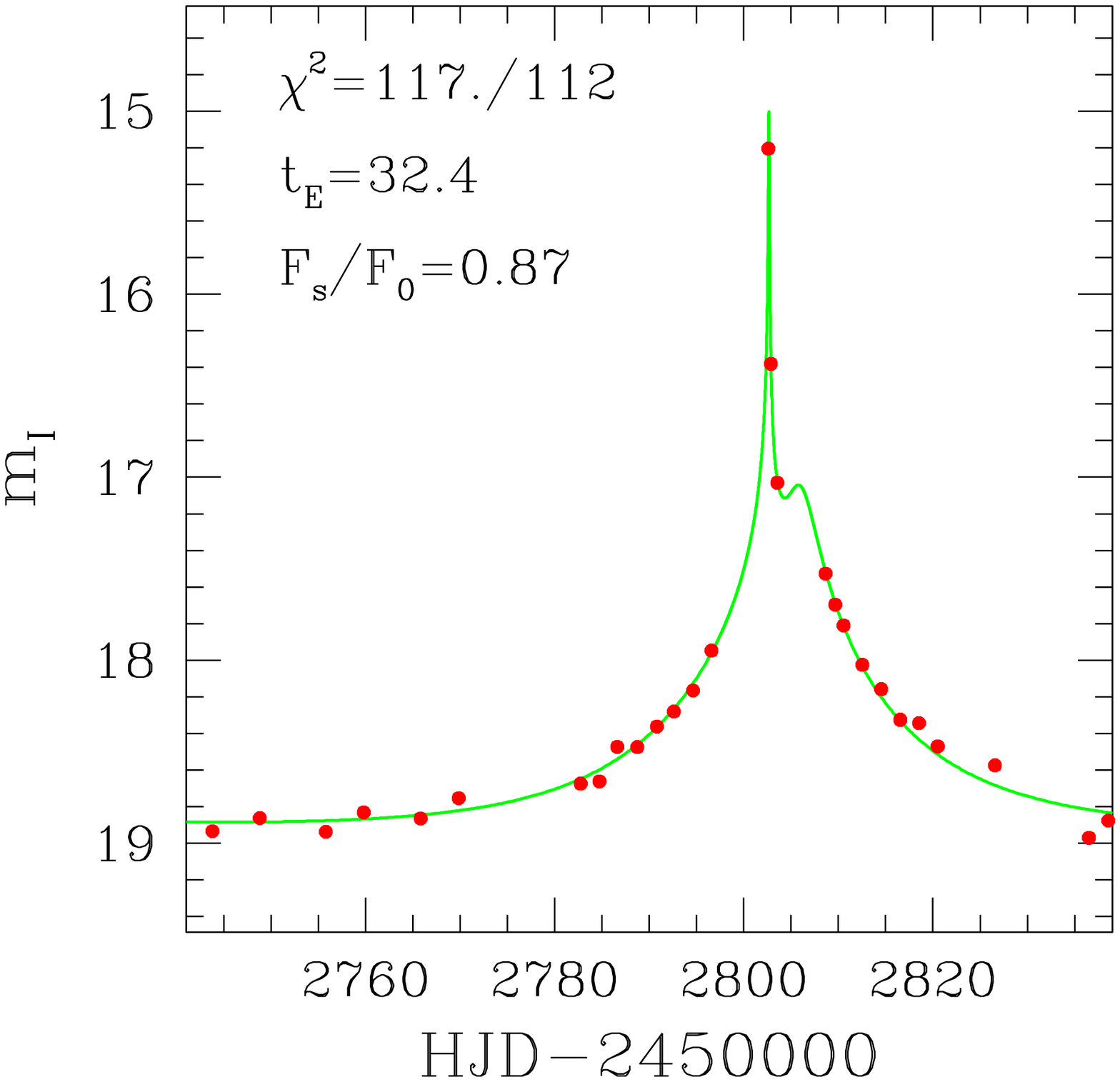}%
 \includegraphics[height=46mm,width=42.5mm]{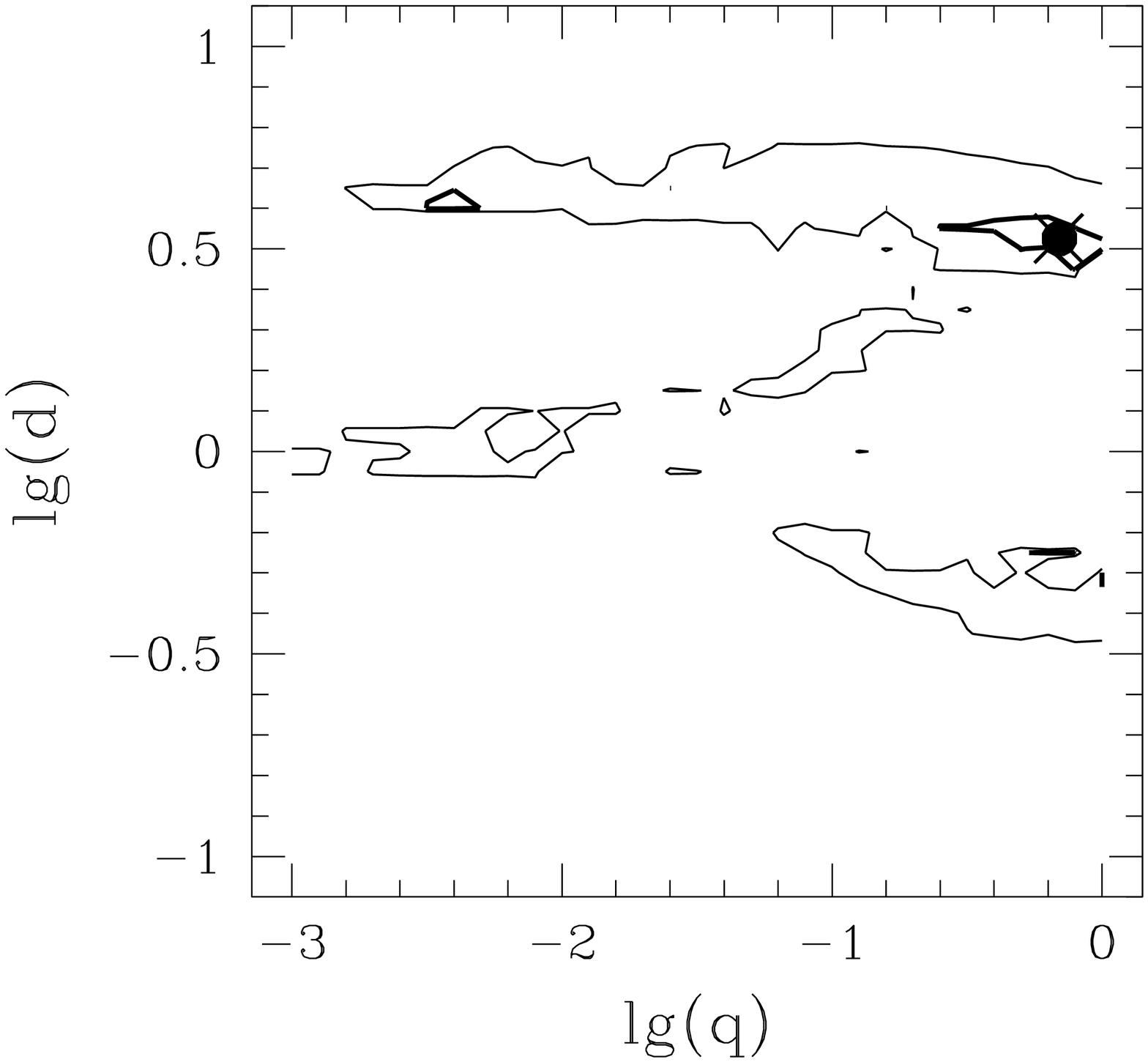}%
}

\noindent\parbox{12.75cm}{
\noindent{\bf OGLE 2003-BLG-194 (2nd model)} 

\vspace*{5pt}

 \includegraphics[height=46mm,width=42.5mm]{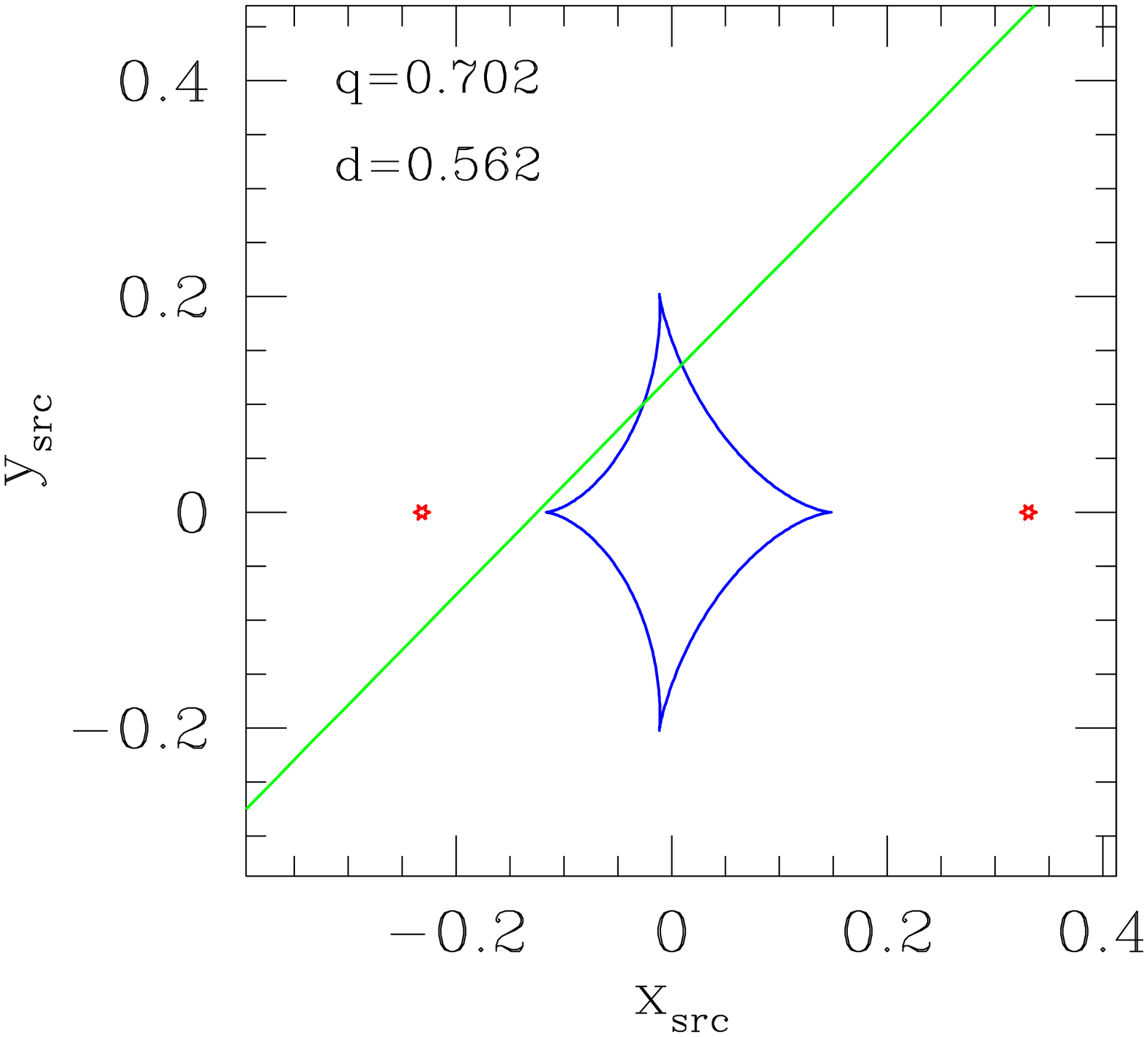}%
 \includegraphics[height=46mm,width=42.5mm]{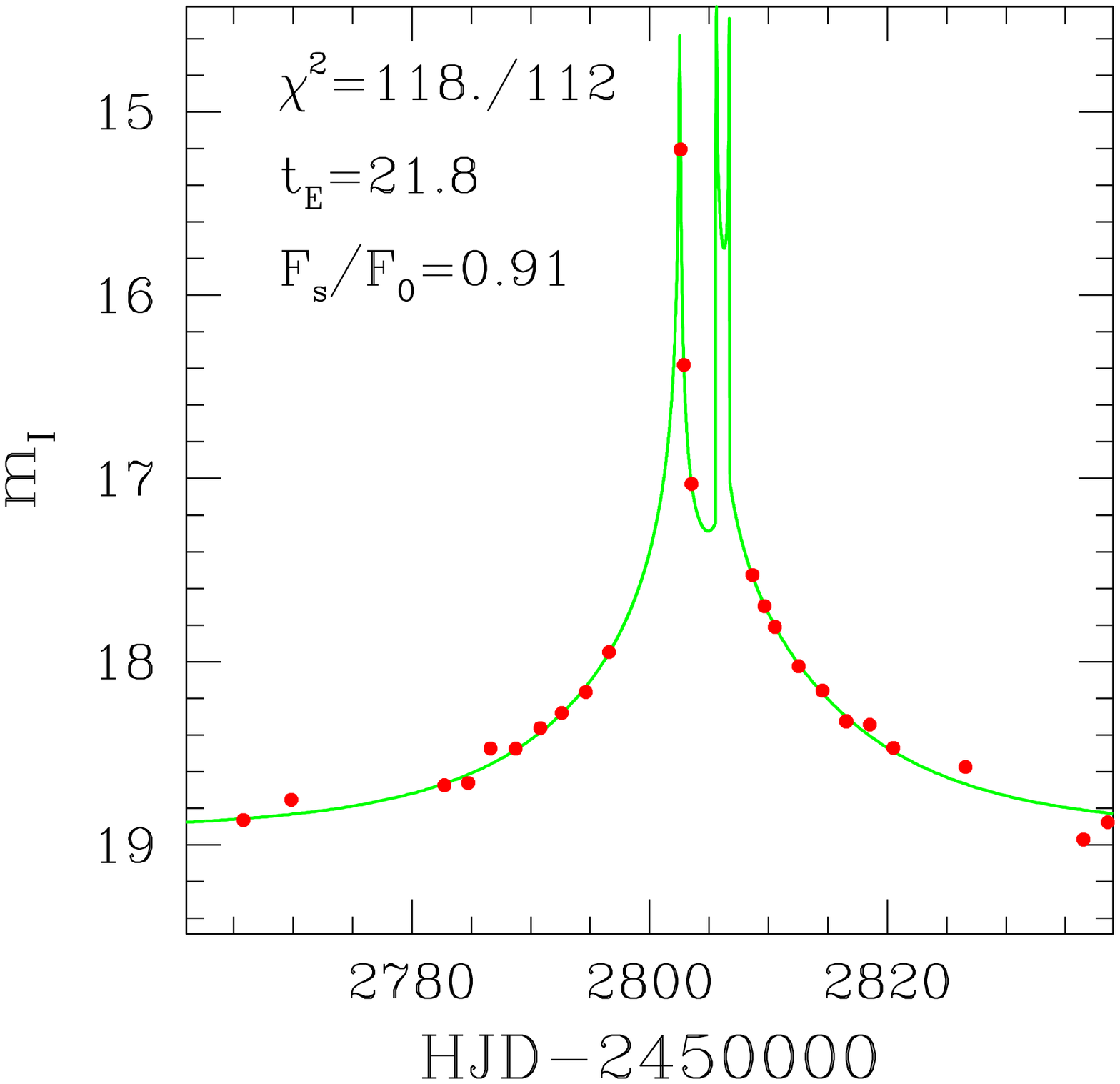}%
 \includegraphics[height=46mm,width=42.5mm]{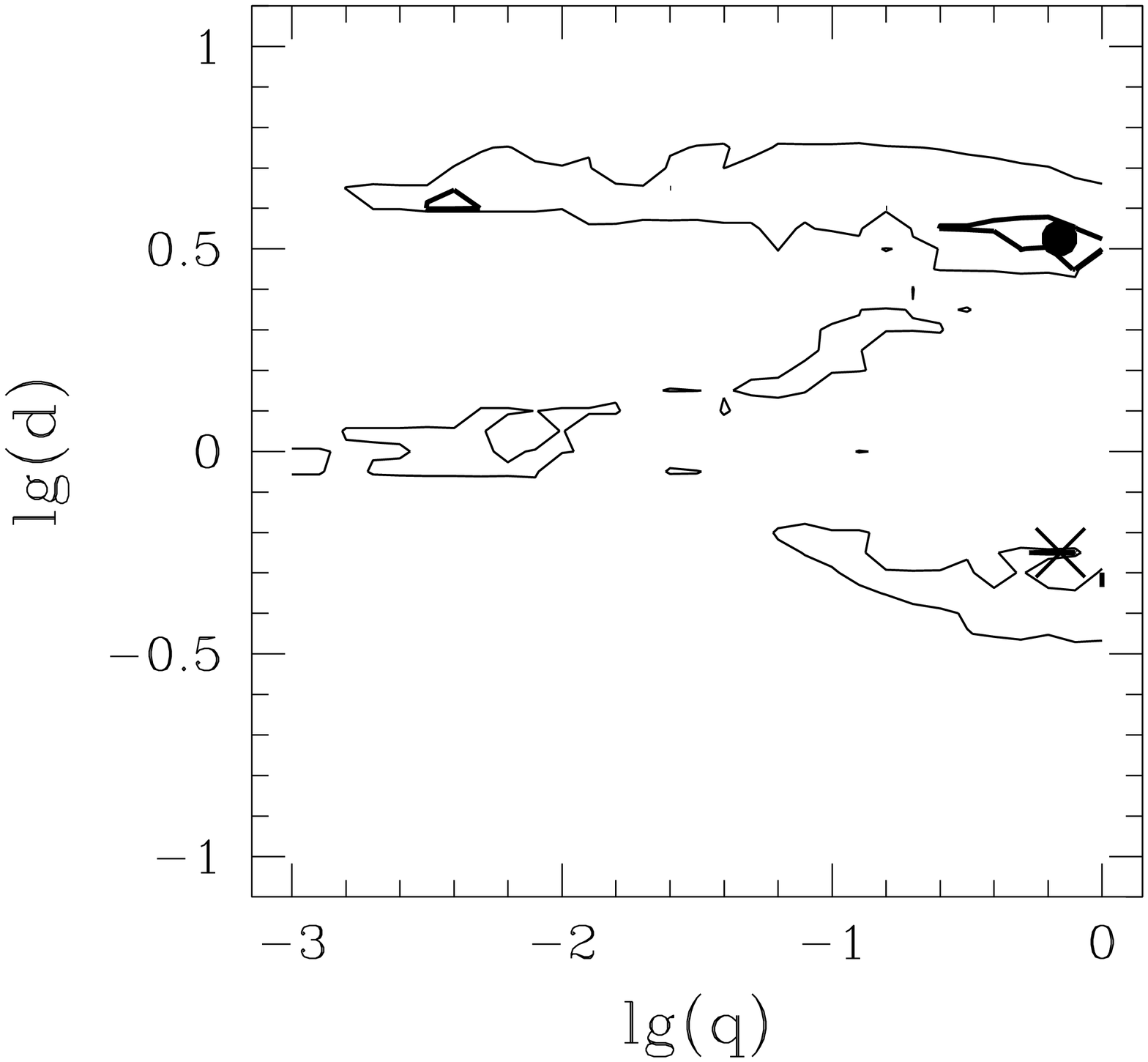}%
}

\noindent\parbox{12.75cm}{
\noindent{\bf OGLE 2003-BLG-200}                           

\vspace*{5pt}

 \includegraphics[height=46mm,width=42.5mm]{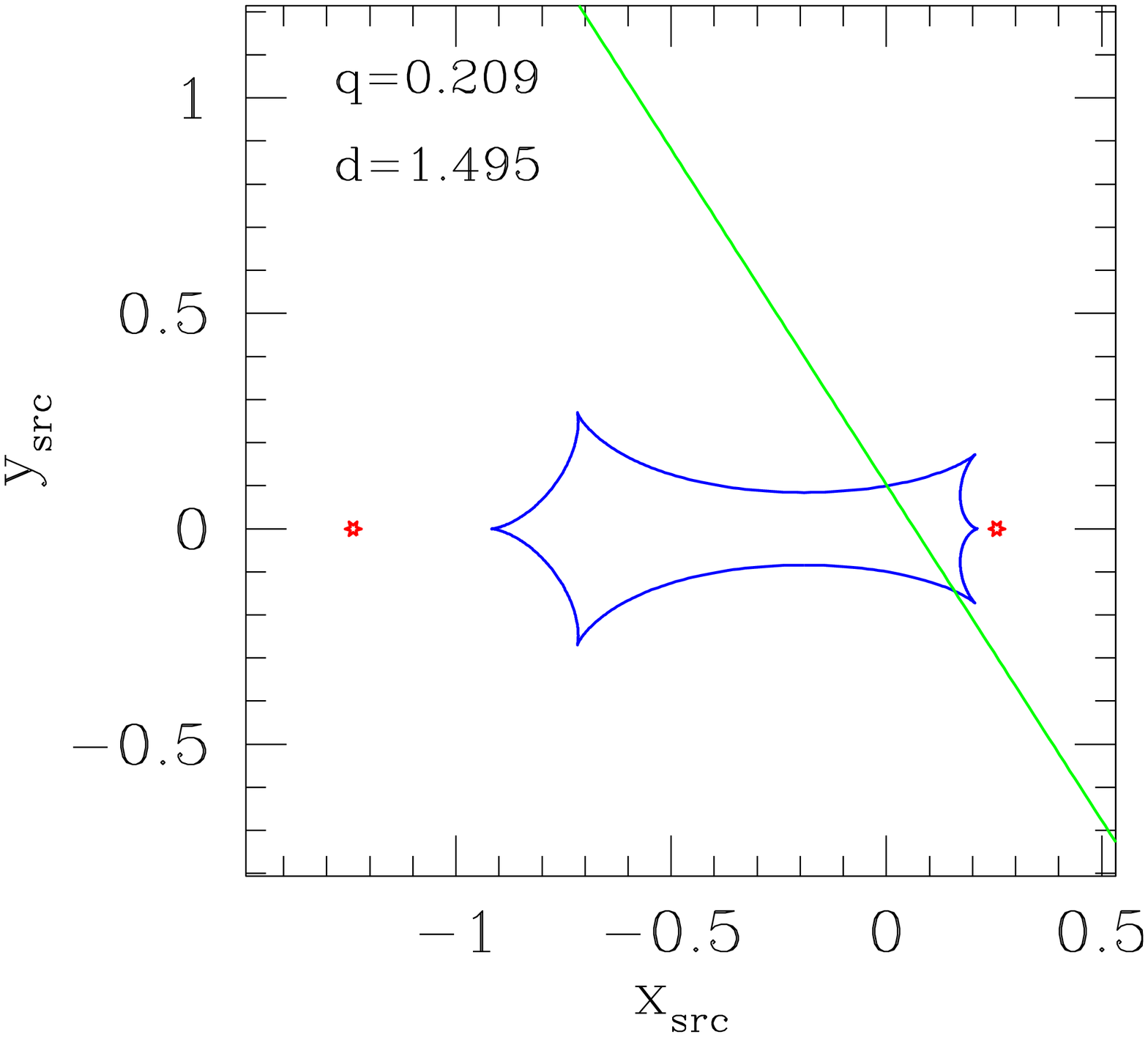}%
 \includegraphics[height=46mm,width=42.5mm]{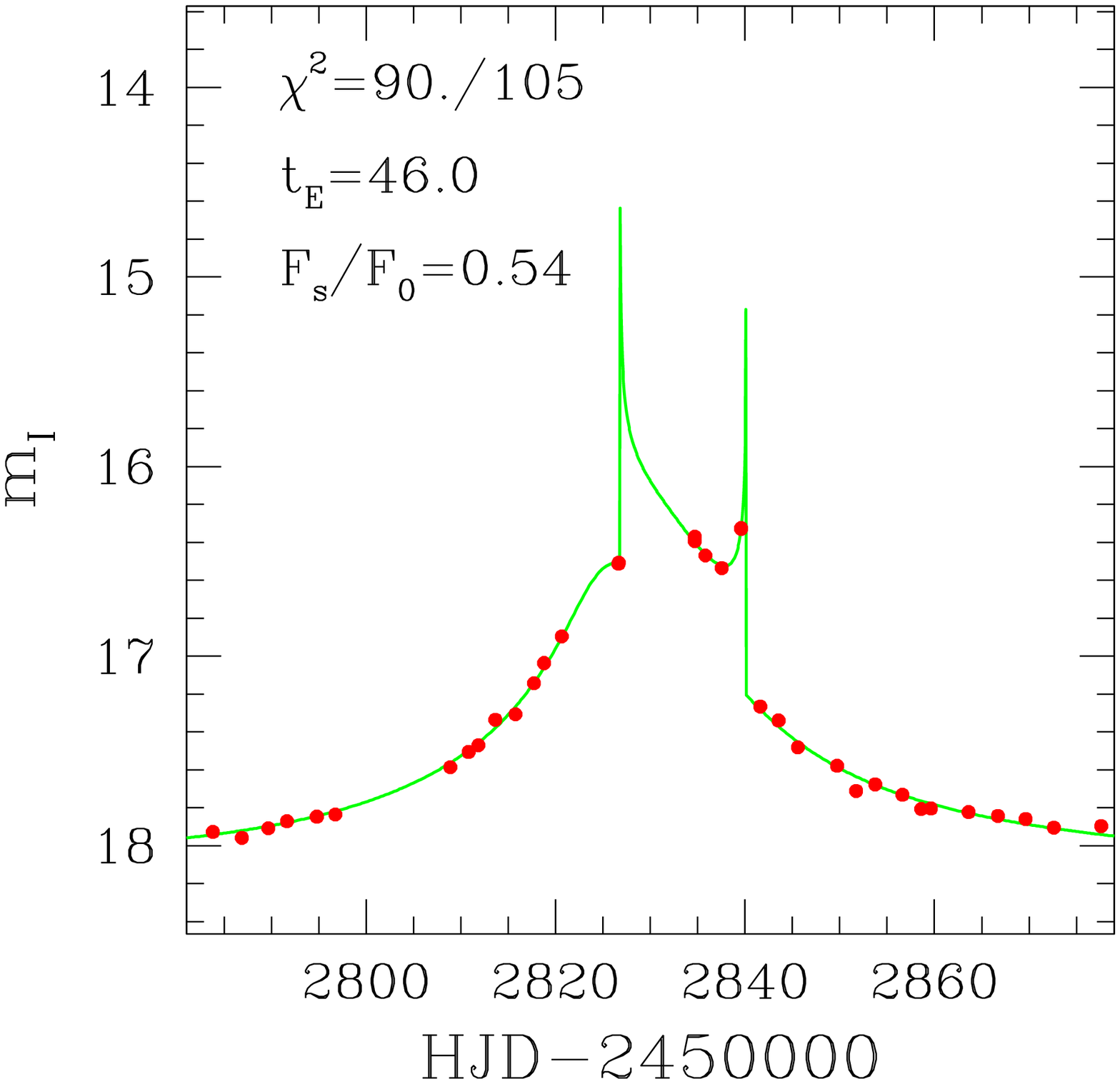}%
 \includegraphics[height=46mm,width=42.5mm]{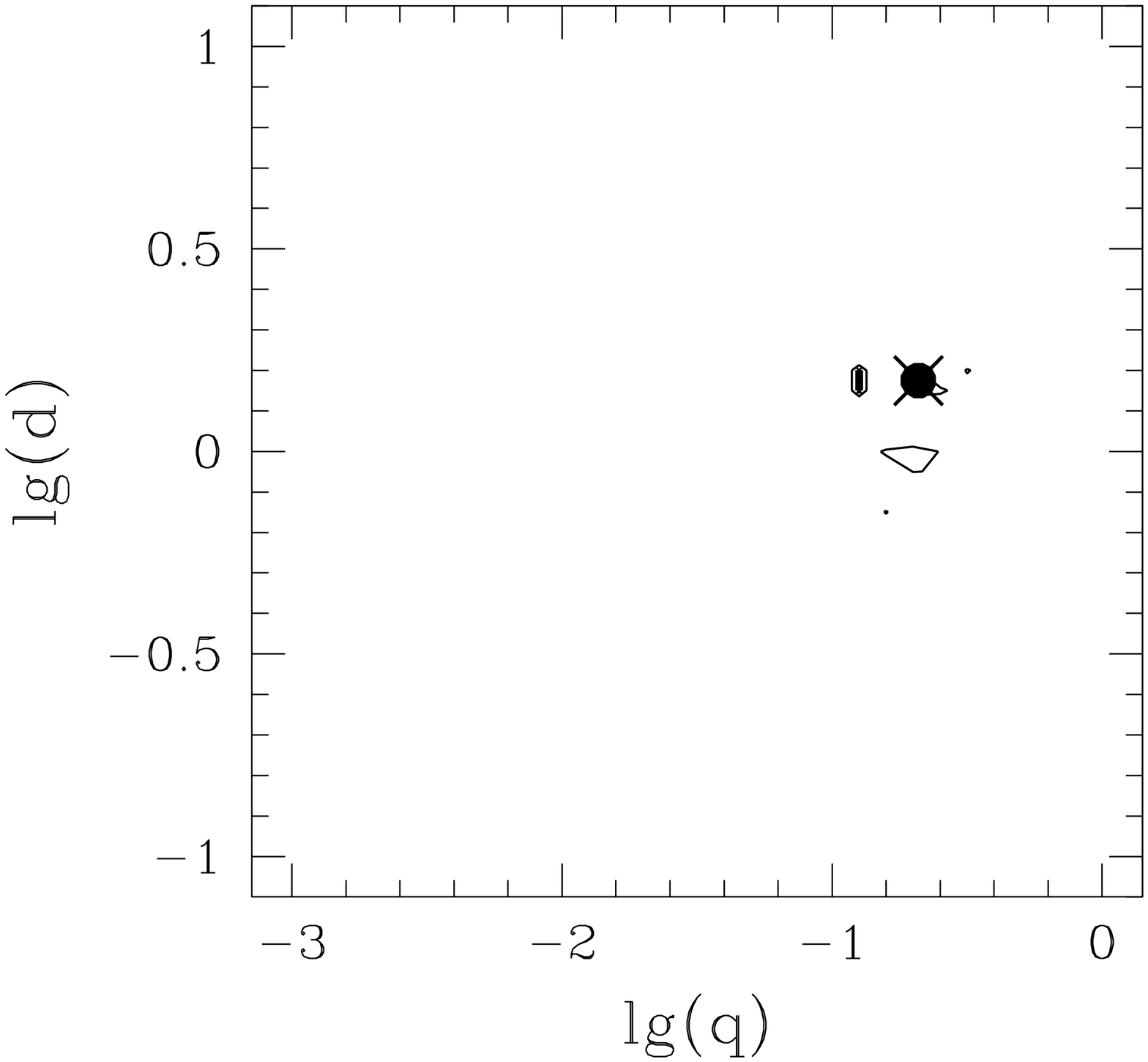}%
}

\noindent\parbox{12.75cm}{
\noindent{\bf OGLE 2003-BLG-235 / MOA 2003-BLG-53}

\vspace*{5pt}

 \includegraphics[height=46mm,width=42.5mm]{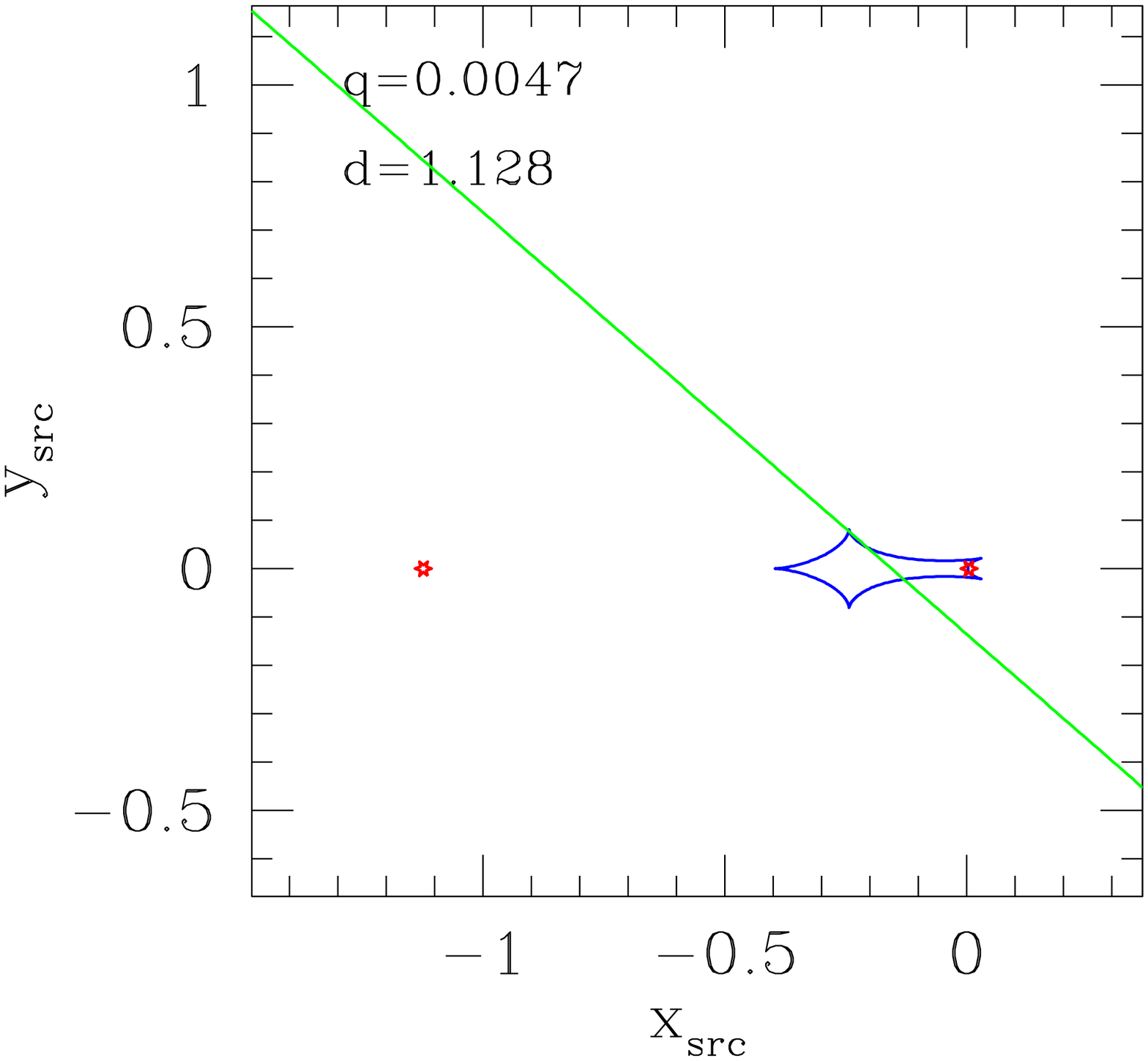}%
 \includegraphics[height=46mm,width=42.5mm]{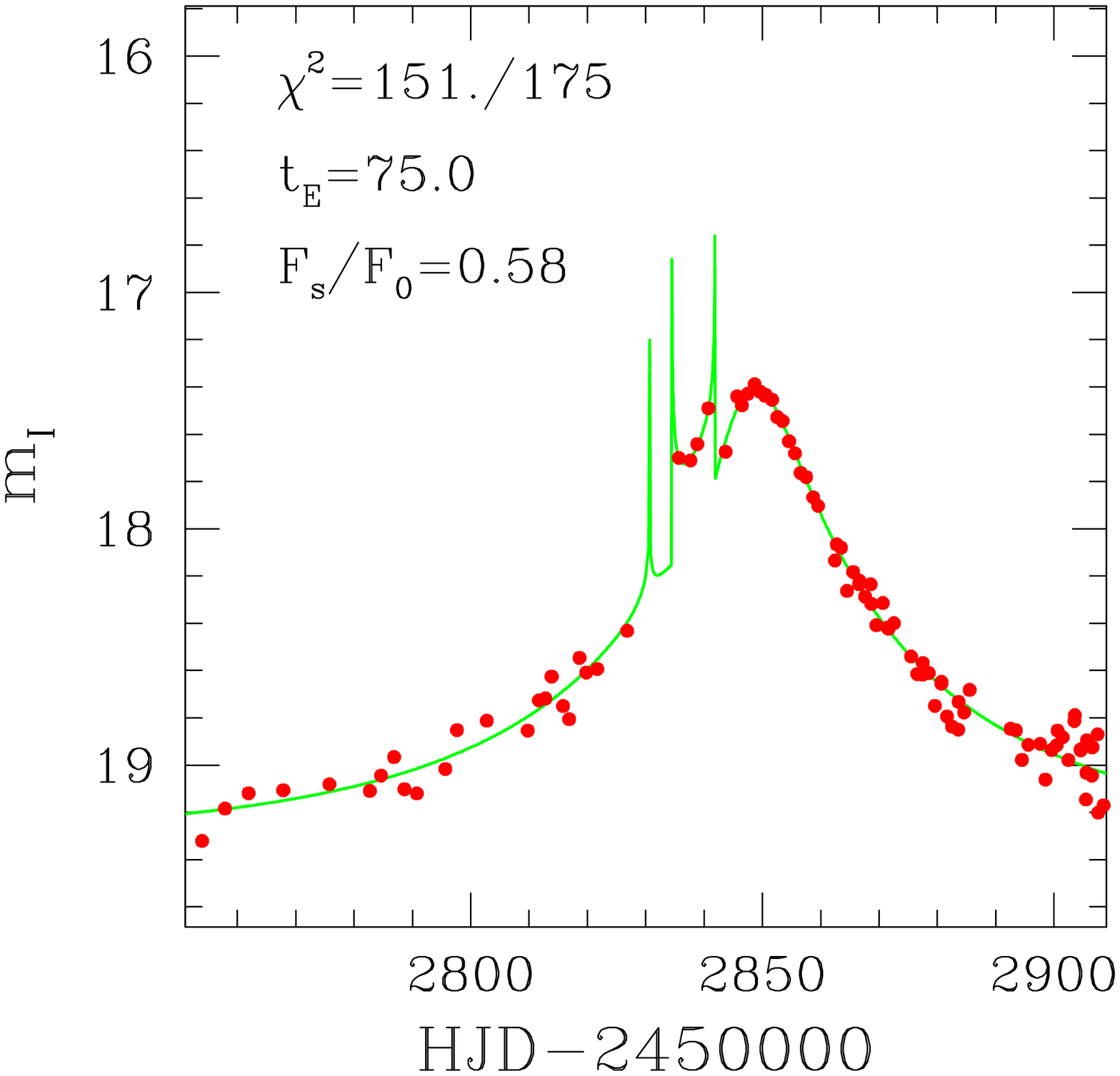}%
 \includegraphics[height=46mm,width=42.5mm]{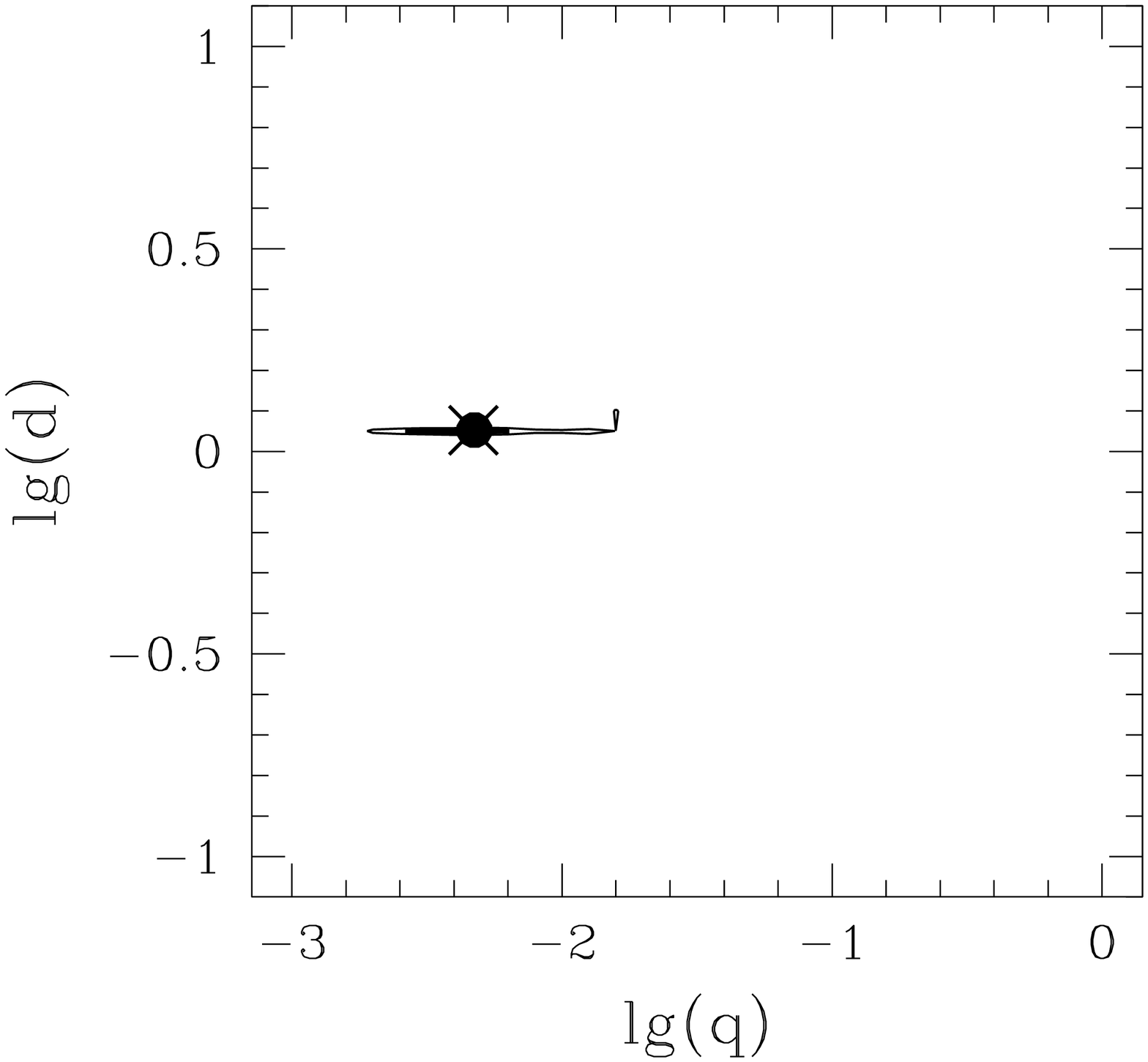}%
}

\noindent\parbox{12.75cm}{
\noindent{\bf OGLE 2003-BLG-236}                           

\vspace*{5pt}

 \includegraphics[height=46mm,width=42.5mm]{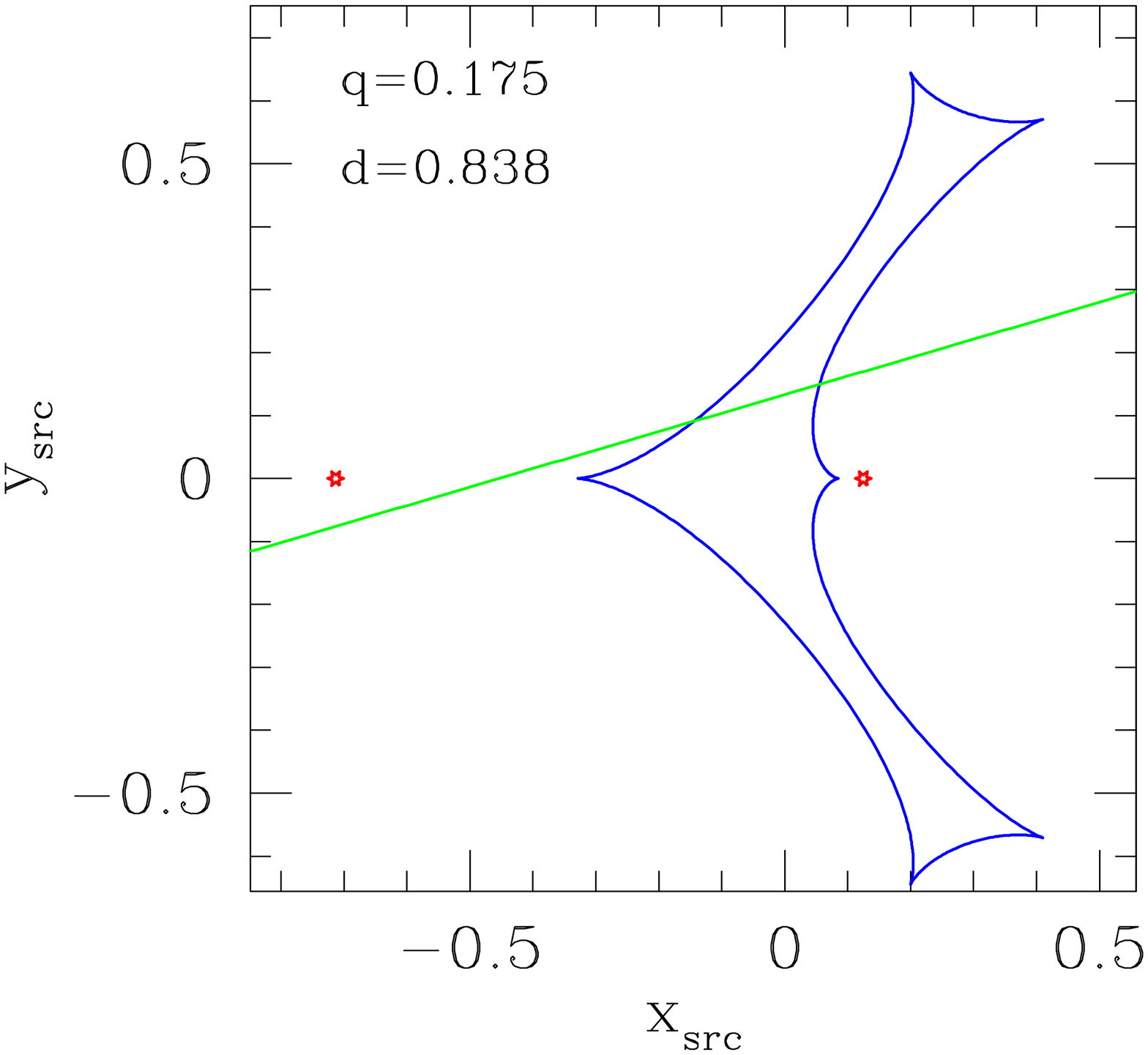}%
 \includegraphics[height=46mm,width=42.5mm]{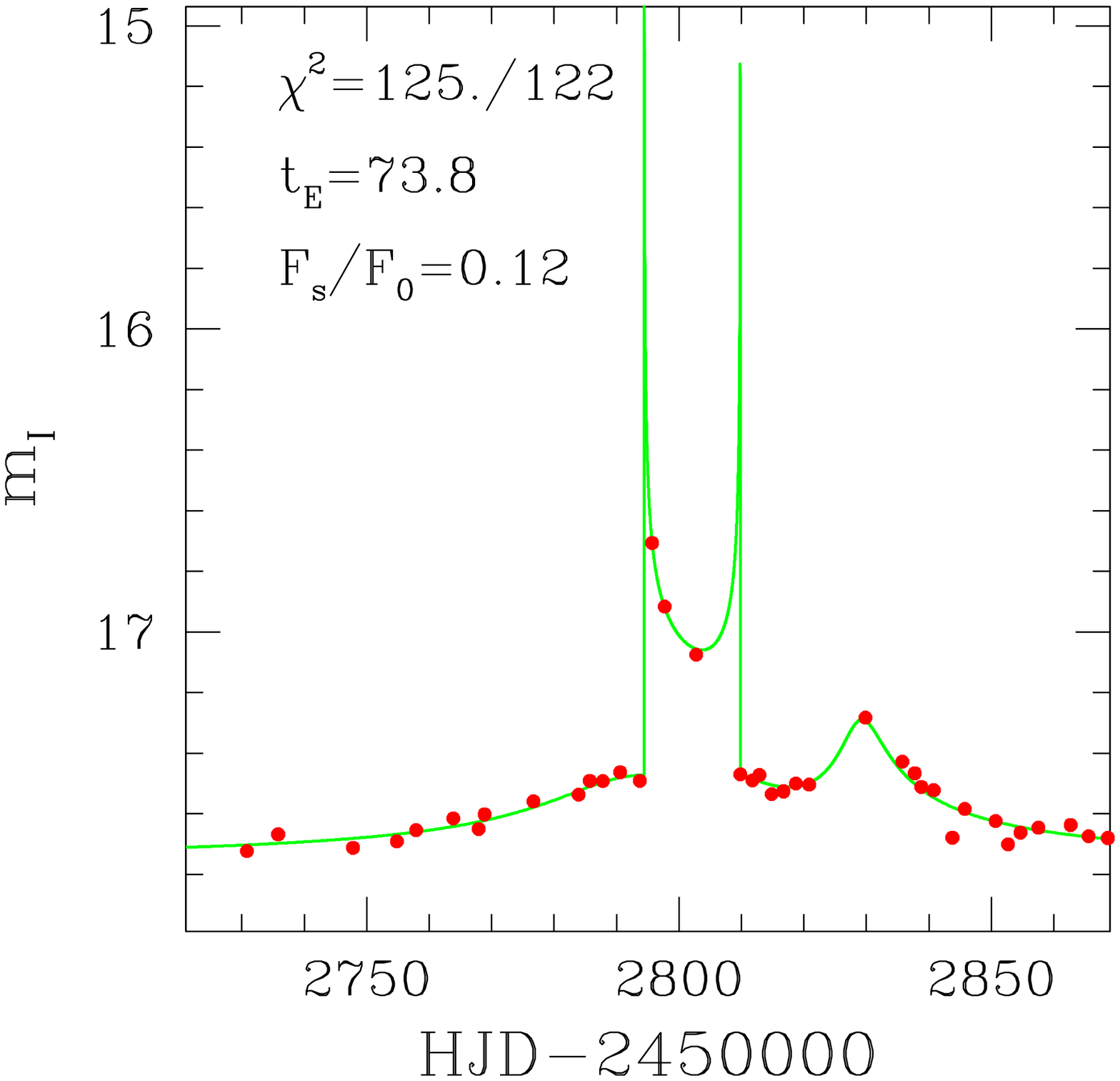}%
 \includegraphics[height=46mm,width=42.5mm]{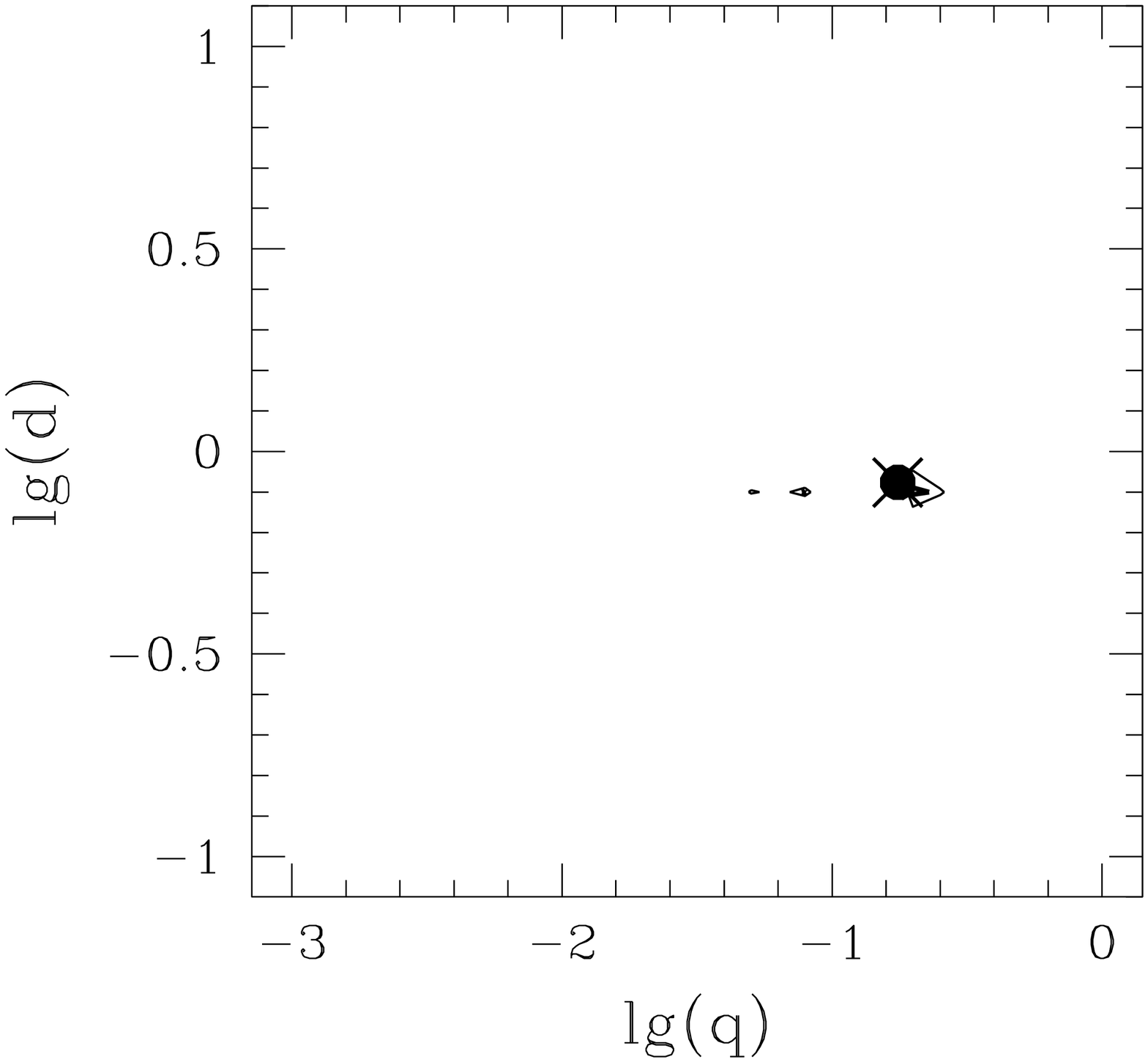}%
}

\noindent\parbox{12.75cm}{
\noindent{\bf OGLE 2003-BLG-260 (1st model)}                     

\vspace*{5pt}

 \includegraphics[height=46mm,width=42.5mm]{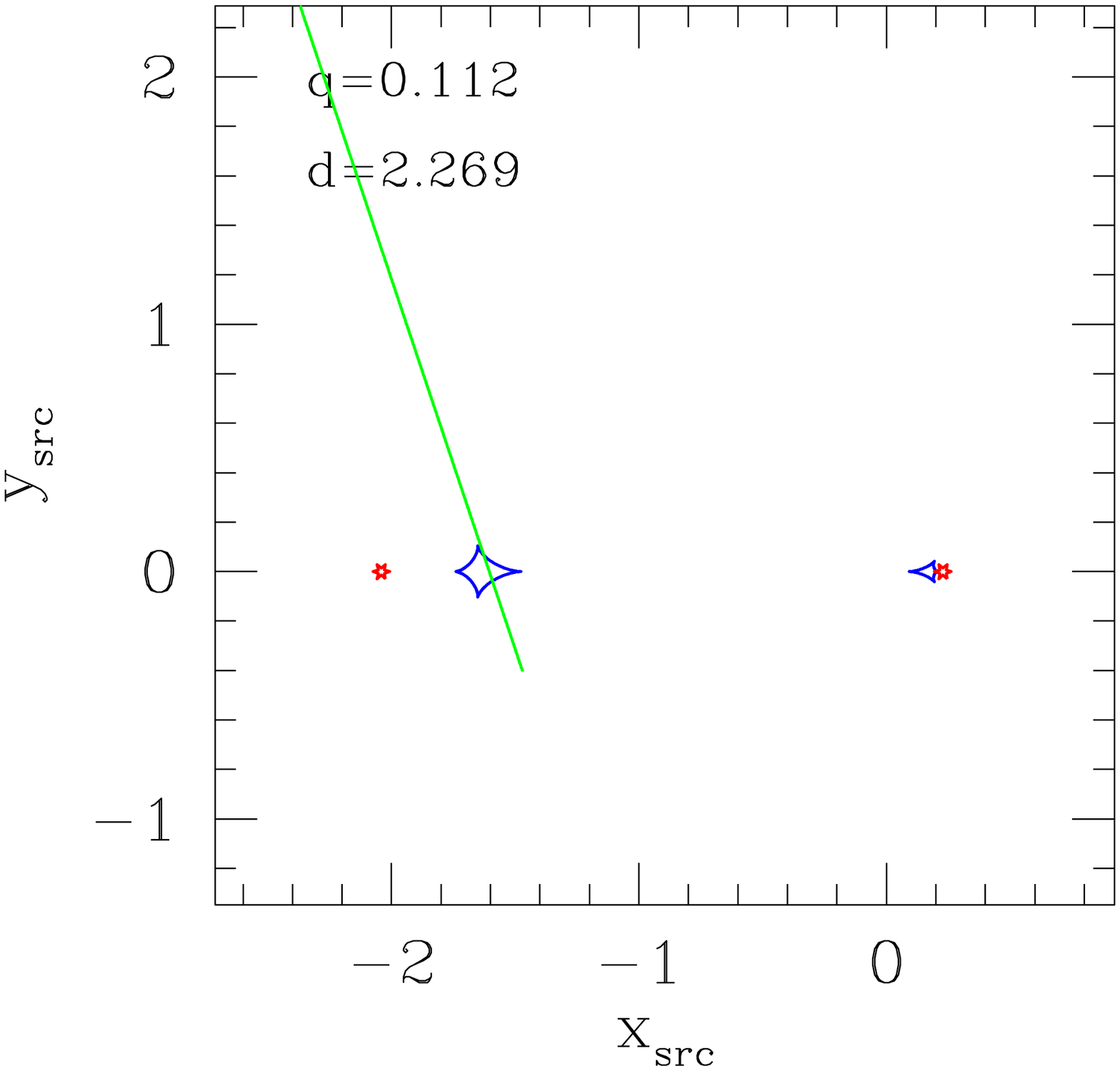}%
 \includegraphics[height=46mm,width=42.5mm]{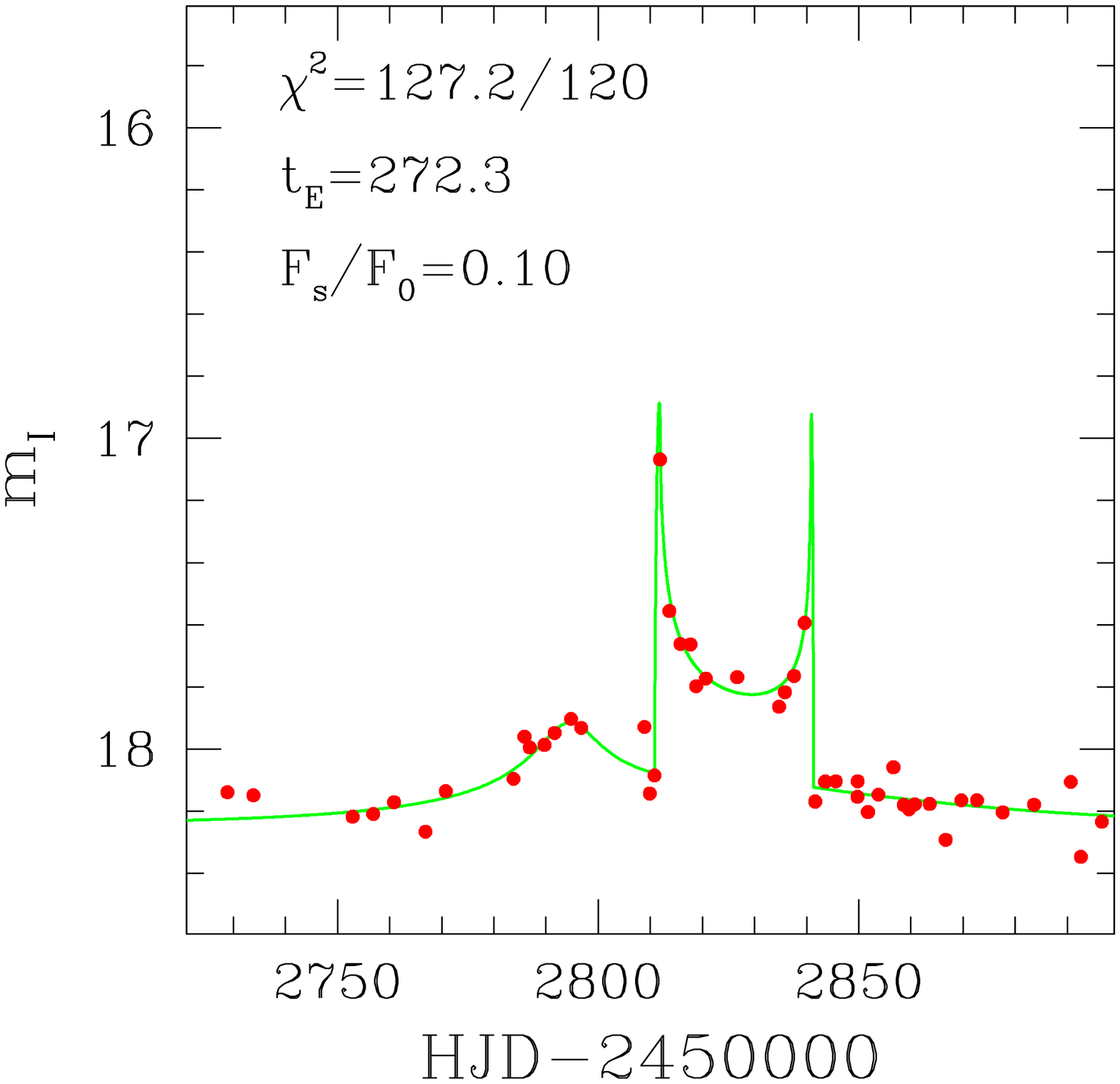}%
 \includegraphics[height=46mm,width=42.5mm]{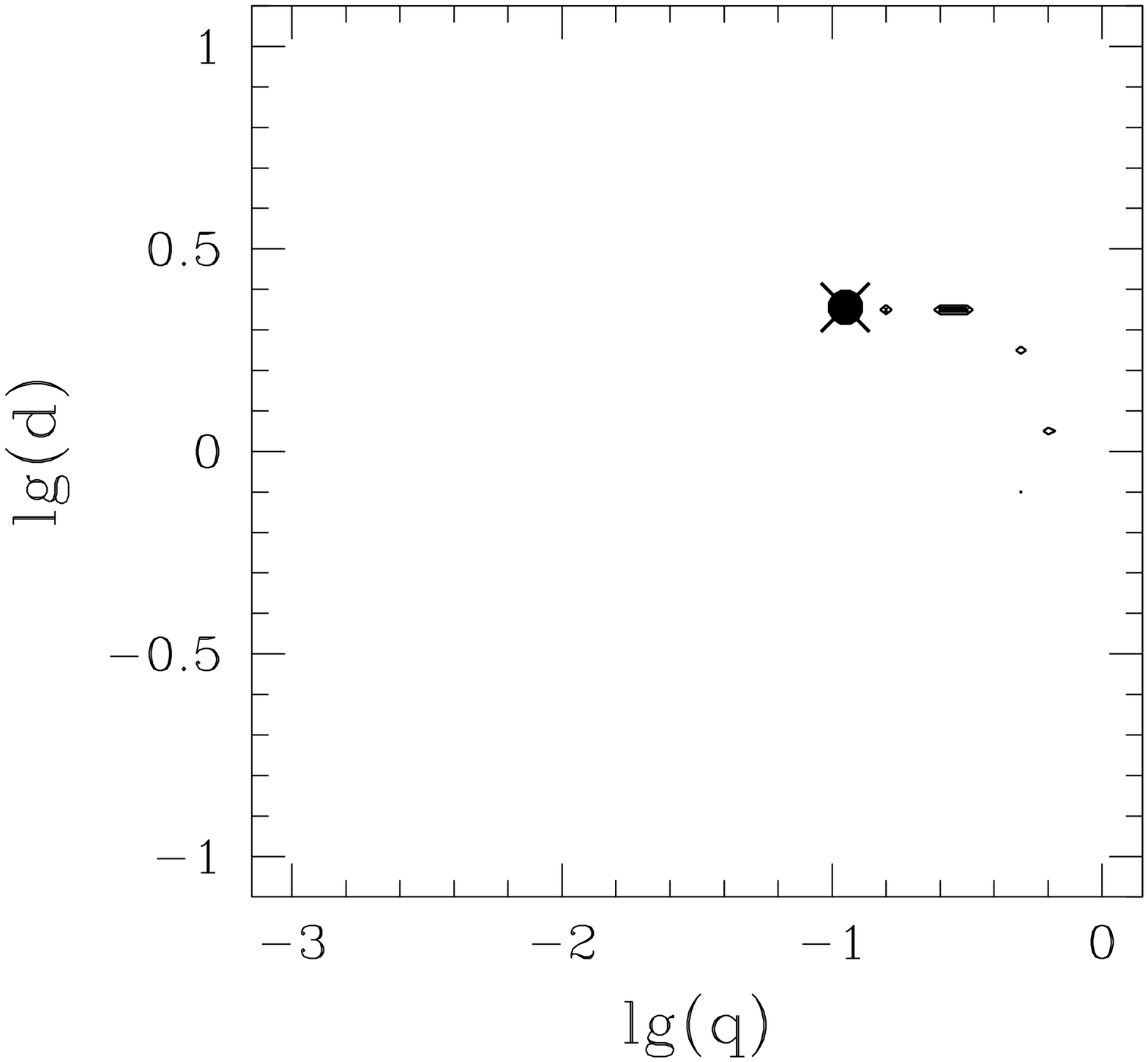}%
}

\noindent\parbox{12.75cm}{
\noindent{\bf OGLE 2003-BLG-260 (2nd model)}                     

\vspace*{5pt}

 \includegraphics[height=46mm,width=42.5mm]{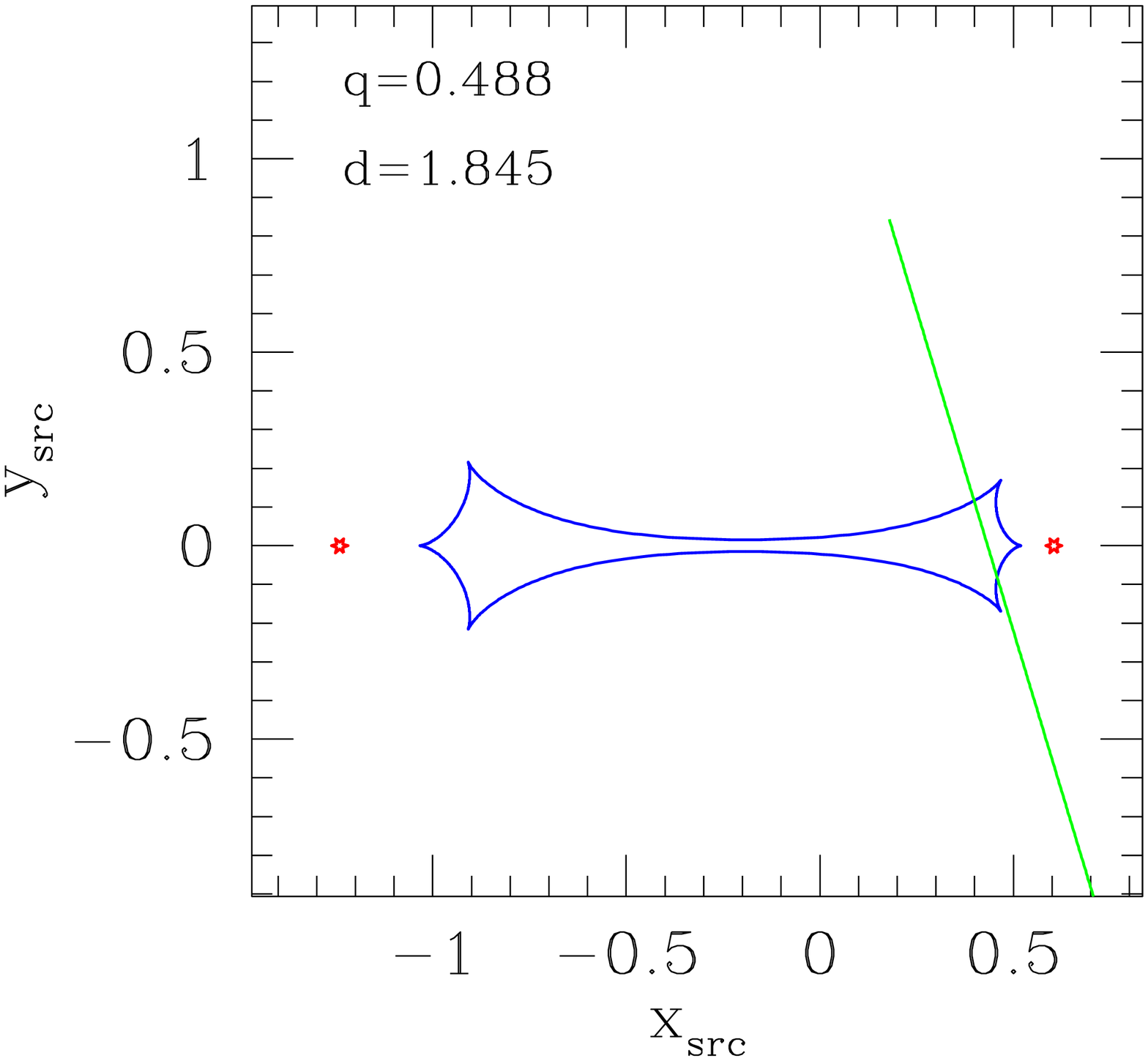}%
 \includegraphics[height=46mm,width=42.5mm]{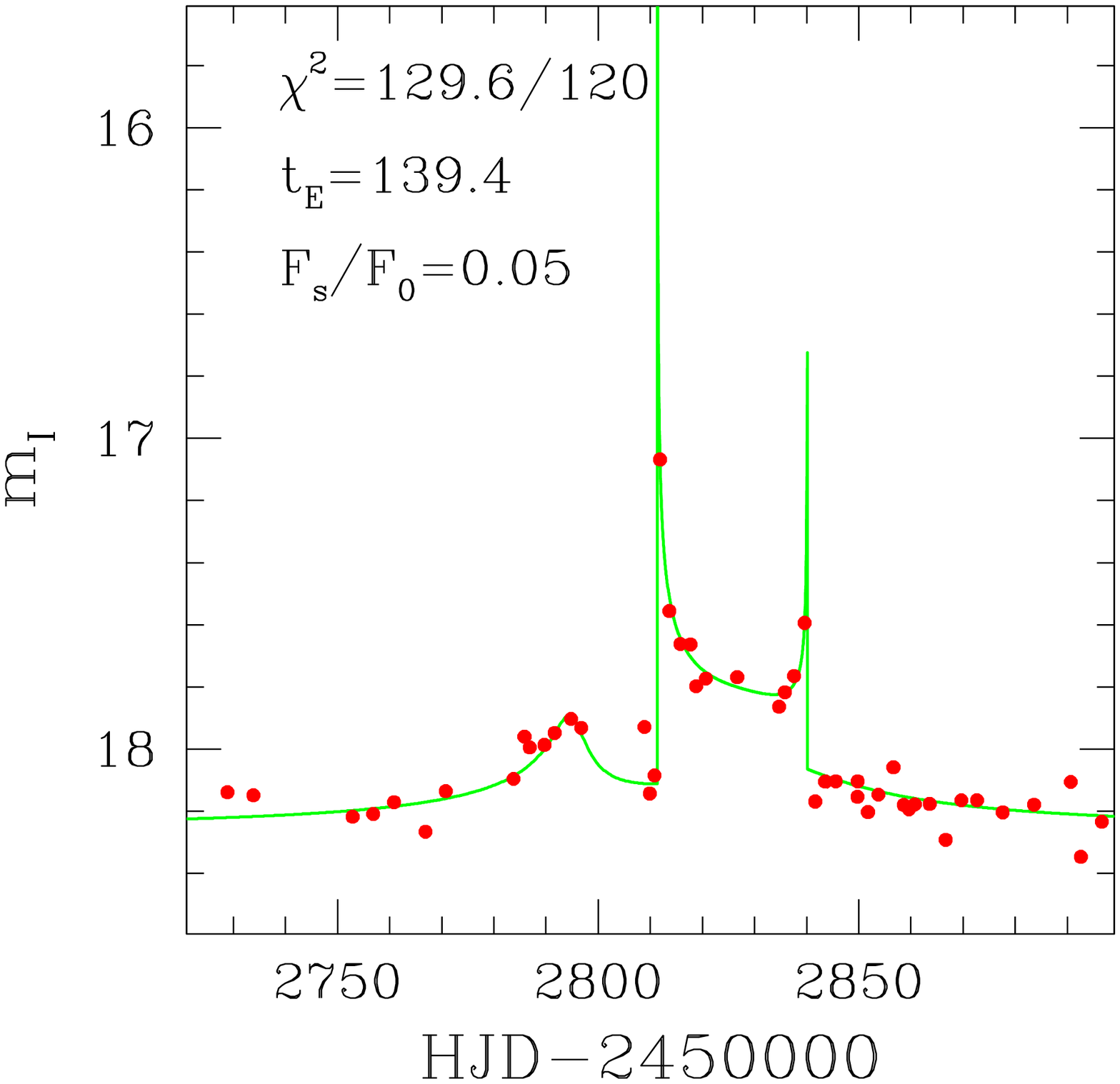}%
 \includegraphics[height=46mm,width=42.5mm]{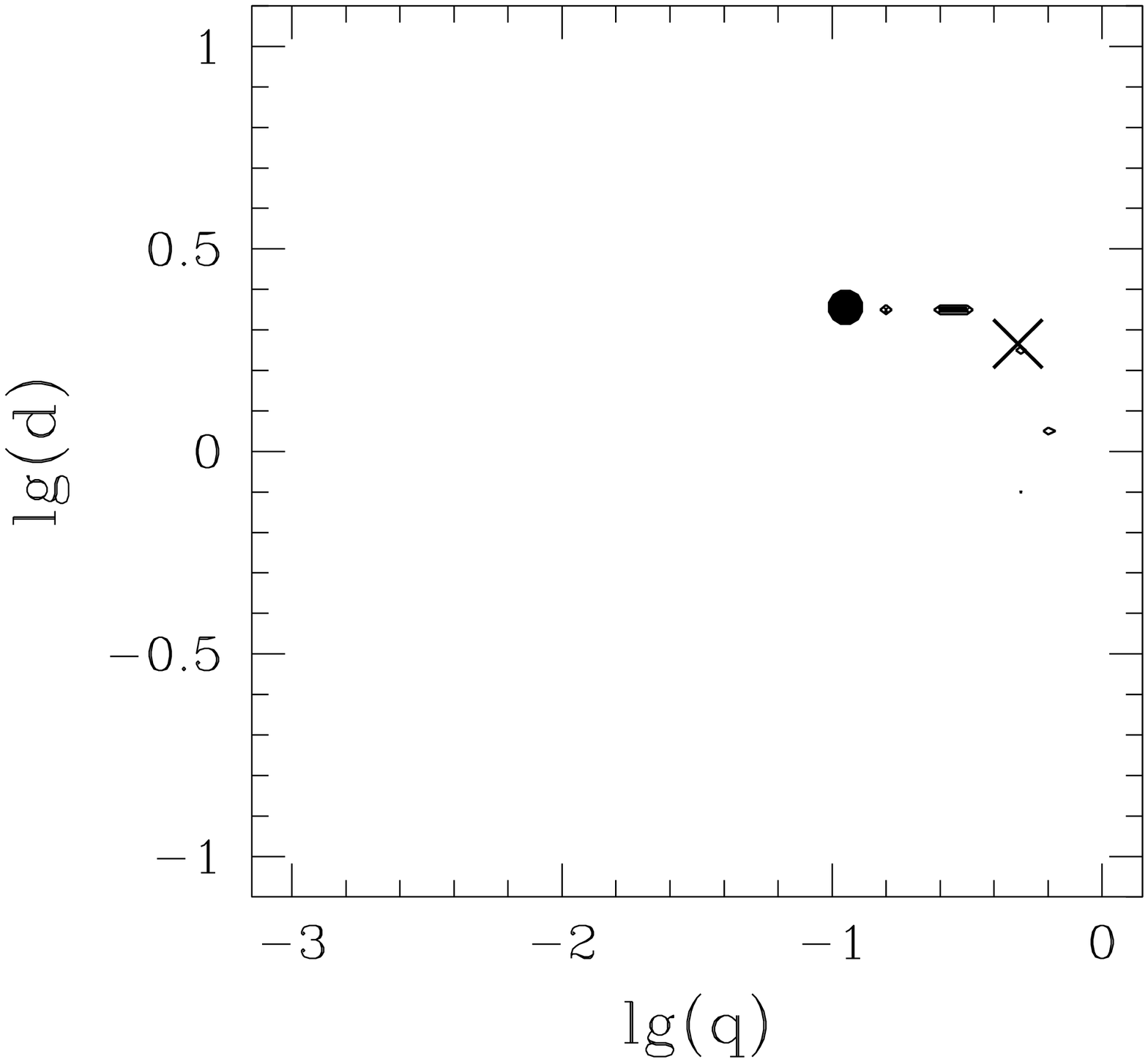}%
}

\noindent\parbox{12.75cm}{
\noindent{\bf OGLE 2003-BLG-260 (3rd model)}                     

\vspace*{5pt}

 \includegraphics[height=46mm,width=42.5mm]{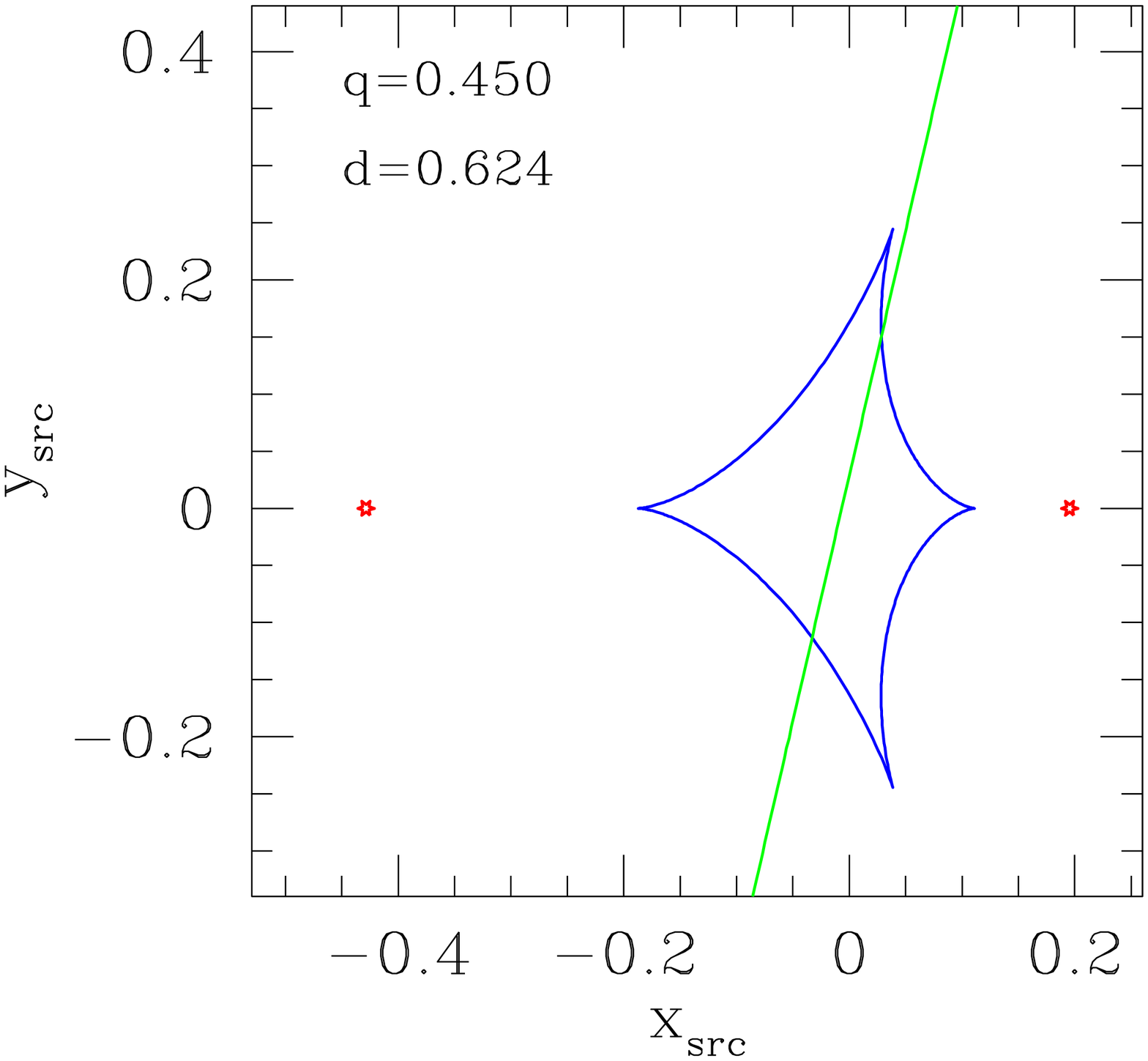}%
 \includegraphics[height=46mm,width=42.5mm]{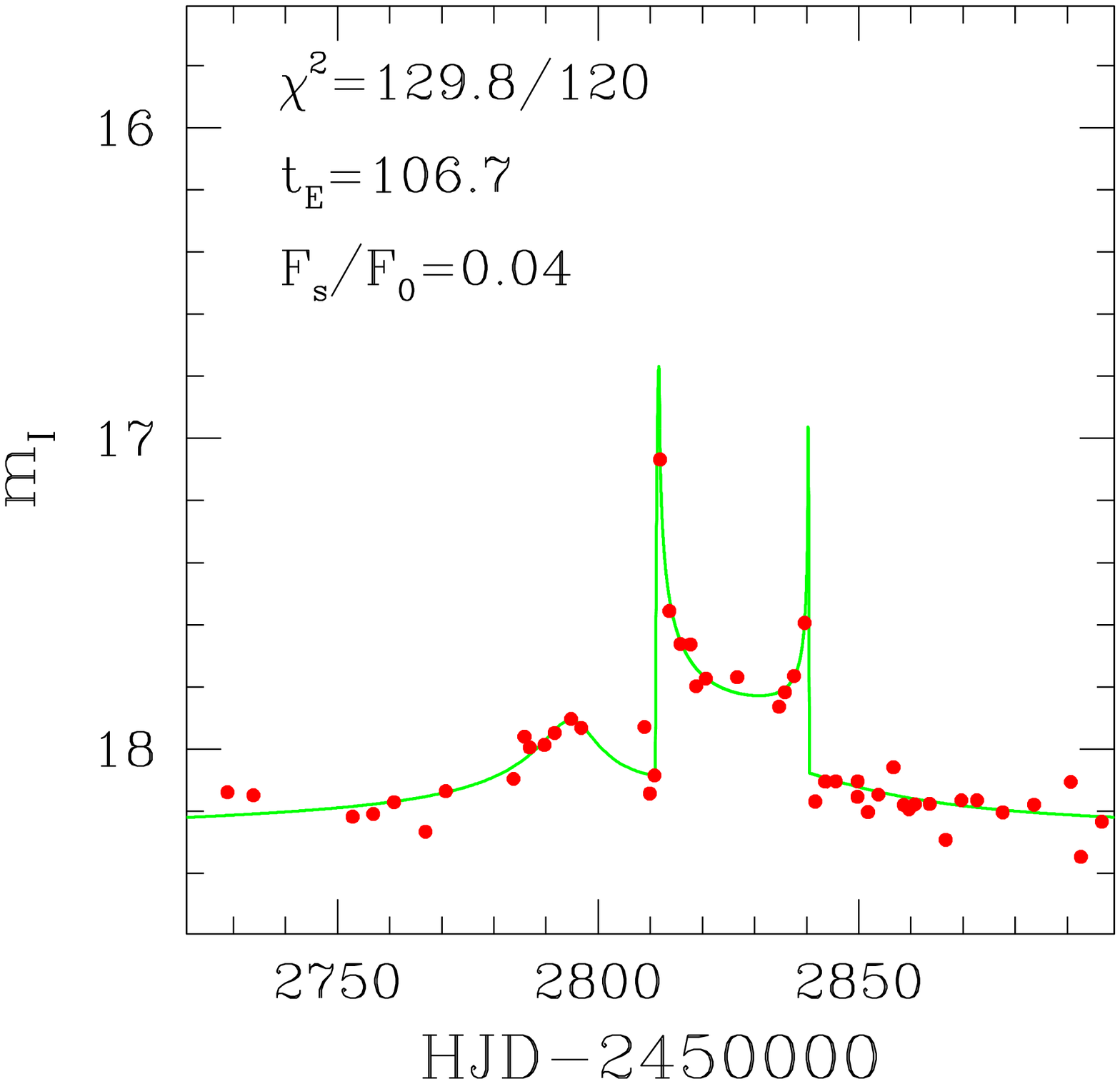}%
 \includegraphics[height=46mm,width=42.5mm]{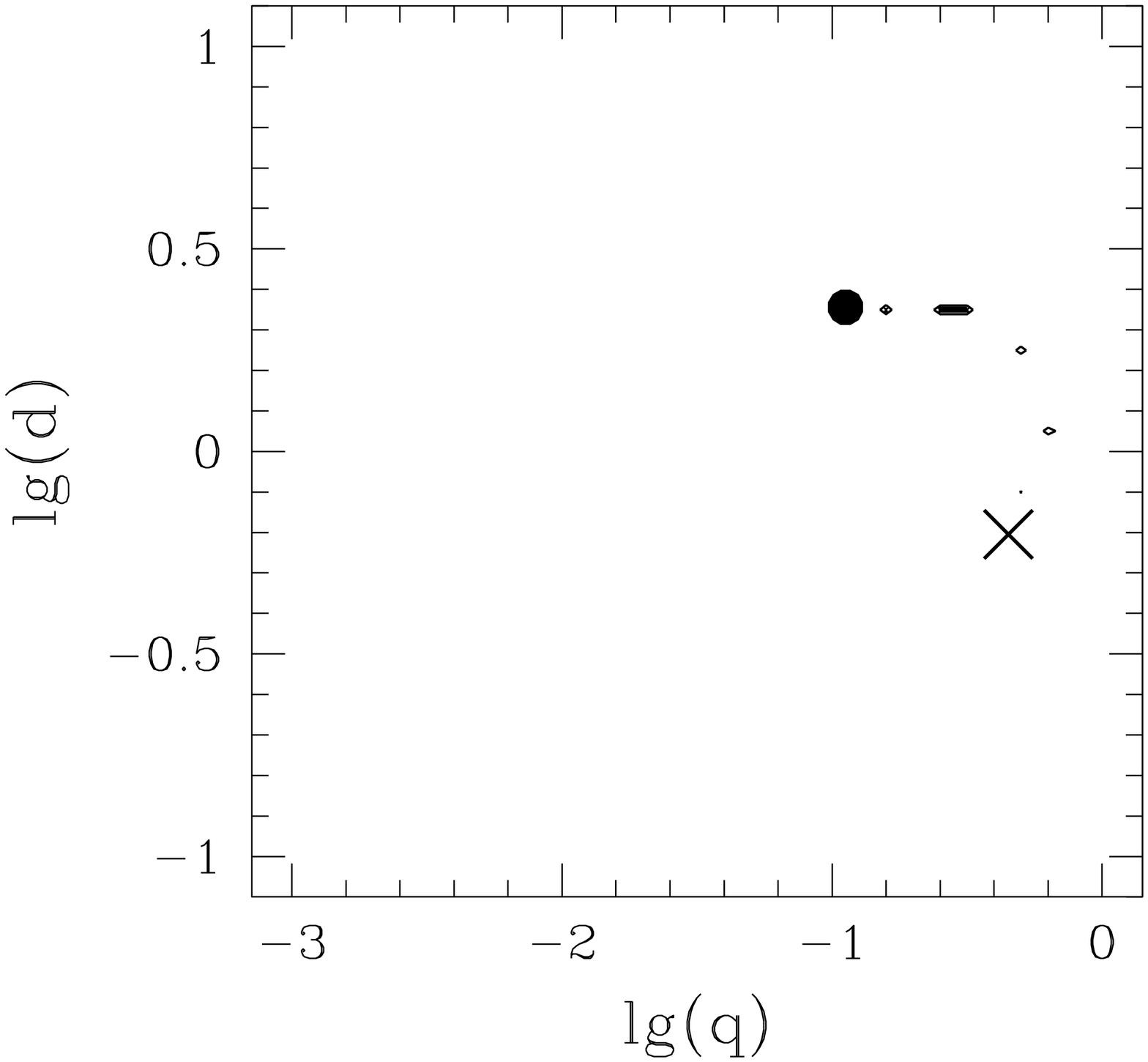}%
}

\noindent\parbox{12.75cm}{
 \noindent{\bf OGLE 2003-BLG-266}                            

\vspace*{5pt}

 \includegraphics[height=46mm,width=42.5mm]{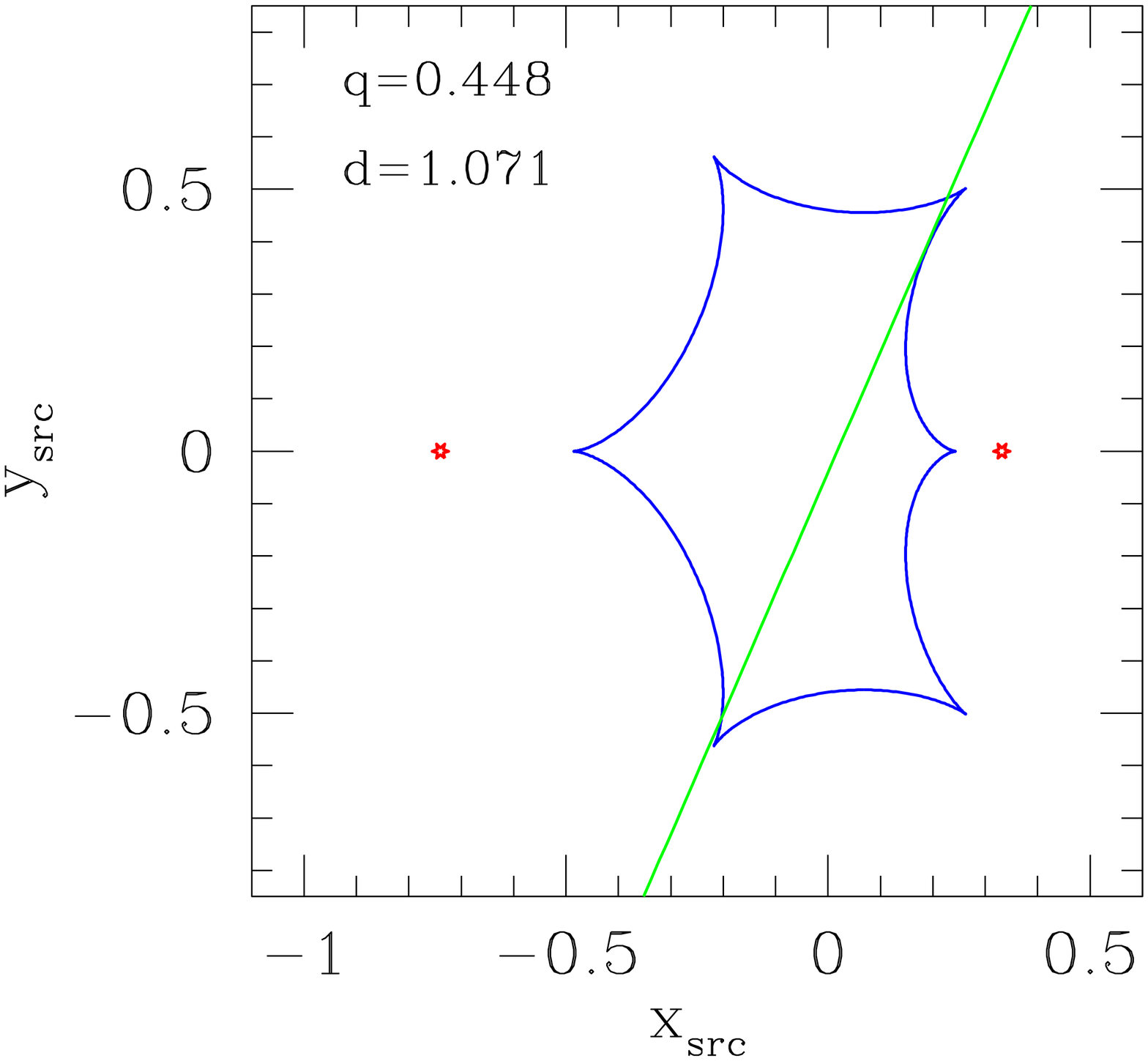}%
 \includegraphics[height=46mm,width=42.5mm]{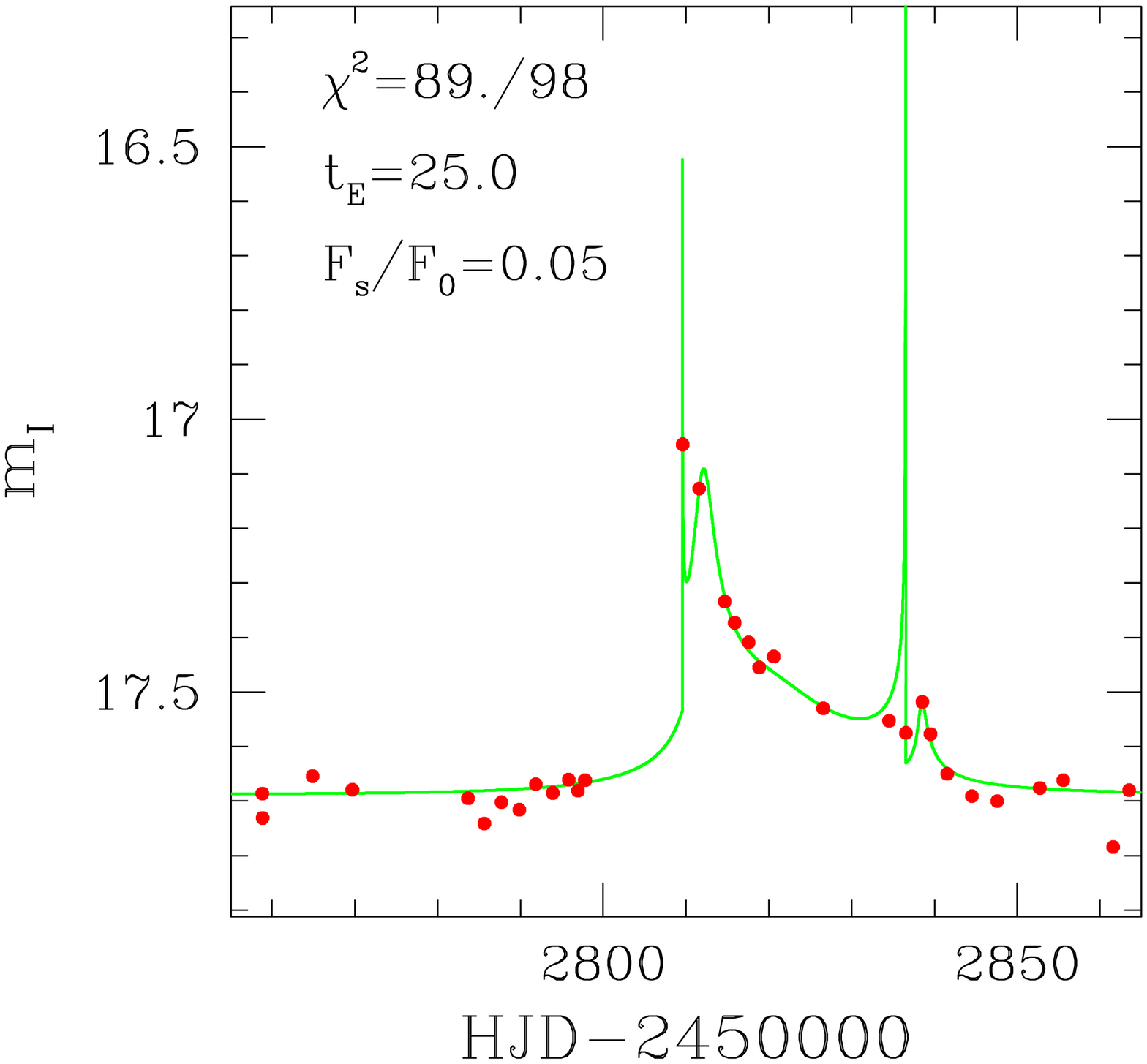}%
 \includegraphics[height=46mm,width=42.5mm]{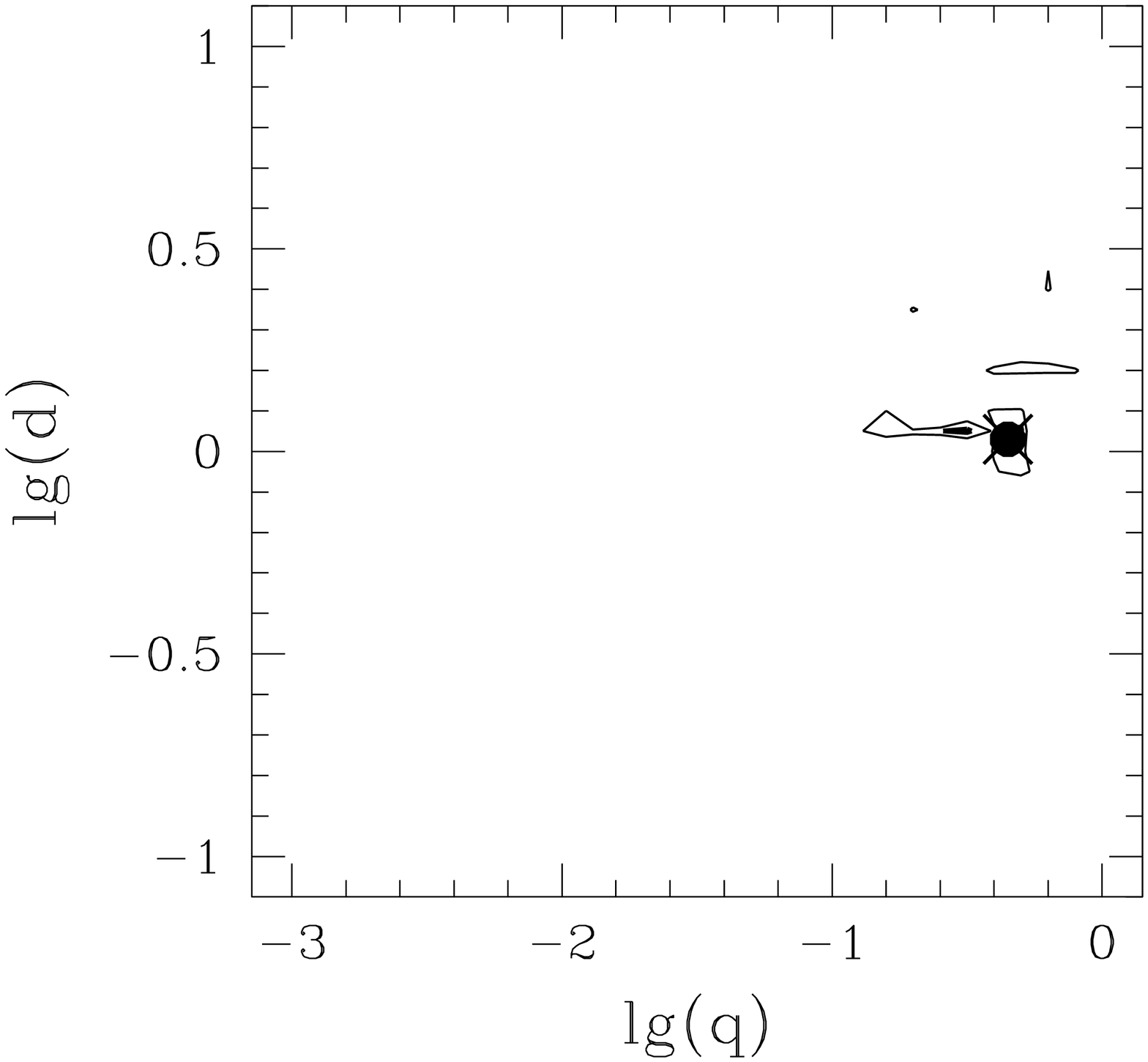}%
}

\noindent\parbox{12.75cm}{
 \noindent{\bf OGLE 2003-BLG-267}                              

\vspace*{5pt}

 \includegraphics[height=46mm,width=42.5mm]{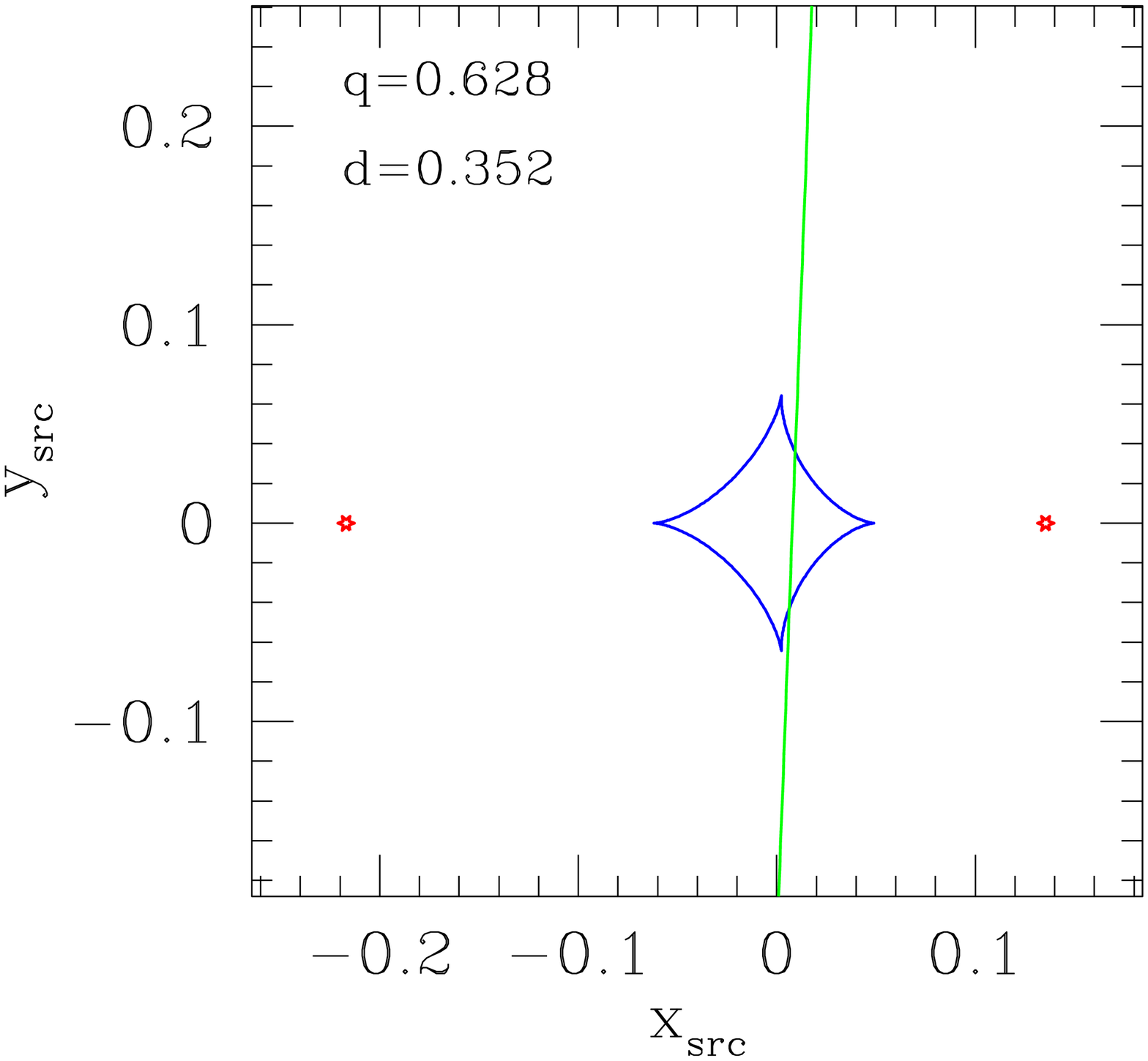}%
 \includegraphics[height=46mm,width=42.5mm]{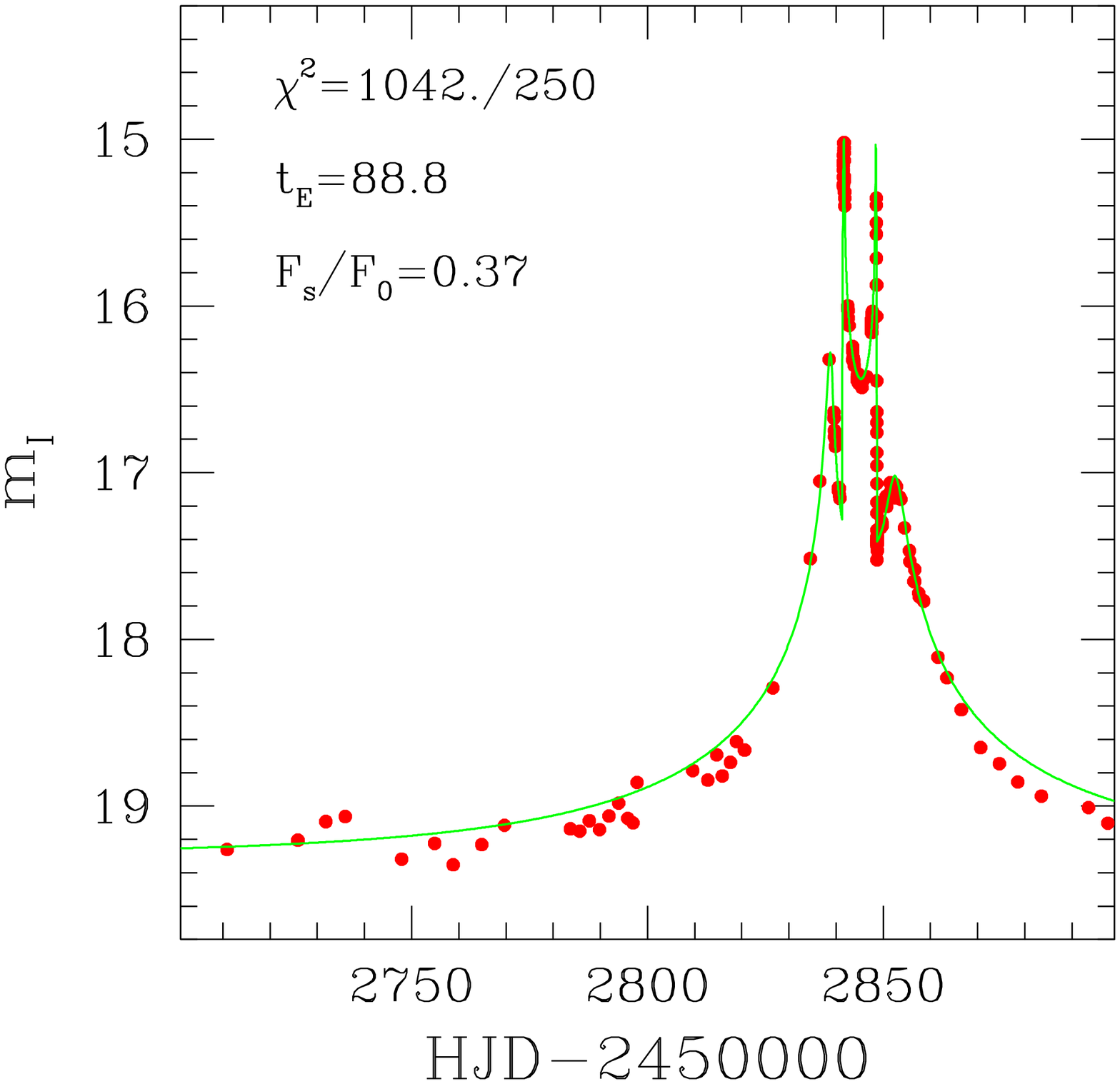}%
 \includegraphics[height=46mm,width=42.5mm]{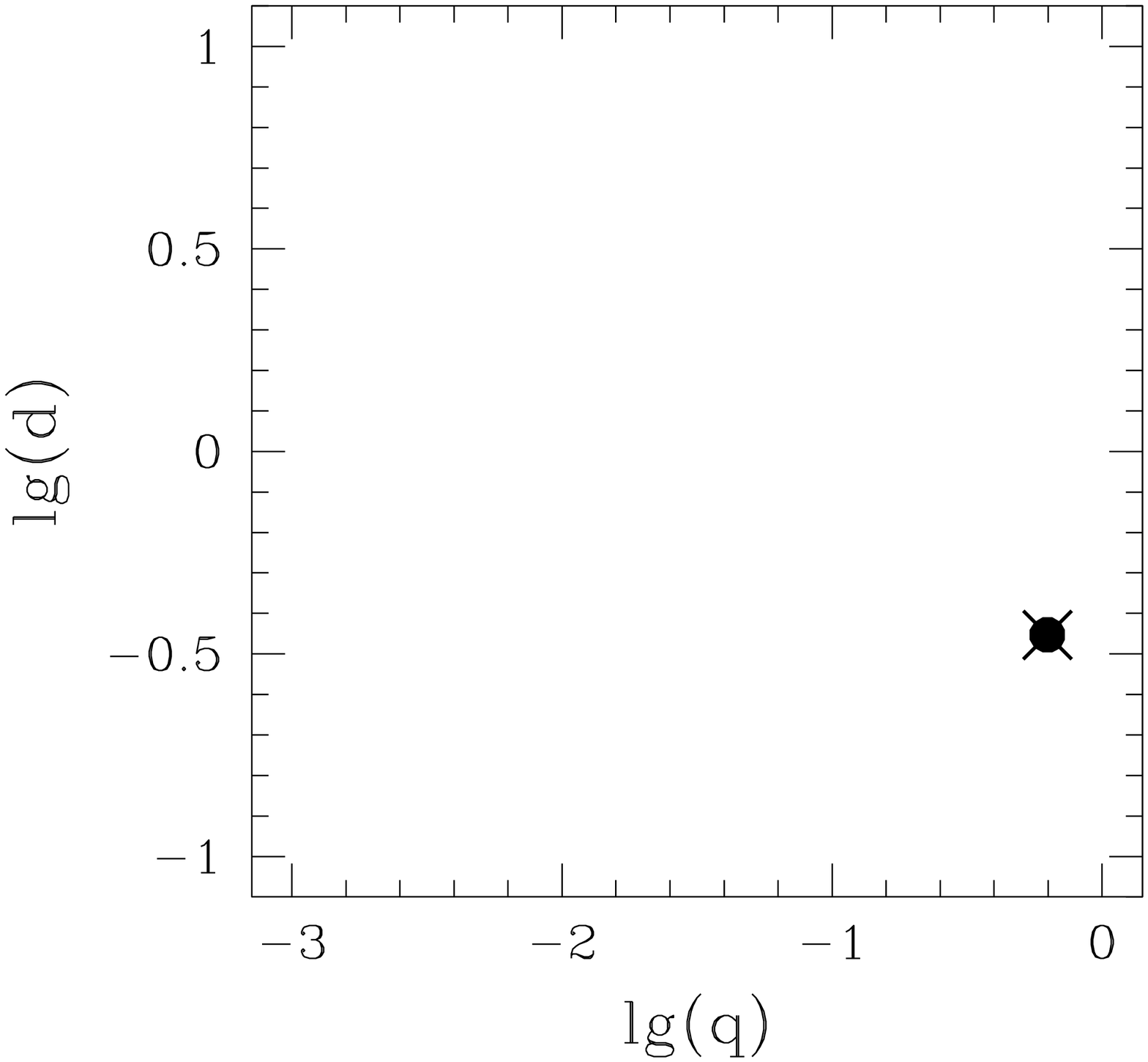}%
}

\noindent\parbox{12.75cm}{
 \noindent{\bf OGLE 2003-BLG-291}           

\vspace*{5pt}

 \includegraphics[height=46mm,width=42.5mm]{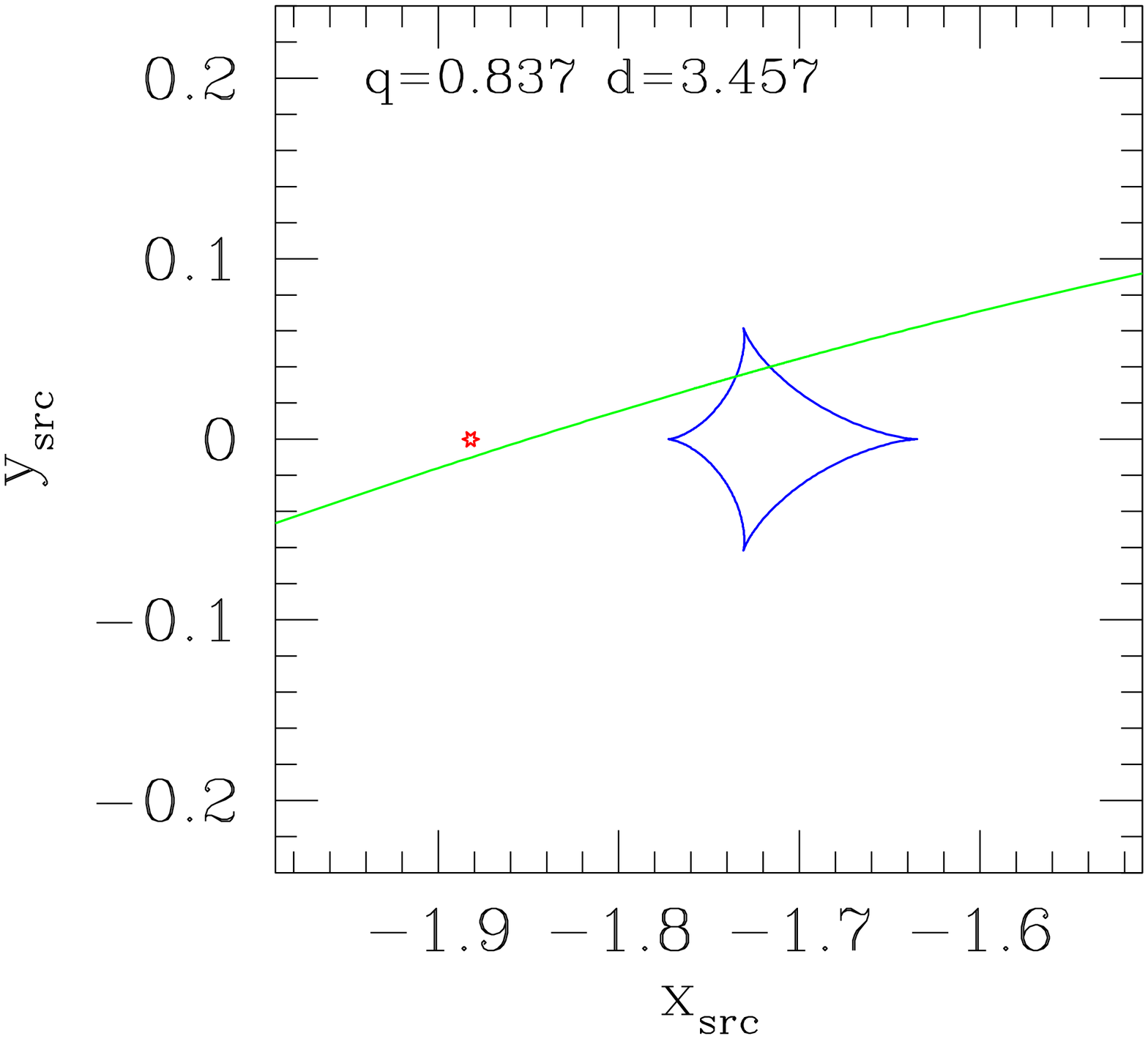}%
 \includegraphics[height=46mm,width=42.5mm]{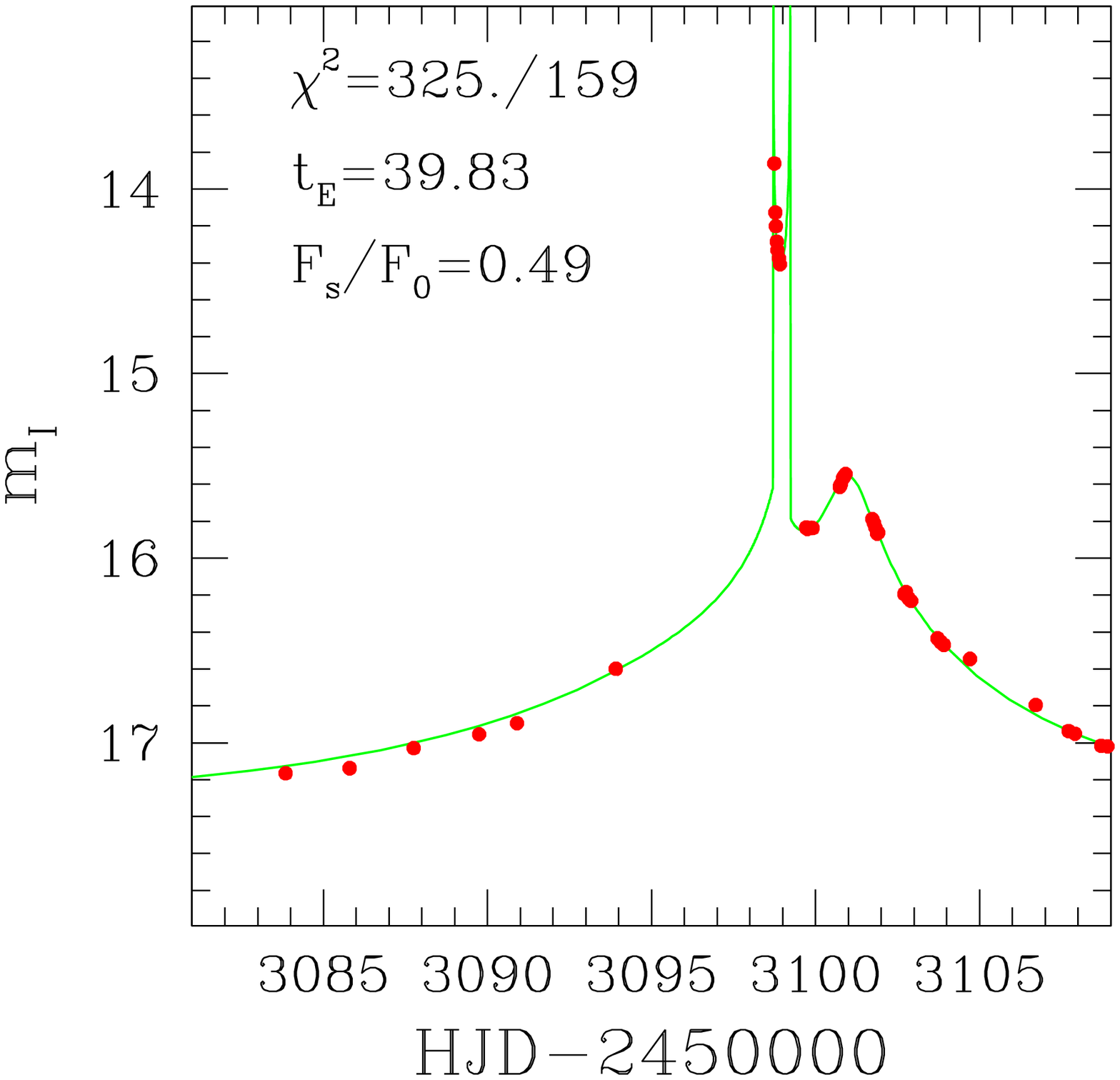}%
 \includegraphics[height=46mm,width=42.5mm]{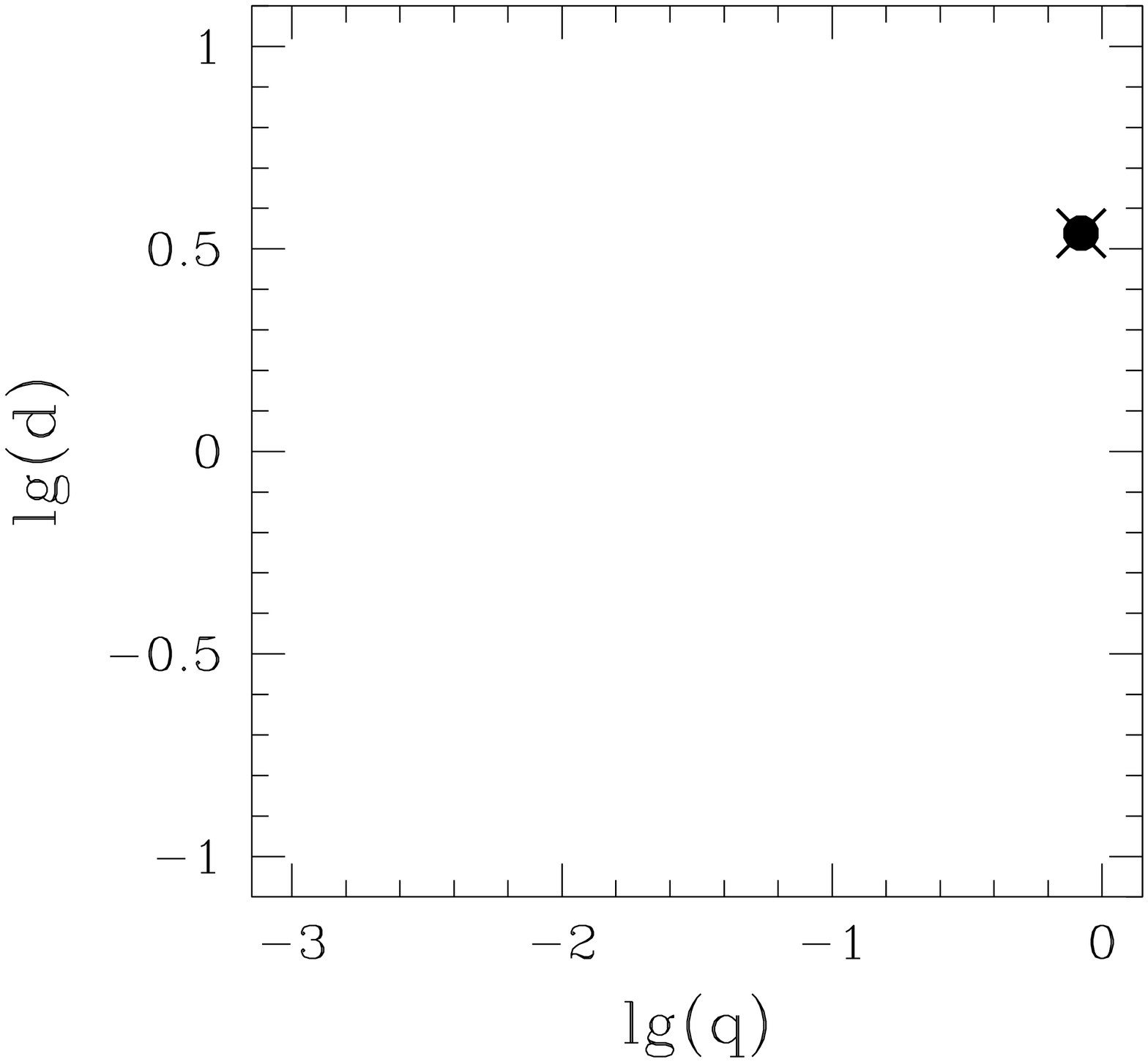}%
}

\noindent\parbox{12.75cm}{
 \noindent{\bf OGLE 2003-BLG-340 (1st model)}   

\vspace*{5pt}

 \includegraphics[height=46mm,width=42.5mm]{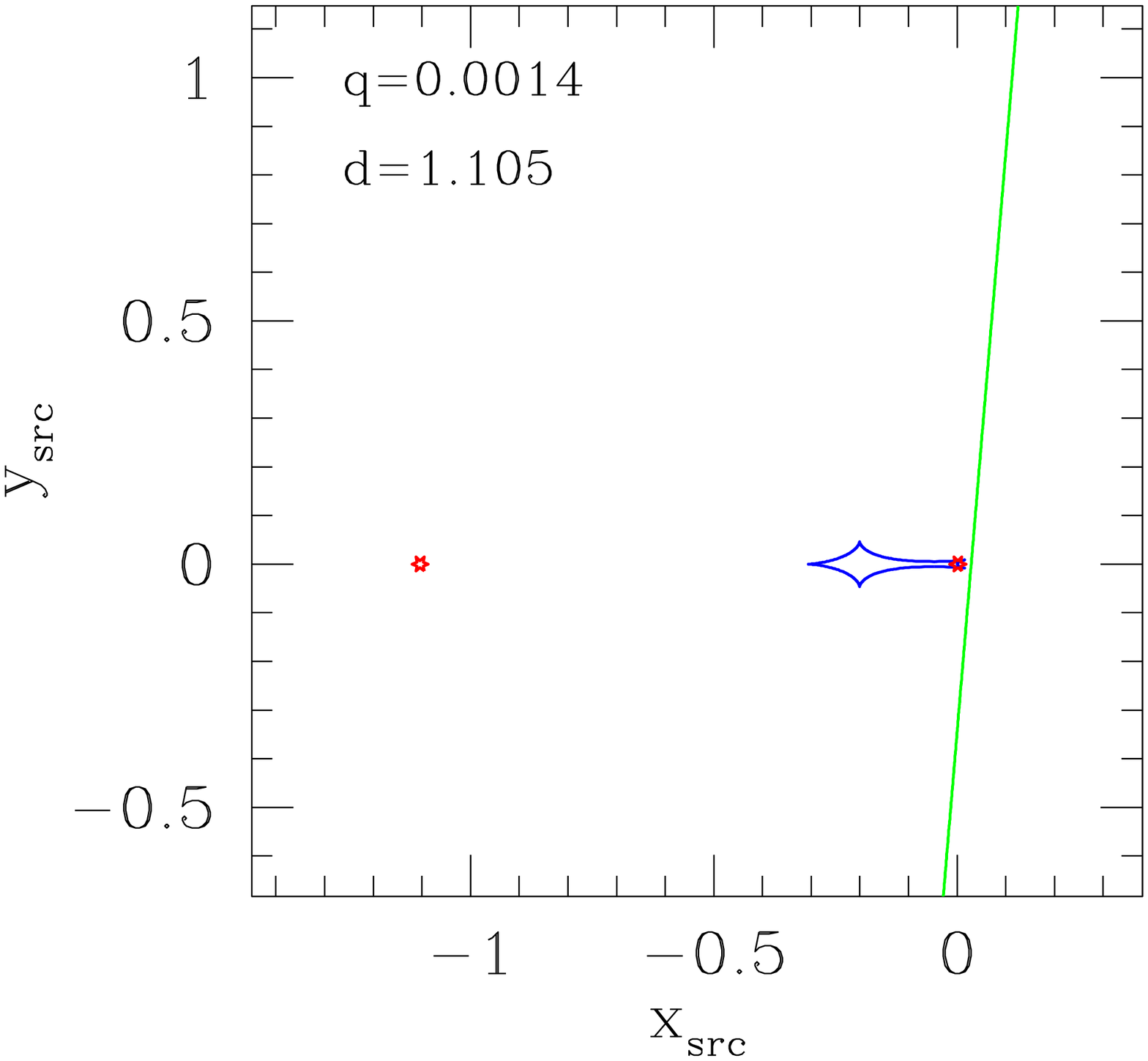}%
 \includegraphics[height=46mm,width=42.5mm]{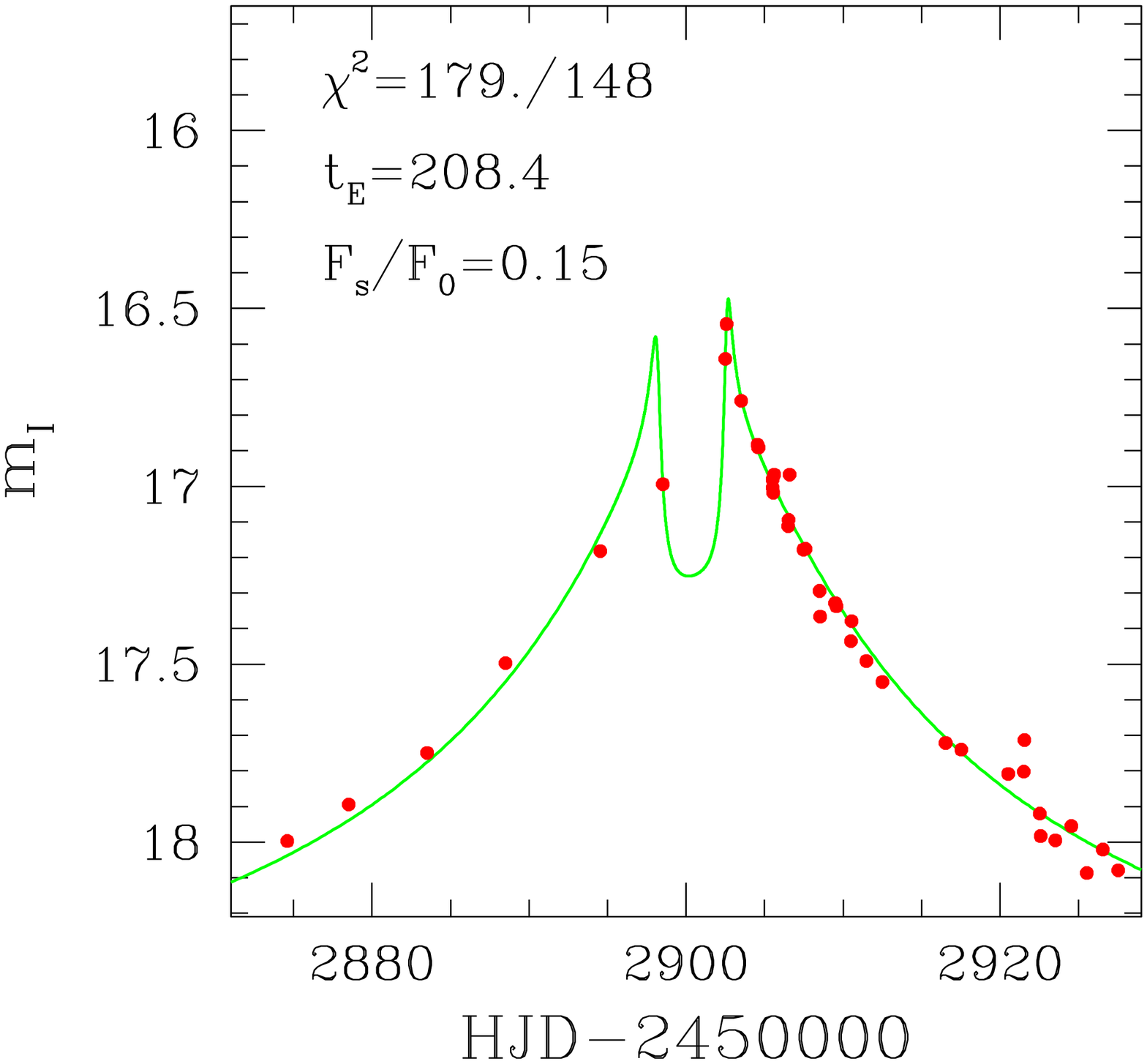}%
 \includegraphics[height=46mm,width=42.5mm]{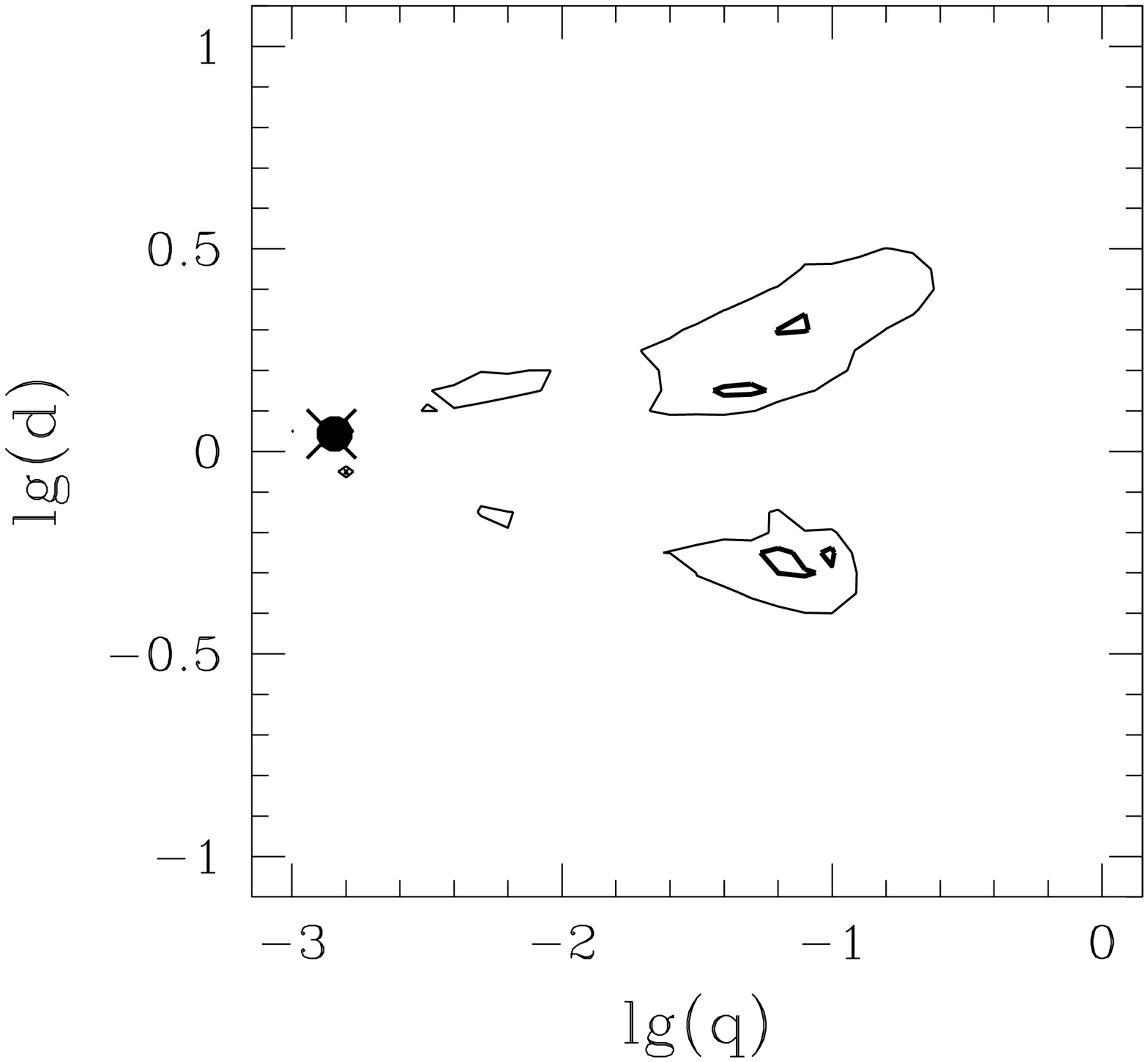}%
}

\noindent\parbox{12.75cm}{
 \noindent{\bf OGLE 2003-BLG-340 (2nd model)}   

\vspace*{5pt}

 \includegraphics[height=46mm,width=42.5mm]{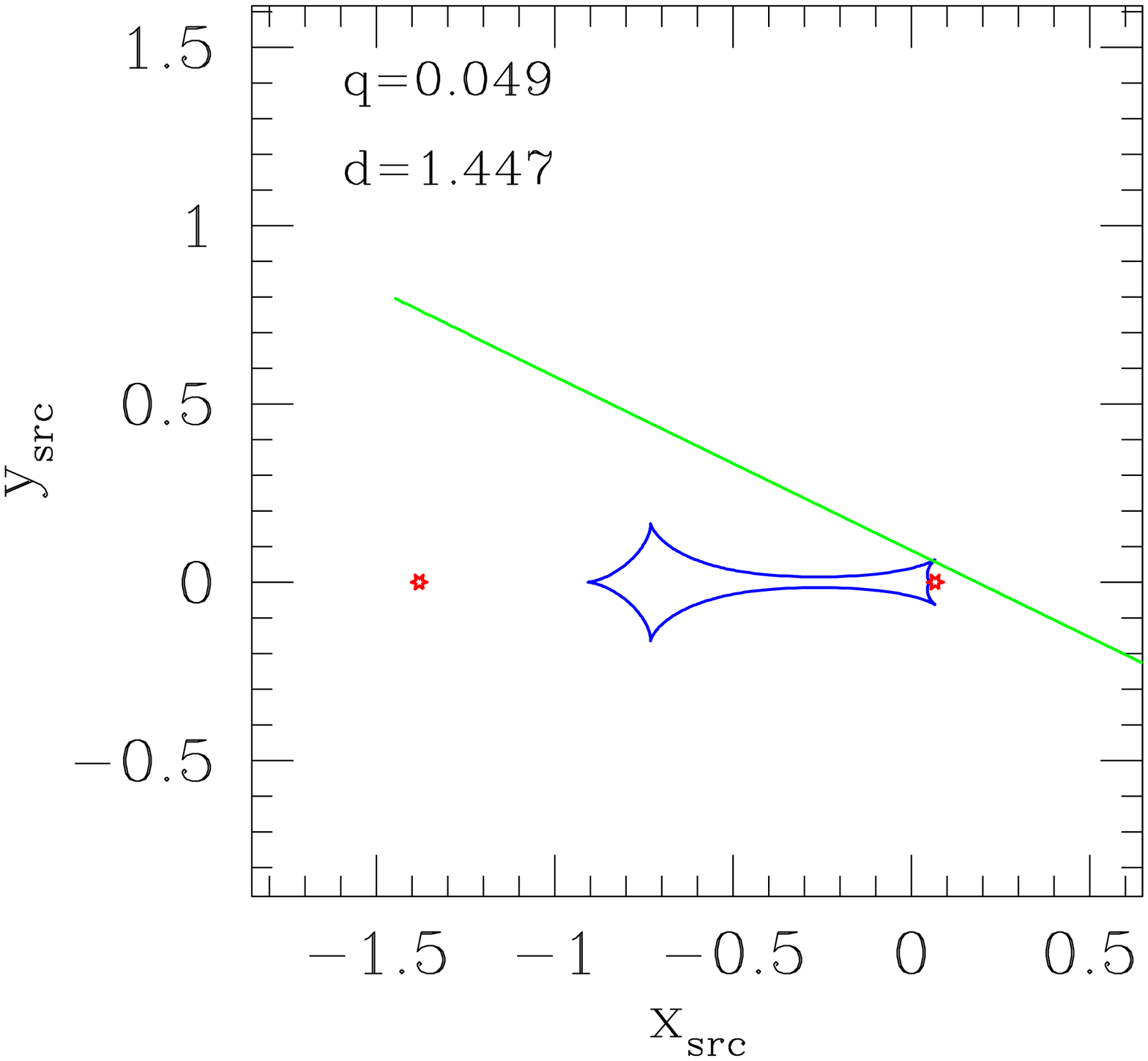}%
 \includegraphics[height=46mm,width=42.5mm]{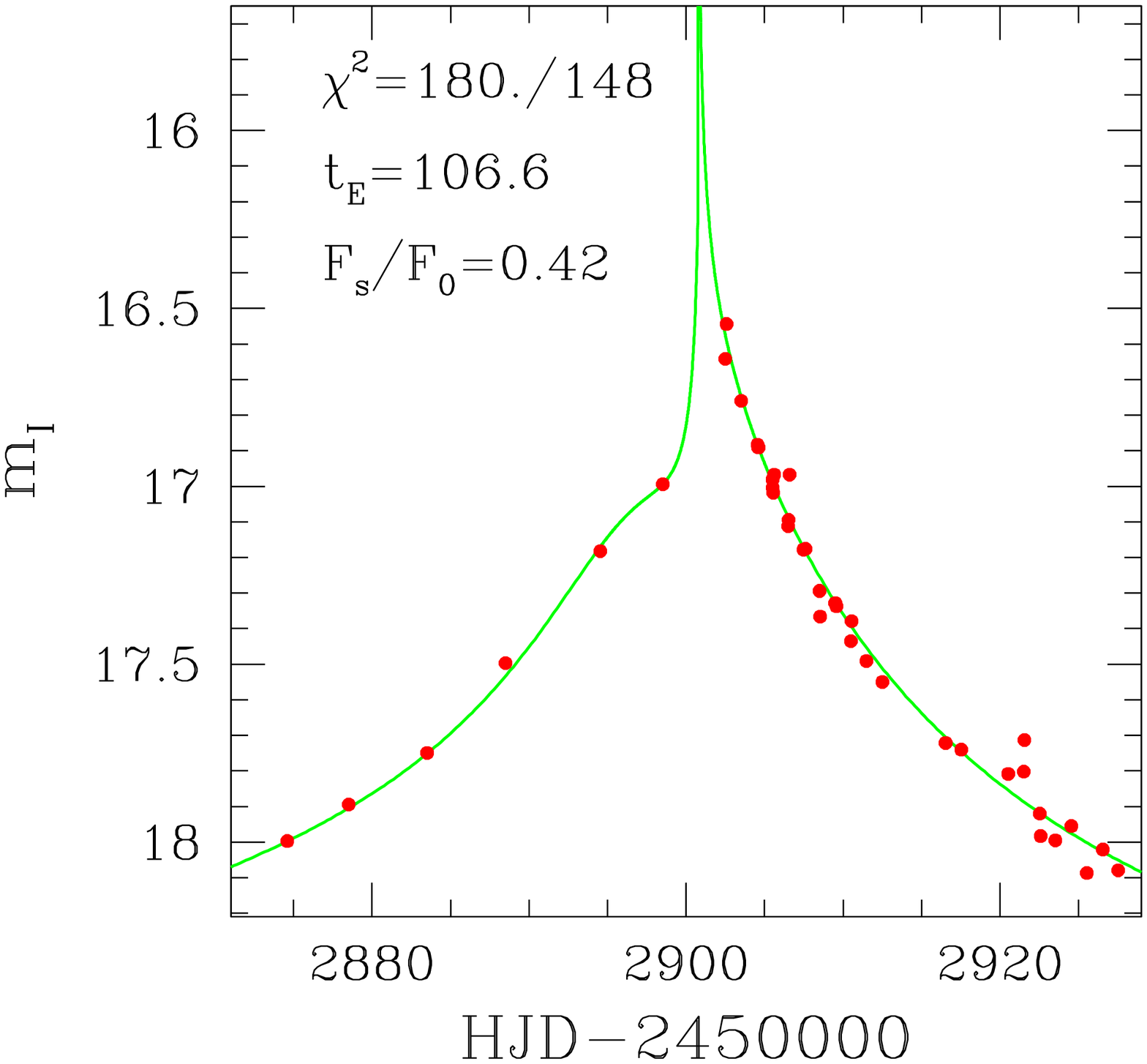}%
 \includegraphics[height=46mm,width=42.5mm]{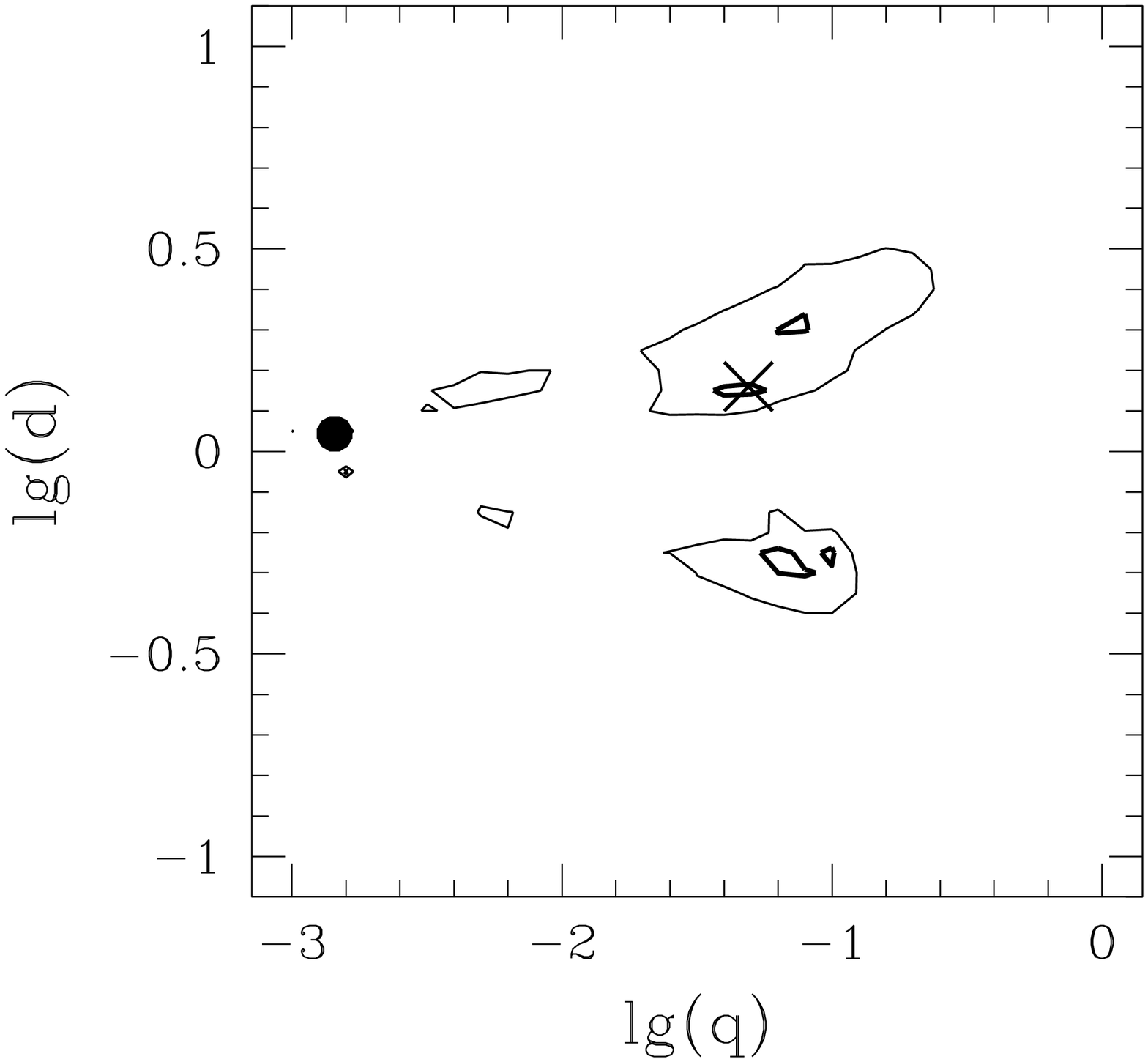}%
}

\noindent\parbox{12.75cm}{
 \noindent{\bf OGLE 2003-BLG-380}                              

\vspace*{5pt}

 \includegraphics[height=46mm,width=42.5mm]{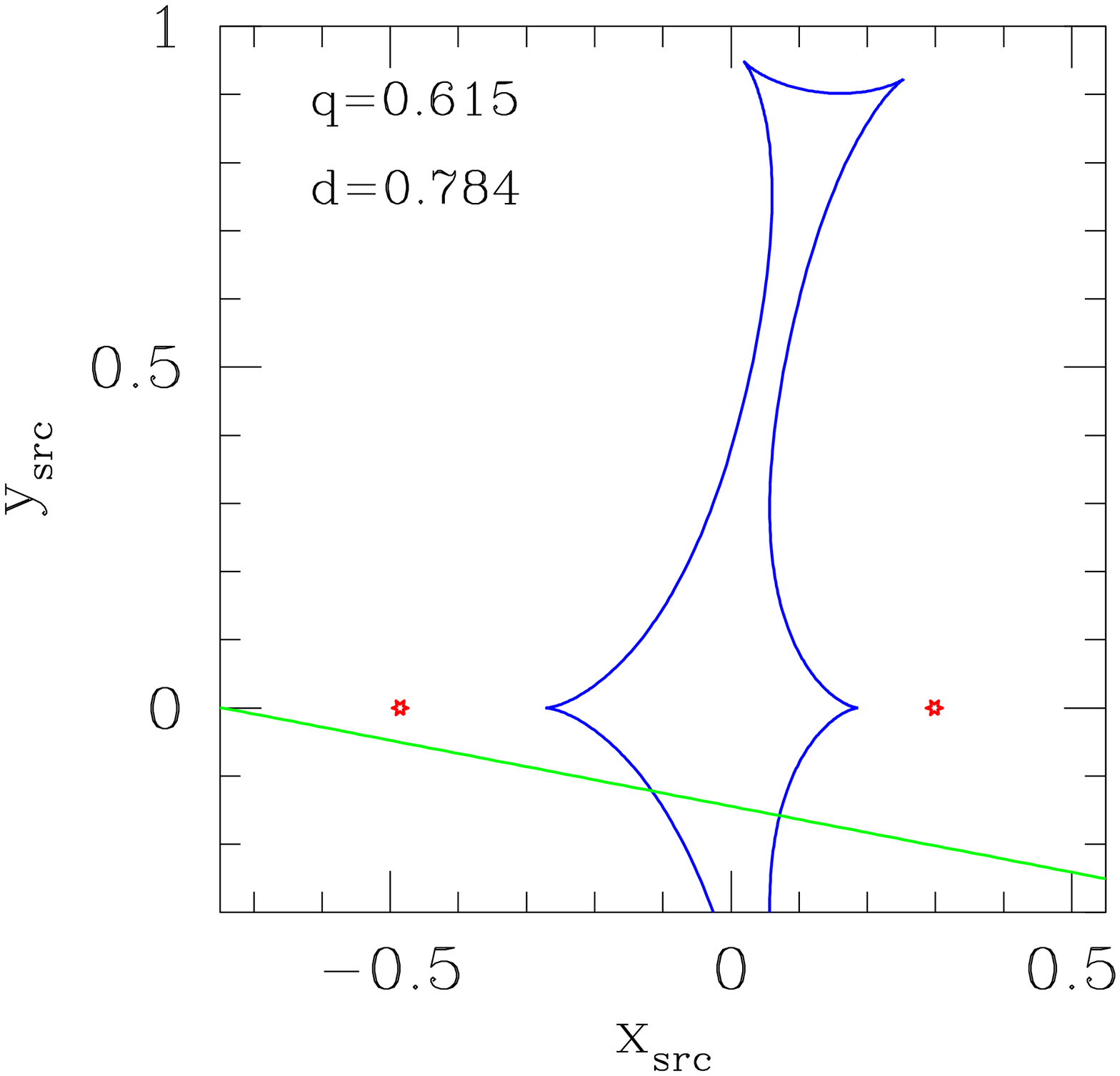}%
 \includegraphics[height=46mm,width=42.5mm]{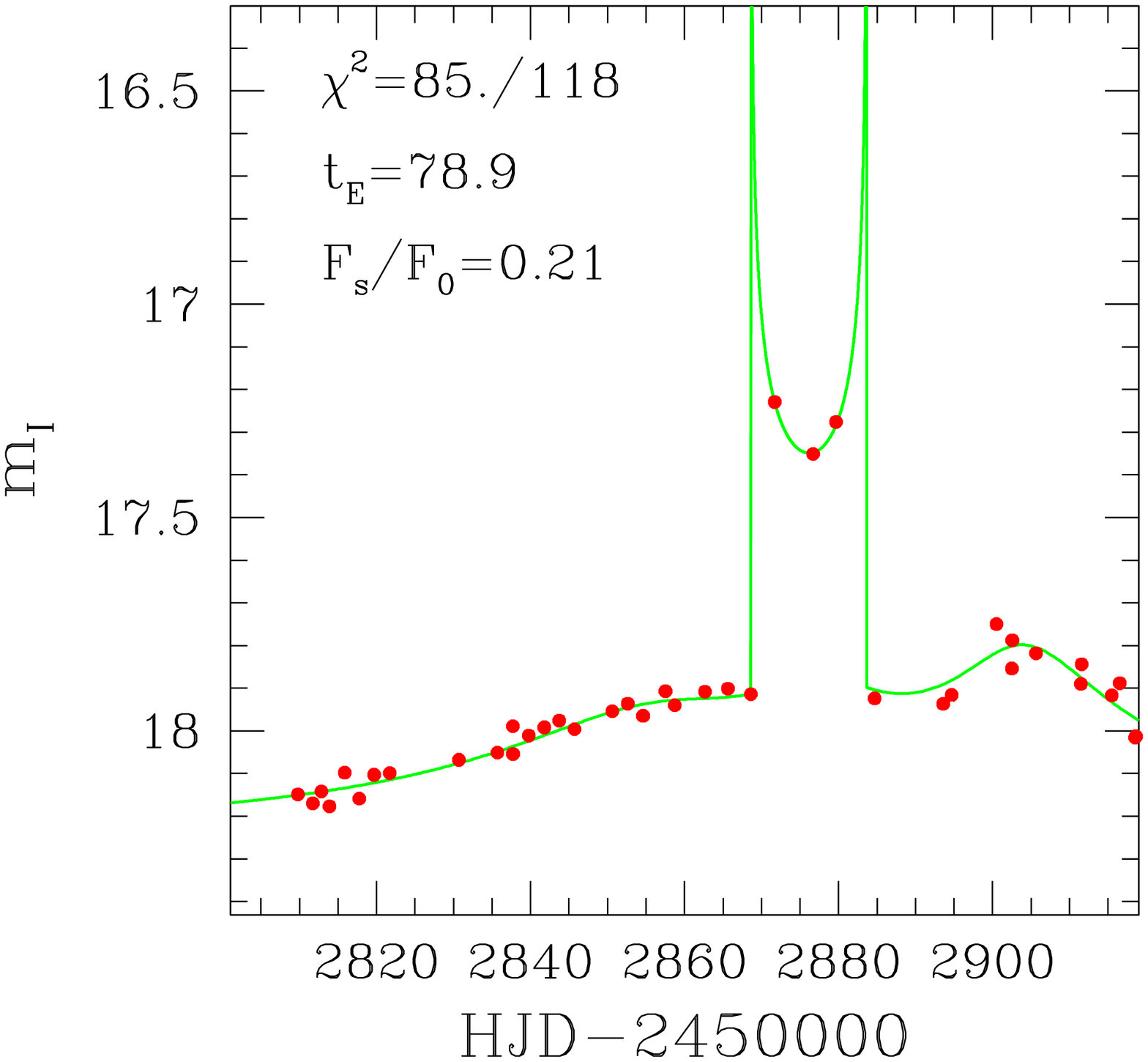}%
 \includegraphics[height=46mm,width=42.5mm]{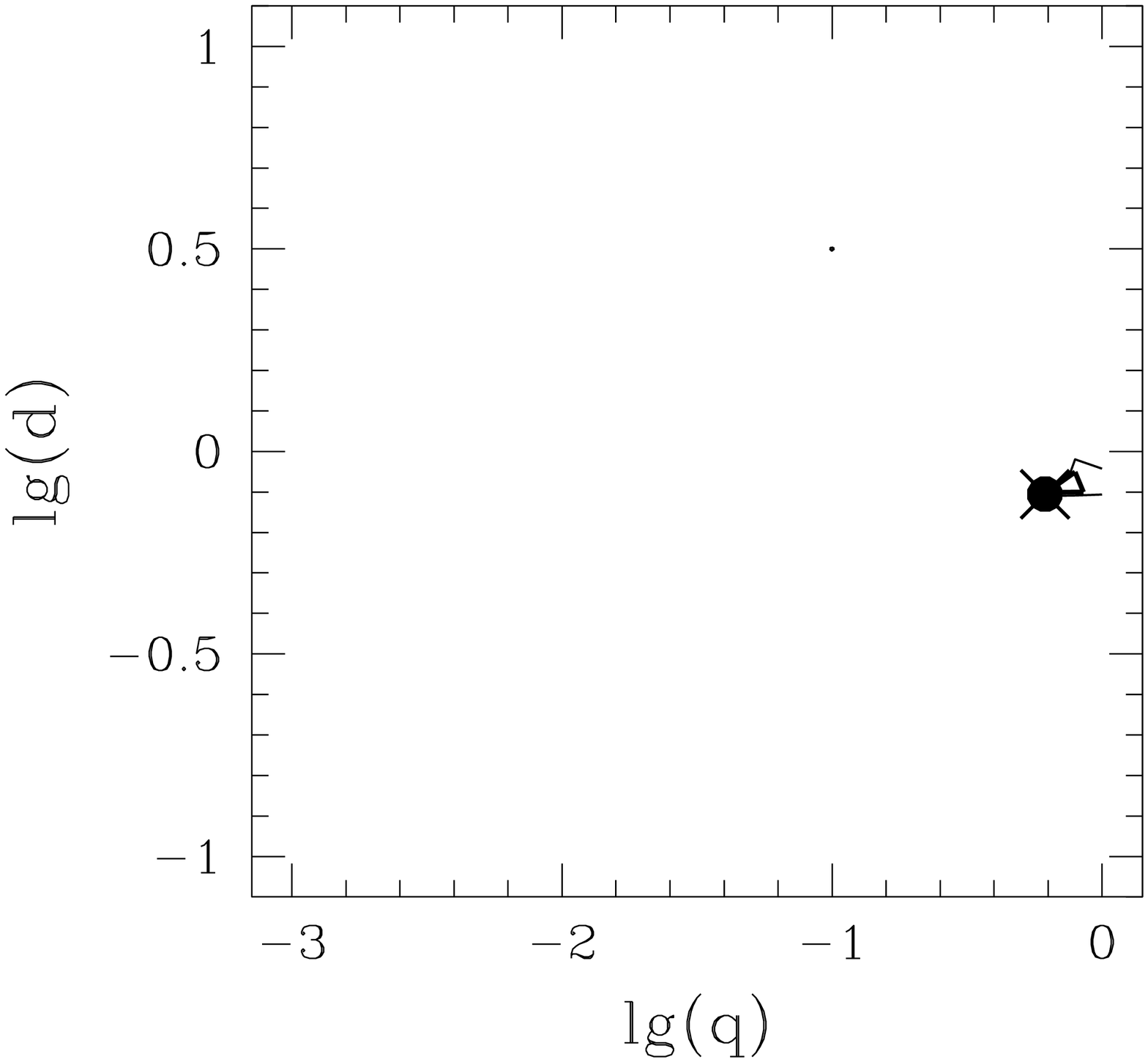}%
}

\newpage
\centerline{{\bf Appendix~2: Ambiguous Binary Lens / Double Source Events}}
\vskip6pt
In the following we show events which have two kinds of models of
comparable quality. The binary lens models are shown on the left and
double source models -- on the right. Since the flux of a double source
is a linear combination of its components and the blend, we use flux
units in the plots. (One flux unit corresponds to ${I=21}$~mag.) The
resulting light curves are shown as thick solid lines, while
observations are marked as dots. For the double source models we also
show the light curves of the contributing components and of the blend
using thin dotted lines.  (More information relating to the binary
models of the events can be found in Appendix~1.)

\vspace{0.5cm}

\noindent\parbox{12.75cm}{
 \noindent{\bf OGLE 2002-BLG-099}

\vspace*{5pt}

 \includegraphics[height=63mm,width=62.0mm]{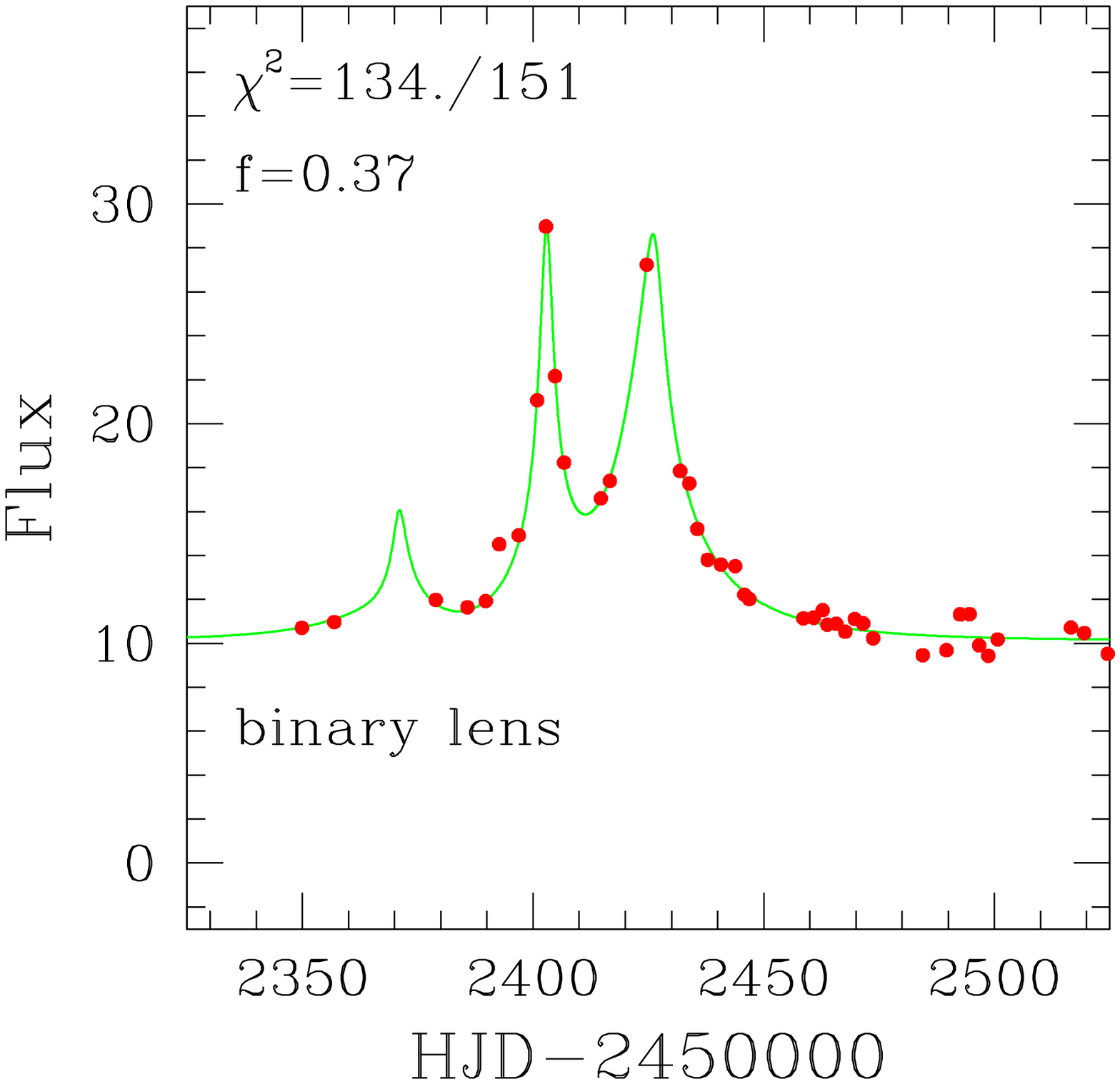}%
 \includegraphics[height=63mm,width=62.0mm]{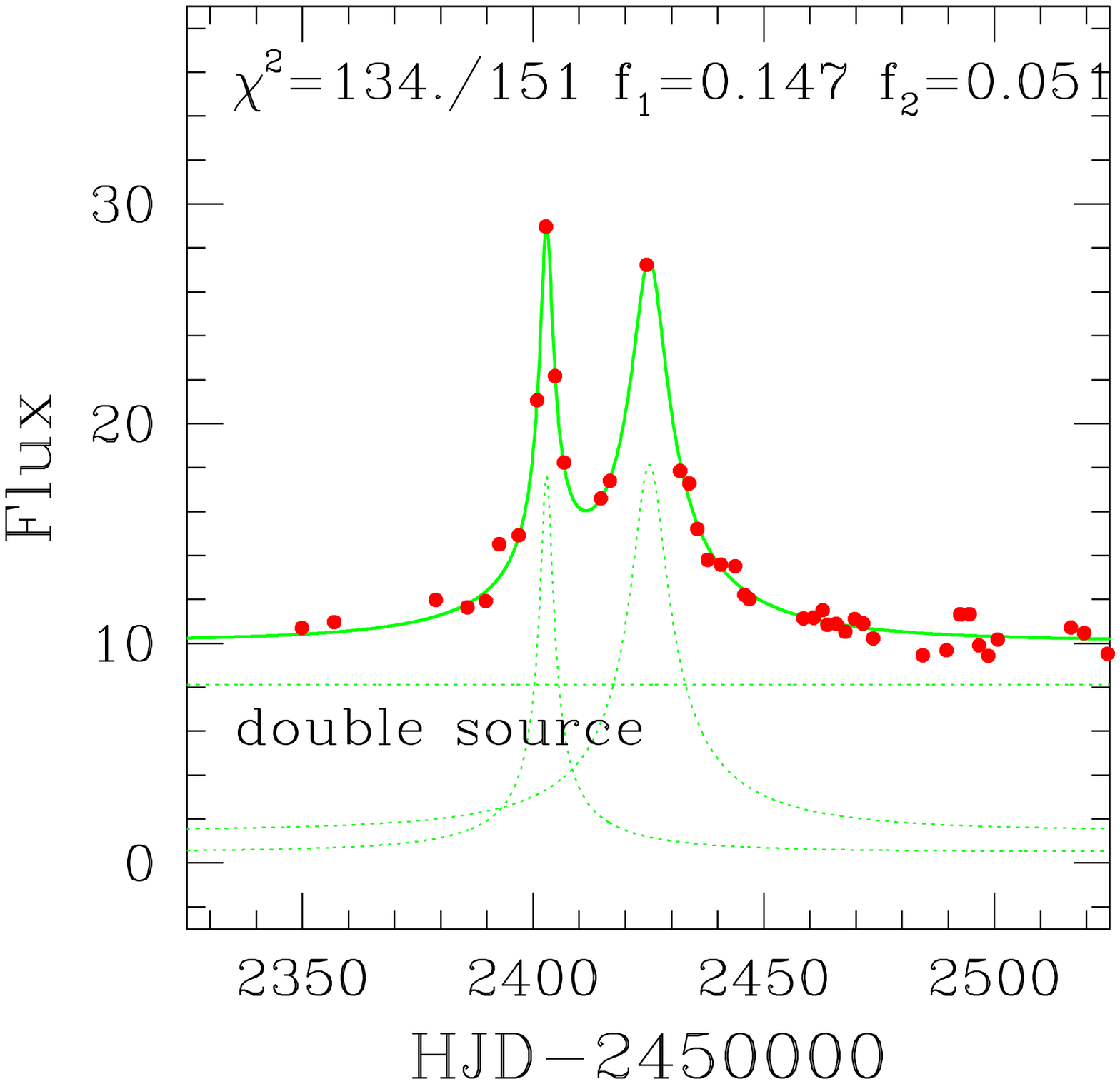}%
}

\noindent\parbox{12.75cm}{
 \noindent{\bf OGLE 2002-BLG-135}

\vspace*{5pt}

 \includegraphics[height=63mm,width=62.0mm]{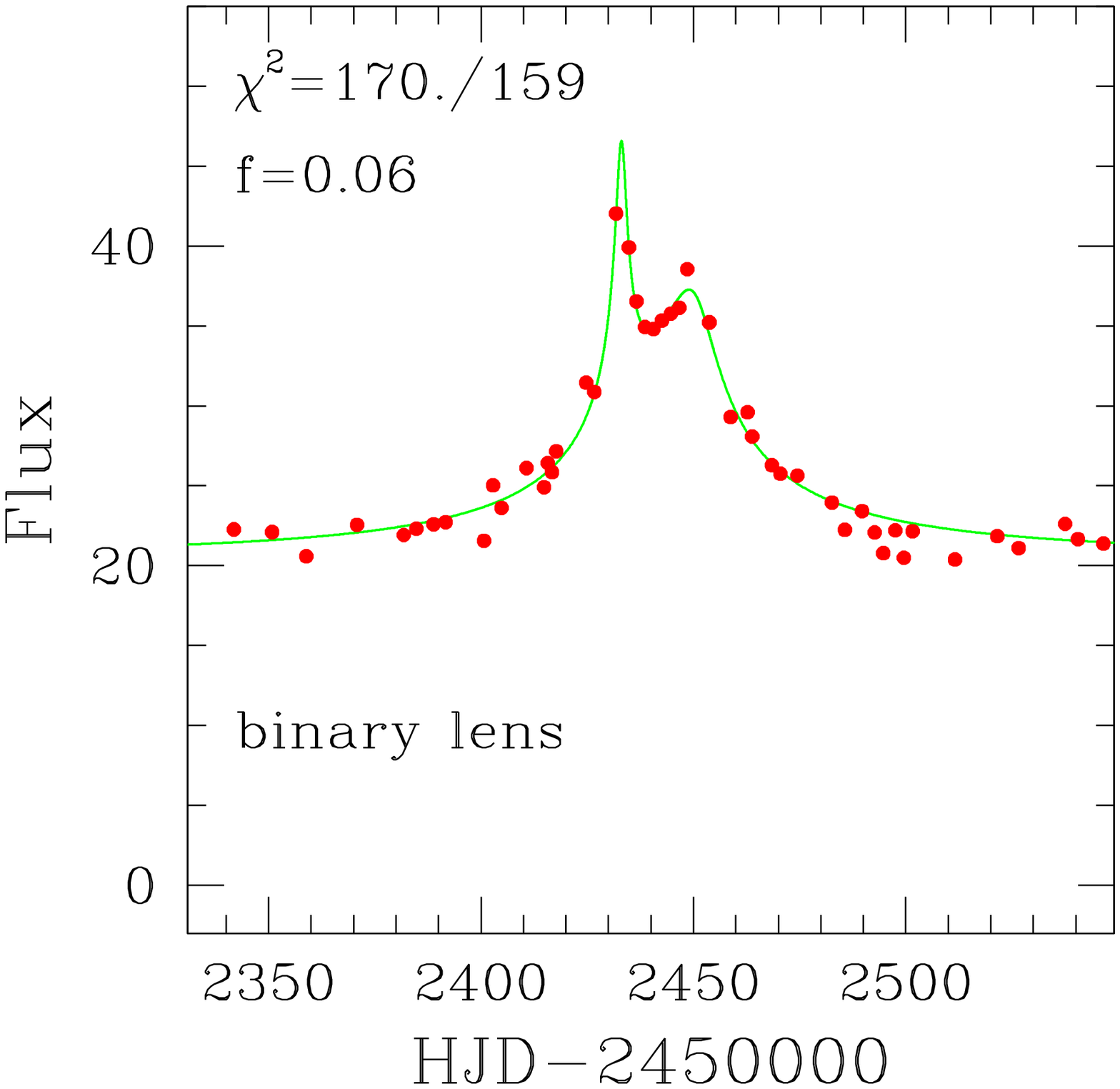}%
 \includegraphics[height=63mm,width=62.0mm]{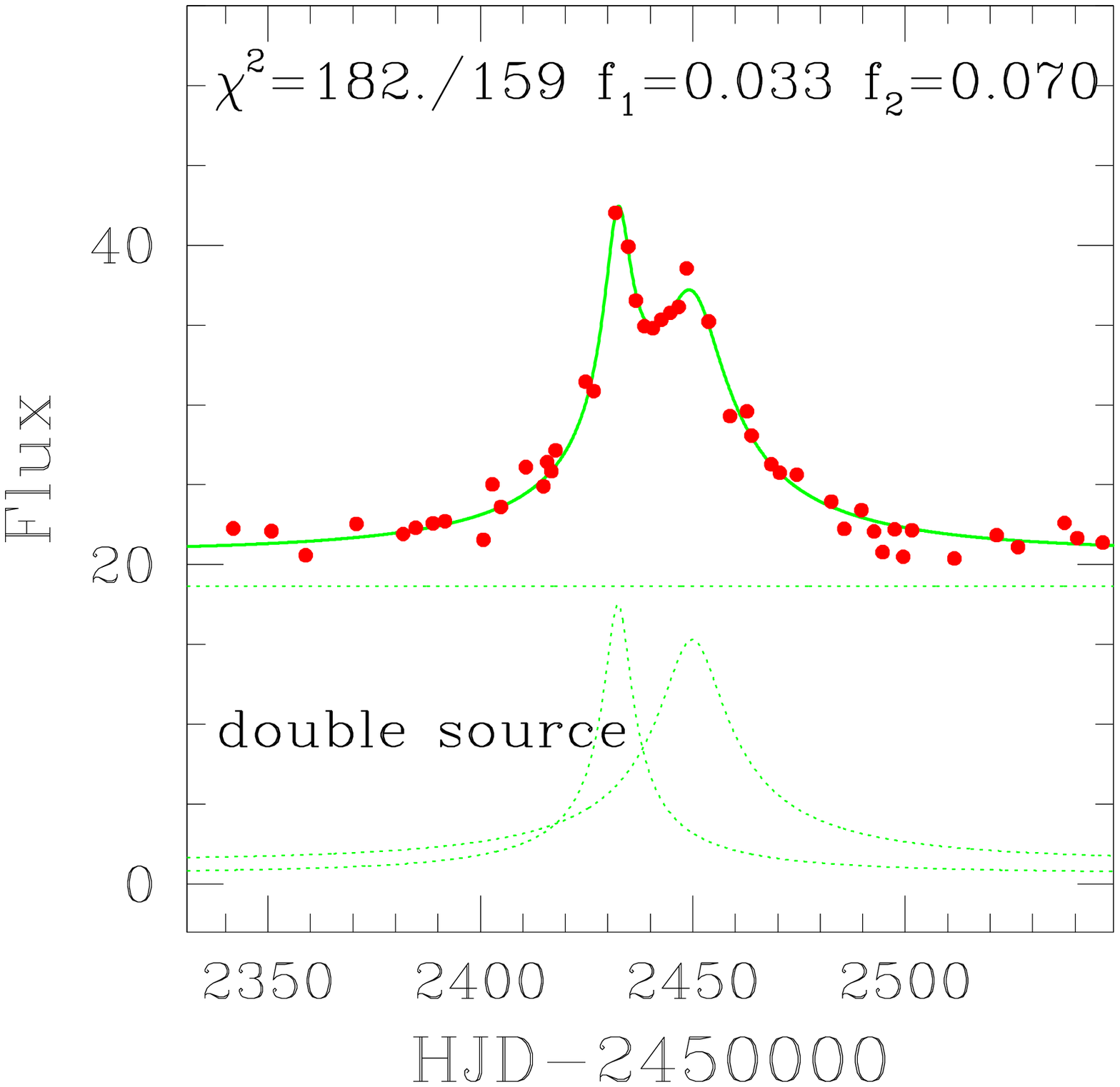}%
}

\noindent\parbox{12.75cm}{
 \noindent{\bf OGLE 2002-BLG-158}

\vspace*{5pt}

 \includegraphics[height=63mm,width=62.0mm]{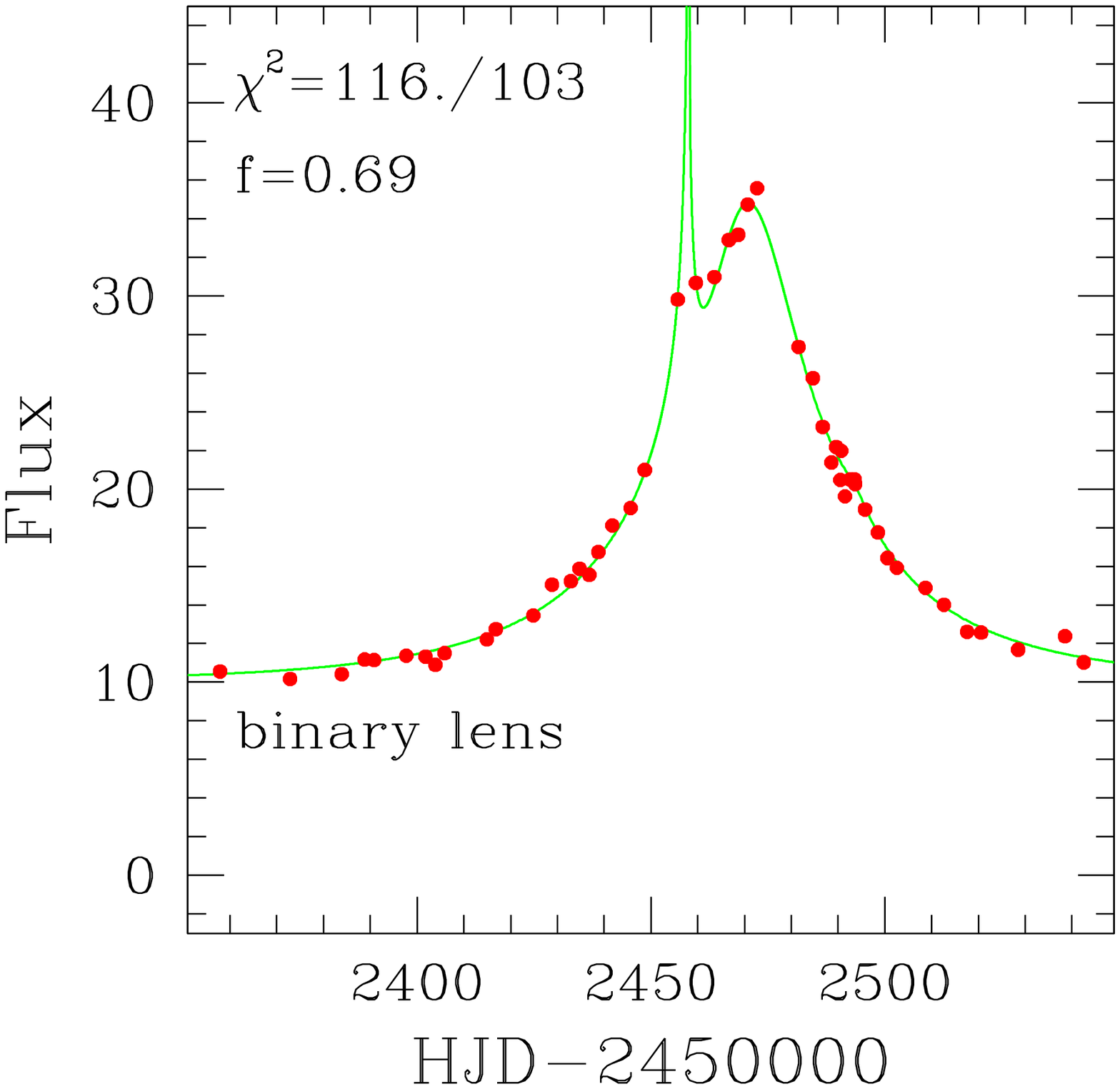}%
 \includegraphics[height=63mm,width=62.0mm]{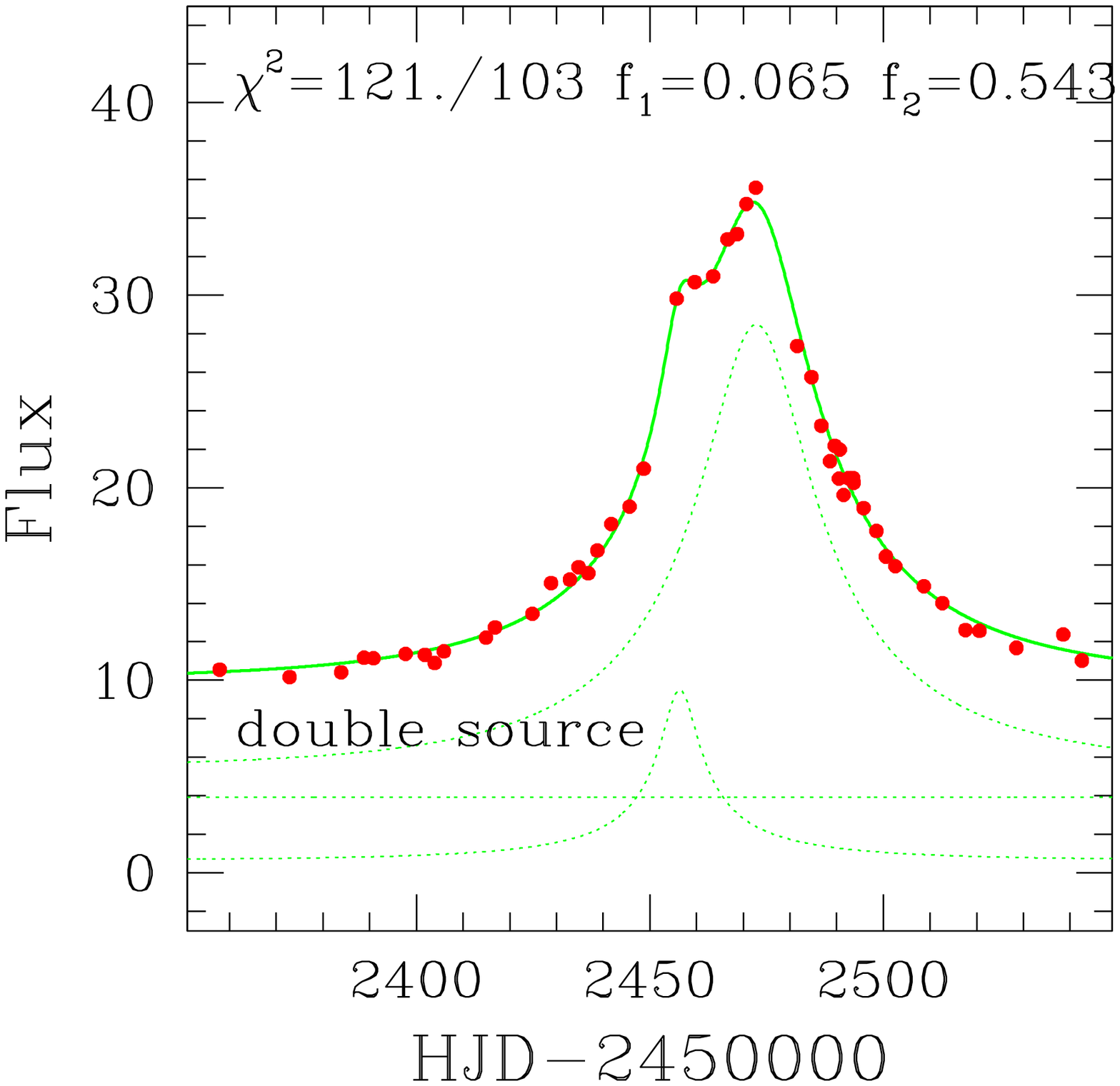}%
}

\noindent\parbox{12.75cm}{
 \noindent{\bf OGLE 2002-BLG-256}

\vspace*{5pt}

 \includegraphics[height=63mm,width=62.0mm]{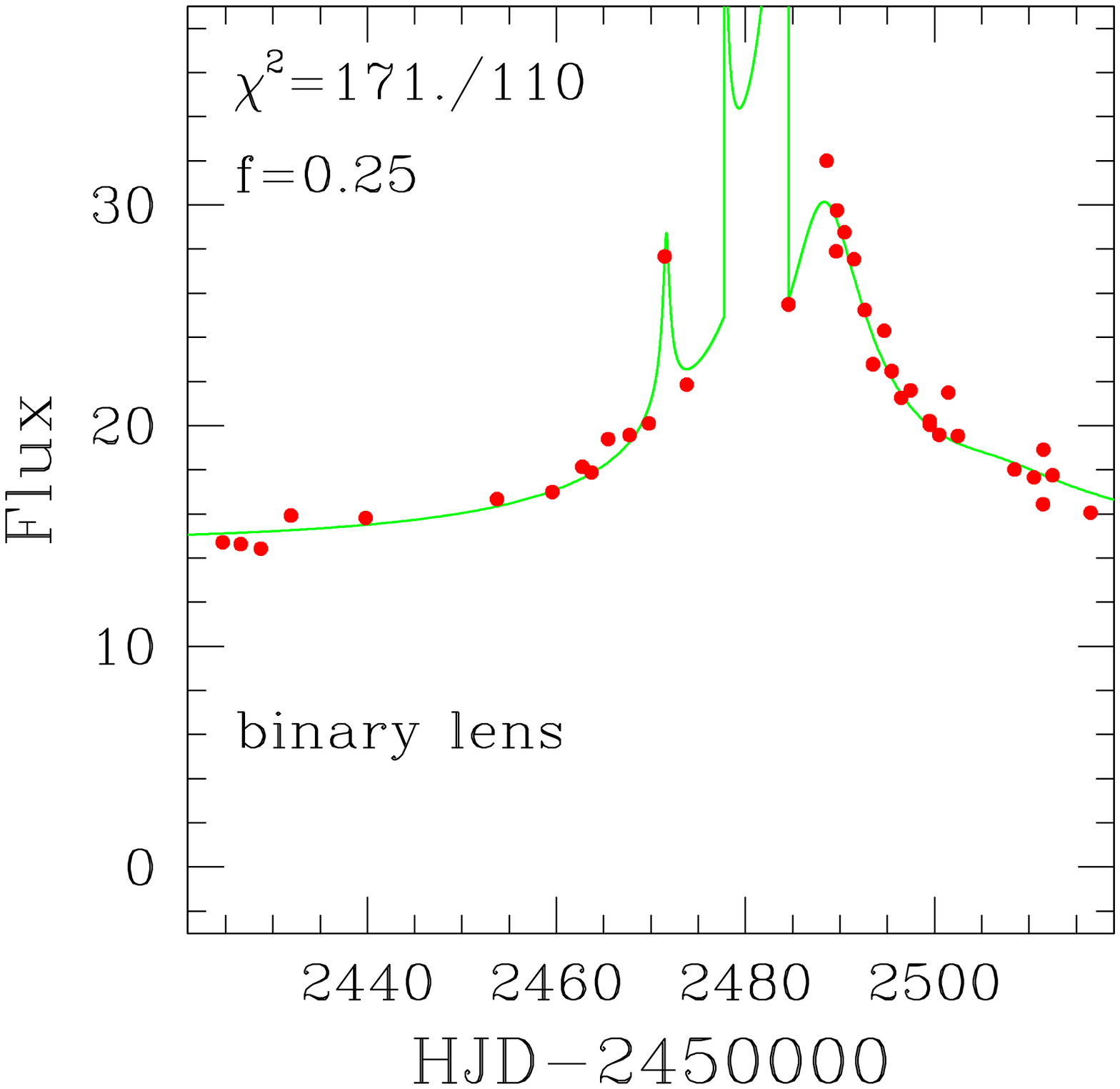}%
 \includegraphics[height=63mm,width=62.0mm]{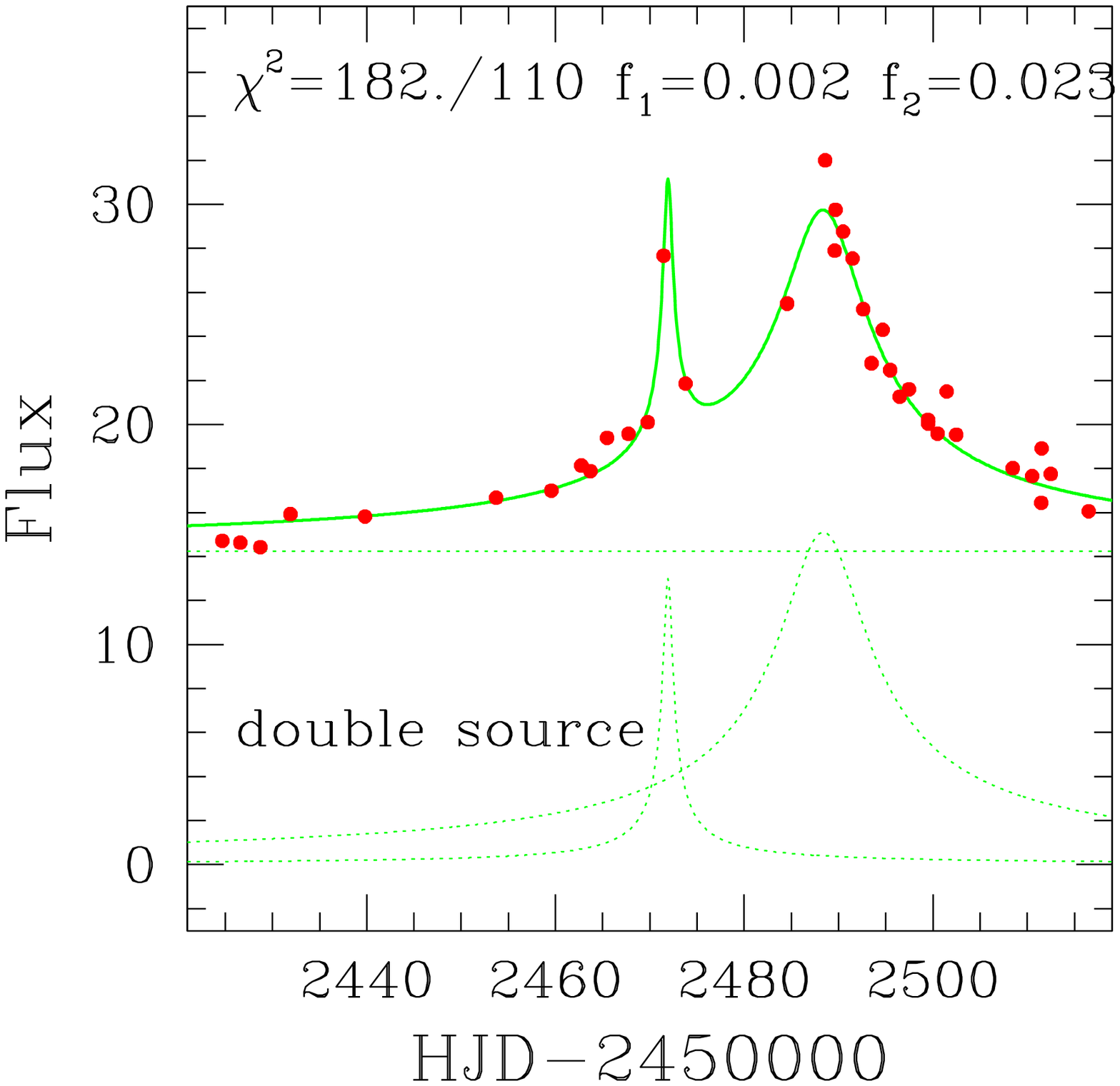}%
}

\noindent\parbox{12.75cm}{
 \noindent{\bf OGLE 2002-BLG-321}

\vspace*{5pt}

 \includegraphics[height=63mm,width=62.0mm]{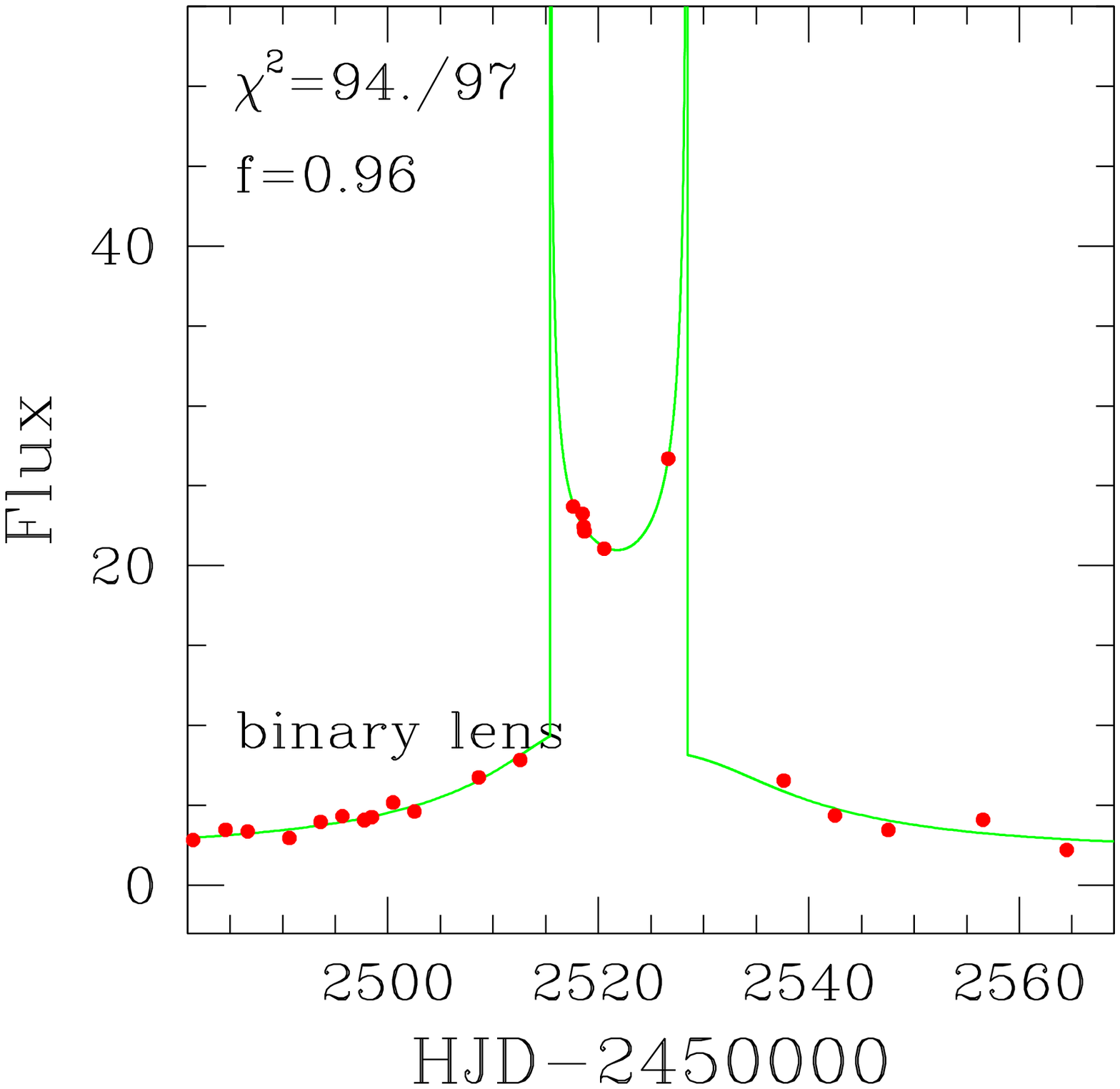}%
 \includegraphics[height=63mm,width=62.0mm]{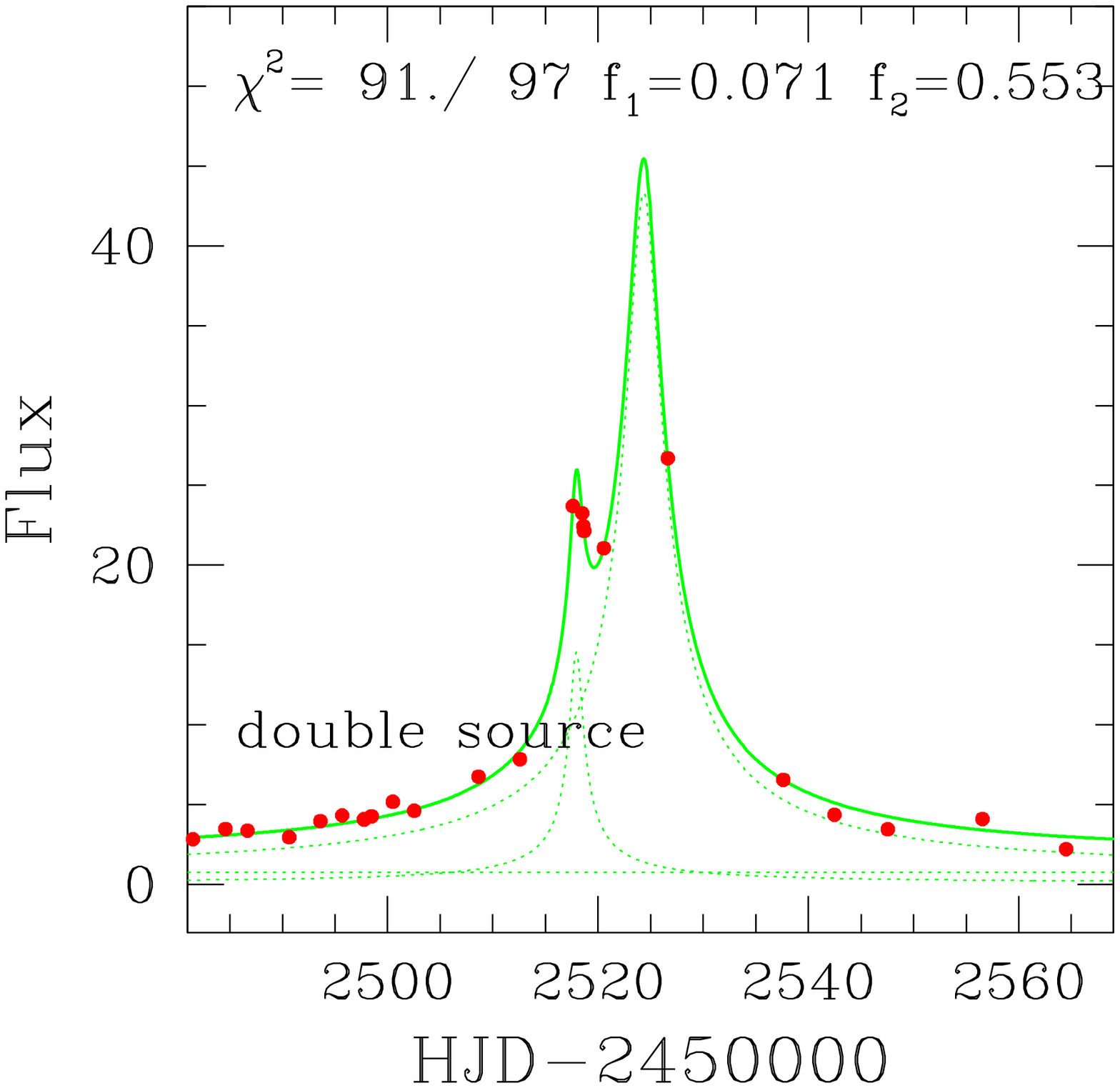}%
}

\noindent\parbox{12.75cm}{
 \noindent{\bf OGLE 2003-BLG-084}

\vspace*{5pt}

 \includegraphics[height=63mm,width=62.0mm]{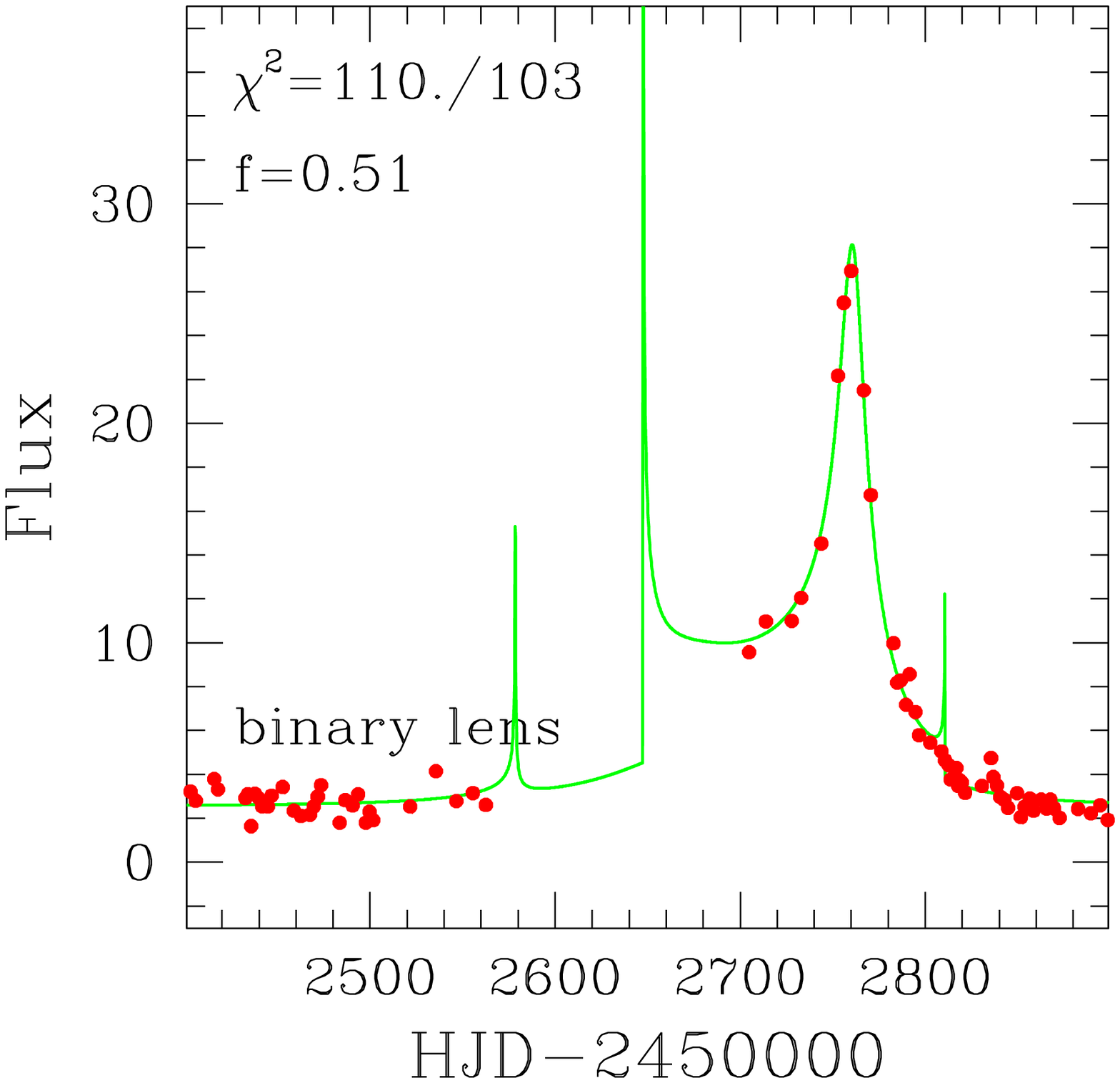}%
 \includegraphics[height=63mm,width=62.0mm]{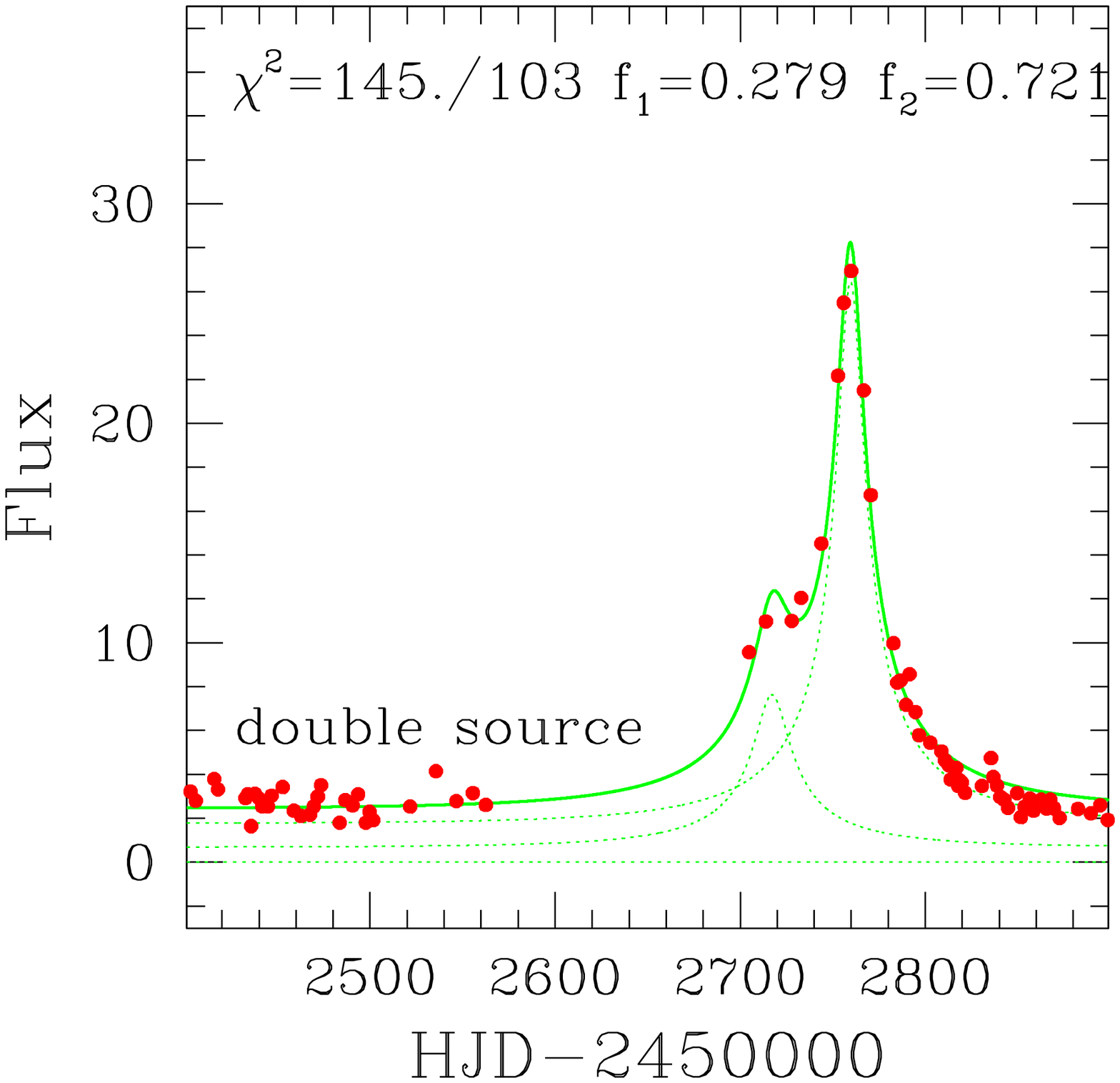}%
}

\noindent\parbox{12.75cm}{
\noindent{\bf OGLE 2003-BLG-194}

\vspace*{5pt}

 \includegraphics[height=63mm,width=62.0mm]{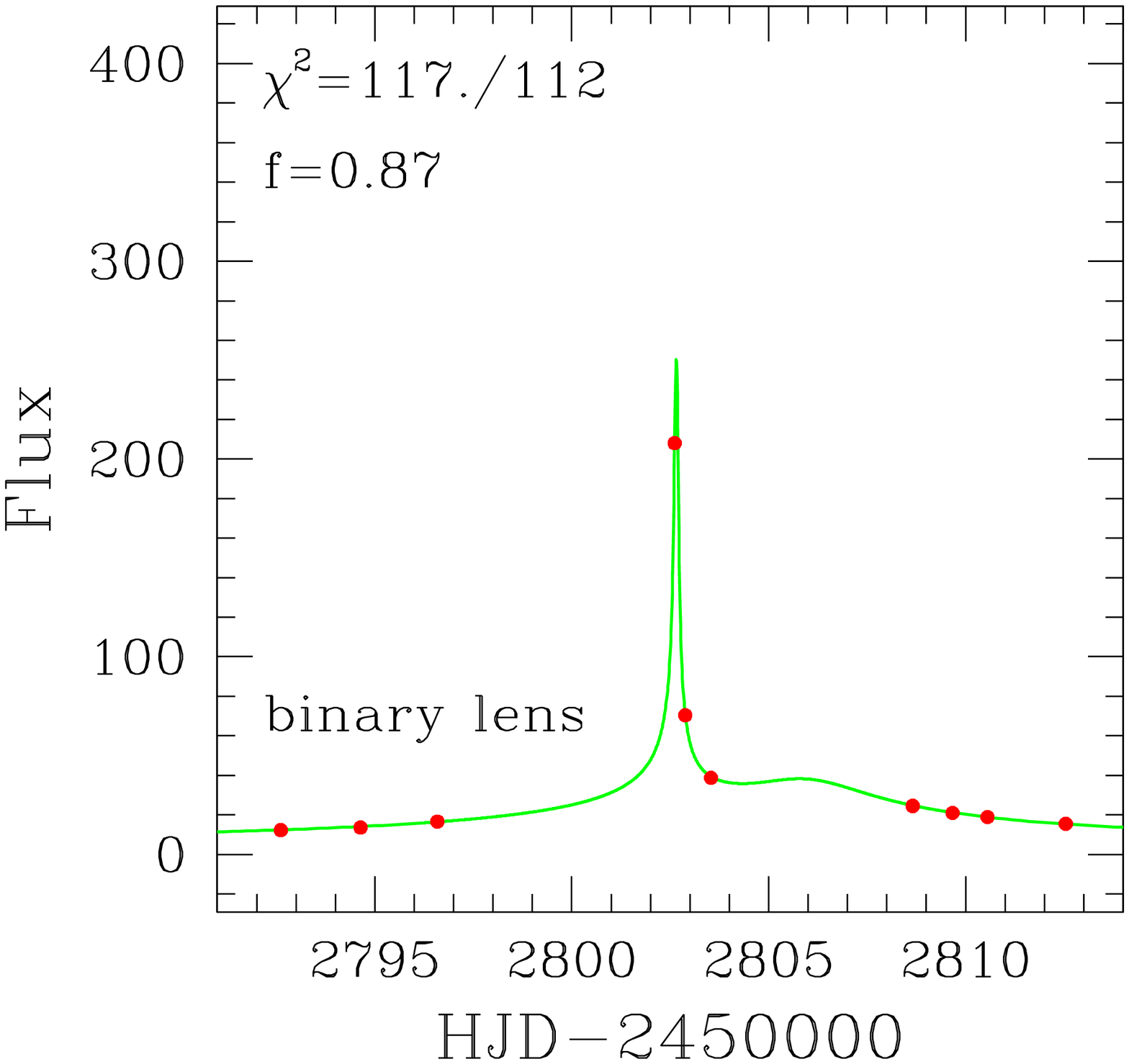}%
 \includegraphics[height=63mm,width=62.0mm]{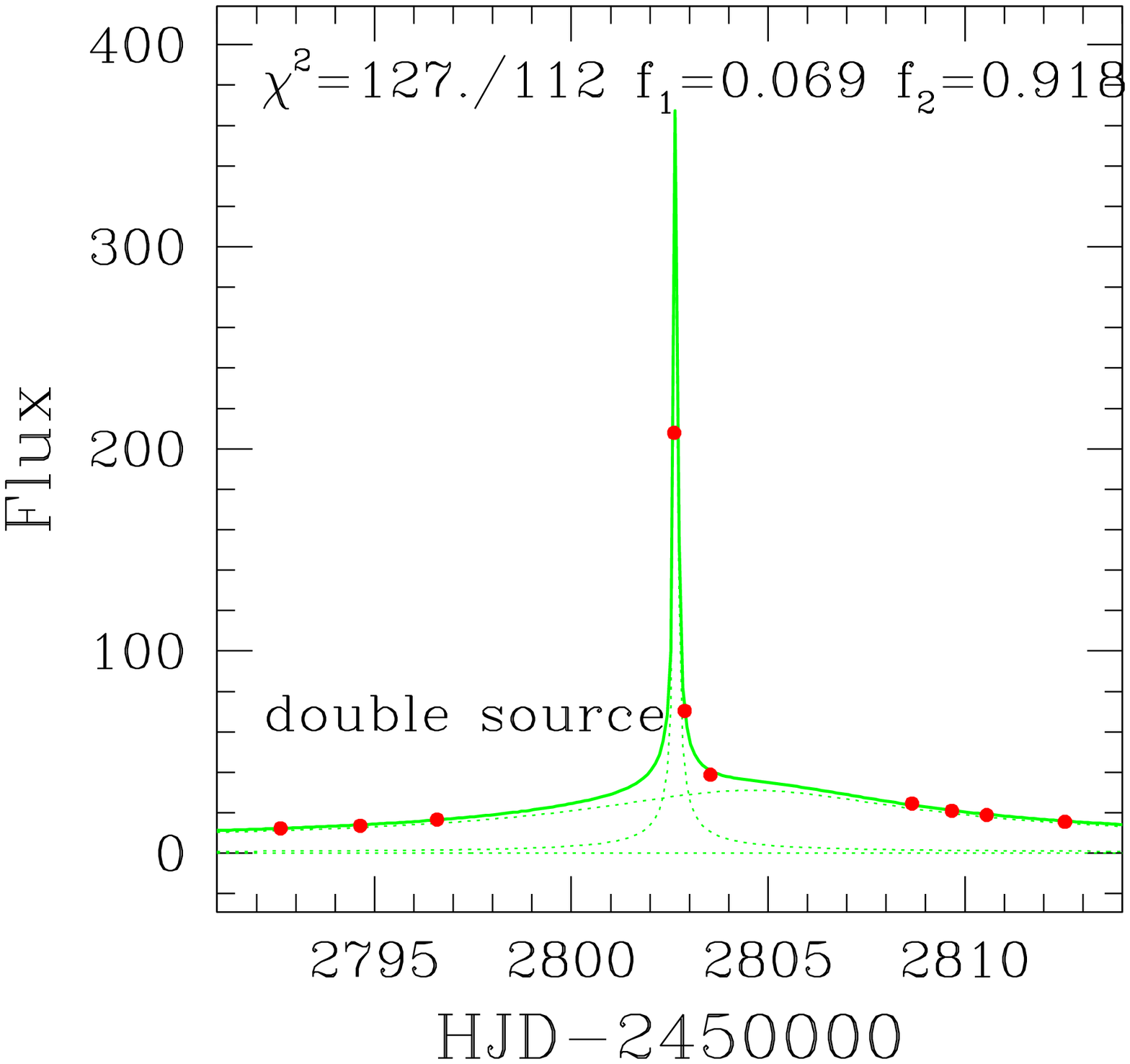}%
}

\noindent\parbox{12.75cm}{
 \noindent{\bf OGLE 2003-BLG-266}

\vspace*{5pt}

 \includegraphics[height=63mm,width=62.0mm]{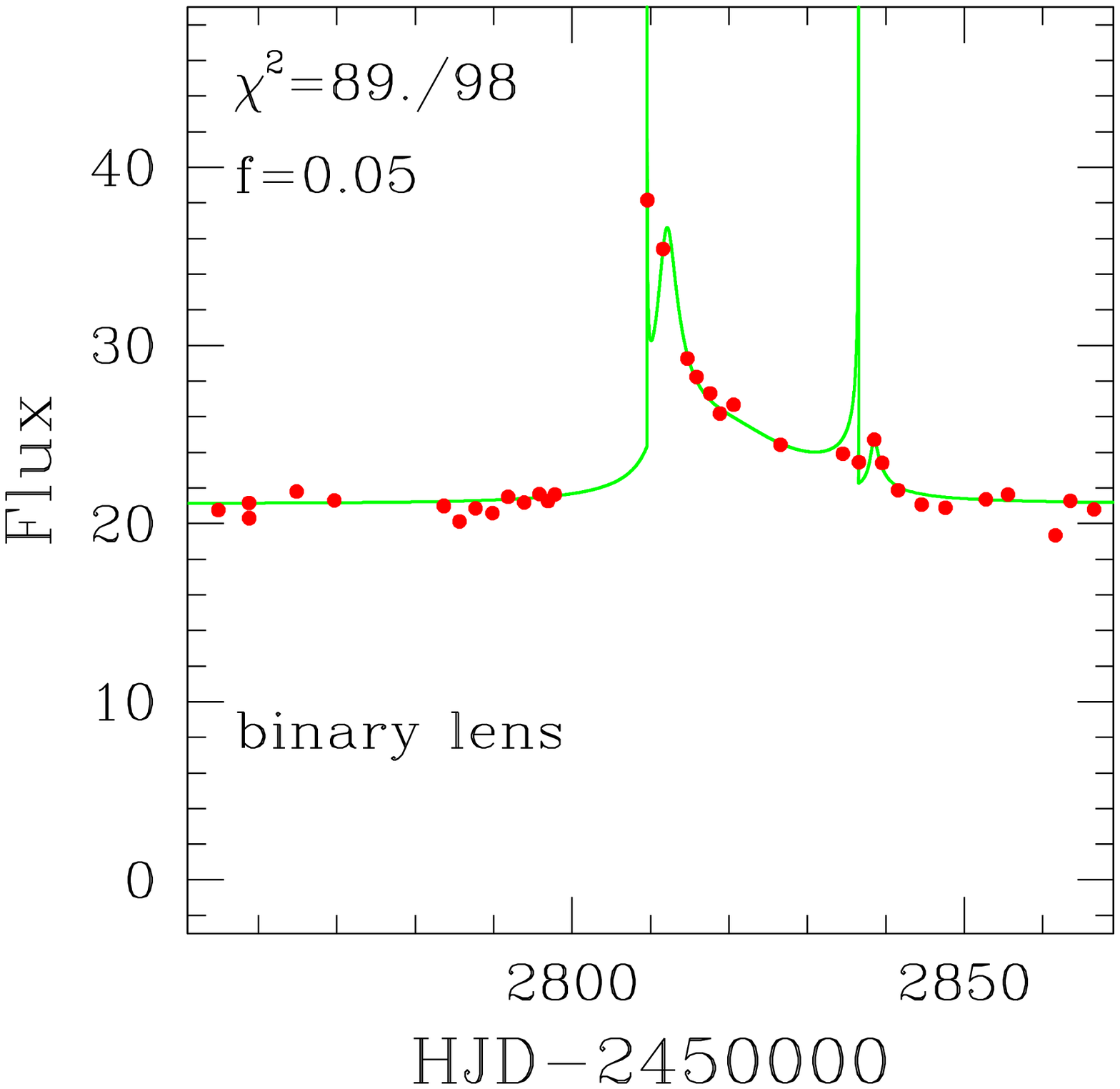}%
 \includegraphics[height=63mm,width=62.0mm]{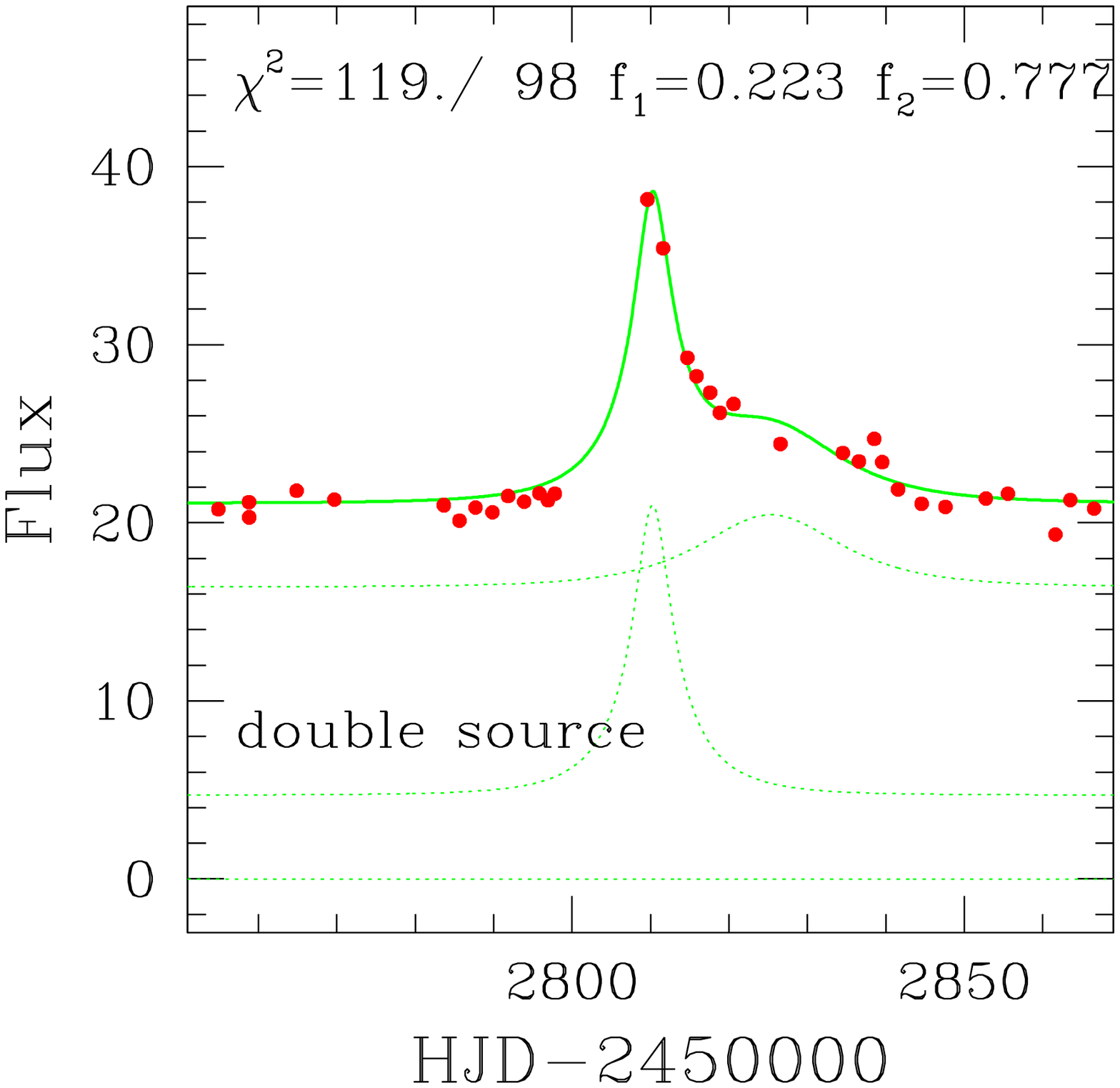}%
}

\noindent\parbox{12.75cm}{
 \noindent{\bf OGLE 2003-BLG-340}

\vspace*{5pt}

 \includegraphics[height=63mm,width=62.0mm]{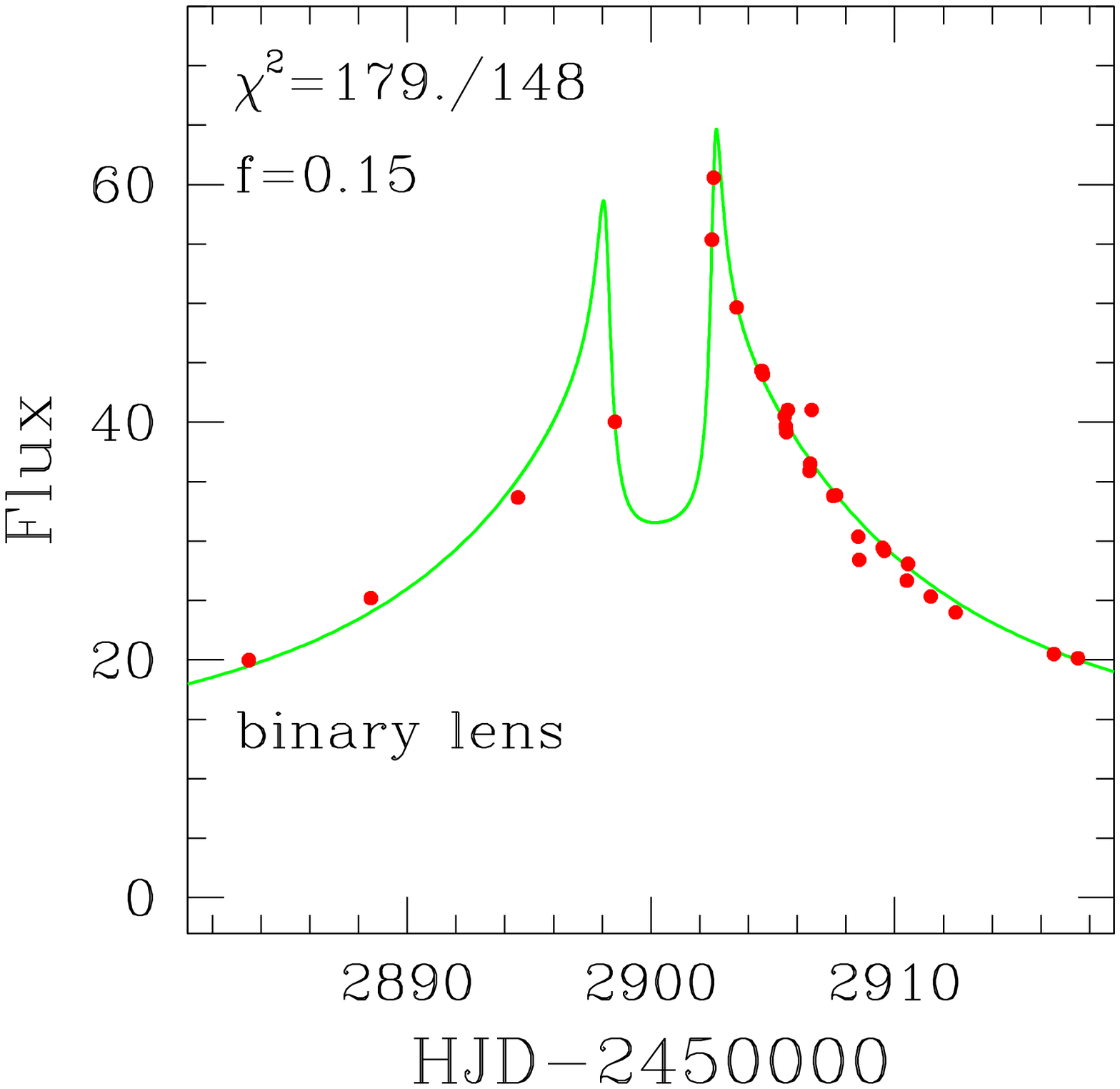}%
 \includegraphics[height=63mm,width=62.0mm]{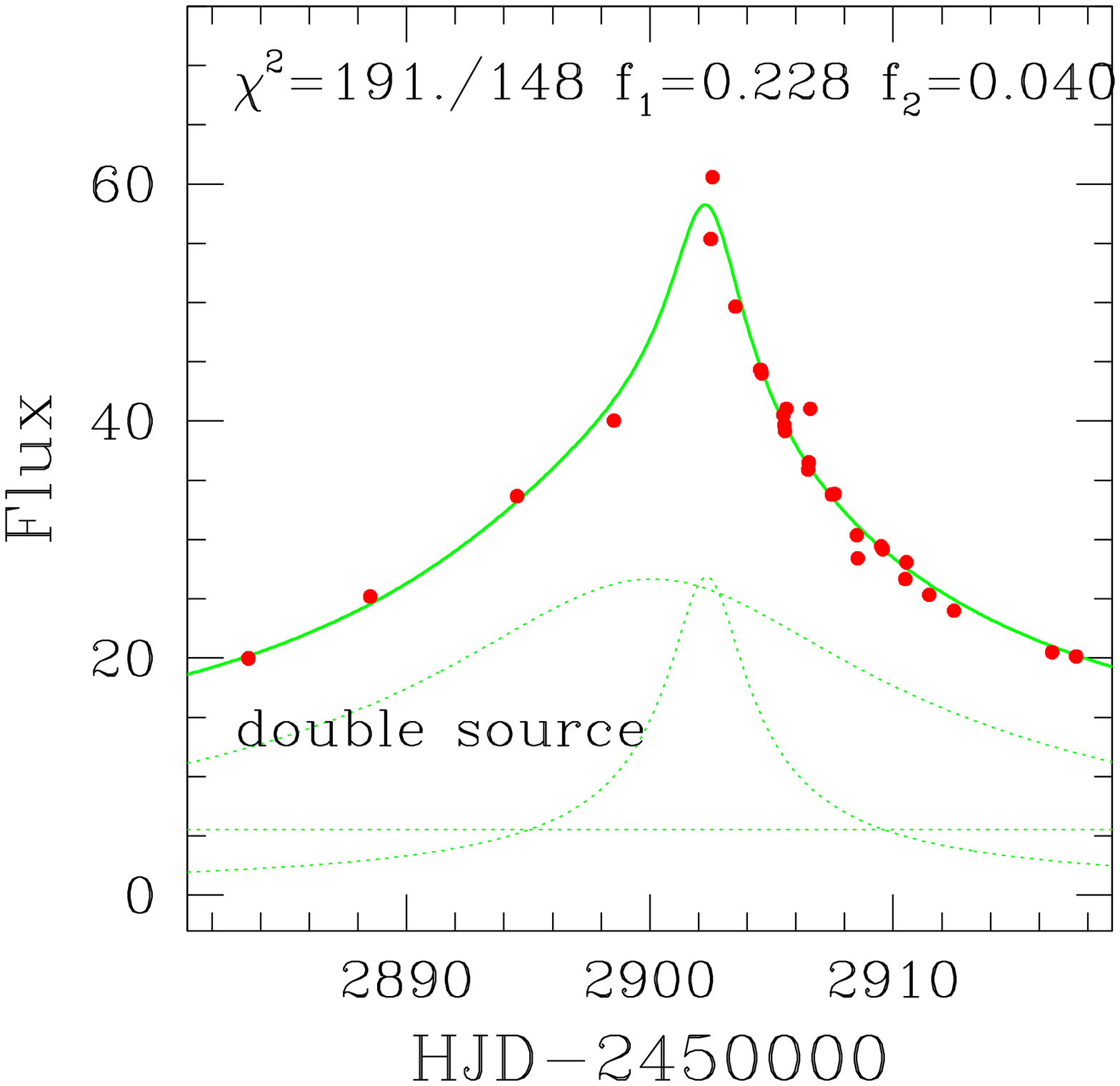}%
}

\vskip4cm
\centerline{{\bf Appendix~3: Well Separated Double Source Events}}
\vskip6pt
We also show a few unambiguous double source events -- cases, where the
source components are well separated which results in two peaks in the
light curves, each resembling closely a single source / single lens
event. 

\noindent\parbox{12.75cm}{
 \noindent{\bf OGLE 2002-BLG-018 \hfill OGLE 2003-BLG-063 \hfil}

\vspace*{5pt}

 \includegraphics[height=63mm,width=62.0mm]{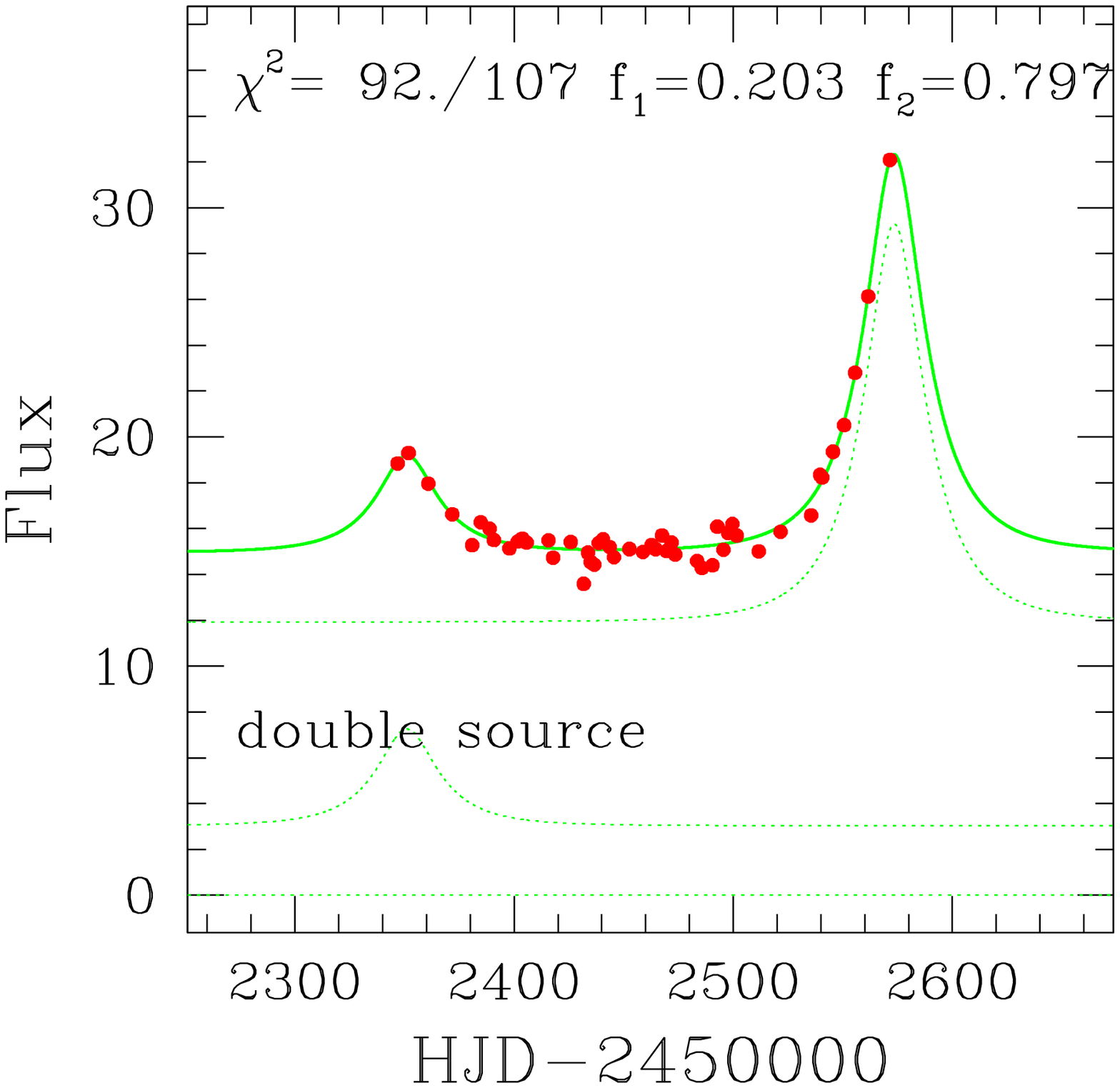}%
 \includegraphics[height=63mm,width=62.0mm]{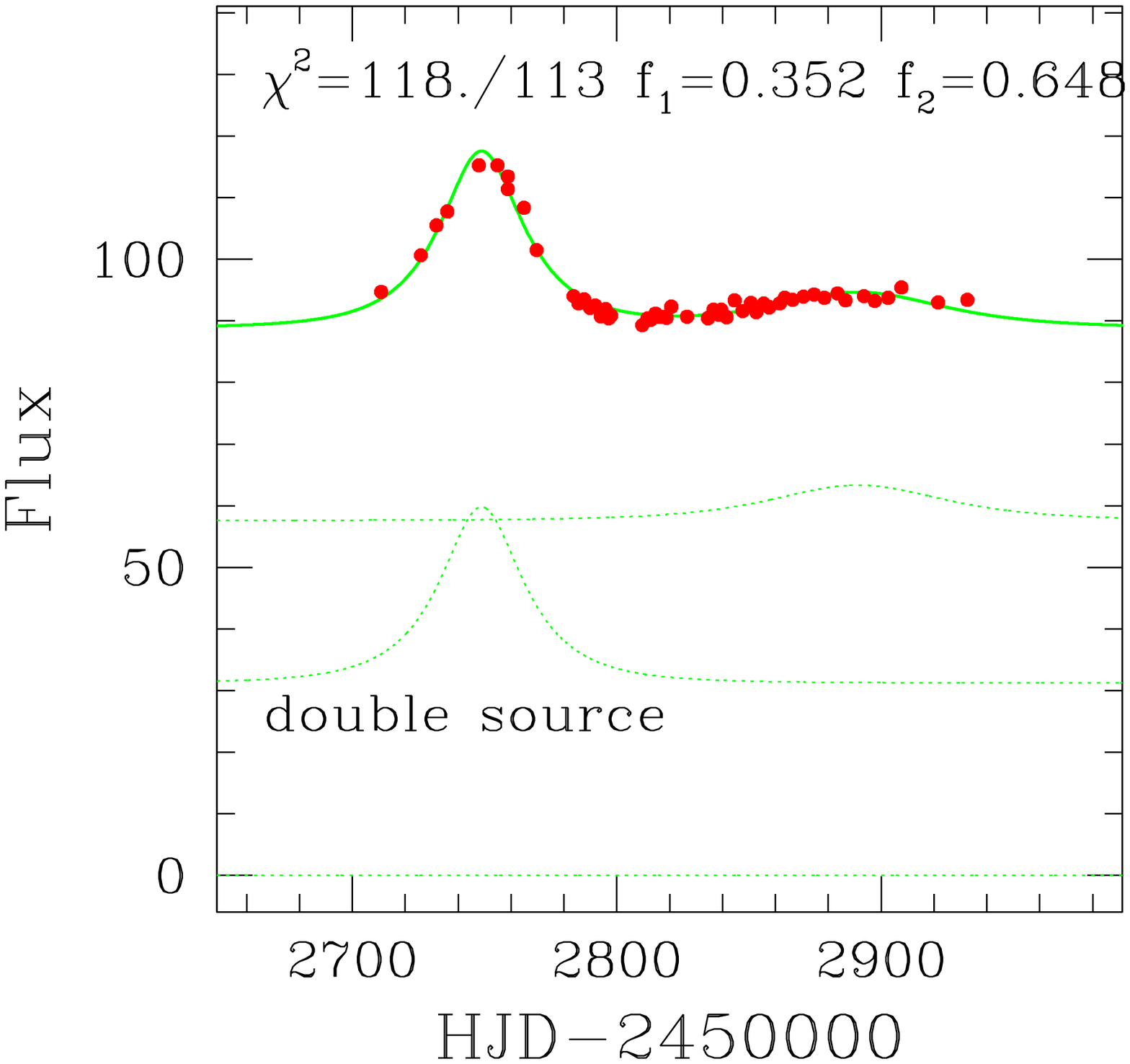}%
}

\noindent\parbox{12.75cm}{
 \noindent{\bf OGLE 2003-BLG-067 \hfill OGLE 2003-BLG-095 \hfil}

\vspace*{4pt}

 \includegraphics[height=63mm,width=62.0mm]{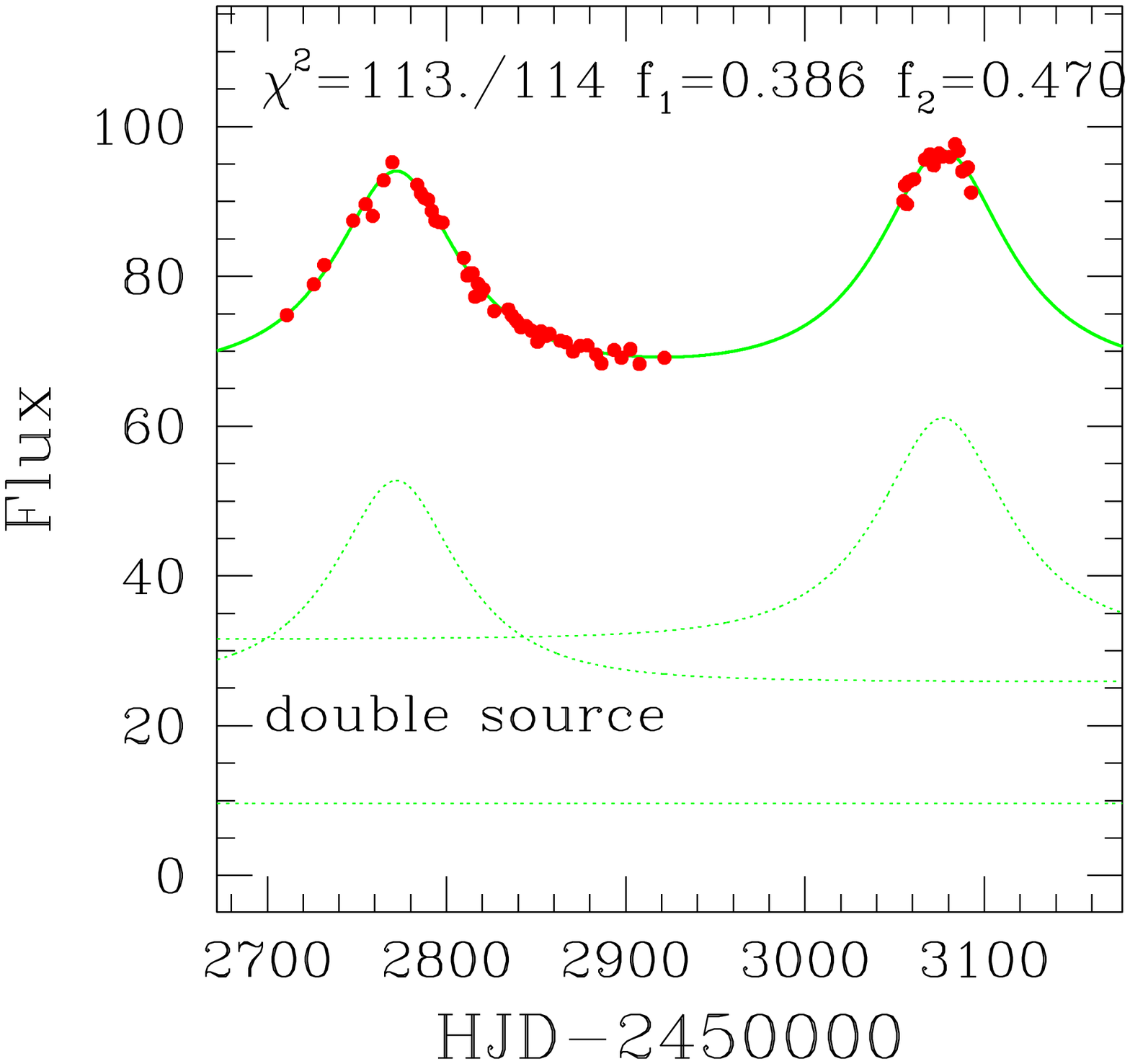}%
 \includegraphics[height=63mm,width=62.0mm]{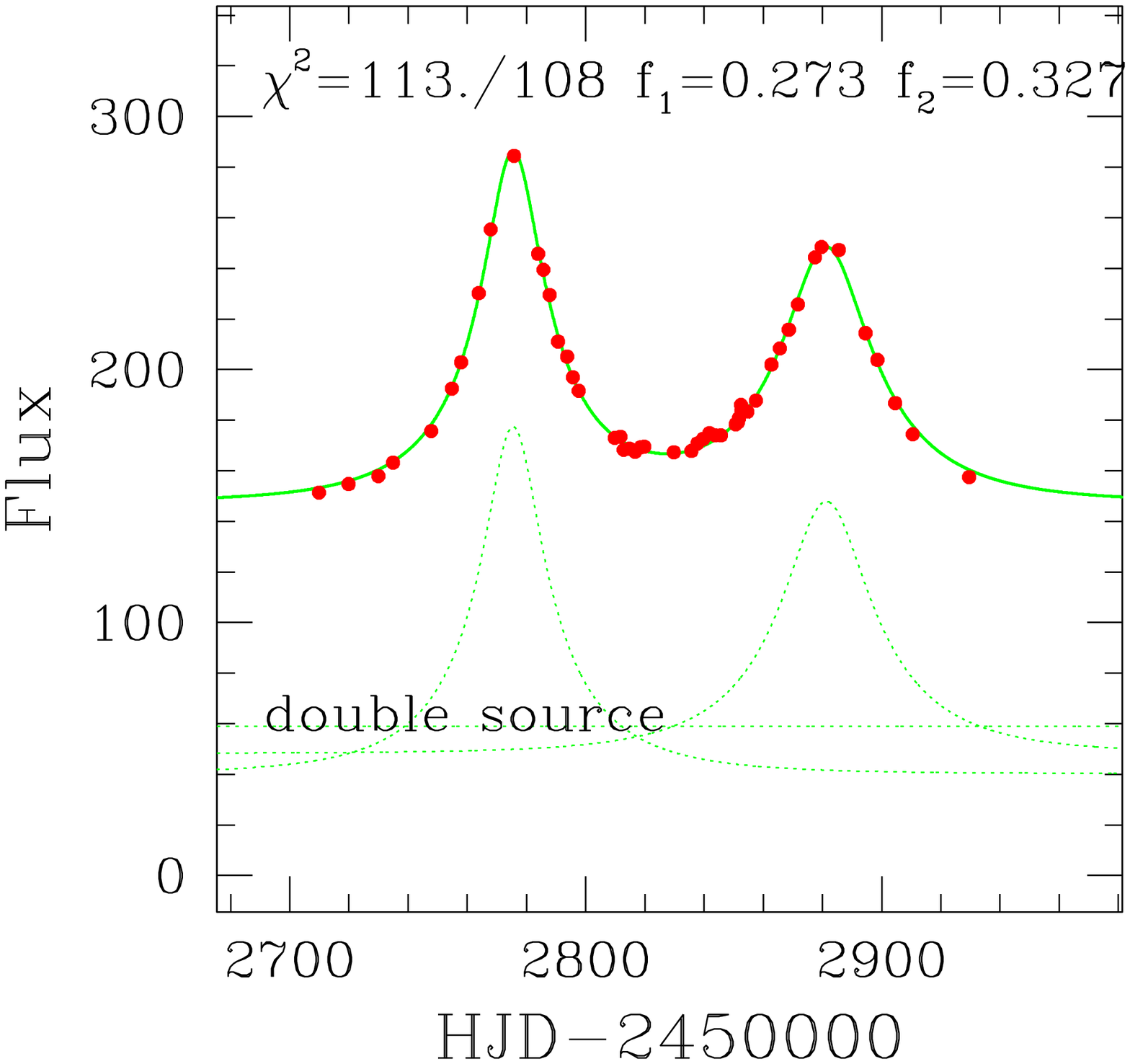}%
}

\noindent\parbox{12.75cm}{
 \noindent{\bf OGLE 2003-BLG-126 \hfill \hfil}

\vspace*{4pt}

 \includegraphics[height=63mm,width=62.0mm]{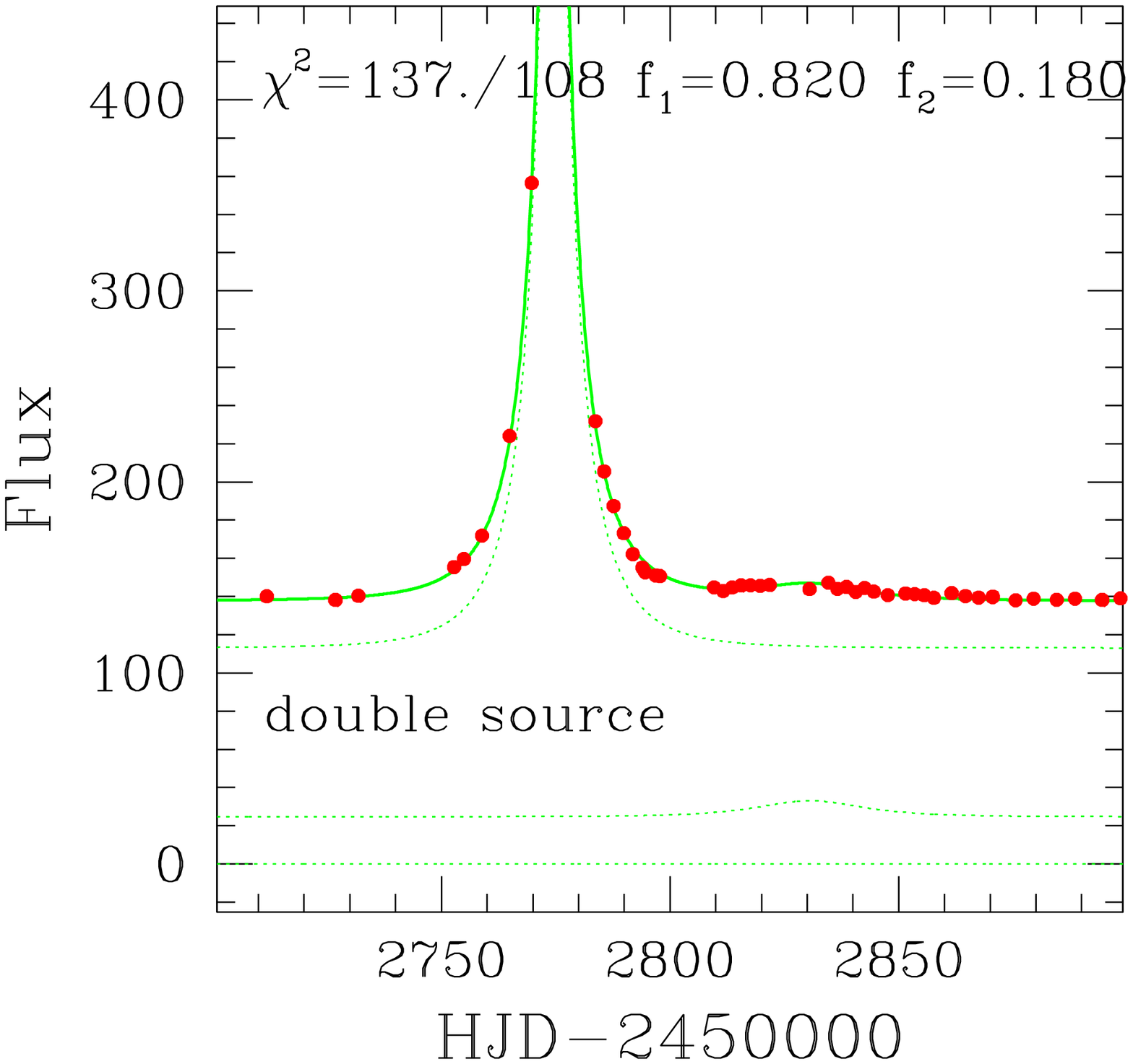}%
}
%\vskip.2cm
\centerline{{\bf Appendix~4: Rejected Event}}
\vskip4pt
The following candidate event has been rejected: we could not find a
binary lens model which would (at least qualitatively) fit its light
curve. 
%\vskip4pt
\vskip8pt
\noindent\parbox{12.75cm}{
 \noindent{\bf OGLE 2003-BLG-303}                           

\vspace*{5pt}

 \includegraphics[height=46mm,width=42.5mm]{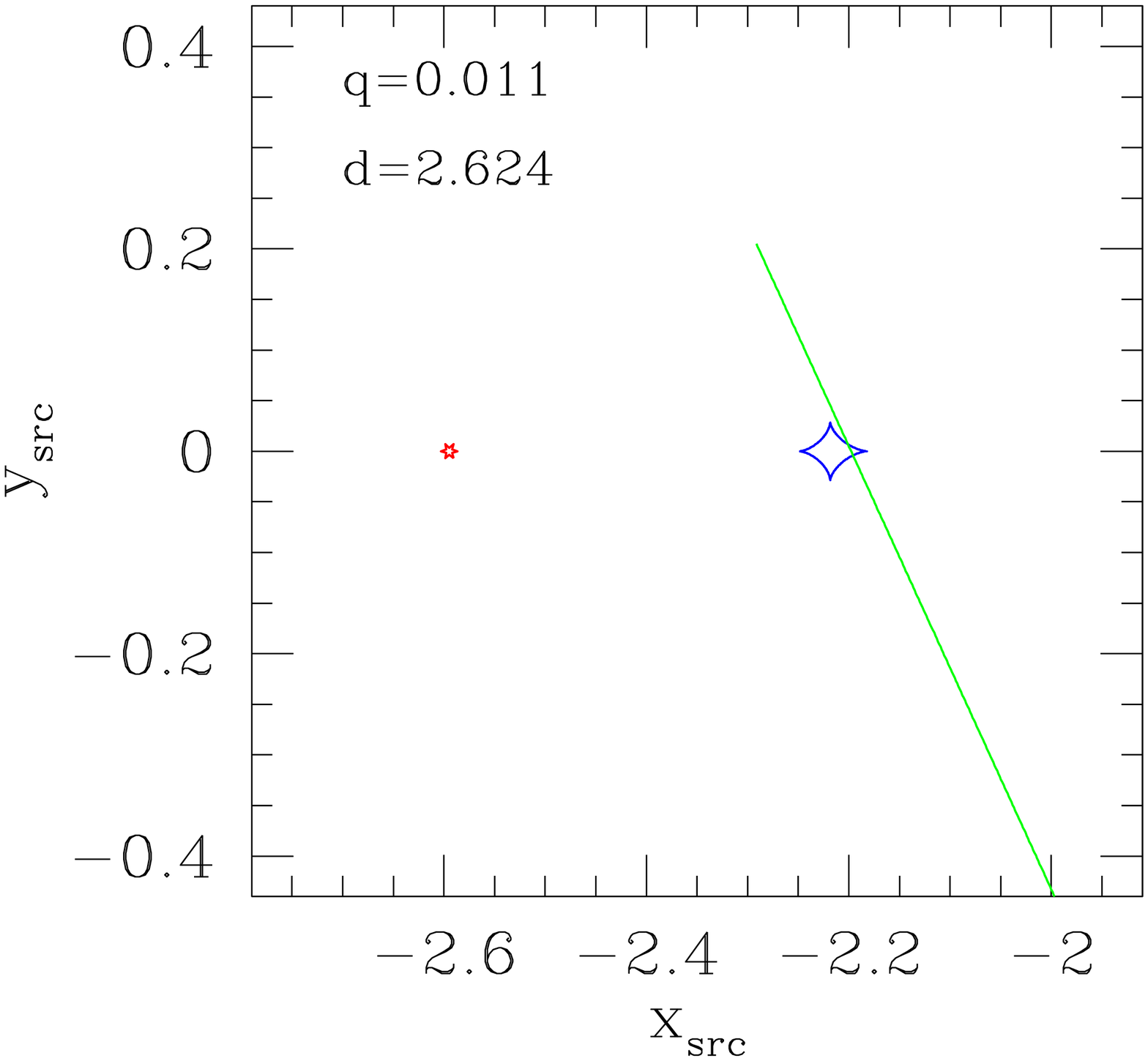}%
 \includegraphics[height=46mm,width=42.5mm]{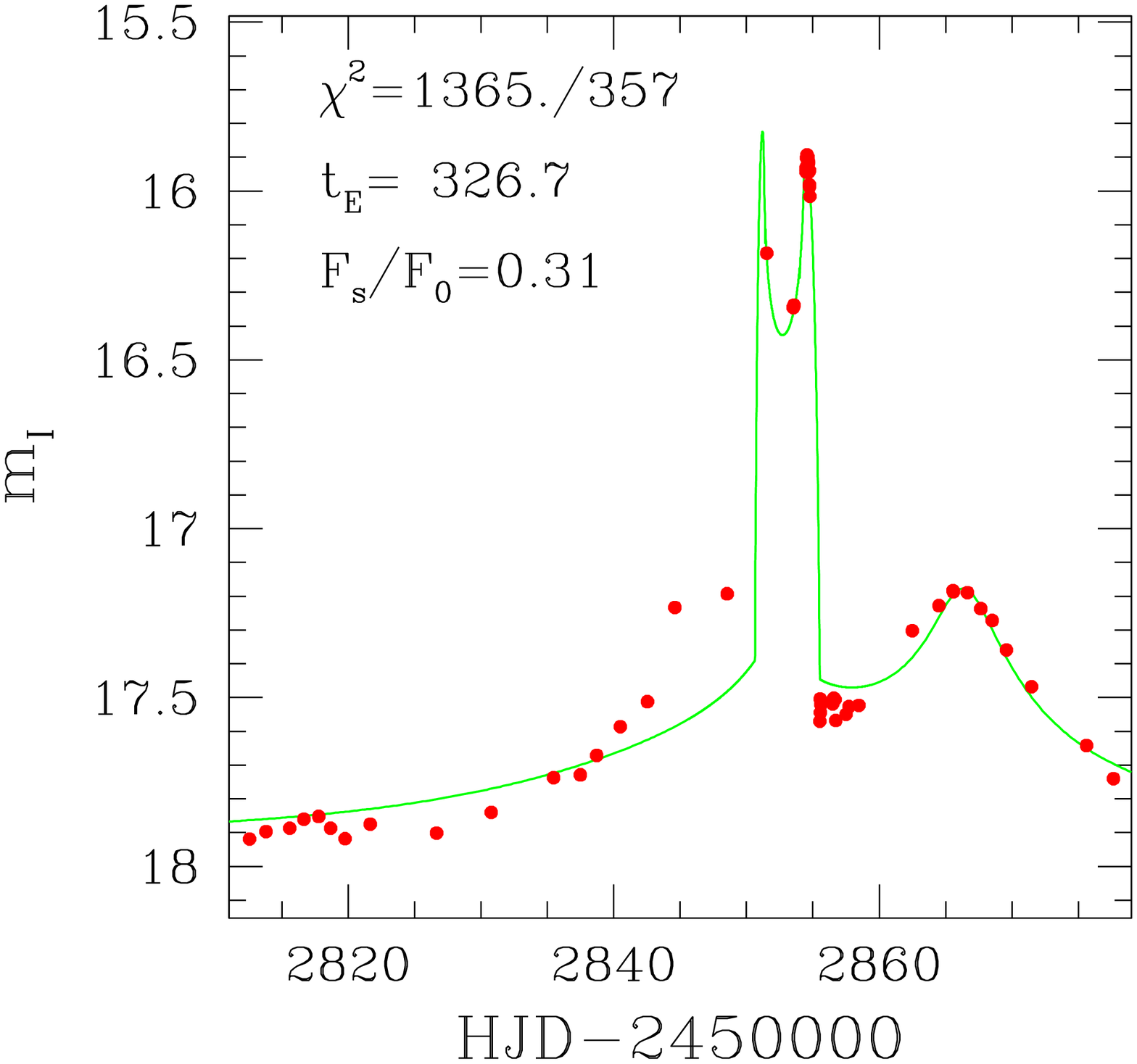}%
 \includegraphics[height=46mm,width=42.5mm]{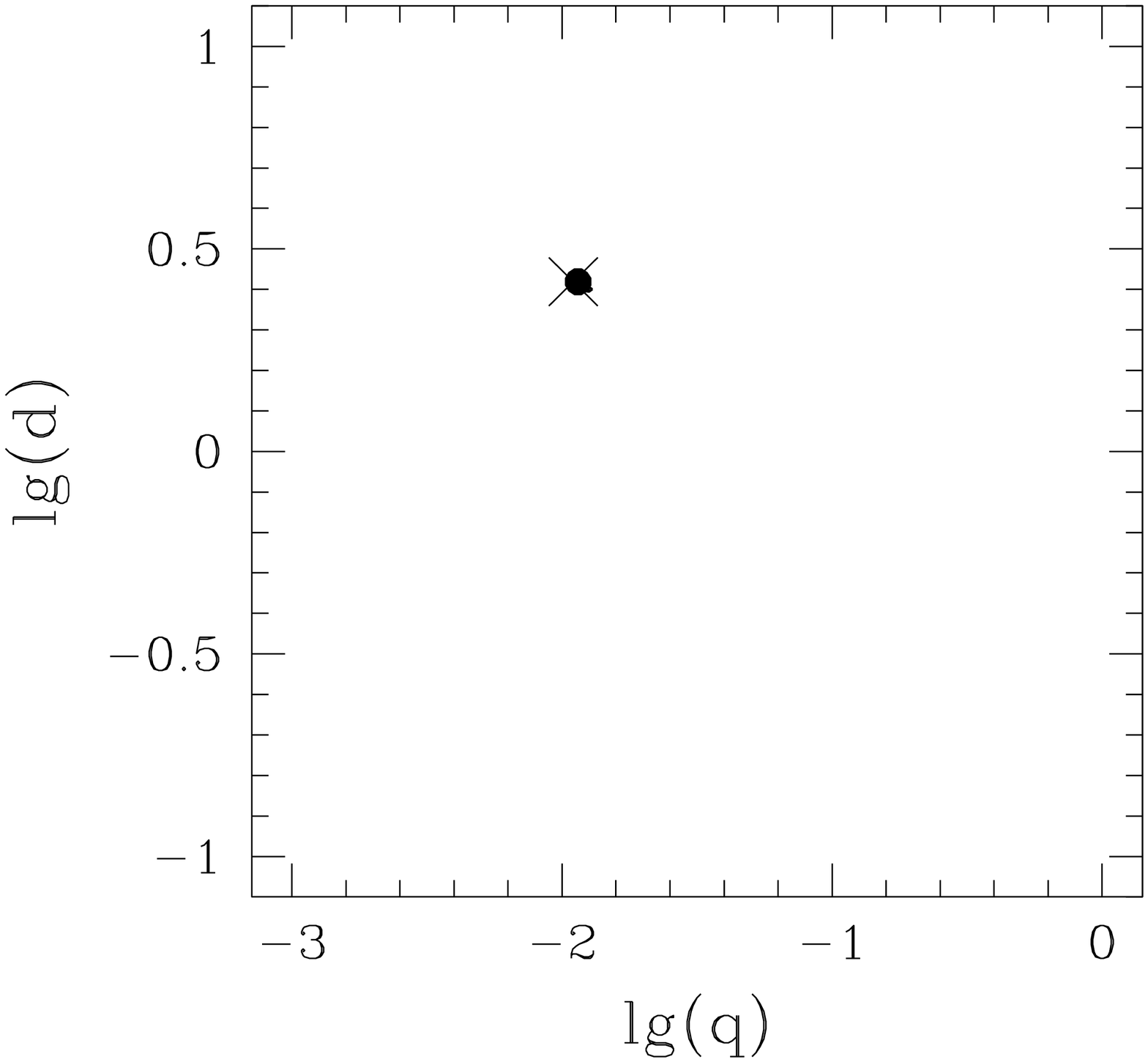}%
}
\end{document}